%% file: main.tex
\documentclass[onecolumn,10pt,a4paper]{article}

\usepackage[ruled,linesnumbered,lined,noend]{algorithm2e}

\usepackage[backend=biber,style=apa]{biblatex}
\usepackage{newfloat}
\usepackage{subcaption}
\usepackage{graphicx}
\usepackage{placeins}
\usepackage{authblk}

\usepackage{amsmath}
\usepackage{amsthm}
\usepackage{stmaryrd}

\usepackage{tikz}
\usetikzlibrary{trees,positioning}

\usepackage{booktabs}
\usepackage{tabularx}
\usepackage{rotating}
\usepackage{makecell}
\usepackage{enumitem}

\usepackage{latexsym}
\usepackage{amssymb}
\usepackage{amsmath}
\usepackage{amsthm}

\usepackage[left=1.5cm,right=2.0cm,top=2cm,bottom=1.5cm]{geometry}

\newtheorem{theorem}{Theorem}
\newtheorem{lemma}{Lemma}
\newtheorem{definition}{Definition}

\usepackage[maxfloats=500]{morefloats}
\input{definitions}

\title{Evolving A$^{*}$ to Efficiently Solve the $\kappa$ Shortest-Path Problem (Extended Version)}
\author[1]{Carlos Linares López}
\author[1]{Ian Herman}
\affil[1]{Computer Science and Engineering Department -- Universidad Carlos III de Madrid}
\date{}

\begin{document}

\maketitle

\begin{abstract}
  The problem of finding the shortest path in a graph $G(V, E)$ has been widely
  studied. However, in many applications it is necessary to compute an arbitrary
  number of them, $\kappa$. Even though the problem has raised a lot of interest
  from different research communities and many applications of it are known, it
  has not been addressed to the same extent as the single shortest path problem.
  The best algorithm known for efficiently solving this task has a time
  complexity of $O (|E| + |V|\log{|V|}+\kappa|V|)$ when computing paths in
  explicit form, and is based on best-first search. This paper introduces a new
  search algorithm with the same time complexity, which results from a natural
  evolution of A$^{*}$ thus, it preserves all its interesting properties, making
  it widely applicable to many different domains. Experiments in various
  testbeds show a significant improvement in performance over the state of the
  art, often by one or two orders of magnitude.
\end{abstract}

\input{supplemental1}

\input{supplemental2}

\input{supplemental3}

\input{supplemental4}

\input{supplemental5}

\input{supplemental6}

\appendix

\input{appendix1}

\FloatBarrier
\printbibliography

\end{document}

%% file: definitions.tex
\newcommand{\astar}{A$^{*}$}
\newcommand{\mAstar}{mA$^{*}$}
\newcommand{\kstar}{K$^{*}$}
\newcommand{\bfk}{K$_{0}$}
\newcommand{\bela}{BELA$^{*}$}
\newcommand{\bfbela}{BELA$_{0}$}

\newcommand{\Z}{\textbf{\textsl{Z}}}

\newcommand{\ze}[2]{$\langle e_{#1, #2},C_{z} \rangle$}


%% file: supplemental1.tex

\section{Introduction}
\label{sec:introduction}

Given a graph $G (V, E)$, the problem of finding the shortest path between two
designated vertices $s$ and $t$ is a long-studied task, and
\astar{}~\parencite{hart.pe.nilsson.nj.ea:formal} is a prominent algorithm used to
solve it. A natural extension consists of computing the best $\kappa$
paths\footnote{The letter $k$, commonly used for referring to the number of
  paths to find, is used throughout this paper as a generic index instead.}
between the same vertices. David Eppstein~\parencite{Eppstein1998} provides a
thorough review in the history of the research on this task, noting that it
dates back as far as 1957. Many variants have been considered, differing on
various criteria, such as whether paths are required to be simple (or loopless)
or whether the graphs considered are directed or undirected. This paper focuses
on the problem of finding the $\kappa$, not necessarily simple, shortest paths
between a start state, $s$, and a goal state, $t$, in directed graphs.

The problem has been already addressed with various heuristic search algorithms,
usually with various derivative versions.
\mAstar{}~\parencite{Dechter2012,Flerova2016} is a straightforward application of
\astar{} which allows the expansion of nodes up to $\kappa$ times. Doing so
clearly allows the discovery of $\kappa$ paths, and the idea can be easily
applied to different domains. In contrast,
\kstar{}~\parencite{aljazzar_directed_2009,Aljazzar2011} expands nodes only once. It
is a heuristic variant of Eppstein's algorithm (EA)~\parencite{Eppstein1998} which,
in addition, can be built on-the-fly significantly improving its running time.
\kstar{} essentially transforms EA to return paths as soon as practical. At the
center of both EA and \kstar{} is the \textit{path graph}, a structure which
stores information from the search allowing the enumeration of paths through a
one-to-one mapping between paths in the path graph and paths in the true graph.
The algorithm swaps between search and enumerating paths from the path graph
based on some swapping criterion, which can lead to the algorithm expanding
nodes unnecessarily. The algorithm has been recently modified~\parencite{katz_k_2023}
with a variety of improvements, including a modification of the swapping
criterion. Still, both EA and \kstar{} have an algorithmic complexity equal to
$O (|E| + |V|\log{|V|}+\kappa)$ when outputting paths in \textit{implicit} form,
i.e., as a sequence of \textit{sidetrack} edges. Usually, however, paths are
required in \textit{explicit} form, i.e., as a sequence of \textit{vertices} and
their algorithm complexity  is then $O (|E| + |V|\log{|V|}+\kappa|V|)$.


In this paper, a novel search algorithm, \bela{} (\textit{Bidirectional Edge
  Labeling A$^{*}$}), is introduced. Some relevant definitions are introduced
first and, among them, a novel use of sidetrack edges is proposed which splits
paths into two components. At the core of our contribution is the notion of a
\textit{centroid} which we then use for the introduction of the brute-force
variant of our algorithm, \bfbela{}. Its theoretical properties are examined and
its algorithmic complexity studied. We then consider the heuristic version of
the algorithm, \bela{}. Afterwards, through empirical evaluation, we show
\bfbela{} and \bela{} outperform both \mAstar{} and \kstar{} (as well as their
brute-force variants), in a wide selection of problems often by one or two
orders of magnitude in running time, and sometimes even more.


%% file: supplemental2.tex

\section{Definitions}
\label{sec:definitions}

Given a \textit{directed} graph $G (V, E)$ characterized by its set of vertices
$v\in V$ and edges $e_{ij}: v_i \rightarrow v_j, e_{ij}\in E$, let $s$ and $t$
denote the \textit{start} and \textit{goal} vertices respectively, between which
an arbitrary number $\kappa$ of shortest-paths has to be found. A path $\pi$ is
defined as a concatenation of vertices
$\pi\langle n_0, n_1, n_2, \ldots, n_k\rangle$ such that
$e_{n_{i-1}, n_i}\in E, 0 < i \leq k$. If $s=n_0$ and $t=n_k$ then $\pi$ is
denoted as a \textit{solution path}. The edges are weighted with non-negative
integers, where $\omega (n_{i-1}, n_i)$ denotes the cost of traversing the edge
$e_{n_{i-1}, n_i}$. Thus, the cost of a path $\pi$ is defined according to the
\textit{additive model} as $C (\pi) =\sum\limits_{i=1}^k \omega (n_{i-1}, n_i)$.
When the path $\pi$ is clear from the context or it is irrelevant, the same cost
can be denoted also as $g (n_i)$ if and only if the path starts at the start
vertex, $s$. Analogously, $g_b (n_i)$ denotes the cost of the path from $n_i$ to
$t$ computed as $g_b (n_i) =\sum\limits_{j=i+1}^k \omega (n_{j-1}, n_j)$ if and
only if $n_k=t$. A path $\pi$ is said to be \textit{optimal}, and is denoted as
$\pi^*$, if and only if $C (\pi^*) \leq C (\pi')$ for every solution path $\pi'$
between $s$ and $t$, and is termed as \textit{suboptimal} otherwise. Following
the previous definitions, $g^* (n_i)$, and $g_b^* (n_i)$ denote the cost of an
optimal path from the start vertex, $s$ to $n_i$, and from $n_i$ to $t$,
respectively.

Heuristic functions are denoted as $h (\cdotp)$. A heuristic function is said to
be \textit{admissible} if and only if $h (n) \leq h^* (n)$ for every node $n$,
where $h^* (n)$ denotes the cost of an optimal solution from $n$ to the goal
$t$. Note that this definition refers to \textit{nodes} instead of vertices,
which are defined in turn, as the representation of a unique path from $s$ to
it, so that the same vertex can be represented with multiple nodes in a search
algorithm. A heuristic function is said to be \textit{consistent} if and only if
$h (n) - h (n_i)\leq \omega (n, n_i)$, for every node $n$, where $n_i$ is any
descendant of it.

The set of all solution paths (either optimal or suboptimal) in $G$ is denoted
as $G_\pi$, and the set of all paths which are suboptimal is denoted as
$G'_{\pi}$, $G'_{\pi}\subset G_{\pi}$.

\begin{definition}
  \label{def:pi}
  Given a directed $G (V, E)$ potentially infinite locally finite graph with
  natural edge weights, and two designated vertices, $s, t\in V$, the
  \emph{single-source $\kappa$ shortest-path problem} consists of finding a set
  of different, not necessarily simple paths
  $\Pi=\{\pi_0, \pi_1, \ldots, \pi_{\kappa-1}\}$ such that:

  \begin{itemize}
    \item If there exists a path $\pi'$ such that $C (\pi') < C (\pi_i), 0 \leq i < \kappa$, then $\pi'\in\Pi$
    \item If $|G_\pi| \leq \kappa$, then $\Pi = G_\pi$
  \end{itemize}

\end{definition}

Every solution path $\pi_{i}\in\Pi$ has a cost possibly different than the cost
of other accepted solution paths. $C^{*}_{0}$ represents the cost of all optimal
solution paths $\pi_{i}\in G_{\pi}\backslash G'_{\pi}$; $C^{*}_{1}$ is the cost
of the cheapest suboptimal solution, $C^{*}_{1}>C^{*}_{0}$. Likewise,
$C^{*}_{i}$ is the cost of all suboptimal solution paths which are the $i$-th
best, and $C^{*}_{\varphi}$ represents the cost of the worst solutions in $\Pi$.
In particular, $C (\pi_{\kappa-1})=C^{*}_{\varphi}$.


\begin{figure}[t]
  \centering
  \begin{tikzpicture}
    \draw (0,0) node [circle, draw] (s) {s};
    \draw (2,0) node [circle, draw] (A) {A};
    \draw (4,0) node [circle, draw] (B) {B};
    \draw (6,0) node [circle, draw] (t) {t};

    \draw (2,1.5) node [circle, draw] (C){C};

    \draw (2,3) node [circle, draw] (D){D};
    \draw (4,3) node [circle, draw] (E){E};

    \draw[-stealth] (s) edge[above] node{1} (A);
    \draw[-stealth] (A) edge[above] node{2}  (B);
    \draw[-stealth] (B) edge[above] node{1}  (t);

    \path [-stealth] (s) edge [bend left] node[above left]{3} (C);
    \path [-stealth, very thick] (C) edge [bend left] node[above right]{2} (B);

    \path [-stealth] (s) edge [bend left] node[above left]{3} (D);
    \draw[-stealth] (D) edge[above] node{2}  (E);
    \path [-stealth, very thick] (E) edge [bend left] node[above right]{2} (t);
  \end{tikzpicture}
  \caption{Examples of sidetrack edges ---shown in thick lines}
  \label{fig:sidetrack-example}
\end{figure}
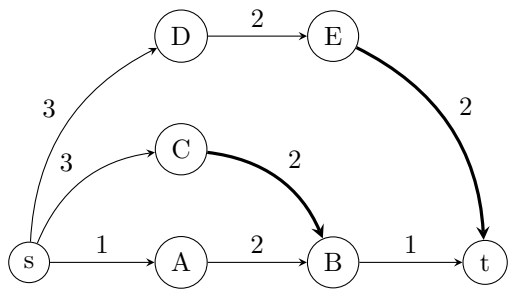

Eppstein's Algorithm (EA) classified all edges in a graph in two different
categories: \textit{tree edges} and \textit{sidetrack edges}, and \kstar{}
slightly modified the definition of the second term. In the following, we adhere
to the definitions used in \kstar:

\begin{definition}
  \label{def:sidetrack}
  An edge $e_{n_{i}, n_{j}}$ is a \emph{tree edge} if and only if
  $g^{*} (n_{j}) = g^{*} (n_{i})+\omega (n_{i}, n_{j})$, and is said to be a
  \emph{sidetrack} edge otherwise, i.e.,
  $g^{*} (n_{j}) < g^{*} (n_{i})+\omega (n_{i}, n_{j})$.
\end{definition}

Clearly, the existence of at least one sidetrack edge is both a necessary and
sufficient condition for a path to be suboptimal. One of our core contributions
is that they also provide a means for distinguishing different components of
any suboptimal solution path:

\begin{definition}
  \label{def:components}
  Given a directed cyclic graph $G (V, E)$ with natural edge weights, and two
  designated vertices, $s, t\in V$, any suboptimal solution path
  $\pi\langle s=n_0, n_1, n_2, \ldots, n_k=t \rangle$ can be decomposed into a
  \emph{prefix} and a \emph{suffix}, via a sidetrack edge
  $e_{n_{i-1}, n_i}\in\pi$, for any $i$ with $0< i\leq k$ as follows: Let $n_i$
  be the \emph{first} node in $\pi$ which verifies that
  $g^* (n_i) < g^* (n_{i-1}) + \omega (n_{i-1}, n_i)$, then:

  \begin{itemize}

    \item $\phi\langle s=n_0, n_1, \ldots, n_{i-1} \rangle$ is the
          \emph{prefix}, possibly empty.
    \item $\sigma\langle n_i, n_{i+1}, \ldots, n_k=t \rangle$ is the
          \emph{suffix}, possibly empty.
    \item The edge $e_{n_{i-1}, n_i}\in\pi$ is a sidetrack edge.

  \end{itemize}

\end{definition}

Figure~\ref{fig:sidetrack-example} shows a graph with two sidetrack edges. In
particular, $e_{C, B}$ decomposes the suboptimal path
$\langle s, C, B, t\rangle$ into its prefix $\phi\langle s, C\rangle$ and its
suffix $\sigma\langle B, t\rangle$ because $g^* (C) = 3$ and
$g^*(B)=3 < g^*(C) + \omega (C, B) = 5$. The other sidetrack exemplifies the
case of sidetrack edges that get to the goal, $t$. Let $\pi$ denote any solution
path (either optimal or suboptimal) such that
$g (n_{i})=g^* (n_{i}), 1\leq i < k$:

\begin{itemize}

  \item If $g(t) > g^*(t)$ then $\pi$ is a suboptimal path, decomposed into its
        prefix and suffix via the sidetrack $e_{n_{k-1}, t}$, because the first
        node verifying $g^* (n_i)< g (n_{i-1}) + \omega (n_{i-1}, n_i)$ is $t$
        indeed.

        Figure~\ref{fig:sidetrack-example} shows this case: the edge $e_{E, t}$
        decomposes the suboptimal path $\langle s, D, E, t\rangle$ into the
        prefix $\phi\langle s, D, E\rangle$ and an empty suffix
        $\sigma =\lambda$.

  \item If $g(t) = g^*(t)$ then $\pi$ is an optimal path, to be denoted as
        $\pi^*$, and thus, there is no vertex $n_i$ verifying that
        $g^* (n_i) < g^* (n_{i-1}) + \omega (n_{i-1}, n_i)$.

        Figure~\ref{fig:sidetrack-example} shows this case: the path
        $\langle s, A, B, t \rangle$ is optimal. Consequently,
        $g(n_i)=g^*(n_i)=g^* (n_{i-1})+\omega(n_{i-1}, n_i), 1\leq i\leq k$ so
        there is no sidetrack edge. It is then said that the prefix of $\pi$ is
        $\pi$ itself.

\end{itemize}

These two types of paths are both considered \emph{direct}:

\begin{definition}
  \label{def:directness}

  A path $\pi\langle s=n_0, n_1, n_2, \ldots, n_k=t \rangle$ is said to be
  \emph{direct} if and only if one of the following conditions hold:

  \begin{itemize}

    \item It is an optimal path or, in other words, it has no sidetrack edge.

    \item It is a suboptimal path with an empty suffix, $\sigma = \lambda$.

  \end{itemize}

  \noindent
  and is called \emph{indirect} otherwise.

\end{definition}

Note there are also paths with an empty prefix, $\phi = \lambda$: if there is a
suboptimal path to get to $t$ from $s$ which consists of only one edge, then it
is a sidetrack edge which makes $\phi = \sigma = \lambda$; any suboptimal path
$\pi\langle s=n_0, n_1, n_2, \ldots, n_k=t\rangle$ with
$g^* (n_1) < \omega (s, n_1)$ is decomposed into an empty prefix,
$\phi=\lambda$, and the suffix $\sigma\langle n_1, n_2, \ldots, n_k=t \rangle$
via the sidetrack $e_{s, n_1}$.








Next, we provide a novel result regarding the relationship between suboptimal
solution paths and adjacent paths via a sidetrack edge:

\begin{lemma}
  \label{tho:relation}
  Let $\pi_{i}\in \Pi$ denote a suboptimal solution path, and let $e_{u, v}$
  denote its first sidetrack edge, then there exists another solution path
  $\pi_{j}\in\Pi, C (\pi_{j}) < C (\pi_{i})$, such that the ending vertex of
   $e_{u, v}$, $v$, belongs to the prefix of $\pi_{j}$.
\end{lemma}

\textbf{Proof}: From the definition of a sidetrack edge it follows that there is
a shorter path to $v$. Let $\phi$ denote it. Denoting the subpath from $v$ to
the end of $\pi_i$ by $\sigma$, the concatenation of $\phi$ and $\sigma$ yields
another solution path, $\pi_{j}$ which is necessarily shorter than $\pi_{i}$,
i.e., $C (\pi_{j}) < C (\pi_{i})$. Note that this is true even if $\pi_{i}$ is a
direct suboptimal path, i.e., $v=t$.\hfill$\Box$







%% file: supplemental3.tex

\section{Centroids}
\label{sec:centroids}


A new definition, which refines the notion of sidetrack edge is proposed first:

\begin{definition}
  A \emph{centroid} $z$ is defined as the association of a sidetrack edge
  $e_{u,v} \in E$ with $u, v \in V$, and an overall cost $C_z$.
\end{definition}

Hence, two centroids are different if they use different sidetrack edges or they
have different overall costs. Clearly, every suboptimal solution path $\pi$
using a centroid is divided into a prefix and suffix, and
$C(\pi)=g^{*} (u)+\omega (u, v)+g_{b} (v)=C_z$. Of course, question is how to
find the paths defined by the cost and sidetrack edge of the centroid. To do
this, we must enumerate all valid suffixes and prefixes for paths of the
centroid. Before returning to this question it is first shown that centroids
create an \textit{equivalence class} over the set of all suboptimal solution
paths, $G'_{\pi}$.

\begin{lemma}
  Any suboptimal path $\pi$ is represented by one and only one centroid $z$.
\end{lemma}

\textbf{Proof}: Indeed, there is a unique combination of an overall cost and a
sidetrack edge that represents any suboptimal solution path $\pi$: It is
trivially observed that the cost of $\pi$ is unique, $C (\pi)$ and thus, its
centroid has to have an overall cost $C_{z}=C (\pi)$; secondly, Definition
(\ref{def:components}) explicitly uses the \textit{first} sidetrack in $\pi$ to
split the path into its prefix and suffix, and hence it has to be unique. To
conclude the second observation, note that every suboptimal path must
necessarily have at least one sidetrack, otherwise it would be an optimal
path.\hfill$\Box$

\begin{lemma}
  The equivalence class induced by the definition of centroids forms a partition
  over the set of all suboptimal solution paths $G'_{\pi}$, i.e., every
  suboptimal solution path $\pi_{i}\in\Pi$ belongs to one and only one
  equivalence class defined by a centroid $z$.
\end{lemma}

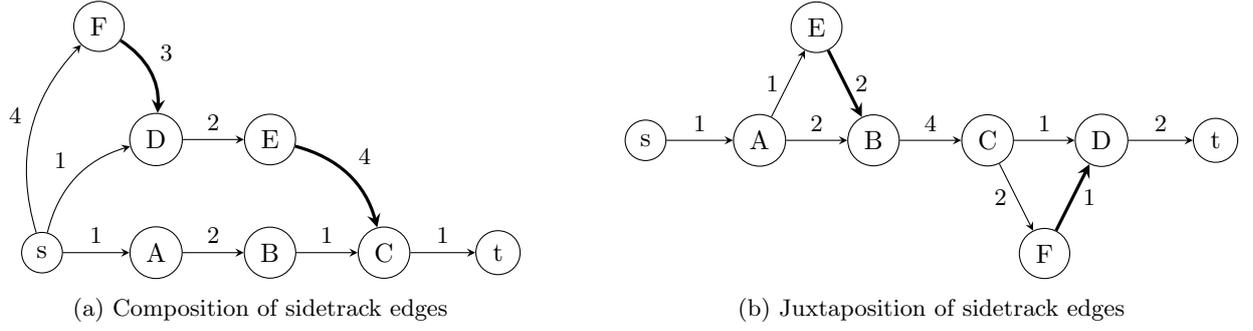
\begin{figure*}[t]
  \begin{subfigure}{.5\linewidth}
    \centering
    \begin{tikzpicture}

      \draw (0.0,0) node [circle, draw] (s) {s};
      \draw (1.5,0) node [circle, draw] (A) {A};
      \draw (3.0,0) node [circle, draw] (B) {B};
      \draw (4.5,0) node [circle, draw] (C) {C};
      \draw (6.0,0) node [circle, draw] (t) {t};

      \draw[-stealth] (s) edge[above] node{\small{1}} (A);
      \draw[-stealth] (A) edge[above] node{\small{2}} (B);
      \draw[-stealth] (B) edge[above] node{\small{1}} (C);
      \draw[-stealth] (C) edge[above] node{\small{1}} (t);

      \draw (1.5,1.5) node [circle, draw] (D) {D};
      \draw (3.0,1.5) node [circle, draw] (E) {E};

      \draw[-stealth] (s) edge[bend left] node[above left]{\small{1}} (D);
      \draw[-stealth] (D) edge[above] node{\small{2}} (E);
      \draw[-stealth, very thick] (E) edge[bend left] node[above right]{\small{4}} (C);

      \draw (0.75,3.0) node [circle, draw] (F) {F};

      \draw[-stealth] (s) edge[bend left] node[above left]{\small{4}} (F);
      \draw[-stealth, very thick] (F) edge[bend left] node[above right]{\small{3}} (D);

    \end{tikzpicture}
    \caption{Composition of sidetrack edges}
    \label{fig:sidetrack-edges:a}
  \end{subfigure}
  \hfill
  \begin{subfigure}{.5\linewidth}
    \centering
    \begin{tikzpicture}

      \draw (0.0,0) node [circle, draw] (s) {s};
      \draw (1.5,0) node [circle, draw] (A) {A};
      \draw (3.0,0) node [circle, draw] (B) {B};
      \draw (4.5,0) node [circle, draw] (C) {C};
      \draw (6.0,0) node [circle, draw] (D) {D};
      \draw (7.5,0) node [circle, draw] (t) {t};

      \draw[-stealth] (s) edge[above] node{\small{1}} (A);
      \draw[-stealth] (A) edge[above] node{\small{2}} (B);
      \draw[-stealth] (B) edge[above] node{\small{4}} (C);
      \draw[-stealth] (C) edge[above] node{\small{1}} (D);
      \draw[-stealth] (D) edge[above] node{\small{2}} (t);

      \draw (2.25,1.5) node [circle, draw] (E) {E};

      \draw[-stealth] (A) edge[left] node{\small{1}} (E);
      \draw[-stealth, very thick] (E) edge[right] node{\small{2}} (B);

      \draw (5.25,-1.5) node [circle, draw] (F) {F};

      \draw[-stealth] (C) edge[left] node{\small{2}} (F);
      \draw[-stealth, very thick] (F) edge[right] node{\small{1}} (D);

    \end{tikzpicture}
    \caption{Juxtaposition of sidetrack edges}
    \label{fig:sidetrack-edges:b}
  \end{subfigure}
  \caption{Example of suboptimal solution paths with various sidetrack edges}
  \label{fig:sidetrack-edges}
\end{figure*}

As a consequence of the preceding Lemma, the computation of all suboptimal
solution paths\footnote{Note that the solution set $\Pi$ is not required to
  contain all suboptimal solution paths with the last cost, $C^{*}_{\varphi}$.}
in the solution set $\Pi$ can be computed from
$\displaystyle\bigcup\limits_{z \in \Z{}}\llbracket z\rrbracket$ where \Z{} is
the set of all centroids of our problem with cost less than or equal to
$C_\varphi^*$, and $\llbracket z\rrbracket$ is the set of all paths that use
centroid $z$. We show next that a unique centroid $z$ represents an arbitrary
number of suboptimal solution paths $\llbracket z\rrbracket$ that can each
contain a different number of sidetrack edges. The only possible cases are shown
in Figure~\ref{fig:sidetrack-edges}. Figure~\ref{fig:sidetrack-edges:a} shows
the case where the sidetrack edges of centroids (shown with thick lines) can be
\textit{composed} to create suboptimal solution paths which are larger. In
particular, the cost of the path $\langle s, D, E, C, t\rangle$, 8, is smaller
than the cost of the solution path $\langle s, F, D, E, C, t \rangle$, 14, which
results of the composition of the sidetrack edge defining its centroid
$\langle e_{F, D}, 14\rangle$ with the defining sidetrack edge of the centroid
representing the former, $\langle e_{E, C}, 8\rangle$.
Figure~\ref{fig:sidetrack-edges:b} shows a more interesting case, where the
sidetrack edges of two centroids (shown with thick lines) can be
\textit{juxtaposed} so that taking them creates a suboptimal solution path which
is larger than the cost of any suboptimal solution path that uses only one of
the two edges. To be clear, the solution set $\Pi$ with $\kappa=4$ is shown
next:

\[
  \begin{array}{rlrlc}
    \pi_{0}: & \langle s, A, B, C, D, t \rangle & C^{*}_{0}: & 10 & - \\
    \pi_{1}: & \langle s, A, E, B, C, D, t \rangle & C^{*}_{1}: & 11 & \langle e_{E,B}, 11\rangle \\
    \pi_{2}: & \langle s, A, B, C, F, D, t \rangle & C^{*}_{2}: & 12 & \langle e_{F,D}, 12\rangle\\
    \pi_{3}: & \langle s, A, E, B, C, F, D, t \rangle & C^{*}_{3}: & 13 &\langle e_{E,B}, 13\rangle \\
  \end{array}
\]

Restricting attention only to the suboptimal paths, $\pi_{1}$ uses the sidetrack
edge $e_{E, B}$ and has an overall cost $C^{*}_{1}=11$, so that the centroid
$\langle e_{E, B}, 11 \rangle$ represents it. Note that this sidetrack splits
$\pi_{1}$ into its prefix $\phi=\langle s, A, E \rangle$ and its suffix
$\sigma=\langle B, C, D, t\rangle$. Because, there is a path from B to $t$ with
cost 7, $g_{b} (B)=7$, then the overall cost of the path can be computed as
$g^{*} (E)+\omega (E, B)+g_{b} (B)=2+2+7=11$. 
However, $\pi_{3}$ contains both sidetracks $e_{E,B}$ and $e_{F,D}$. According
to Definition (\ref{def:components}), it is the first sidetrack edge in
$\pi_{3}$, $e_{E,B}$ which splits $\pi_{3}$ into its prefix
$\phi=\langle s, A, E \rangle$ and its suffix
$\sigma=\langle B, C, F, D, t \rangle$. The cost of the suffix is the $g_{b}$
cost of node $B$ for constructing $\pi_{3}$, which is equal to $g_{b} (B)=9$,
hence $C^{*}_{3}=C (\pi_{3})=g^{*} (E)+\omega (E,B)+g_{b} (B)=2+2+9=13$. As a
result, the node B has two $g_{b}$ costs, 7 and 9 used respectively for
$\pi_{1}$ and $\pi_{3}$. In conclusion, the juxtaposition of several sidetrack
edges in the same suboptimal solution path is represented with different $g_{b}$
values in the same node. The last column above shows the centroid of each
suboptimal path, where it can be seen that the sidetrack $e_{E,B}$ has been used
twice to generate $\pi_{1}$ and $\pi_{3}$.

To conclude, any suboptimal solution path results from either the consideration
of solely the defining sidetrack edge of a centroid, or the combination of an
arbitrary number of them, either composed, juxtaposed or a combination of both.


%% file: supplemental4.tex

\section{BELA$_0$}
\label{sec:bela0}

We consider first the uninformed variant of our search algorithm, \bfbela{},
where heuristics are not available. From the preceding Section, the computation
of the $\kappa$ shortest-paths can be computed from the union of all centroids
with cost less than or equal to $C^*_{\varphi}$, where every centroid is defined
as the association of a sidetrack edge and an overall cost. As indicated in the
Definitions, $C^{*}_{0}$ is the cost of all the optimal solution paths and an
ordinary application of Dijkstra's can be used to compute all of them. For the
case of a centroid $z$ such that $C_z = C^*_i$, $i\geq 1$, we will soon show how
to compute its set of paths from its cost and sidetrack edge.

The first extension that we propose to Dijkstra's algorithm consists of storing
all edges traversed in the \textsc{closed} list. When a duplicate is found
(e.g., node $D$ in Figure~\ref{fig:sidetrack-edges:a}), the edge to it (i.e.,
$e_{F, D}$) is stored in \textsc{closed}, and the node is not re-expanded. This
way, all existing sidetrack edges can be easily distinguished from tree edges:
Given a node $n$ in \textsc{closed}, one of its incoming edges $e_{m, n}$ is a
sidetrack edge if and only if $g^{*} (n) < g^{*} (m)+\omega (m, n)$. This
operation can be performed in $O (1)$ because Dijkstra's algorithm already
stores in \textsc{closed} the optimal cost to reach each node from the start
state, $g^{*} (\cdotp)$, and so does \bfbela{}. According to the normal
operation of Dijkstra's, when the goal state is about to be expanded it knows
that a new direct (either optimal or suboptimal) path has been found. Because it
also knows the cost of the new path and its parent is already in
\textsc{closed}, it can output the new solution path by following all
backpointers. Conducting a depth-first search in \textsc{closed} where the next
node is a parent of the current one with a $g^{*}$-value equal to the
$g^{*}$-value of the current node minus the cost of the edge that joins them,
delivers all the direct solution paths with a cost equal to the desired one. We
call this process \textit{prefix construction}.

More specifically, prefix construction, shown in
Algorithm~\ref{algo:getprefixes}, is the process of computing all optimal paths
from the start state to a designated node in \textsc{closed} by following the
backpointers from it, and also discovering new centroids if any exist. When
using a centroid \ze{u}{v} only one specific $g_{b}$-value is used among all in
the ending vertex, $g_{b}=C_{z}-g^{*} (u)-\omega (u, v)$. Thus, the
$g_{b}$-value of any node selected when enumerating a prefix can be computed as
the sum of the $g_{b}$-value of its descendant plus the cost of the edge as
shown in Lines~\ref{algo:getprefixes:gb:0}--\ref{algo:getprefixes:insert} of
Algorithm~\ref{algo:getprefixes}. Moreover, once a new $g_{b}$-value is
discovered, the existence of new centroids can be verified by checking the
condition given in Definition (\ref{def:sidetrack}), see
Line~\ref{algo:getprefixes:sidetrack}. Finally, the enumeration of prefixes is
done recursively in
Lines~\ref{algo:getprefixes:normal:0}--\ref{algo:getprefixes:normal:n}, where
$\otimes$ denotes the cross-product of its arguments.


\IncMargin{1em}
\begin{algorithm}
  \LinesNumbered
  \DontPrintSemicolon
  \SetKwData{Z}{Z}
  \SetKwFunction{Insert}{insert}
  \SetKwFunction{GetPrefixes}{GetPrefixes}
  \KwData{A node $n$ and a backward $g$-cost, $g_{b}$}
  \KwResult{All optimal paths from $s$ to $n$. In addition, it creates new centroids if any is found.}
  \BlankLine
  \If{$g_{b}\notin g_{b} (n)$}{\label{algo:getprefixes:gb:0}
    \Insert ($g_{b} (n)$, $g_{b}$)\;\label{algo:getprefixes:insert}
    \For{every parent $p_{n}$ of node $n$}{
      \If{$g^{*} (n) < g^{*} (p_{n})+\omega (p_{n}, n)$}{\label{algo:getprefixes:sidetrack}
        \Insert{\Z, $g^{*} (p_{n})+\omega (p_{n}, n)+g_{b}$}\;\label{algo:getprefixes:gb:n}
      }
    }
  }
  \If{$n=s$}{\label{algo:getprefixes:normal:0}
    return $\{s\}$\;
  }
  \For{every parent $p_{n}$ of node $n$}{\label{algo:getprefixes:normal:0}
    \If{$g^{*} (p_{n})+\omega (p_{n}, n)=g^{*} (n)$}{
      $\phi\leftarrow\phi\cup$\GetPrefixes{$p_{n}$, $g_{b}+\omega (p_{n}, n)$}$\otimes\{n\}$;\label{algo:getprefixes:normal:n}
    }
  }
  return $\phi$\;
  \BlankLine
  \caption{Pseudocode of \texttt{GetPrefixes}}
  \label{algo:getprefixes}
\end{algorithm}
\DecMargin{1em}

Given a centroid $z$, i.e., a sidetrack edge and a known overall cost $C_z$, all
solution paths that go from $s$ to the starting vertex of the sidetrack edge,
traverse it and then arrive at the goal state with overall cost $C_z$, or
simply, the paths of the centroid, are computed as the cross-product of all
prefixes and suffixes corresponding to the centroid. As in the case of prefix
construction, all suffixes with a given cost can be found by conducting a
depth-first search in \textsc{closed} where the next node selected is a child of
the current one with a $g_{b}$-value equal to the $g_{b}$-value of its parent
minus the cost of the edge that joins them, until the goal state is reached.
This procedure is known as \textit{suffix construction}.

\IncMargin{1em}
\begin{algorithm}[tb]
  \small
  \LinesNumbered
  \DontPrintSemicolon
  \SetKwData{Open}{open}
  \SetKwData{Closed}{closed}
  \SetKwData{Z}{Z}
  \SetKwFunction{GetPaths}{GetPaths}
  \SetKwFunction{Pop}{pop}
  \SetKwFunction{Add}{add}
  \SetKwFunction{Insert}{insert}
  \SetKwFunction{Children}{children}
  \KwData{A graph $G (V, E)$ and two designated vertices $s, t\in V$}
  \KwResult{Solution set $\Pi$ with $\kappa$ shortest paths}
  \BlankLine
  \Closed$\leftarrow\varnothing$, \Z$\leftarrow\varnothing$\;
  \Open$\leftarrow\{s\}$ with $g (s)=0$\;
  \While{\Open$\neq \varnothing$}{
    $n\leftarrow$\Pop{\Open}\;\label{algo:bela:n}
    \While{$\exists z\in$\Z, $C_z \leq f (n)$}{\label{algo:bela:z}
      $z\leftarrow$\Pop (\Z)\;
      $\Pi$ = $\Pi\cup$\GetPaths ($z$)\;
      \If{$|\Pi|\geq\kappa$}{
        return $\Pi$\;\label{algo:bela:return:0}
      }
    }
    \If{$n=t$}{\label{algo:bela:t}
      \Insert{\Z, $\langle e_{p_{t}, t}, g (n) \rangle$} in increasing order of cost\;\label{algo:insert:t}
      continue\;
    }
    \If{$n\in$\Closed}{
      \Add (\Closed, $e_{p_{n}, n}$)\;\label{algo:bela:edge}
      \For{every $g_{b}$-value in $n$}{\label{algo:bela:check-centroid}
        \Insert{\Z, $\langle e_{p_{n}, n}, g^{*} (p_{n}) + \omega (p_{n}, n) + g_{b}\rangle$} in increasing order of cost\;\label{algo:new-centroid}
      }
      continue\;
    }
    \Insert{\Closed, $n$}\;
    \For{$c_i\in$\Children{n}}{\label{algo:bela:gb:0}
      $g (c_i)=g (n)+\omega (n, c_i)$\;
      \Insert{\Open, $c_i$} in increasing order of $f (\cdotp)$ \;\label{algo:bela:gb:n}
    }
  }
  \While{\Z$\neq\varnothing$}{\label{algo:bela:out}
    $z\leftarrow$\Pop (\Z)\;
    $\Pi$ = $\Pi\cup$\GetPaths ($z$)\;\label{algo:bela:getpaths-2}
    \If{$|\Pi|\geq\kappa$}{
      return $\Pi$\;\label{algo:bela:end-2}
    }
  }
  return $\Pi$\;
  \BlankLine
  \normalsize
  \caption{Pseudocode of \bfbela{}/\bela{}}
  \label{algo:bela}
\end{algorithm}
\DecMargin{1em}

Algorithm~\ref{algo:bela} shows the pseudocode of \bfbela{}/\bela{}, where \Z{}
is the set of the current centroids and $p_{n}$ represents the parent of a node
$n$. The function \texttt{GetPaths} computes all paths represented by the
centroid given to it as the cross-product of all its prefixes and suffixes. To
guarantee \textit{admissibility} the algorithm first checks whether there is a
centroid with a cost strictly less or equal than the current $f (n)$ value, with
$n$ being the node just popped out from \textsc{open}. If so, all its paths are
added to the solution path until no more centroids can be used, or $\kappa$
shortest paths have been found, in which case the algorithm returns. Otherwise,
if a direct path to the goal has been found, then a new centroid with the cost
of this solution is added and the current iteration is skipped. Because Dijkstra
expands nodes in ascending order of cost, this guarantees that the next
iteration will start by outputting all paths corresponding to the centroid just
added. In case the current node, $n$, has been already expanded, its edge is
added to the \textsc{closed} list and, before skipping the current iteration, it
is verified whether there are known $g_{b}$-values of it in \textsc{closed}. If
so, this node has known suffixes, therefore, new centroids have been discovered,
so we add them to the set of centroids, \Z{}. Finally, the current node is
expanded and its children are added to the \textsc{open} list in increasing
order of their $f$-value.

Note that the \textsc{open} list can be exhausted without having found $\kappa$
shortest paths. In such case, the algorithm considers all centroids in
increasing order of their cost adding their solution paths to $\Pi$. While
computing these paths, new centroids might be discovered and so the loop
proceeds until $\kappa$ paths have been found. If after considering all
centroids, $\kappa$ paths are not found, the algorithm simply returns those that
were found.

It is noted that Algorithm~\ref{algo:bela} follows the same mechanics as
Dijkstra's/\astar{}, the only difference being that it uses information in
\textsc{closed} to reconstruct the $\kappa$ paths and, for this, it uses an
ordered set of centroids to generate paths from, \Z{}. Hence, its theoretical
properties naturally derive from those of Dijkstra's/\astar{}.

\begin{lemma}[Sufficient condition for expansion]
  \label{lma:sufficient}
  \bfbela{} expands all nodes with $f (n)<C^{*}_{\varphi}$.
\end{lemma}

\textbf{Proof}: Once \bfbela{} discovers a centroid, it is stored in \Z{} for
consideration only once the $f (n)$ value of the current node from \textsc{open}
is greater or equal than the cost of the cheapest centroid in \Z{} ---see
Line~\ref{algo:bela:z} in Algorithm~\ref{algo:bela}. Because $f (n)=g (n)$ is
monotonically increasing and nodes from \textsc{open} are expanded in ascending
order of $f (n)$, the $\kappa$ paths can be discovered only once
$C^{*}_{\varphi}\leq f (n)$, thus, after expanding all nodes with
$f (n)<C^{*}_{\varphi}$.\hfill$\Box$

\begin{lemma}[Necessary condition for expansion]
  \label{lma:necessary}
  \bfbela{} never expands nodes with $f (n)>C^{*}_{\varphi}$. Thus, a necessary
  condition for expansion in \bfbela{} is $f (n)\leq C^{*}_{\varphi}$.
\end{lemma}

\textbf{Proof}: Once a centroid $z$ is considered and $\kappa$ paths are
generated, the algorithm halts execution ---see Line~\ref{algo:bela:return:0}.
From the proof of the preceding Lemma, it is observed that this happens as soon
as $C^{*}_{\varphi}\leq f (n)$ and the considered centroid yields the necessary
number of paths to complete the search of $\kappa$ shortest paths, i.e., no node
with $f (n)>C^{*}_{\varphi}$ is ever expanded.\hfill$\Box$

\begin{theorem}[Commpleteness and Admissibility]
  \label{tho:completeness}
  \bfbela{} finds all paths in $\Pi$ as given in Definition (\ref{def:pi}).
\end{theorem}

\textbf{Proof}: Let $\pi_{i}$ denote a path in $\Pi$, and let us consider two
different cases: Either $\pi_{i}$ is a direct path or it is indirect ---see
Definition (\ref{def:directness}).

If it is a direct path, its node corresponding to $t$ will eventually be
expanded, and thus inserted in \Z{} in Line~\ref{algo:insert:t}. Because nodes
in \textsc{open} are considered in increasing order of cost, the next node to
expand will have a cost equal or greater than the cost of $\pi_{i}$ and thus,
all paths represented by the latest centroid will be discovered in the loop of
Lines \ref{algo:bela:z}--\ref{algo:bela:return:0}.

If it is an indirect path, then it must contain at least one sidetrack edge, see
Definition (\ref{def:components}). Let $e_{u, v}$ denote its first sidetrack
edge and let $C_{z}$ denote its cost. There are two cases to consider: Either
the ending vertex of the sidetrack edge, $v$, has at least one $g_{b}$-value
when reached from $u$, or it does not. In the case it has at least one
$g_{b}$-value, then a new centroid $z$, with
$C_z=g^{*} (u)+\omega (u, v)+g_{b} (v)$, is created for every $g_{b}$-value of
$v$ in Line~\ref{algo:new-centroid}. Once the centroid representing $\pi_{i}$ is
created, it will be eventually considered and $\pi_{i}$ found as soon as the
first node $n$ with $f (n)\geq C_{z}$ is popped out from \textsc{open}. Note
that this has to happen because $\pi_{i}$ is assumed to have a cost strictly
less or equal than $C^{*}_{\varphi}$, and the sufficient and necessary
conditions for expansion guarantee that a node $n$ with
$f (n)\geq C^{*}_{\varphi}\geq C_{z}$ should be eventually popped out from
\textsc{open}. In case that vertex $v$ has no $g_{b}$-values when reached from
$u$, then according to Lemma (\ref{tho:relation}) there shall exist a path
$\pi_{j}\in\Pi$, such that $C (\pi_{j}) \leq C (\pi_{i})$, and $v$ belongs to
the prefix of $\pi_{j}$. Upon discovery of the path $\pi_{j}$, the prefix
construction procedure necessarily will set a $g_{b}$-value for vertex $v$ and
it will discover $e_{u, v}$, and thus a centroid $z$ will be created that
represents $\pi_{i}$. As in the previous case, this centroid will be eventually
considered and $\pi_{i}$ will be found.\hfill$\Box$

The following result ensures that \bfbela{} preserves the best known asymptotic
worst-case complexity:

\begin{theorem}[Algorithmic complexity]
  \label{tho:complexity}
  \bfbela{} runs in $O (|E| + |V|\log{|V|}+\kappa|V|)$.
\end{theorem}

\textbf{Proof}: Algorithm~\ref{algo:bela} has a complexity at least as bad as
Dijkstra's, $O (|E| + |V|\log{|V|})$\parencite{dechter.r.pearl.j:generalized}.
On top of this, it has to output $\kappa$ paths in explicit form, which come
from the cross-product of all prefixes and suffixes of each centroid. In the
worst case every centroid yields a single prefix and suffix, and thus, the added
complexity is $O (\kappa|V|)$, with $|V|$ being the length of the shortest paths
in the worst case, resulting in a worst-case time complexity of
$O (|E| + |V|\log{|V|}+\kappa|V|)$.\hfill$\Box$

To conclude the presentation of the brute-force variant of our algorithm, the
first example introduced in the description of
\kstar{}\parencite{aljazzar_directed_2009,Aljazzar2011} is solved using
\bfbela{}. The example considered is shown in Figure~\ref{fig:example-1}, where it is
requested to find three shortest paths between $s_{0}$ and $s_{4}$ in the graph
shown in Figure~\ref{fig:example-1:0}, i.e., $\kappa=3$. The algorithm shown in
Pseudocode~\ref{algo:bela} is considered next using $f (n)=g (n)$, i.e.,
ignoring any heuristic guidance.

\begin{figure*}
  \centering
  \begin{subfigure}{0.3\textwidth}
    \begin{center}
      \begin{tikzpicture}

        \draw (0,0) node [circle, draw] (s0) {$s_{0}$};
        \draw (2,1.5) node [circle, draw] (s1) {$s_{1}$};
        \draw (2,0) node [circle, draw] (s2) {$s_{2}$};
        \draw (2,-1.5) node [circle, draw] (s3) {$s_{3}$};
        \draw (4,0) node [circle, draw] (s4) {$s_{4}$};

        \draw[-stealth] (s0) edge[above] node{\footnotesize 3} (s1);
        \draw[-stealth] (s0) edge[above] node{\footnotesize 2}  (s2);

        \draw[-stealth] (s1) edge[right] node{\footnotesize 1} (s2);
        \draw[-stealth] (s1) edge[right] node{\footnotesize 1} (s4);
        \path (s1) edge [loop above] node {\footnotesize 2} (s1);

        \draw[-stealth] (s2) edge[bend right] node[left]{\footnotesize 1} (s3);
        \draw[-stealth] (s2) edge[above] node{\footnotesize 3} (s4);

        \draw[-stealth] (s3) edge[bend right] node[right]{\footnotesize 2} (s2);

      \end{tikzpicture}
    \end{center}
    \caption{Simple example of \bfbela{}. $s_{0}$ is the start vertex and $s_{4}$ is
      the goal vertex.}
    \label{fig:example-1:0}
  \end{subfigure}
  \begin{subfigure}{0.3\textwidth}
    \begin{center}
      \begin{tikzpicture}

        \draw (0,0) node [circle, draw] (s0) {$s_{0}$};
        \draw (2,1.5) node [circle, draw] (s1) {$s_{1}$};
        \draw (2,0) node [circle, draw] (s2) {$s_{2}$};
        \draw (2,-1.5) node [circle, draw] (s3) {$s_{3}$};
        \draw (4,0) node [circle, draw] (s4) {$s_{4}$};

        \draw[-stealth] (s0) edge[above] node{\footnotesize 3} (s1);
        \draw[-stealth] (s0) edge[above] node{\footnotesize 2}  (s2);

        \draw[-stealth] (s1) edge[right] node{\footnotesize 1} (s2);
        \draw[dashed,-stealth] (s1) edge[right] node{\footnotesize 1} (s4);
        \path (s1) edge [dashed,loop above] node {\footnotesize 2} (s1);

        \draw[-stealth] (s2) edge[bend right] node[left]{\footnotesize 1} (s3);
        \draw[dashed,-stealth] (s2) edge[above] node{\footnotesize 3} (s4);

        \draw[dashed,-stealth] (s3) edge[bend right] node[right]{\footnotesize 2} (s2);

      \end{tikzpicture}
    \end{center}
    \caption{The result of the first five iterations of \bfbela{}. The solid lines represent edges stored in \textsc{closed}, whereas dotted lines represent edges whose start nodes have only been generated, but not expanded and are therefore not in \textsc{closed}.}
    \label{fig:example-1:a}
  \end{subfigure}
  \begin{subfigure}{0.3\textwidth}
    \begin{center}
      \begin{tikzpicture}

        \draw (0,0) node [circle, draw] (s0) {$s_{0}$};
        \draw (2,1.5) node [circle, draw] (s1) {$s_{1}$};
        \draw (2,0) node [circle, draw] (s2) {$s_{2}$};
        \draw (2,-1.5) node [circle, draw] (s3) {$s_{3}$};
        \draw (4,0) node [circle, draw] (s4) {$s_{4}$};

        \draw[-stealth] (s0) edge[above] node{\footnotesize 3} (s1);
        \draw[-stealth] (s0) edge[above] node{\footnotesize 2}  (s2);

        \draw[-stealth] (s1) edge[right] node{\footnotesize 1} (s2);
        \draw[-stealth] (s1) edge[right] node{\footnotesize 1} (s4);
        \path (s1) edge [dashed,loop above] node {\footnotesize 2} (s1);

        \draw[-stealth] (s2) edge[bend right] node[left]{\footnotesize 1} (s3);
        \draw[dashed,-stealth] (s2) edge[above] node{\footnotesize 3} (s4);

        \draw[dashed,-stealth] (s3) edge[bend right] node[right]{\footnotesize 2} (s2);

      \end{tikzpicture}
    \end{center}
    \caption{Result of the fifth iteration of \bfbela{}. $s_{4}$ is chosen
      for expansion next, at the beginning of the sixth iteration}
    \label{fig:example-1:b}
  \end{subfigure}
  \begin{subfigure}{0.3\textwidth}
    \begin{center}
      \begin{tikzpicture}

        \draw (0,0) node [circle, draw] (s0) {$s_{0}$};
        \draw (2,1.5) node [circle, draw] (s1) {$s_{1}$};
        \draw (2,0) node [circle, draw] (s2) {$s_{2}$};
        \draw (2,-1.5) node [circle, draw] (s3) {$s_{3}$};
        \draw (4,0) node [circle, draw] (s4) {$s_{4}$};

        \draw[-stealth] (s0) edge[above] node{\footnotesize 3} (s1);
        \draw[-stealth] (s0) edge[above] node{\footnotesize 2}  (s2);

        \draw[-stealth] (s1) edge[right] node{\footnotesize 1} (s2);
        \draw[-stealth] (s1) edge[right] node{\footnotesize 1} (s4);
        \path (s1) edge [loop above] node {\footnotesize 2} (s1);

        \draw[-stealth] (s2) edge[bend right] node[left]{\footnotesize 1} (s3);
        \draw[-stealth] (s2) edge[above] node{\footnotesize 3} (s4);

        \draw[dashed,-stealth] (s3) edge[bend right] node[right]{\footnotesize 2} (s2);

      \end{tikzpicture}
    \end{center}
    \caption{Beginning of the ninth iteration of \bfbela{}.}
    \label{fig:example-1:c}
  \end{subfigure}
  \begin{subfigure}{0.3\textwidth}
    \begin{center}
      \begin{tikzpicture}

        \draw (0,0) node [circle, draw] (s0) {$s_{0}$};
        \draw (2,1.5) node [circle, draw] (s1) {$s_{1}$};
        \draw (2,0) node [circle, draw] (s2) {$s_{2}$};
        \draw (2,-1.5) node [circle, draw] (s3) {$s_{3}$};
        \draw (4,0) node [circle, draw] (s4) {$s_{4}$};

        \draw[-stealth] (s0) edge[above] node{\footnotesize 3} (s1);
        \draw[-stealth] (s0) edge[above] node{\footnotesize 2}  (s2);

        \draw[-stealth] (s1) edge[right] node{\footnotesize 1} (s2);
        \draw[-stealth] (s1) edge[right] node{\footnotesize 1} (s4);
        \path (s1) edge [loop above] node {\footnotesize 2} (s1);

        \draw[-stealth] (s2) edge[bend right] node[left]{\footnotesize 1} (s3);
        \draw[-stealth] (s2) edge[above] node{\footnotesize 3} (s4);

        \draw[-stealth] (s3) edge[bend right] node[right]{\footnotesize 2} (s2);

      \end{tikzpicture}
    \end{center}
    \caption{End of the ninth iteration of \bfbela{}. All nodes have been
      expanded and \textsc{open} is empty}
    \label{fig:example-1:d}
  \end{subfigure}
  \caption{Example of \bfbela{}}
  \label{fig:example-1}
\end{figure*}

Figure~\ref{fig:example-1:a} shows the first five iterations of
\bfbela{}. Because no centroids have been discovered yet, the algorithm
proceeds in exactly the same fashion as Dijkstra's. The expansion order (with
the $g^{*}$-values shown between parenthesis) is $s_{0} (g=0)$, $s_{2} (g=2)$,
$s_{1} (g=3)$, $s_{3} (g=3)$, and $s_{2} (g=4)$. Here, ties are broken favoring
nodes which enter \textsc{open} earlier. A solid line in
Figure~\ref{fig:example-1:a} indicates that the start vertex of the edge has been
expanded, and thus, it is present in \textsc{closed}. At this point, the
contents of \textsc{open} (with the $g$-values shown between parenthesis) are:
$s_{4} (g=4)$, $s_{4} (g=5)$, $s_{1} (g=5)$, $s_{2} (g=5)$. The dotted lines
are incoming edges of nodes that have only been generated, but not expanded, and are thus only in \textsc{open} and not \textsc{closed}.

In the sixth iteration, $s_{4} (g=4)$ is popped out from \textsc{open}, which is
detected to be the goal on Line~\ref{algo:bela:t} of Algorithm~\ref{algo:bela}.
As a consequence, the first centroid, $\langle e_{s_{1}, s_{4}}, 4\rangle$ is
added to the set of centroids \Z{}, and the current iteration is skipped without
generating any children.

Next, the node $s_{4} (g=5)$ is popped from \textsc{open}, but before it is
examined it is observed that there is a centroid with cost 4, being less or
equal than the $g$-value of the current node, 5, on Line~\ref{algo:bela:z}.
Thus, the centroid $\langle e_{s_{1}, s_{4}}, 4\rangle$ is popped from \Z{} (so
that it becomes empty) and all paths represented by this centroid are computed
by \texttt{GetPaths}. The prefixes of the starting vertex of the centroid,
$s_{1}$ are all the optimal paths from the start state, $s_{0}$ to it. There is
only one such prefix, namely $\langle s_{0}, s_{1} \rangle$ with a cost equal to
3. Recall that while prefixes are computed, all visited nodes are checked for incoming sidetrack edges, which give us new centroids. Currently,
$s_{1}$ only has one incoming edge which is a tree edge, and $s_{0}$ has no
incoming edges, so no new centroids are found. Because the cost of the prefix
plus the cost of the defining edge of the centroid is equal to the overall cost of the centroid, 4, there are no suffixes to compute. Therefore, the first path found is
$\pi_{1}: \langle s_{0}, s_{1}, s_{4}\rangle$.

After examining all paths represented by the centroid considered, the next
iteration proceeds as usual. Next, $s_{4} (g=5)$ is popped from open
\textsc{open}. $s_4$ is found to represent the goal vertex on
Line~\ref{algo:bela:t}, so the centroid $\langle e_{s_{2}, s_{4}}, 5 \rangle$ is
added to \Z{}, and the current iteration is terminated without expanding $s_4$.

At the beginning of the eighth iteration, the contents of \textsc{closed} are
still the same as those shown in Figure~\ref{fig:example-1:b}, but the
\textsc{open} list has shrunk. Now, it only contains $s_{1} (g=5)$ and $s_{2} (g=5)$.
Thus, the next node chosen for expansion on Line~\ref{algo:bela:n} is $s_{1}$.
However, before proceeding, it is observed that there is currently a centroid,
$\langle e_{s_{2}, s_{4}}, 5 \rangle$ with a cost less than or equal to
the $g$-value of the node just popped from \textsc{open}, so
\textsc{GetPaths} is invoked again on this centroid.

\textsc{GetPaths} starts by assigning a $g_{b}$-value of 3 to node $s_{2}$,
because that is the cost of the centroid under consideration minus the cost of
the prefix. It then starts computing the prefixes from the starting vertex of
the centroid, $s_{2}$. First, it discovers an incoming sidetrack edge of
$s_{2}$, $e_{s_{1}, s_{2}}$. Because $s_2$ has a $g_{b}$-value 3, the
$g^{*}$-value of $s_{1}$ is 3, and the edge cost of the sidetrack edge is 1, a
new centroid, $\langle e_{s_{1}, s_{2}}, 7\rangle$ is added to \Z{} which now
only contains this centroid. It then continues following backpointers, reaching
the start state, which has no incoming edges. Thus, the full list of prefixes
just contains $\langle s_{0}, s_{2}\rangle$. Like earlier, the list of suffixes
is empty because the ending vertex of the centroid currently under consideration
is the goal state itself, and thus has a $g_{b}$-value equal to 0. Hence, the
only path returned is $\pi_{2}: \langle s_{0}, s_{2}, s_{4}\rangle$.

We still have not returned $\kappa=3$ paths, so we continue with the eighth iteration. The current node, $s_{1}$ is already in
\textsc{closed}, so the new edge (the self-loop, $e_{s_{1}, s_{1}}$) is
added to \textsc{closed} on Line~\ref{algo:bela:edge}. Before continuing to the next iteration, it is observed that $s_{1}$ contains one $g_{b}$-value, 1,
and thus, a new centroid $\langle e_{s_{1}, s_{1}}, 6\rangle$ is added to \Z{}.
The cost of the new centroid, 6, is computed as the
sum of the $g^{*}$-value of node $s_{1}$, plus the cost of the sidetrack edge,
2, plus the $g_{b}$-value, 1. Currently, the list of centroids stored is:
$\langle e_{s_{1}, s_{1}}, 6\rangle$, $\langle e_{s_{1}, s_{2}}, 7\rangle$.
Because the node popped out from \textsc{open} was a duplicate, the current
iteration is terminated before expansion.

In the ninth iteration, there is only one node in \textsc{open}, $s_{2} (g=5)$,
and the \textsc{closed} list has been updated to contain the self-loop of node
$s_{1}$ as shown in Figure~\ref{fig:example-1:c}. At this point, \Z{} contains
two centroids, but none of them have a cost less than or equal to the $g$-value
of the node popped from \textsc{open}, so they are ignored. Because $s_{2}$ is
not the goal state, it is looked up in \textsc{closed} and it is found to be a
duplicate, so the edge $e_{s_{3}, s_{2}}$ is added to \textsc{closed}. Recall
that node $s_{2}$ had a known $g_{b}$-value equal to 3, and thus the loop on
lines \ref{algo:bela:gb:0}--\ref{algo:bela:gb:n} adds a new centroid,
$\langle e_{s_{3}, s_{2}}, 8 \rangle$, where the cost of the centroid is
computed as follows:
$g^{*} (s_{3}) + \omega (s_{3}, s_{2}) + g_{b} (s_{2}) = 3 + 2 + 3 = 8$.
Currently, \Z{}=$\{\langle e_{s_{1}, s_{1}}, 6\rangle$,
$\langle e_{s_{1}, s_{2}}, 7\rangle$, $\langle e_{s_{3}, s_{2}}, 8 \rangle\}$.
Because this node has been found in \textsc{closed} the current iteration ends.
The result of this iteration is shown in Figure~\ref{fig:example-1:d}

At the end of the ninth iteration, the \textsc{open} list is empty, meaning all nodes have already been expanded, so execution continues in the loop
starting at Line~\ref{algo:bela:out}. From this point on, \bfbela{} only uses
information from the \textsc{closed} list to find new paths. Note it can update
\textsc{closed} with new centroids if any are found during prefix computation.

The first centroid popped is $\langle e_{s_{1}, s_{1}}, 6\rangle$. Exactly as it
happened in the eighth iteration, the prefix computation starts by setting a new
$g_{b}$-value of 3 for the starting vertex of this centroid, $s_{1}$, equal to the
cost of the current centroid, 6, minus the cost of the prefix, 3. While
computing the prefixes from $s_{1}$ it realizes (again) that the self-loop
$e_{s_{1}, s_{1}}$ is a sidetrack edge, as it did in the eighth iteration, so a new
centroid $\langle e_{s_{1}, s_{1}}, 8 \rangle$ is added to \Z{} with the cost of
the centroid computed like so:
$g^{*} (s_{1}) + \omega (s_{1}, s_{1})+g_{b} (s_{1})=3 + 2 + 3 =8$. The next
node visited is $s_{0}$, which has no incoming edges, so the prefix
computation ends, returning the path $\langle s_{0}, s_{1}\rangle$. This time,
the suffix computation produces a non-empty path. Starting at $s_{1}$ with a
$g_{b}$-value equal to 1, it chooses all descendants $n'$ with a $g_{b}$-value
equal to $g_{b} (s_{1})-\omega (s_{1}, n')$. There is only one such descendant,
$s_{4}$, so the only suffix produced is $\langle s_{1}, s_{4}\rangle$. Finally,
\texttt{GetPaths} computes the cross-product of all prefixes and suffixes
giving us the path $\pi_{3}:\langle s_{0}, s_{1}, s_{1}, s_{4}\rangle$,
with a cost equal to 6. Because the number of paths produced so far, 3, is equal to the
desired number of paths, $\kappa=3$, the algorithm terminates on
Line~\ref{algo:bela:end-2}, returning the following paths:

\[
  \begin{array}{lll}
    \pi_{1} & \langle s_{0}, s_{1}, s_{4}\rangle & C^{*}_{0} = 4\\
    \pi_{2} & \langle s_{0}, s_{2}, s_{4}\rangle & C^{*}_{1} = 5\\
    \pi_{3} & \langle s_{0}, s_{1}, s_{1}, s_{4}\rangle & C^{*}_{2} = 6\\
  \end{array}
\]


%% file: supplemental5.tex

\section{BELA$^*$}
\label{sec:belastar}

This Section considers the availability of a heuristic function, $h (\cdotp)$,
which is assumed to be \textit{consistent} and thus, \textit{admissible}. As a
matter of fact, all of the discussion from the previous Section apply to
this one and only a few novel remarks are necessary. Indeed, Algorithm~\ref{algo:bela}
becomes \bela{} when using $f (n)=g (n)+h (n)$.

The first observation is that $f (\cdotp)$ is monotonically increasing, provided
that $h (\cdotp)$ is consistent, as assumed. From this, Lemmas
(\ref{lma:sufficient}) and (\ref{lma:necessary}) are still valid for \bela{}.

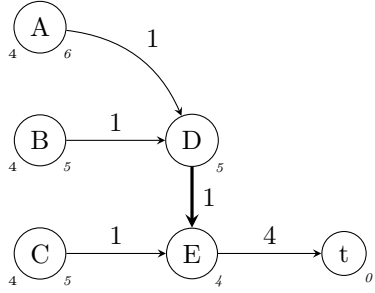
\begin{figure}
  \begin{center}
    \begin{tikzpicture}

      \node [draw, circle] (A) at (2,3) {A};
      \draw (2,1.5) node [circle, draw] (B) {B};
      \draw (2,0) node [circle, draw] (C) {C};
      \draw (4,1.5) node [circle, draw] (D) {D};
      \draw (4,0) node [circle, draw] (E) {E};
      \draw (6,0) node [circle, draw] (t) {t};

      \node[below left=-0.10cm of A] {\tiny{4}};
      \node[below left=-0.10cm of B] {\tiny{4}};
      \node[below left=-0.10cm of C] {\tiny{4}};

      \node[below right=-0.10cm of A] {\textit{\tiny{6}}};
      \node[below right=-0.10cm of B] {\textit{\tiny{5}}};
      \node[below right=-0.10cm of C] {\textit{\tiny{5}}};
      \node[below right=-0.10cm of D] {\textit{\tiny{5}}};
      \node[below right=-0.10cm of E] {\textit{\tiny{4}}};
      \node[below right=-0.10cm of t] {\textit{\tiny{0}}};

      \path [-stealth] (A) edge [bend left] node[above right]{1} (D);
      \path [-stealth] (B) edge [above] node{1} (D);
      \draw[-stealth] (C) edge[above] node{1}  (E);
      \draw[-stealth, very thick] (D) edge[right] node{1}  (E);
      \draw[-stealth] (E) edge[above] node{4}  (t);

    \end{tikzpicture}
  \end{center}
  \caption{Expansion order of \bela{}}
  \label{fig:bela:expansion}
\end{figure}

The case of Theorem (\ref{tho:completeness}) deserves further consideration,
though. Consider Figure~\ref{fig:bela:expansion}, where the $g$-value of each
node is shown below it to the left, and its $h$-value appears below it to the
right in italics. Consider next the order expansion of \bfbela{} and how it
contrasts with the order expansion of \bela{}. \bfbela{} first expands node C
($g^{*}$=4), generating node E; next it expands nodes B ($g^{*}=4$) and A
($g^{*}=4$), which generate two copies of node D with $g^{*}=5$. After expanding
node E ($g^{*}=5$), the goal is generated with a cost equal to 9. The first copy
of node D is expanded immediately after, adding a new copy of node E ($g^{*}=6$)
to the \textsc{open} list. The next node in open is the second copy of node D
($g^{*}=5$) which is found to be a duplicate and thus, it is skipped for
expansion after adding the edge $\langle A, D\rangle$ to the \textsc{closed}
list. The next node popped from \textsc{open} is the node E ($g^{*}=6$) which,
as in the previous case, is found to be a duplicate so that the edge
$\langle D, E \rangle$ is added to \textsc{closed}. The last node in
\textsc{open} is $t$, which generates a new centroid with a cost equal to 9. In
the next iteration, the centroid just created is considered and, while
constructing its prefixes, node E gets a $g_{b}$-value equal to 4 which is
propagated backwards from the goal as the sum of all cost edges traversed so
far. At this point, the prefix construction procedure notes that there is an
incoming sidetrack edge, $\langle D, E\rangle$, and hence a new centroid with
cost $C=g^{*} (D)+\omega (D, E)+g_{b}$ = 5 + 1 + 4 = 10 is added. The prefix
construction will continue moving backwards to C and from there eventually
reaching the start state $s$, setting $g_{b}$ values for all nodes traversed.
The important aspect to note is that, if $\kappa$ nodes have not been produced
yet, \bfbela{} will consider the centroid $\langle e_{D, E}, 10\rangle$ created
in the last prefix construction. Proceeding as in the previous case, it will
search both prefixes starting from node D, one through A and the other one
through B. As it can be seen, \bfbela{} provides a \textit{one to many} mapping
of centroids to solution paths, in contrast to \kstar{} which provides a
\textit{one to one} mapping from paths in the \textit{path graph} and paths in
the true graph.

Still, the same observation is true for \bela{}, but the cardinality of this
mapping might shrink as a result of \textit{tie-breaking} rules of $f$-values.
When using \bela{} nodes are expanded in increasing order of their $f$-value.
Succinctly, C ($f=9$) and B ($f=9$) are expanded first generating E ($f=9$) and
D ($f=10$), respectively. After expanding E ($f=9$), the goal is generated with
cost 9 which becomes the first node in \textsc{open}, so it is selected for
expansion in the next iteration, triggering the prefix construction, which
returns all optimal paths from $s$ to $t$ with a cost equal to 9 that use the
edge $\langle E, t\rangle$. Note, however, that node $D$ is still in
\textsc{open} and thus, the sidetrack edge $\langle e_{D, E}, 10\rangle$ has not
been discovered yet, as with \bfbela{}. There are two nodes in \textsc{open} at
this point, A ($f=10$) and D ($f=10$). The expansion order matters indeed and if
A is expanded before D, paths will be discovered in the same order as \bfbela{}.
Assuming the opposite, the expansion of D ($f=10$) generates E ($f=10$), which
is found to be a duplicate, and the edge $\langle D, E \rangle$ is added to
\textsc{closed}. This time, before skipping its expansion,
Lines~\ref{algo:bela:check-centroid}--\ref{algo:new-centroid} in
Algorithm~\ref{algo:bela} add a new centroid because node E was already given a
$g_{b}$-value equal to 4. Because the new centroid has a cost equal to 10 units,
which is equal to the $f$-value of the next node in \textsc{open}, it is
considered before expanding node A. Because A has not yet been expanded, the
incoming edge $\langle A, D\rangle$ is not discovered in prefix construction.
\bela{} will have to wait for the expansion of node A to realize that a new
centroid can be created and thus, the existence of new paths. Consequently, only
one prefix will be considered even though there are two.

There are two important consequences of the expansion order of \bela{}: On one
hand, the consideration of a centroid might yield less paths than the number of
paths returned by \bfbela{} when considering the same centroid; On the other
hand, and only in the context of \bela{}, centroids can be constructed from a
tree edge! Note that in the last example, the eventual expansion of node A
($f=10$) generates D ($f=10$) which, when being expanded is observed to have a
$g_{b}$-value equal to 4, and thus a new centroid is generated. However, the
edge $\langle A, D\rangle$ is a tree edge. There remains a third consequence. In
spite of the effects of \textit{tie-breaking} policies on the expansion order of
\bela{}, all cases considered in the Proof of Theorem (\ref{tho:completeness})
are still valid and thus, \bela{} is both complete and admissible. Moreover,
like \astar{} when contrasted with Dijkstra's, \bela{} should considerably
reduce the number of expansions in comparison with \bfbela{} by avoiding the
consideration of all nodes with $f(n)>C^{*}_{\varphi}$. The accuracy of the
heuristic function plays a major role in the level of reduction. The better the
heuristic, the larger the improvement in the number of necessary expansions.

To conclude, note that when using a consistent heuristic, duplicates are never
expanded, thus Theorem (\ref{tho:complexity}) still applies.


%% file: supplemental6.tex

\section{Empirical evaluation}
\label{sec:empirical-evaluation}

This last section provides all relevant details of the experiments described in
the main paper. All of the source code, along with documentation, unit tests,
and various scripts for running the experiments and generating figures and
tables are available on
github\footnote{\url{https://github.com/clinaresl/ksearch}}. All the instances for
all experiments are stored in Zenodo~\parencite{linares_lopez_2024_13293103}.
The selection of domains considers both map-like and combinatorial domains, with
branching factors ranging from slightly above 2 (in the roadmap domain), to
two-digit branching factors in the $N$-Pancake domain; depths ranging from
dozens of vertices (as in the $N$-Puzzle or the $N$-Pancake domains) to several
hundreds, often exceeding 1,000 ---as in the Random Maps and the Roadmap
domains. We also consider both unit cost and non-unit cost versions (the
definition of non-unit costs is domain dependent). The selection of $\kappa$
values has been always from 1 to 10, from 10 to 100 in steps of 10, next getting
to 1,000 in steps of 100 and, finally, to 10,000 in steps of 1,000, unless
inferior values were enough to compare the selected algorithms, or too hard to
solve. The benchmarking suite has been configured so that every algorithm is
able to solve all instances for all the selected values of $\kappa$.

In each domain, we measure runtime, number of expansions, and memory usage for
each algorithm. Importantly, memory usage is simply the memory measured at the
termination of the algorithm, with the memory needed for storing solutions
subtracted. Data is provided first, as figures, and also in tabular form
in Appendix~\ref{cha:tables}.

All the experiments have been executed on a machine with 8 core i7 and 32 Gb of
RAM. All algorithms have been implemented in c++-17.

\input{supplemental6-roadmap}

\input{supplemental6-maps}

\input{supplemental6-npancake}

\input{supplemental6-npuzzle}


%% file: supplemental6-roadmap.tex

\subsection{Roadmap}
\label{sec:empirical-evaluation:roadmap}

The roadmap domain was used in the empirical evaluation of \kstar{} in
\parencite{Aljazzar2011} and thus, it is considered in this section. It is taken from
the 9th DIMACS Shortest-Path Challenge. Two variants are considered, dimacs and
unit. The first uses the provided edge costs. The latter considers all edges to
have cost 1.

\subsubsection{9th DIMACS Challenge}
\label{sec:empirical-evaluation:roadmap:9th-dimacs-challenge}

Figures~\ref{fig:roadmap:dimacs:brute-force:runtime}--\ref{fig:roadmap:dimacs:heuristic:expansions}
show the results of running \bela{}, \mAstar{}, \kstar{}, and their brute-force
variants over a selection of maps from the 9th DIMACS Shortest Paths Challenge.
In the empirical evaluation of \kstar{} only NY and E were used.
Table~\ref{tab:dimacs} shows all of the available maps and their size measured
in the number of vertices and edges. The figures show the runtime (in seconds),
memory usage (in Mbytes) and number of expansions of each algorithm. Every point
has been averaged over 100 instances randomly generated where, as in the
original evaluation of \kstar{} a random pair $s-t$ was accepted if and only if
the distance between them was at least 50 km measured as the great-circle
distance using the haversine function, as described in \parencite{Aljazzar2011}.

\begin{table}[h]
  \centering
  \begin{tabular}{llrr}
    \multicolumn{1}{c}{\small{Name}} & \multicolumn{1}{c}{\small{Description}} & \multicolumn{1}{c}{\small{\# Vertices}} & \multicolumn{1}{c}{\small{\# Edges}} \\ \toprule
    USA  & Full USA & 23,947,347 & 58,333,344 \\
    CTR  & Central USA & 14,081,816 & 34,292,496 \\
    W    & Western USA & 6,262,104 & 15,248,146 \\
    E    & Eastern USA & 3,598,623 & 8,778,114 \\
    LKS  & Great Lakes & 2,758,119 & 6,885,658 \\
    CAL  & California and Nevada & 1,890,815 & 4,657,742 \\
    NE   & Northeast USA & 1,524,453 & 3,897,636 \\
    NW   & Northwest USA & 1,207,945 & 2,840,208 \\
    FLA  & Florida & 1,070,376 & 2,712,798 \\
    COL  & Colorado & 435,666 & 1,057,066 \\
    BAY  & San Francisco Bay Area & 321,270 & 800,172 \\
    NY   & New York City & 264,346 & 733,846 \\ \bottomrule
  \end{tabular}
  \caption{9th DIMACS Shortest Path Challenge}
  \label{tab:dimacs}
\end{table}

Figure~\ref{fig:roadmap:dimacs:brute-force:runtime} compares only \bfbela{},
\bfk{} and mDijkstra, and it shows a clear trend. Even if \bfk{} is faster than
\bfbela{} for large values of $\kappa$, this only occurs in the smallest graphs,
NY and BAY. In larger graphs, the margin of improvement in runtime provided by
\bfbela{} increases with the $\kappa$ --- see
Figures~\ref{fig:roadmap:dimacs:brute-force:runtime:d}
and~\ref{fig:roadmap:dimacs:brute-force:runtime:f}. Note that mDijkstra, the
brute-force variant of \mAstar{} performs so poorly that it was not practically
possible to compute more than $\kappa=10$ paths with it, while either \bfbela{}
of \bfk{} output 10,000 paths in roughly the same amount of time.

Regarding memory usage, Figure~\ref{fig:roadmap:dimacs:brute-force:mem:a} shows
an effect that will be seen in other experiments as well, i.e., that memory
usage in either \bfbela{} or \bela{} can decrease when increasing the number of
paths to seek, $\kappa$. This phenomena is attributed to the fact that the
number of centroids can decrease when looking for more paths and thus, less
memory is required to store all the necessary information, as shown in the
example discussed in Section~\ref{sec:bela0}.

Figure~\ref{fig:roadmap:dimacs:heuristic:runtime} shows the runtime (in seconds)
of \bela{}, \mAstar{} and \kstar{}. Again, the same trend we observed before is
seen here. Even though \kstar{} performs better than \bela{} in smaller graphs,
this margin shrinks as the size of the graph increases, and, eventually, it
performs worse in the larger map, E. The relatively good performance of \kstar{}
in this domain is attributed to a variety of factors. On one hand, the maximum
number of paths requested, 10,000 does not require expanding the whole graph as
Figure~\ref{fig:roadmap:dimacs:heuristic:expansions} shows. Second, all of the
graphs where \kstar{} performs better than \bela{} are rather small (the largest
one with less than 2 million vertices). Third, of all the benchmarks tried, this
is the one with the lowest branching factor. Most importantly, the heuristic
function suggested in \parencite{Aljazzar2011} is very poor.
Figure~\ref{fig:roadmap:dimacs:mixed:expansions} (where \mAstar{} has been
removed due to its poor performance) shows that the reduction in the number of
expansions is always around 10\% both for \bela{} and \kstar{}, which is too
small to pay off for the extra work at each node for computing the heuristic
function. In fact, the brute-force variants of both \kstar{} and \bela{}, i.e.,
\bfk{} and \bfbela{}, outperform their heuristic counterparts in all maps, as
shown in Figure~\ref{fig:roadmap:dimacs:mixed:runtime}, with the only exception
being the smallest graphs, NY and BAY. As a consequence of the poor performance
of the heuristic function, \mAstar{} performs the worst, as it expands nodes
near the start state many times.

In the end, \bfbela{} is the fastest among all algorithms tried in this domain
in the majority of cases.

\input{roadmap.dimacs.runtime.brute-force.tex}
\input{roadmap.dimacs.mem.brute-force.tex}
\input{roadmap.dimacs.expansions.brute-force.tex}

\input{roadmap.dimacs.runtime.heuristic.tex}
\input{roadmap.dimacs.mem.heuristic.tex}
\input{roadmap.dimacs.expansions.heuristic.tex}

\input{roadmap.dimacs.runtime.mixed.tex}
\input{roadmap.dimacs.mem.mixed.tex}
\input{roadmap.dimacs.expansions.mixed.tex}

\subsubsection{Unit variant}
\label{sec:empirical-evaluation:roadmap:unit}

The edge costs found in the dimacs variant of the roadmap domain vary quite a
lot and are decently large. Weighting every edge with these costs makes the
mapping between centroids and solution paths provided by \bela{} to be very
poor, because each centroid can only be expected to represent a few paths. For
example, \bela{} exploits about 1,800 centroids to generate 10,000 paths in the
NY map, whereas in the E map (which is larger), on average, it uses almost 1,100
centroids to generate the same number of solution paths. Thus, every centroid
generated in the dimacs variant of this domain approximately represents 5 to 9
paths. Simply using unit costs produces a dramatic change in these figures. Of
course, doing so invalidates the heuristic function used in the dimacs variant
and thus, only the brute-force versions are considered in this case. In the unit
variant of the roadmap domain, \bfbela{} needs 15 centroids on average to
generate 10,000 paths in the NY map, and a little bit more than 18 in the E map
to generate the same number of paths, improving the number of paths per centroid
by about two orders of magnitude.

The experiments conducted in the unit variant aim to demonstrate how
\bfbela{} can benefit from this increase in the number of paths per centroid.
Figure~\ref{fig:roadmap:unit:brute-force:runtime} shows the runtime (in seconds)
of all brute-force search algorithms in the unit variant. Again, mDijkstra
performs so poorly that only $\kappa=10$ paths can be computed in the time used
by \bfk{} and \bfbela{} to find 10,000 distinct paths. As
Figure~\ref{fig:roadmap:unit:brute-force:runtime} shows, the difference in
running time between \bfk{} and \bfbela{} increases with larger values of
$\kappa$ in all maps, regardless of their size.

Thus, \bfbela{} strongly dominates both \bfk{} and mDijkstra in the unit variant of the
roadmap domain, being three or four times
faster than \bfk{}.

\input{roadmap.unit.runtime.brute-force.tex}
\input{roadmap.unit.mem.brute-force.tex}
\input{roadmap.unit.expansions.brute-force.tex}


%% file: roadmap.dimacs.runtime.brute-force.tex
\begin{figure*}
  \centering
  \begin{subfigure}{0.3\textwidth}
    \begin{center}
        \includegraphics[width=\textwidth]{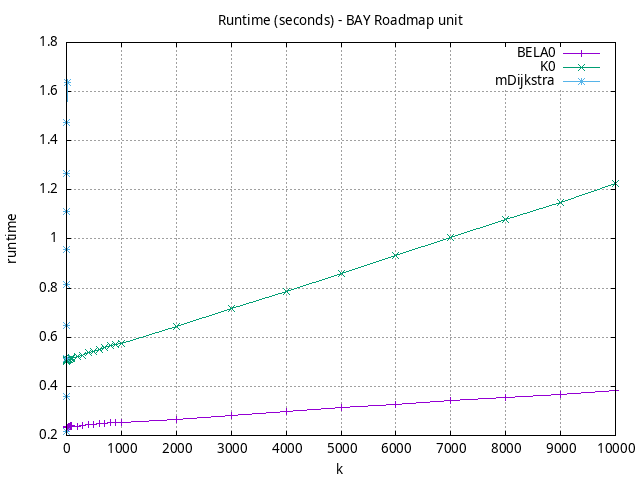}
    \end{center}
    \caption{}
    \label{fig:roadmap:dimacs:brute-force:runtime:a}
  \end{subfigure}
  \begin{subfigure}{0.3\textwidth}
    \begin{center}
        \includegraphics[width=\textwidth]{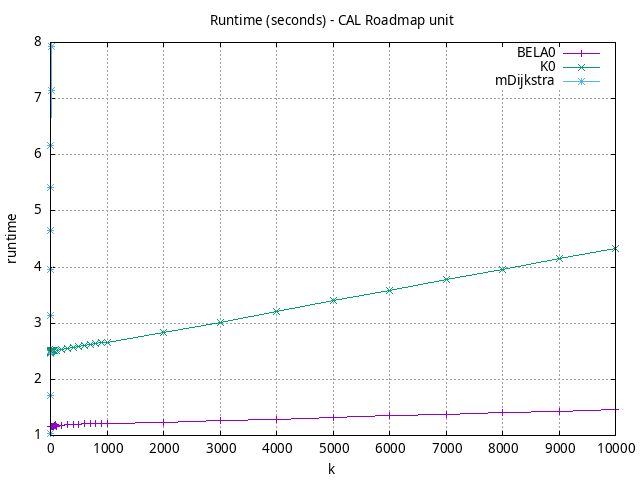}
    \end{center}
    \caption{}
    \label{fig:roadmap:dimacs:brute-force:runtime:b}
  \end{subfigure}
  \begin{subfigure}{0.3\textwidth}
    \begin{center}
        \includegraphics[width=\textwidth]{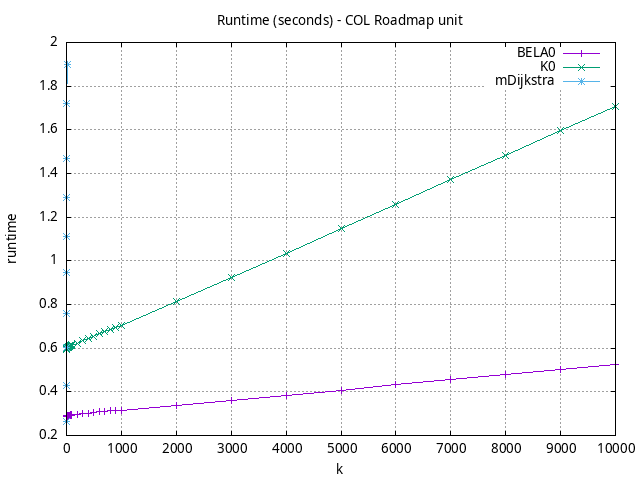}
    \end{center}
    \caption{}
    \label{fig:roadmap:dimacs:brute-force:runtime:c}
  \end{subfigure}
  \begin{subfigure}{0.3\textwidth}
    \begin{center}
        \includegraphics[width=\textwidth]{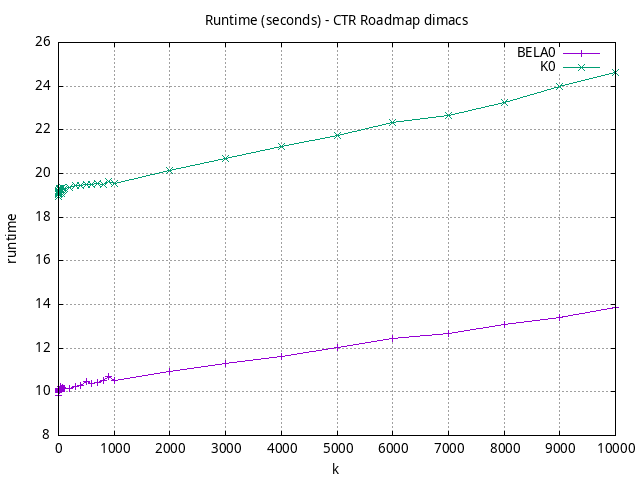}
    \end{center}
    \caption{}
    \label{fig:roadmap:dimacs:brute-force:runtime:d}
  \end{subfigure}
  \begin{subfigure}{0.3\textwidth}
    \begin{center}
        \includegraphics[width=\textwidth]{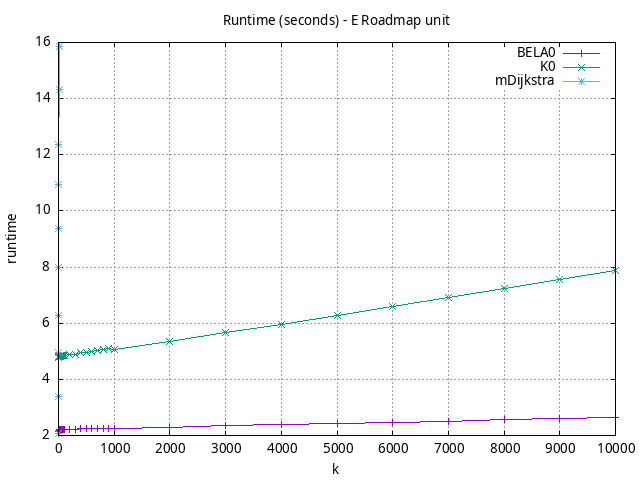}
    \end{center}
    \caption{}
    \label{fig:roadmap:dimacs:brute-force:runtime:e}
  \end{subfigure}
  \begin{subfigure}{0.3\textwidth}
    \begin{center}
        \includegraphics[width=\textwidth]{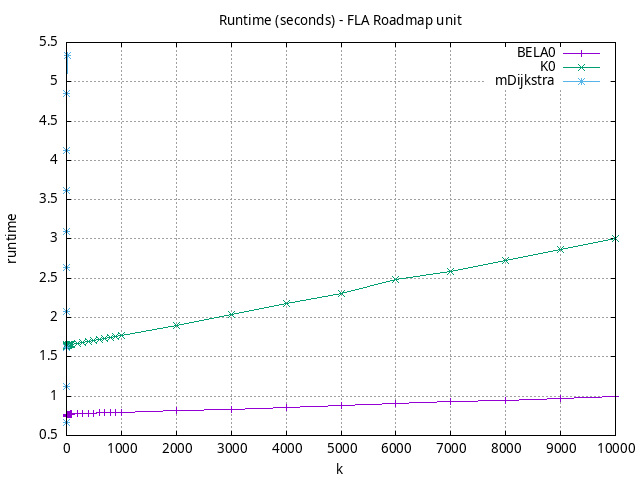}
    \end{center}
    \caption{}
    \label{fig:roadmap:dimacs:brute-force:runtime:f}
  \end{subfigure}
  \begin{subfigure}{0.3\textwidth}
    \begin{center}
        \includegraphics[width=\textwidth]{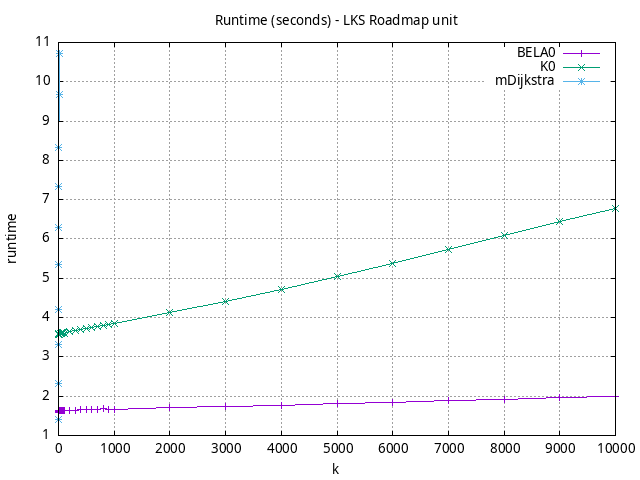}
    \end{center}
    \caption{}
    \label{fig:roadmap:dimacs:brute-force:runtime:g}
  \end{subfigure}
  \begin{subfigure}{0.3\textwidth}
    \begin{center}
        \includegraphics[width=\textwidth]{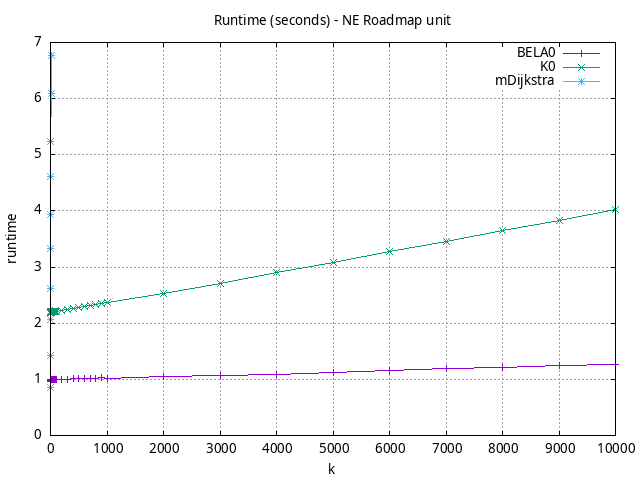}
    \end{center}
    \caption{}
    \label{fig:roadmap:dimacs:brute-force:runtime:h}
  \end{subfigure}
  \begin{subfigure}{0.3\textwidth}
    \begin{center}
        \includegraphics[width=\textwidth]{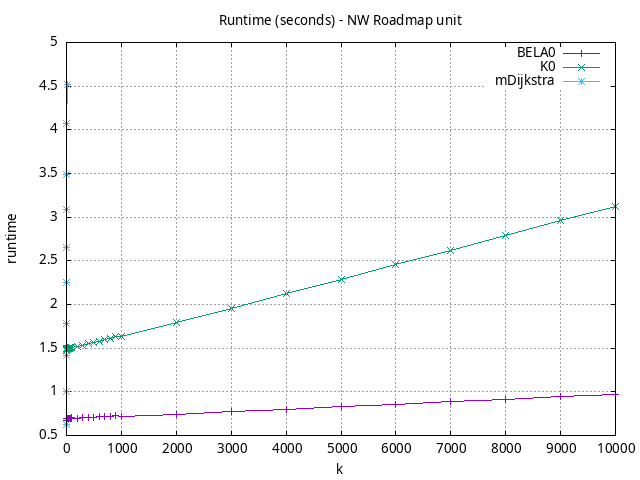}
    \end{center}
    \caption{}
    \label{fig:roadmap:dimacs:brute-force:runtime:i}
  \end{subfigure}
  \begin{subfigure}{0.3\textwidth}
    \begin{center}
        \includegraphics[width=\textwidth]{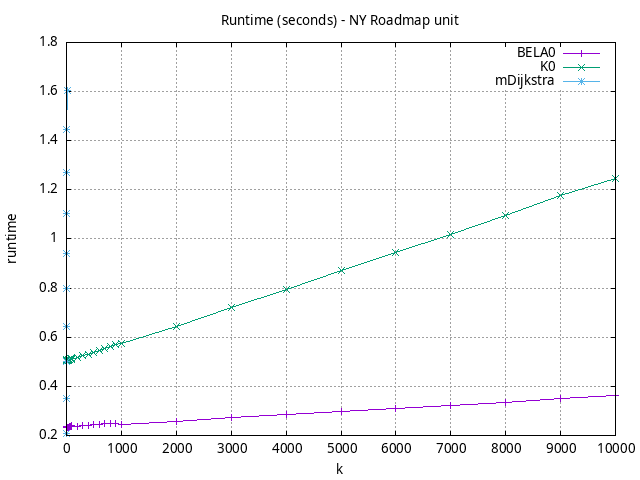}
    \end{center}
    \caption{}
    \label{fig:roadmap:dimacs:brute-force:runtime:j}
  \end{subfigure}
  \begin{subfigure}{0.3\textwidth}
    \begin{center}
        \includegraphics[width=\textwidth]{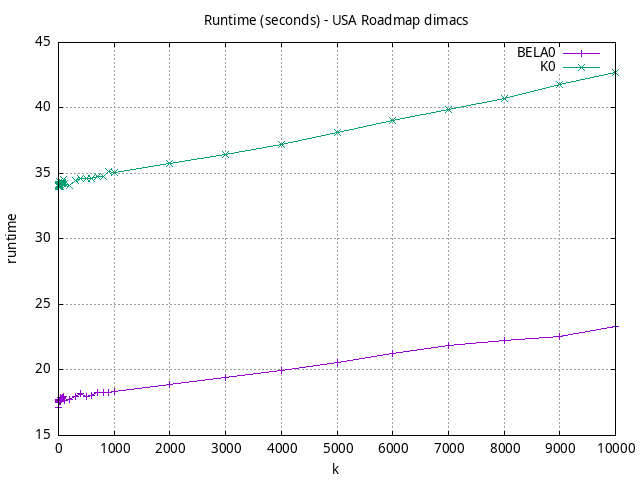}
    \end{center}
    \caption{}
    \label{fig:roadmap:dimacs:brute-force:runtime:k}
  \end{subfigure}
  \begin{subfigure}{0.3\textwidth}
    \begin{center}
        \includegraphics[width=\textwidth]{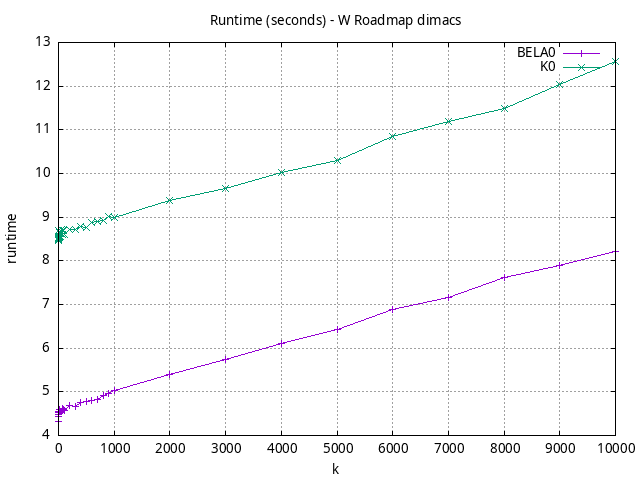}
    \end{center}
    \caption{}
    \label{fig:roadmap:dimacs:brute-force:runtime:l}
  \end{subfigure}
  \caption{Runtime (in seconds) in the roadmap (dimacs) domain with brute-force search algorithms}
  \label{fig:roadmap:dimacs:brute-force:runtime}
\end{figure*}

%% file: roadmap.dimacs.mem.brute-force.tex
\begin{figure*}
  \centering
  \begin{subfigure}{0.3\textwidth}
    \begin{center}
        \includegraphics[width=\textwidth]{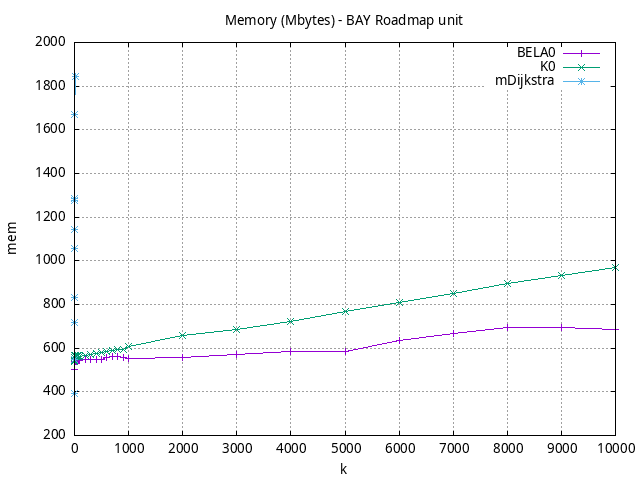}
    \end{center}
    \caption{}
    \label{fig:roadmap:dimacs:brute-force:mem:a}
  \end{subfigure}
  \begin{subfigure}{0.3\textwidth}
    \begin{center}
        \includegraphics[width=\textwidth]{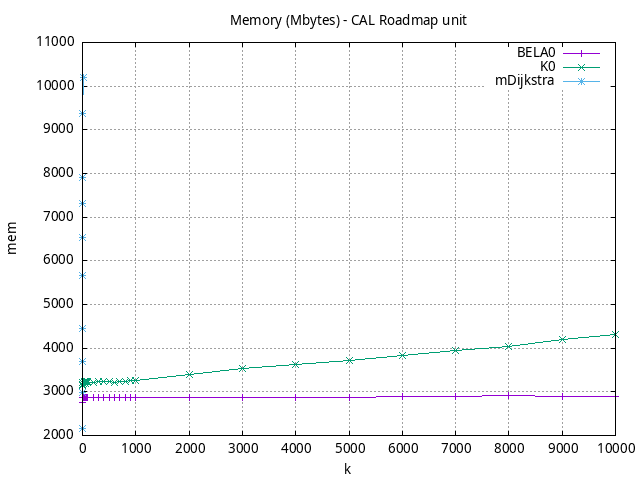}
    \end{center}
    \caption{}
    \label{fig:roadmap:dimacs:brute-force:mem:b}
  \end{subfigure}
  \begin{subfigure}{0.3\textwidth}
    \begin{center}
        \includegraphics[width=\textwidth]{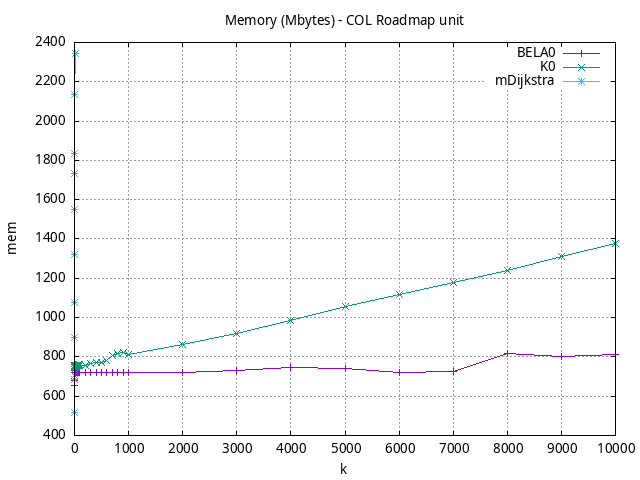}
    \end{center}
    \caption{}
    \label{fig:roadmap:dimacs:brute-force:mem:c}
  \end{subfigure}
  \begin{subfigure}{0.3\textwidth}
    \begin{center}
        \includegraphics[width=\textwidth]{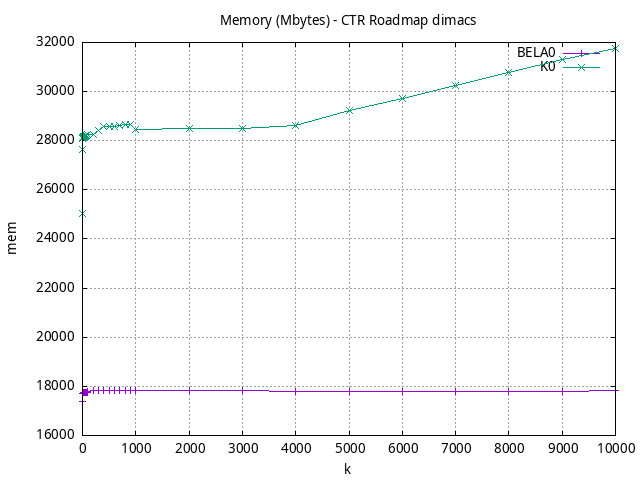}
    \end{center}
    \caption{}
    \label{fig:roadmap:dimacs:brute-force:mem:d}
  \end{subfigure}
  \begin{subfigure}{0.3\textwidth}
    \begin{center}
        \includegraphics[width=\textwidth]{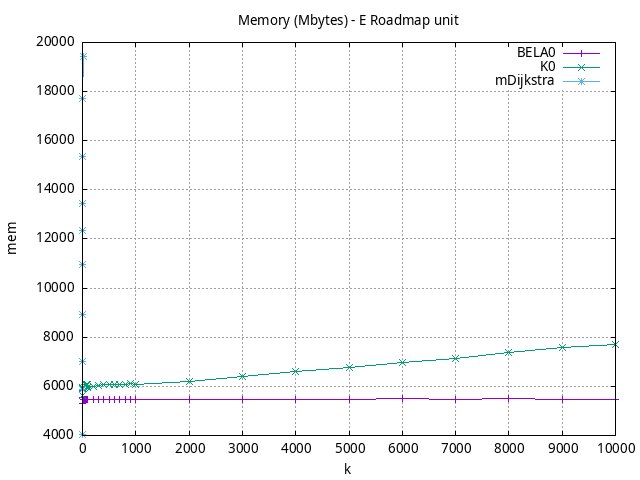}
    \end{center}
    \caption{}
    \label{fig:roadmap:dimacs:brute-force:mem:e}
  \end{subfigure}
  \begin{subfigure}{0.3\textwidth}
    \begin{center}
        \includegraphics[width=\textwidth]{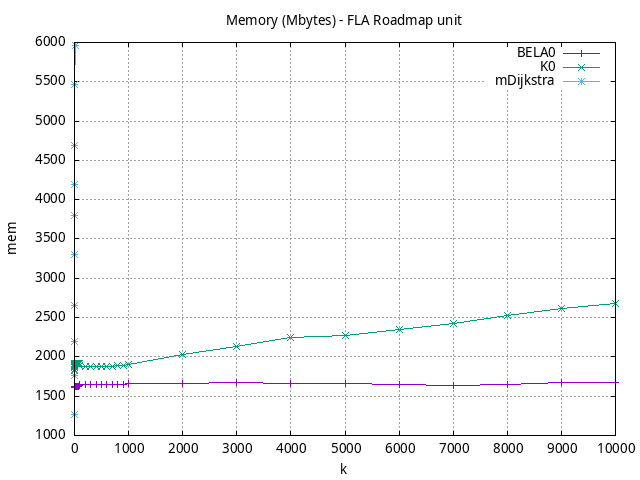}
    \end{center}
    \caption{}
    \label{fig:roadmap:dimacs:brute-force:mem:f}
  \end{subfigure}
  \begin{subfigure}{0.3\textwidth}
    \begin{center}
        \includegraphics[width=\textwidth]{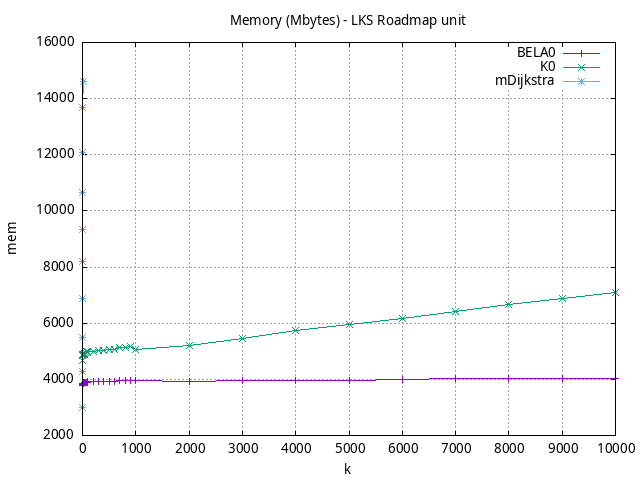}
    \end{center}
    \caption{}
    \label{fig:roadmap:dimacs:brute-force:mem:g}
  \end{subfigure}
  \begin{subfigure}{0.3\textwidth}
    \begin{center}
        \includegraphics[width=\textwidth]{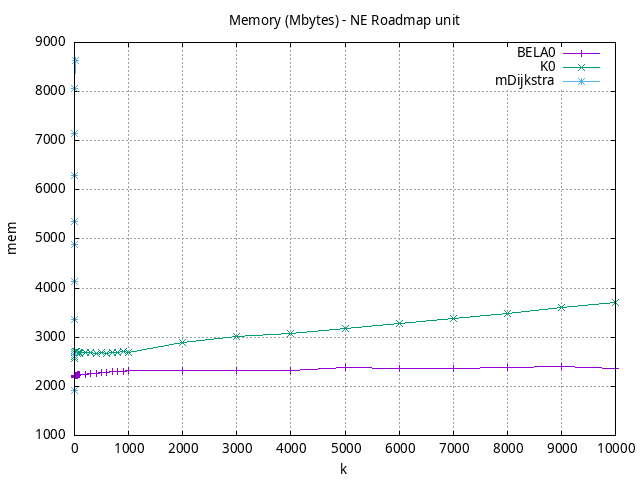}
    \end{center}
    \caption{}
    \label{fig:roadmap:dimacs:brute-force:mem:h}
  \end{subfigure}
  \begin{subfigure}{0.3\textwidth}
    \begin{center}
        \includegraphics[width=\textwidth]{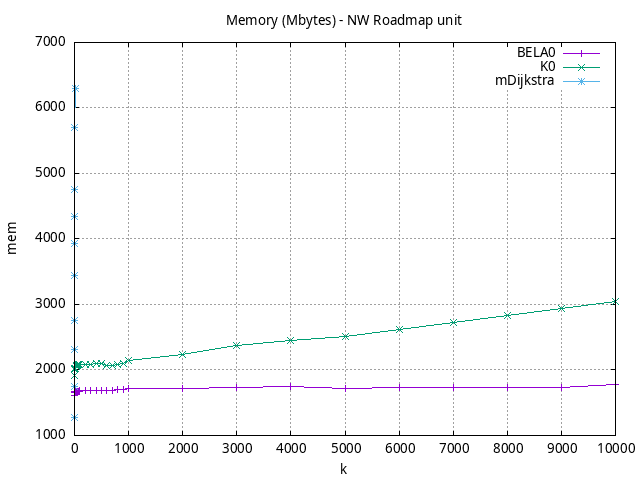}
    \end{center}
    \caption{}
    \label{fig:roadmap:dimacs:brute-force:mem:i}
  \end{subfigure}
  \begin{subfigure}{0.3\textwidth}
    \begin{center}
        \includegraphics[width=\textwidth]{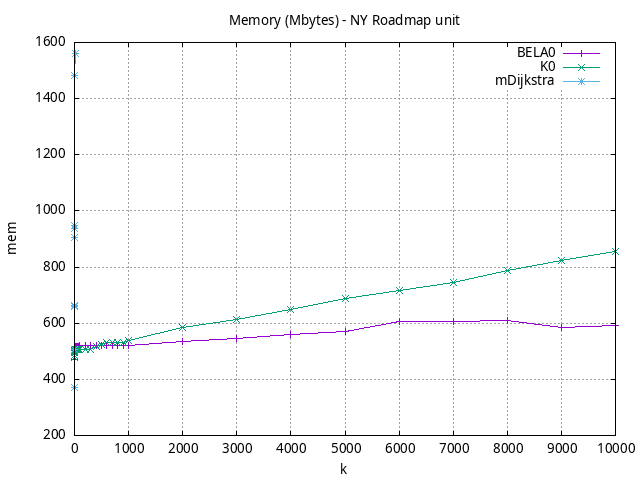}
    \end{center}
    \caption{}
    \label{fig:roadmap:dimacs:brute-force:mem:j}
  \end{subfigure}
  \begin{subfigure}{0.3\textwidth}
    \begin{center}
        \includegraphics[width=\textwidth]{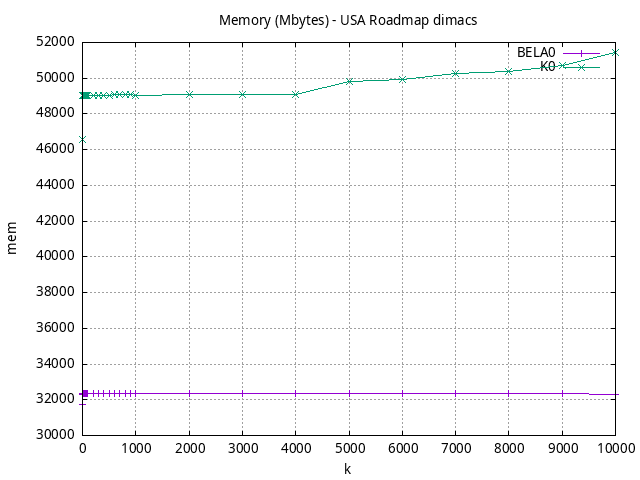}
    \end{center}
    \caption{}
    \label{fig:roadmap:dimacs:brute-force:mem:k}
  \end{subfigure}
  \begin{subfigure}{0.3\textwidth}
    \begin{center}
        \includegraphics[width=\textwidth]{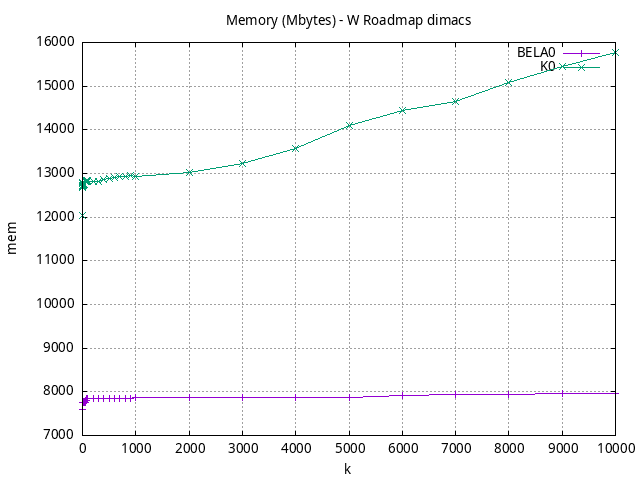}
    \end{center}
    \caption{}
    \label{fig:roadmap:dimacs:brute-force:mem:l}
  \end{subfigure}
  \caption{Memory usage (in Mbytes) in the roadmap (dimacs) domain with brute-force search algorithms}
  \label{fig:roadmap:dimacs:brute-force:mem}
\end{figure*}

%% file: roadmap.dimacs.expansions.brute-force.tex
\begin{figure*}
  \centering
  \begin{subfigure}{0.3\textwidth}
    \begin{center}
        \includegraphics[width=\textwidth]{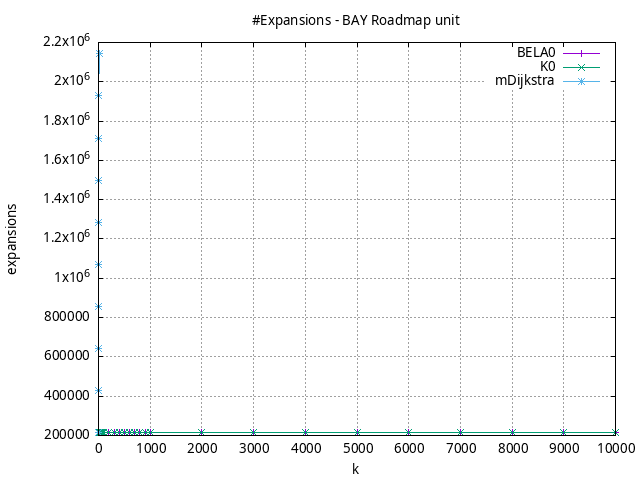}
    \end{center}
    \caption{}
    \label{fig:roadmap:dimacs:brute-force:expansions:a}
  \end{subfigure}
  \begin{subfigure}{0.3\textwidth}
    \begin{center}
        \includegraphics[width=\textwidth]{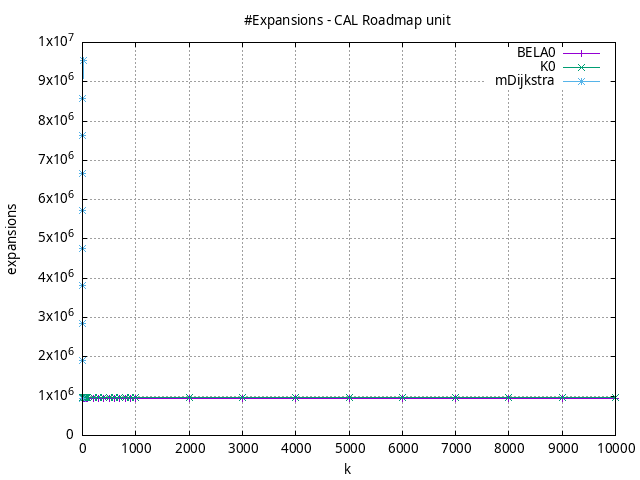}
    \end{center}
    \caption{}
    \label{fig:roadmap:dimacs:brute-force:expansions:b}
  \end{subfigure}
  \begin{subfigure}{0.3\textwidth}
    \begin{center}
        \includegraphics[width=\textwidth]{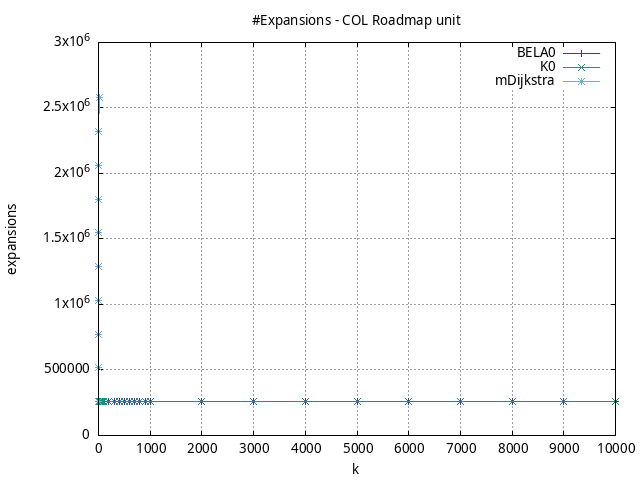}
    \end{center}
    \caption{}
    \label{fig:roadmap:dimacs:brute-force:expansions:c}
  \end{subfigure}
  \begin{subfigure}{0.3\textwidth}
    \begin{center}
        \includegraphics[width=\textwidth]{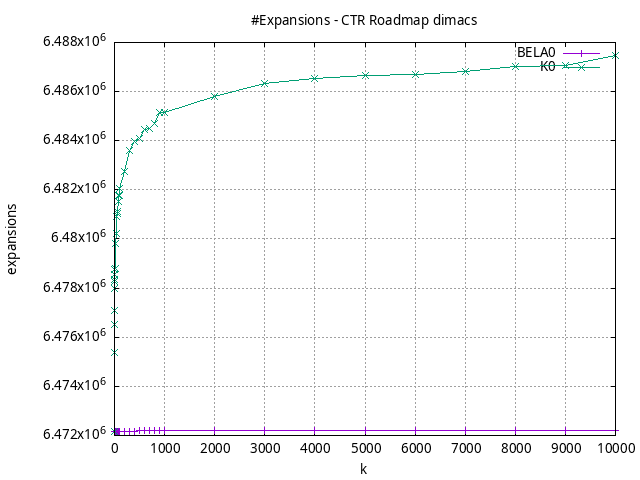}
    \end{center}
    \caption{}
    \label{fig:roadmap:dimacs:brute-force:expansions:d}
  \end{subfigure}
  \begin{subfigure}{0.3\textwidth}
    \begin{center}
        \includegraphics[width=\textwidth]{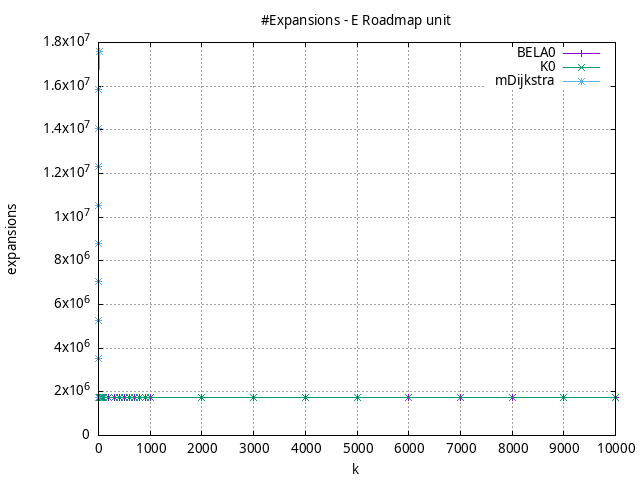}
    \end{center}
    \caption{}
    \label{fig:roadmap:dimacs:brute-force:expansions:e}
  \end{subfigure}
  \begin{subfigure}{0.3\textwidth}
    \begin{center}
        \includegraphics[width=\textwidth]{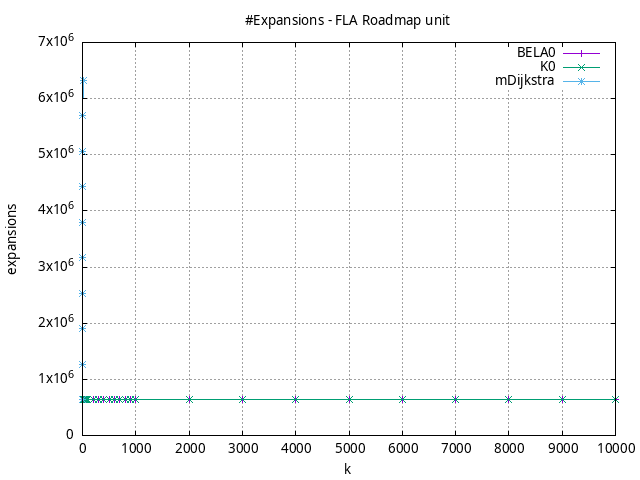}
    \end{center}
    \caption{}
    \label{fig:roadmap:dimacs:brute-force:expansions:f}
  \end{subfigure}
  \begin{subfigure}{0.3\textwidth}
    \begin{center}
        \includegraphics[width=\textwidth]{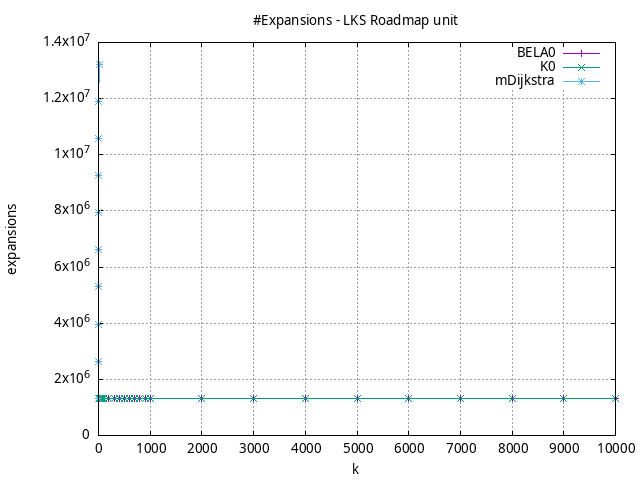}
    \end{center}
    \caption{}
    \label{fig:roadmap:dimacs:brute-force:expansions:g}
  \end{subfigure}
  \begin{subfigure}{0.3\textwidth}
    \begin{center}
        \includegraphics[width=\textwidth]{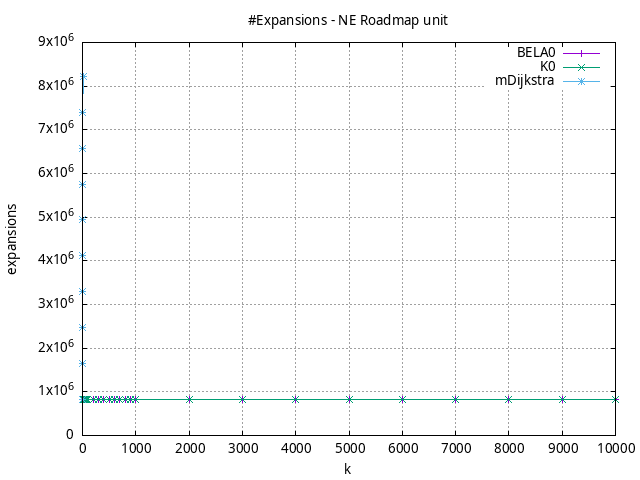}
    \end{center}
    \caption{}
    \label{fig:roadmap:dimacs:brute-force:expansions:h}
  \end{subfigure}
  \begin{subfigure}{0.3\textwidth}
    \begin{center}
        \includegraphics[width=\textwidth]{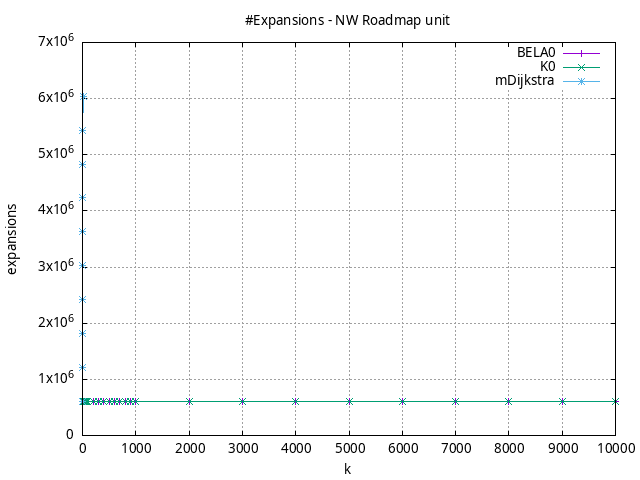}
    \end{center}
    \caption{}
    \label{fig:roadmap:dimacs:brute-force:expansions:i}
  \end{subfigure}
  \begin{subfigure}{0.3\textwidth}
    \begin{center}
        \includegraphics[width=\textwidth]{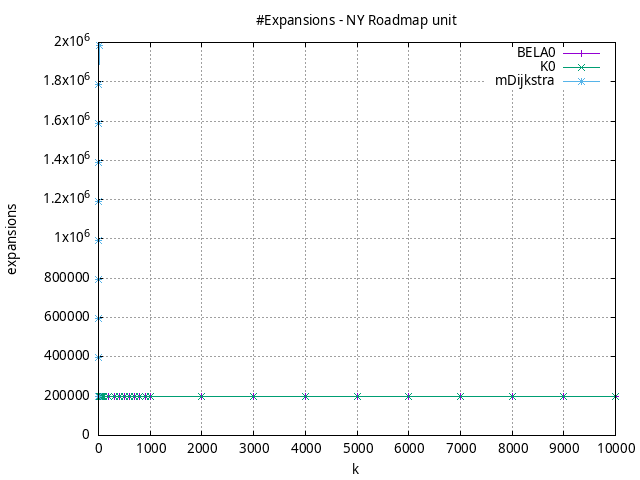}
    \end{center}
    \caption{}
    \label{fig:roadmap:dimacs:brute-force:expansions:j}
  \end{subfigure}
  \begin{subfigure}{0.3\textwidth}
    \begin{center}
        \includegraphics[width=\textwidth]{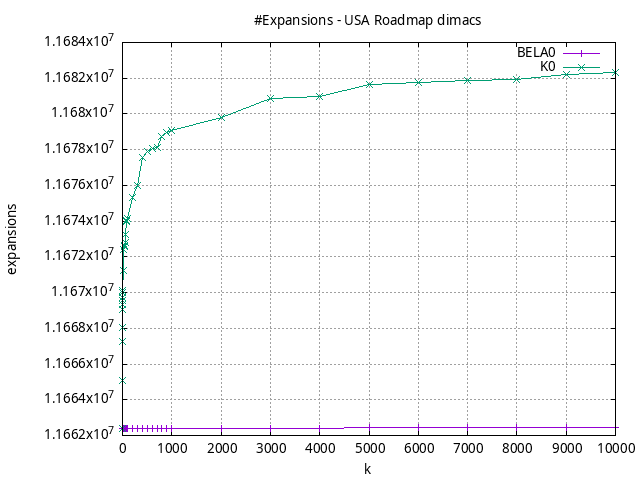}
    \end{center}
    \caption{}
    \label{fig:roadmap:dimacs:brute-force:expansions:k}
  \end{subfigure}
  \begin{subfigure}{0.3\textwidth}
    \begin{center}
        \includegraphics[width=\textwidth]{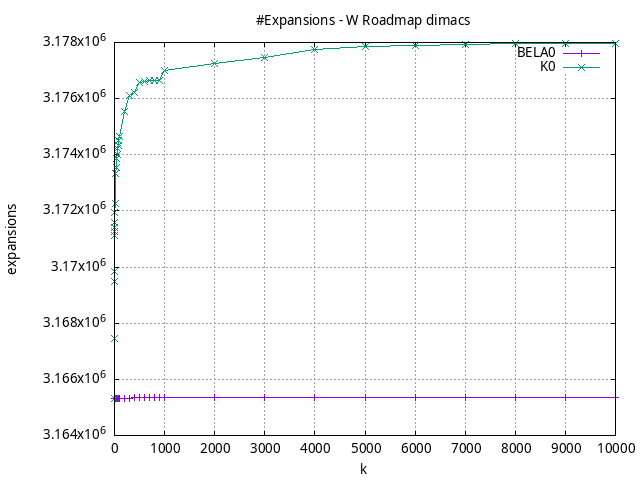}
    \end{center}
    \caption{}
    \label{fig:roadmap:dimacs:brute-force:expansions:l}
  \end{subfigure}
  \caption{Number of expansions in the roadmap (dimacs) domain with brute-force search algorithms}
  \label{fig:roadmap:dimacs:brute-force:expansions}
\end{figure*}

%% file: roadmap.dimacs.runtime.heuristic.tex
\begin{figure*}
  \centering
  \begin{subfigure}{0.3\textwidth}
    \begin{center}
        \includegraphics[width=\textwidth]{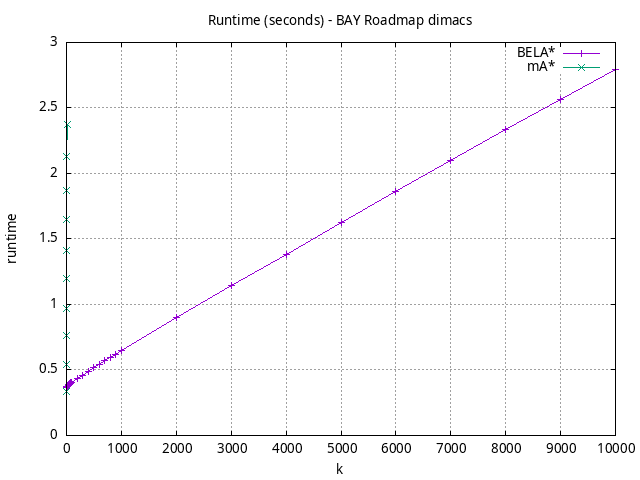}
    \end{center}
    \caption{}
    \label{fig:roadmap:dimacs:heuristic:runtime:a}
  \end{subfigure}
  \begin{subfigure}{0.3\textwidth}
    \begin{center}
        \includegraphics[width=\textwidth]{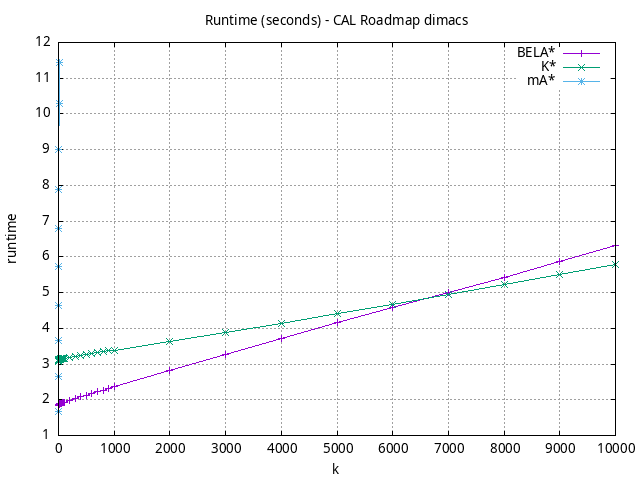}
    \end{center}
    \caption{}
    \label{fig:roadmap:dimacs:heuristic:runtime:b}
  \end{subfigure}
  \begin{subfigure}{0.3\textwidth}
    \begin{center}
        \includegraphics[width=\textwidth]{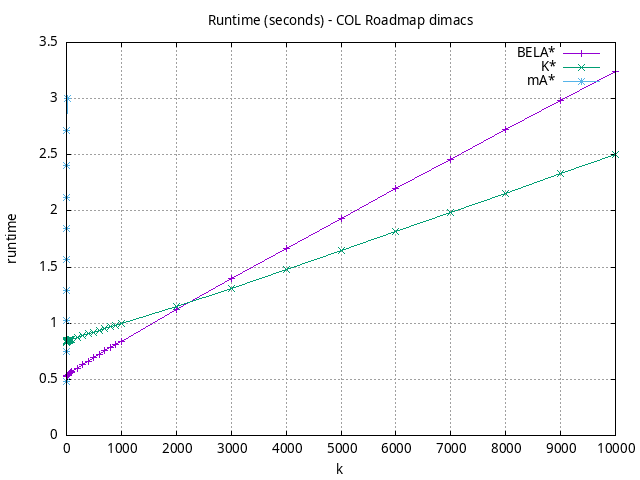}
    \end{center}
    \caption{}
    \label{fig:roadmap:dimacs:heuristic:runtime:c}
  \end{subfigure}
  \begin{subfigure}{0.3\textwidth}
    \begin{center}
        \includegraphics[width=\textwidth]{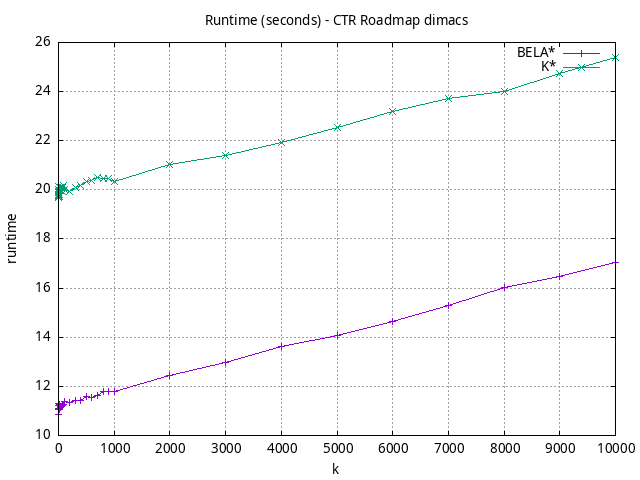}
    \end{center}
    \caption{}
    \label{fig:roadmap:dimacs:heuristic:runtime:d}
  \end{subfigure}
  \begin{subfigure}{0.3\textwidth}
    \begin{center}
        \includegraphics[width=\textwidth]{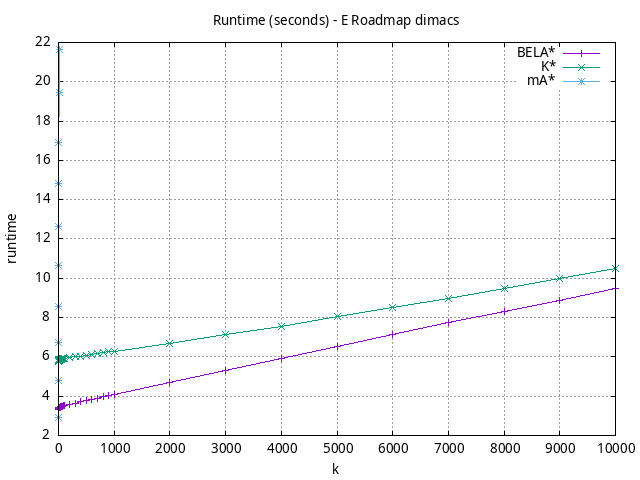}
    \end{center}
    \caption{}
    \label{fig:roadmap:dimacs:heuristic:runtime:e}
  \end{subfigure}
  \begin{subfigure}{0.3\textwidth}
    \begin{center}
        \includegraphics[width=\textwidth]{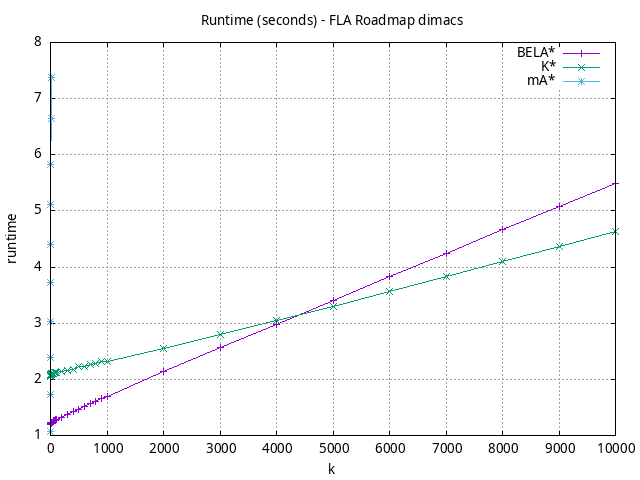}
    \end{center}
    \caption{}
    \label{fig:roadmap:dimacs:heuristic:runtime:f}
  \end{subfigure}
  \begin{subfigure}{0.3\textwidth}
    \begin{center}
        \includegraphics[width=\textwidth]{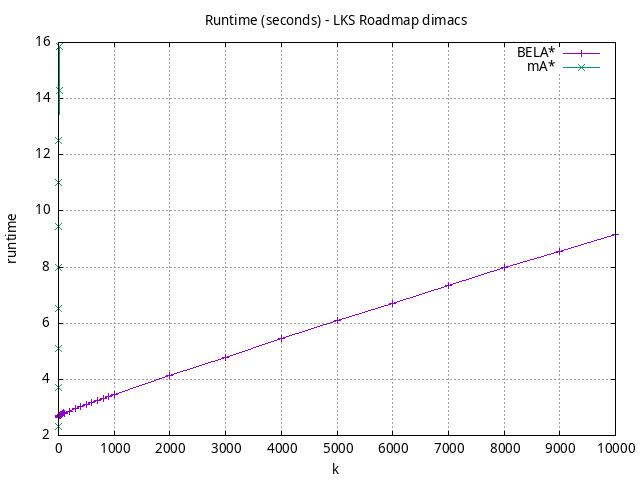}
    \end{center}
    \caption{}
    \label{fig:roadmap:dimacs:heuristic:runtime:g}
  \end{subfigure}
  \begin{subfigure}{0.3\textwidth}
    \begin{center}
        \includegraphics[width=\textwidth]{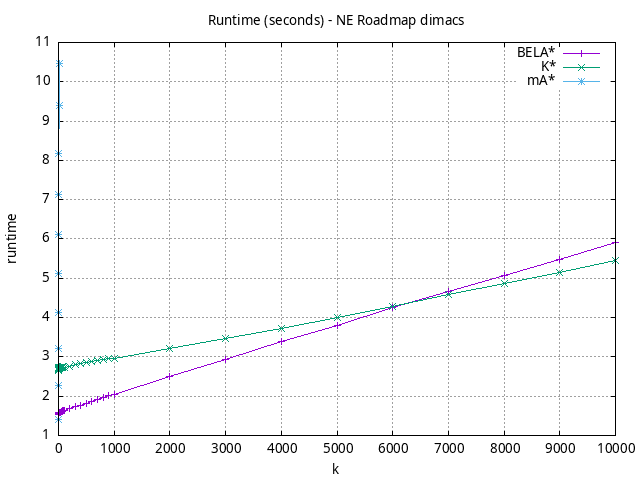}
    \end{center}
    \caption{}
    \label{fig:roadmap:dimacs:heuristic:runtime:h}
  \end{subfigure}
  \begin{subfigure}{0.3\textwidth}
    \begin{center}
        \includegraphics[width=\textwidth]{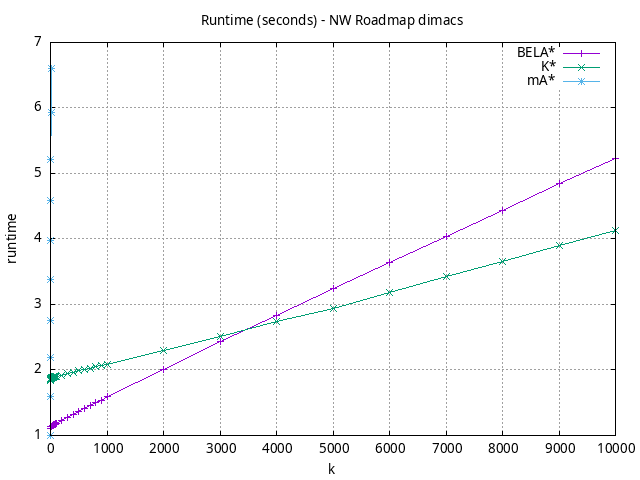}
    \end{center}
    \caption{}
    \label{fig:roadmap:dimacs:heuristic:runtime:i}
  \end{subfigure}
  \begin{subfigure}{0.3\textwidth}
    \begin{center}
        \includegraphics[width=\textwidth]{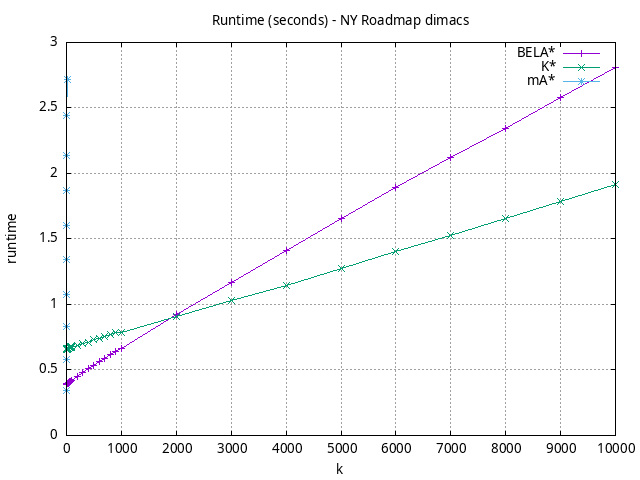}
    \end{center}
    \caption{}
    \label{fig:roadmap:dimacs:heuristic:runtime:j}
  \end{subfigure}
  \begin{subfigure}{0.3\textwidth}
    \begin{center}
        \includegraphics[width=\textwidth]{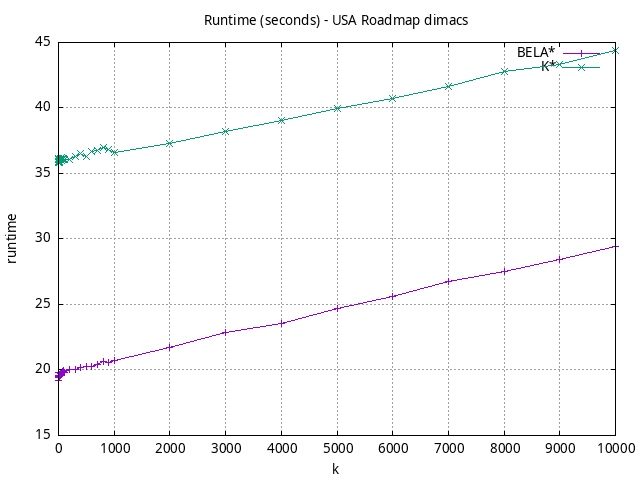}
    \end{center}
    \caption{}
    \label{fig:roadmap:dimacs:heuristic:runtime:k}
  \end{subfigure}
  \begin{subfigure}{0.3\textwidth}
    \begin{center}
        \includegraphics[width=\textwidth]{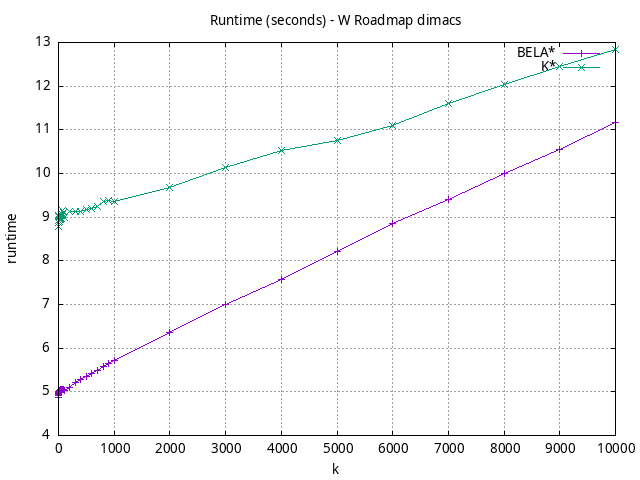}
    \end{center}
    \caption{}
    \label{fig:roadmap:dimacs:heuristic:runtime:l}
  \end{subfigure}
  \caption{Runtime (in seconds) in the roadmap (dimacs) domain with heuristic search algorithms}
  \label{fig:roadmap:dimacs:heuristic:runtime}
\end{figure*}

%% file: roadmap.dimacs.mem.heuristic.tex
\begin{figure*}
  \centering
  \begin{subfigure}{0.3\textwidth}
    \begin{center}
        \includegraphics[width=\textwidth]{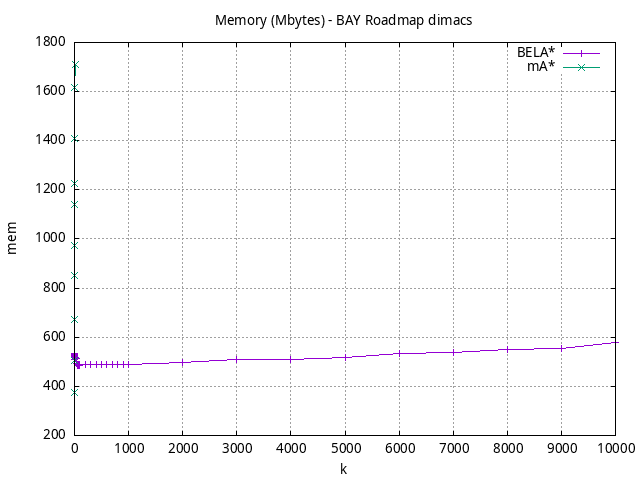}
    \end{center}
    \caption{}
    \label{fig:roadmap:dimacs:heuristic:mem:a}
  \end{subfigure}
  \begin{subfigure}{0.3\textwidth}
    \begin{center}
        \includegraphics[width=\textwidth]{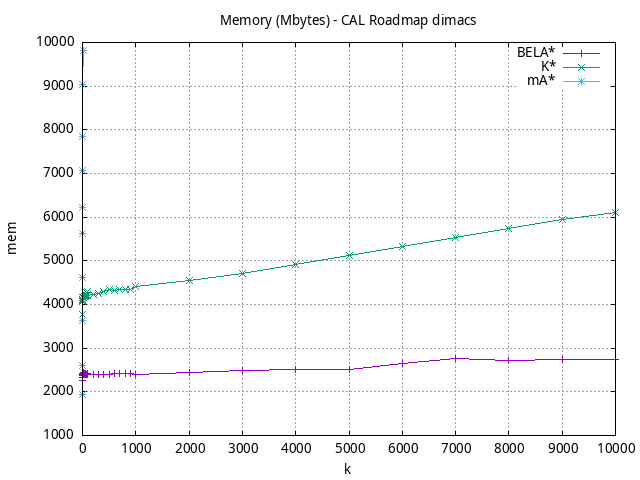}
    \end{center}
    \caption{}
    \label{fig:roadmap:dimacs:heuristic:mem:b}
  \end{subfigure}
  \begin{subfigure}{0.3\textwidth}
    \begin{center}
        \includegraphics[width=\textwidth]{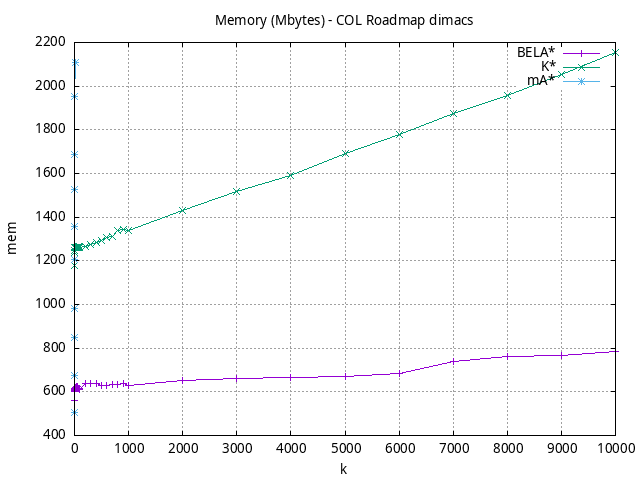}
    \end{center}
    \caption{}
    \label{fig:roadmap:dimacs:heuristic:mem:c}
  \end{subfigure}
  \begin{subfigure}{0.3\textwidth}
    \begin{center}
        \includegraphics[width=\textwidth]{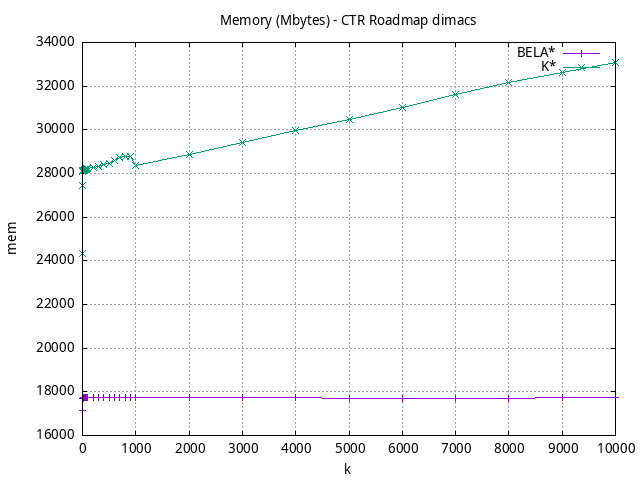}
    \end{center}
    \caption{}
    \label{fig:roadmap:dimacs:heuristic:mem:d}
  \end{subfigure}
  \begin{subfigure}{0.3\textwidth}
    \begin{center}
        \includegraphics[width=\textwidth]{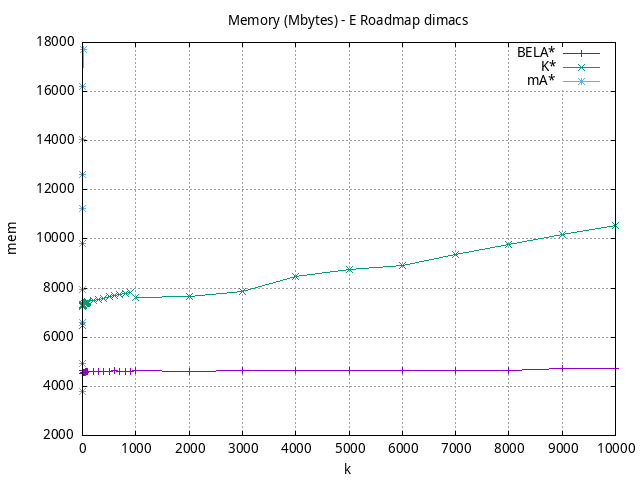}
    \end{center}
    \caption{}
    \label{fig:roadmap:dimacs:heuristic:mem:e}
  \end{subfigure}
  \begin{subfigure}{0.3\textwidth}
    \begin{center}
        \includegraphics[width=\textwidth]{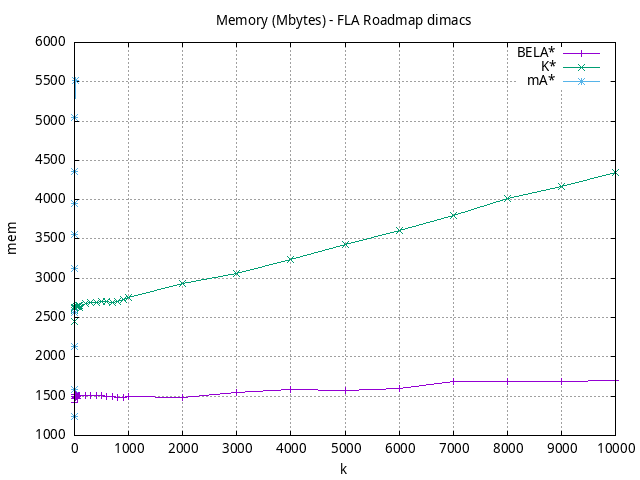}
    \end{center}
    \caption{}
    \label{fig:roadmap:dimacs:heuristic:mem:f}
  \end{subfigure}
  \begin{subfigure}{0.3\textwidth}
    \begin{center}
        \includegraphics[width=\textwidth]{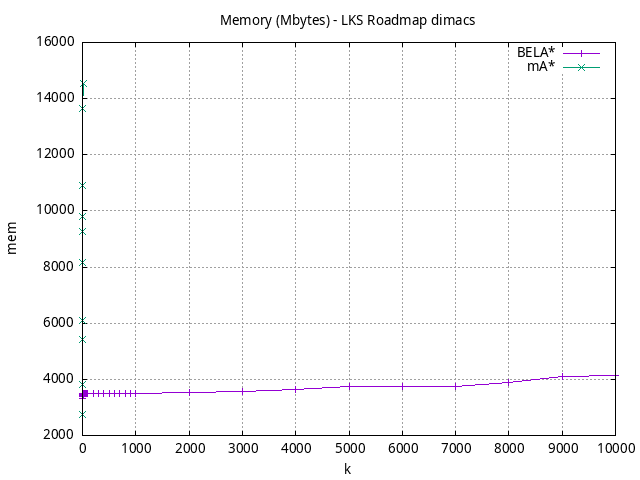}
    \end{center}
    \caption{}
    \label{fig:roadmap:dimacs:heuristic:mem:g}
  \end{subfigure}
  \begin{subfigure}{0.3\textwidth}
    \begin{center}
        \includegraphics[width=\textwidth]{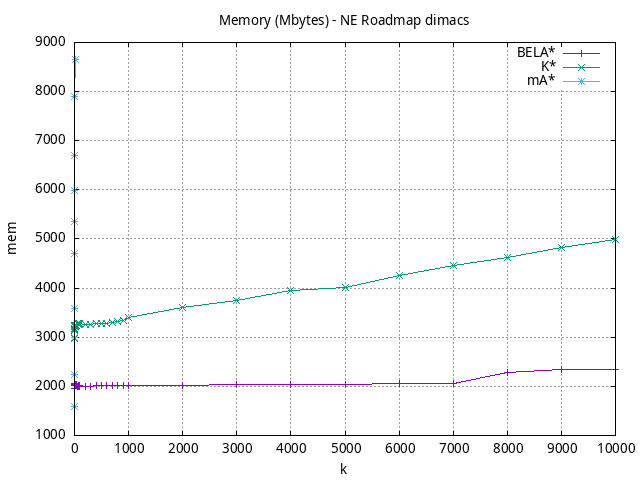}
    \end{center}
    \caption{}
    \label{fig:roadmap:dimacs:heuristic:mem:h}
  \end{subfigure}
  \begin{subfigure}{0.3\textwidth}
    \begin{center}
        \includegraphics[width=\textwidth]{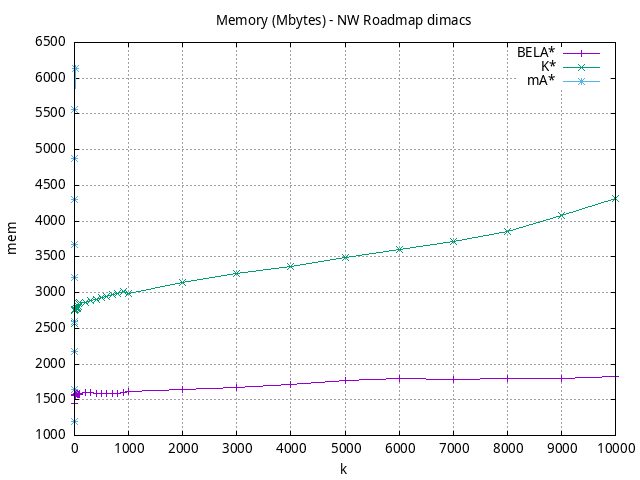}
    \end{center}
    \caption{}
    \label{fig:roadmap:dimacs:heuristic:mem:i}
  \end{subfigure}
  \begin{subfigure}{0.3\textwidth}
    \begin{center}
        \includegraphics[width=\textwidth]{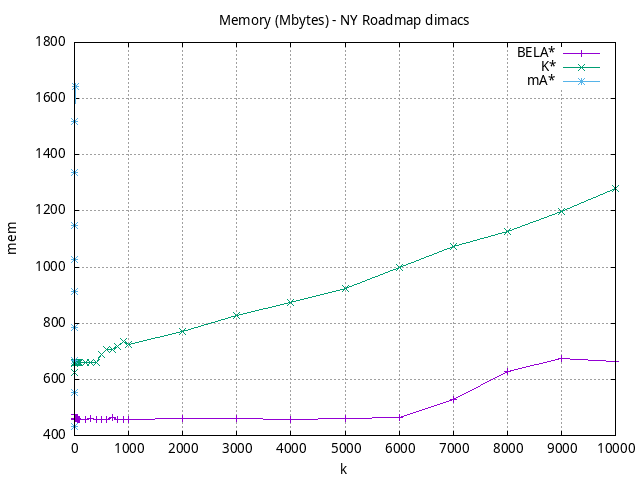}
    \end{center}
    \caption{}
    \label{fig:roadmap:dimacs:heuristic:mem:j}
  \end{subfigure}
  \begin{subfigure}{0.3\textwidth}
    \begin{center}
        \includegraphics[width=\textwidth]{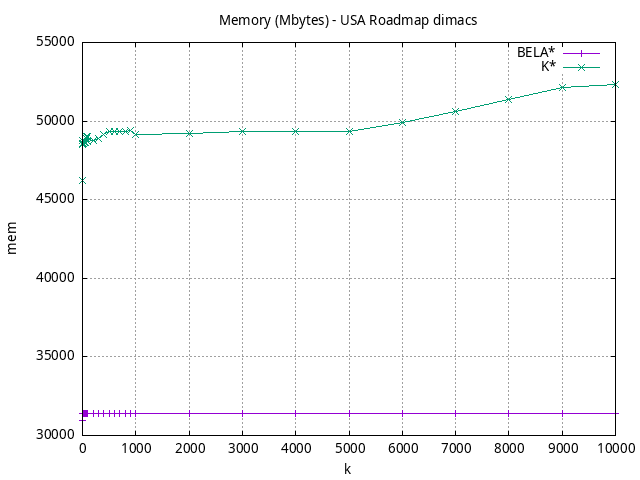}
    \end{center}
    \caption{}
    \label{fig:roadmap:dimacs:heuristic:mem:k}
  \end{subfigure}
  \begin{subfigure}{0.3\textwidth}
    \begin{center}
        \includegraphics[width=\textwidth]{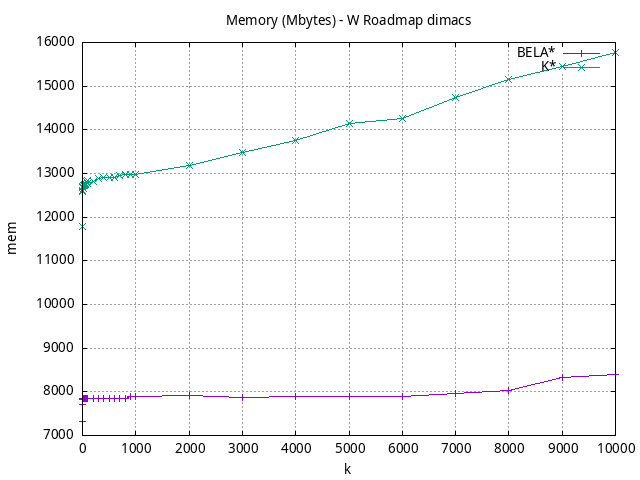}
    \end{center}
    \caption{}
    \label{fig:roadmap:dimacs:heuristic:mem:l}
  \end{subfigure}
  \caption{Memory usage (in Mbytes) in the roadmap (dimacs) domain with heuristic search algorithms}
  \label{fig:roadmap:dimacs:heuristic:mem}
\end{figure*}

%% file: roadmap.dimacs.expansions.heuristic.tex
\begin{figure*}
  \centering
  \begin{subfigure}{0.3\textwidth}
    \begin{center}
        \includegraphics[width=\textwidth]{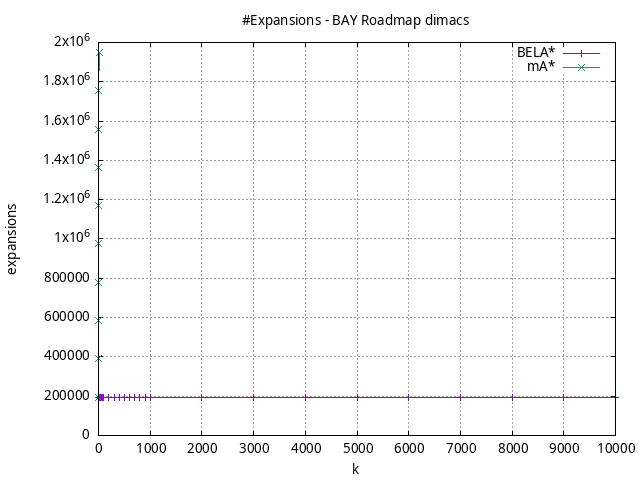}
    \end{center}
    \caption{}
    \label{fig:roadmap:dimacs:heuristic:expansions:a}
  \end{subfigure}
  \begin{subfigure}{0.3\textwidth}
    \begin{center}
        \includegraphics[width=\textwidth]{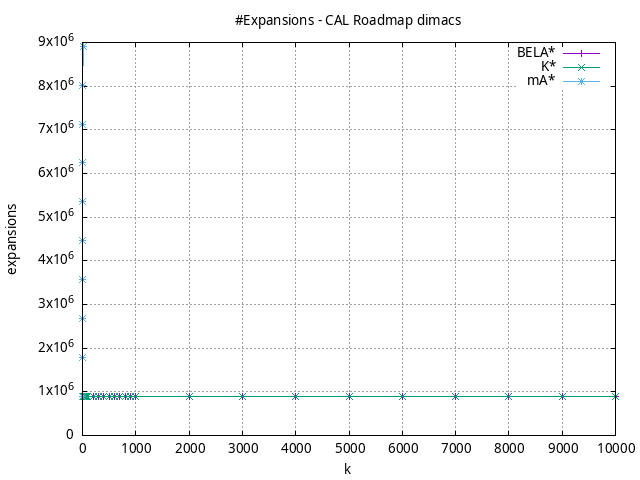}
    \end{center}
    \caption{}
    \label{fig:roadmap:dimacs:heuristic:expansions:b}
  \end{subfigure}
  \begin{subfigure}{0.3\textwidth}
    \begin{center}
        \includegraphics[width=\textwidth]{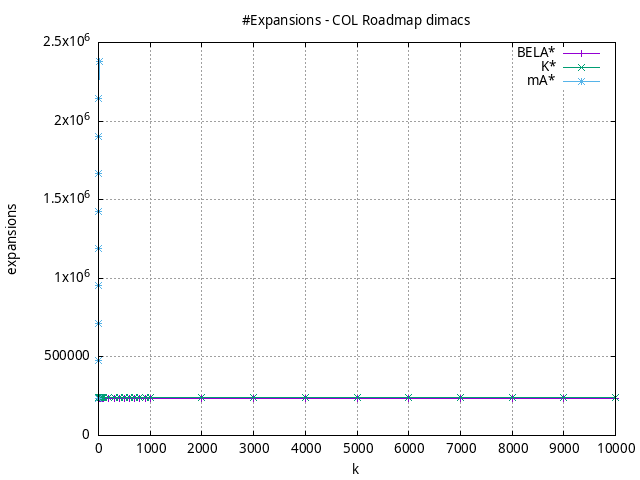}
    \end{center}
    \caption{}
    \label{fig:roadmap:dimacs:heuristic:expansions:c}
  \end{subfigure}
  \begin{subfigure}{0.3\textwidth}
    \begin{center}
        \includegraphics[width=\textwidth]{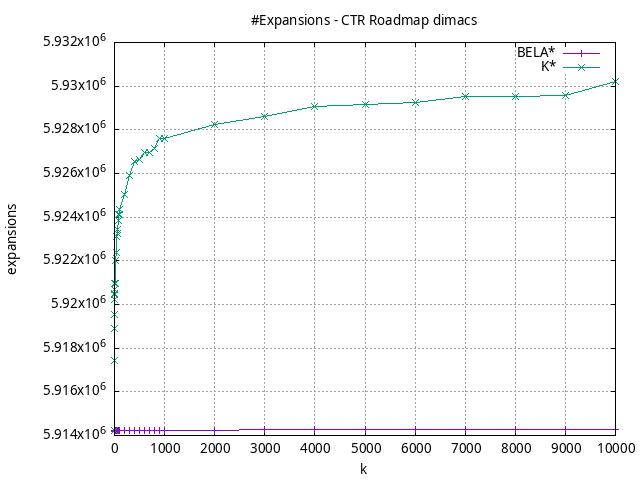}
    \end{center}
    \caption{}
    \label{fig:roadmap:dimacs:heuristic:expansions:d}
  \end{subfigure}
  \begin{subfigure}{0.3\textwidth}
    \begin{center}
        \includegraphics[width=\textwidth]{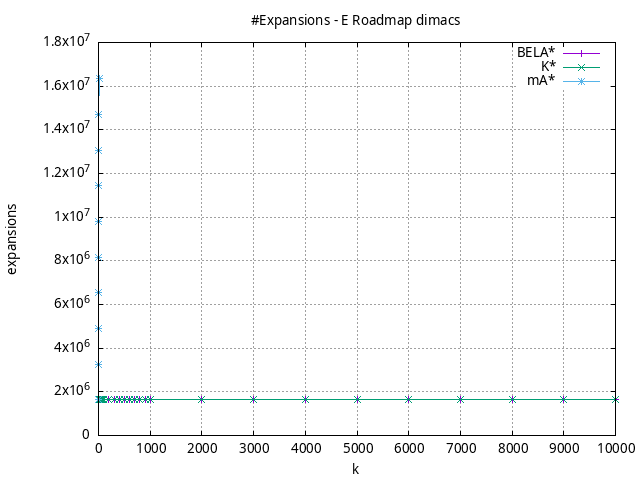}
    \end{center}
    \caption{}
    \label{fig:roadmap:dimacs:heuristic:expansions:e}
  \end{subfigure}
  \begin{subfigure}{0.3\textwidth}
    \begin{center}
        \includegraphics[width=\textwidth]{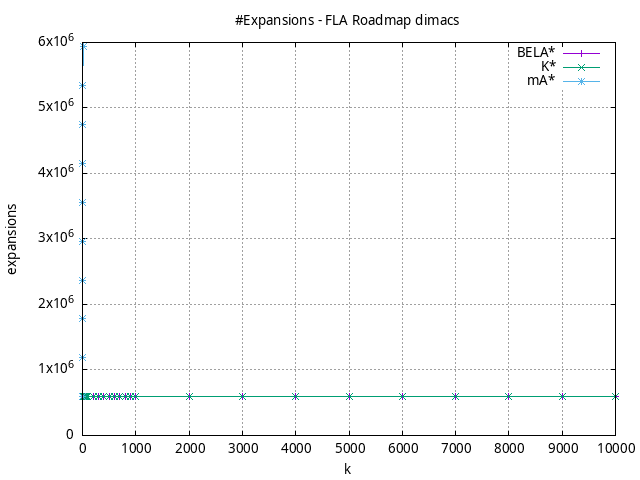}
    \end{center}
    \caption{}
    \label{fig:roadmap:dimacs:heuristic:expansions:f}
  \end{subfigure}
  \begin{subfigure}{0.3\textwidth}
    \begin{center}
        \includegraphics[width=\textwidth]{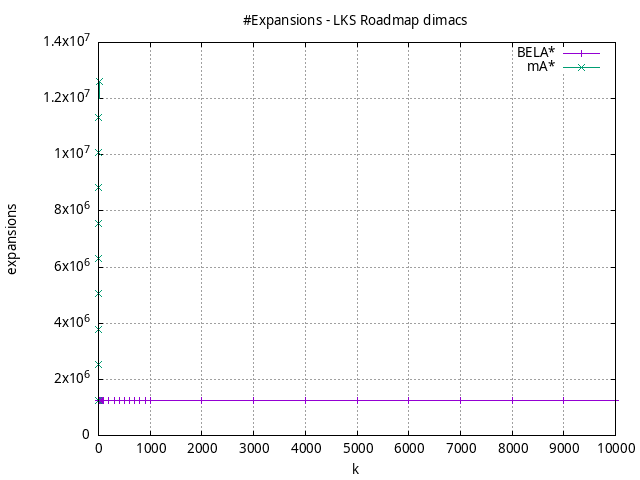}
    \end{center}
    \caption{}
    \label{fig:roadmap:dimacs:heuristic:expansions:g}
  \end{subfigure}
  \begin{subfigure}{0.3\textwidth}
    \begin{center}
        \includegraphics[width=\textwidth]{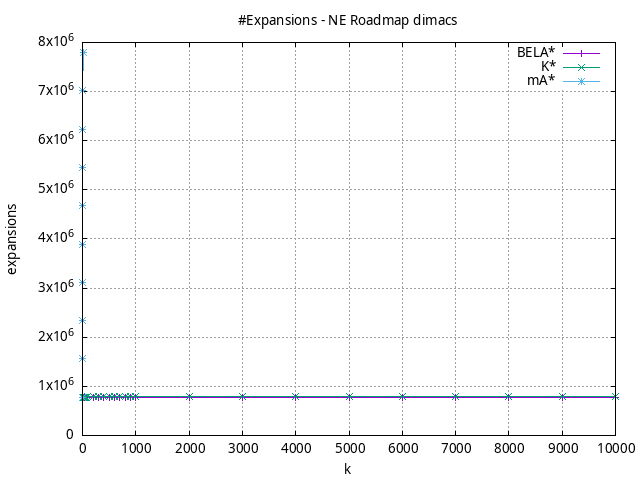}
    \end{center}
    \caption{}
    \label{fig:roadmap:dimacs:heuristic:expansions:h}
  \end{subfigure}
  \begin{subfigure}{0.3\textwidth}
    \begin{center}
        \includegraphics[width=\textwidth]{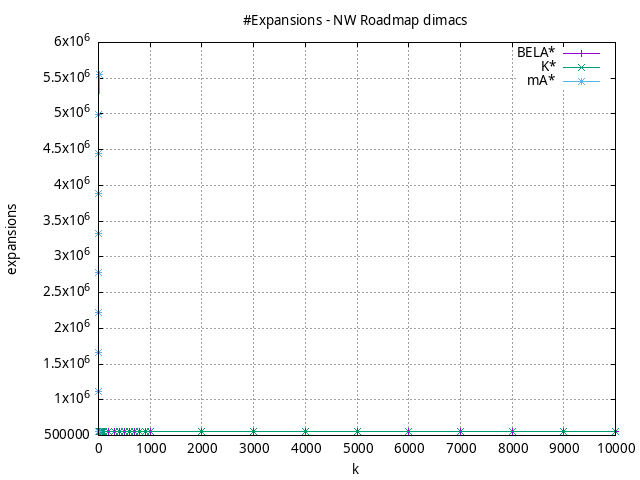}
    \end{center}
    \caption{}
    \label{fig:roadmap:dimacs:heuristic:expansions:i}
  \end{subfigure}
  \begin{subfigure}{0.3\textwidth}
    \begin{center}
        \includegraphics[width=\textwidth]{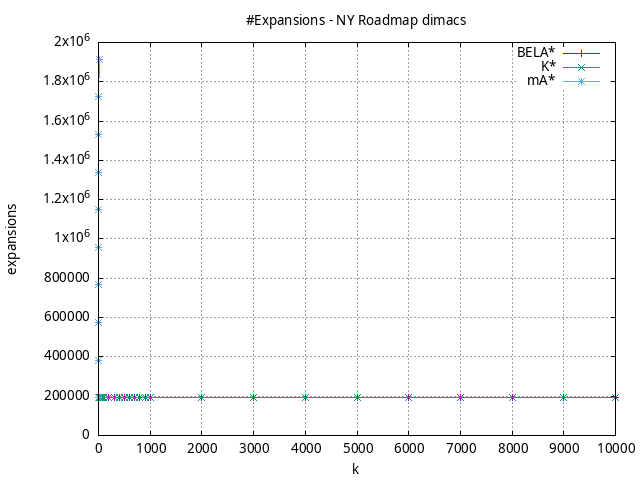}
    \end{center}
    \caption{}
    \label{fig:roadmap:dimacs:heuristic:expansions:j}
  \end{subfigure}
  \begin{subfigure}{0.3\textwidth}
    \begin{center}
        \includegraphics[width=\textwidth]{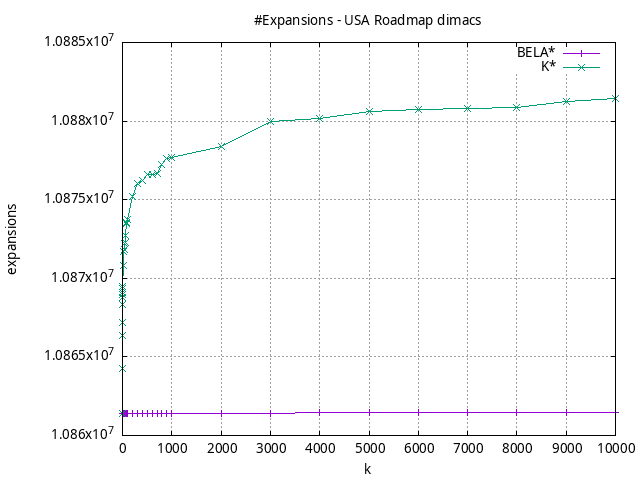}
    \end{center}
    \caption{}
    \label{fig:roadmap:dimacs:heuristic:expansions:k}
  \end{subfigure}
  \begin{subfigure}{0.3\textwidth}
    \begin{center}
        \includegraphics[width=\textwidth]{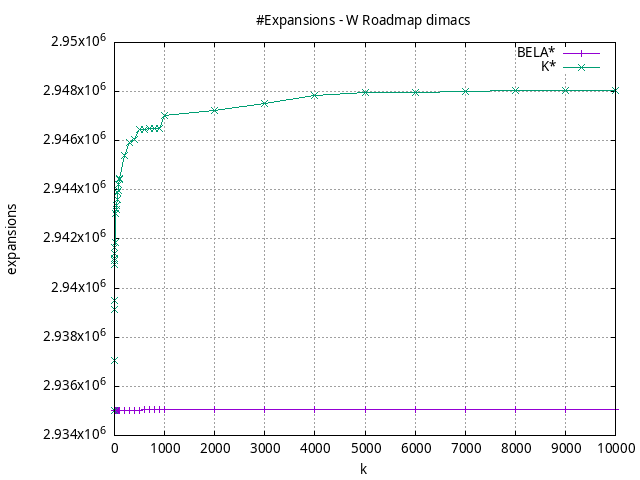}
    \end{center}
    \caption{}
    \label{fig:roadmap:dimacs:heuristic:expansions:l}
  \end{subfigure}
  \caption{Number of expansions in the roadmap (dimacs) domain with heuristic search algorithms}
  \label{fig:roadmap:dimacs:heuristic:expansions}
\end{figure*}

%% file: roadmap.dimacs.runtime.mixed.tex
\begin{figure*}
  \centering
  \begin{subfigure}{0.3\textwidth}
    \begin{center}
        \includegraphics[width=\textwidth]{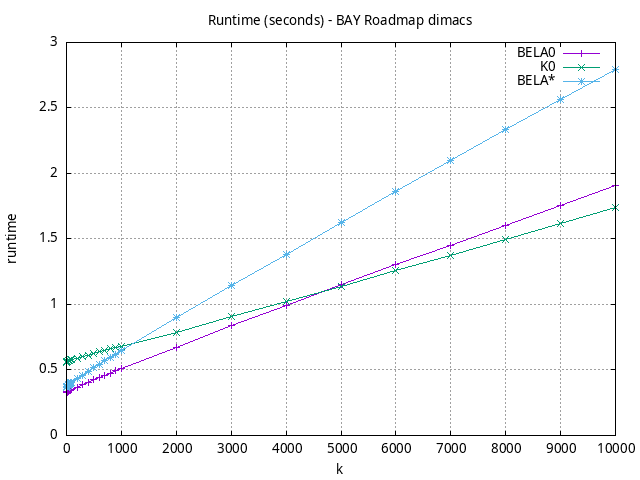}
    \end{center}
    \caption{}
    \label{fig:roadmap:dimacs:mixed:runtime:a}
  \end{subfigure}
  \begin{subfigure}{0.3\textwidth}
    \begin{center}
        \includegraphics[width=\textwidth]{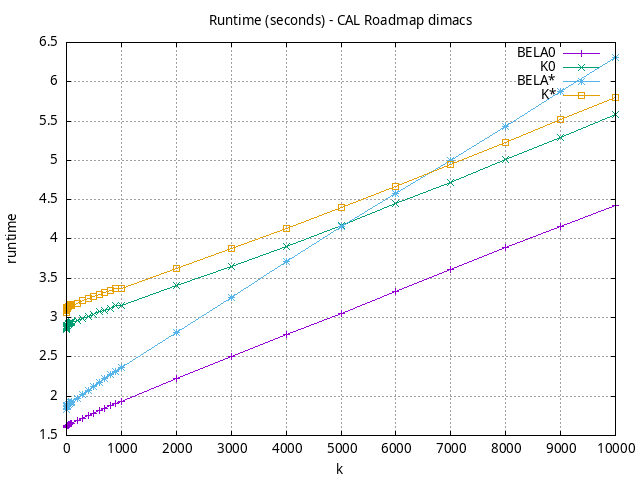}
    \end{center}
    \caption{}
    \label{fig:roadmap:dimacs:mixed:runtime:b}
  \end{subfigure}
  \begin{subfigure}{0.3\textwidth}
    \begin{center}
        \includegraphics[width=\textwidth]{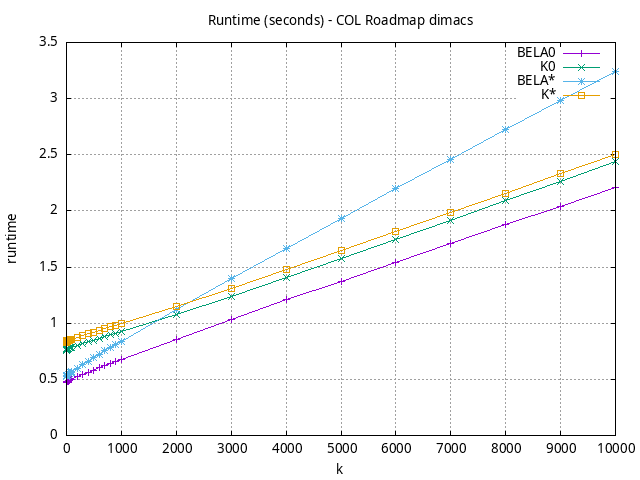}
    \end{center}
    \caption{}
    \label{fig:roadmap:dimacs:mixed:runtime:c}
  \end{subfigure}
  \begin{subfigure}{0.3\textwidth}
    \begin{center}
        \includegraphics[width=\textwidth]{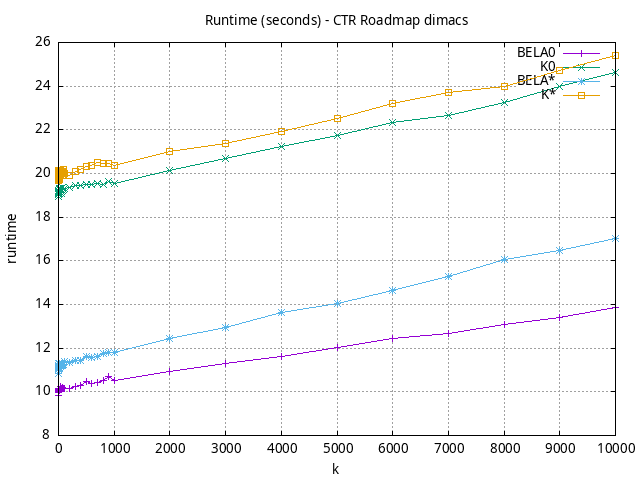}
    \end{center}
    \caption{}
    \label{fig:roadmap:dimacs:mixed:runtime:d}
  \end{subfigure}
  \begin{subfigure}{0.3\textwidth}
    \begin{center}
        \includegraphics[width=\textwidth]{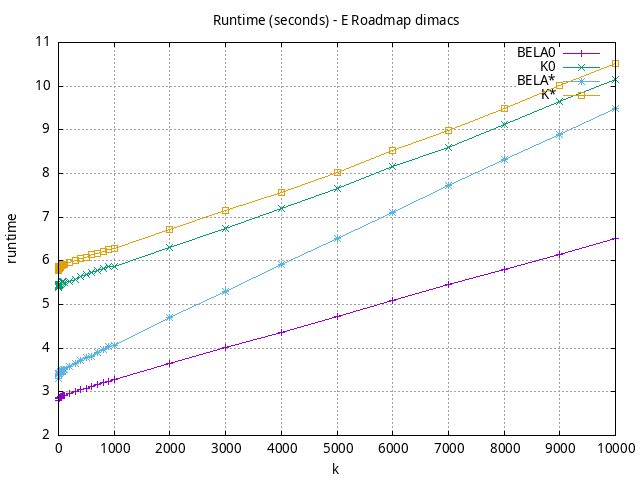}
    \end{center}
    \caption{}
    \label{fig:roadmap:dimacs:mixed:runtime:e}
  \end{subfigure}
  \begin{subfigure}{0.3\textwidth}
    \begin{center}
        \includegraphics[width=\textwidth]{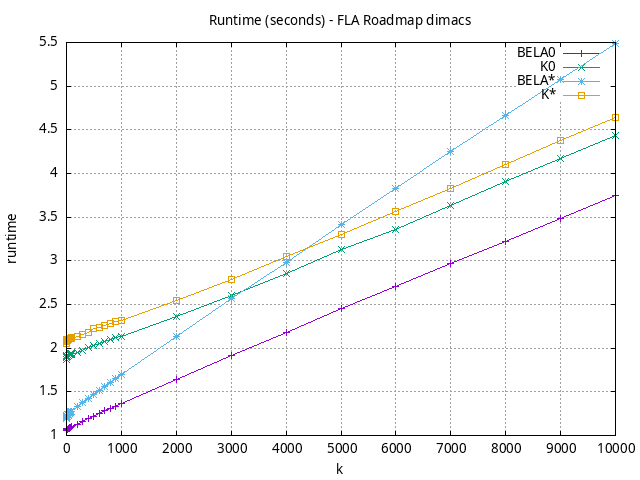}
    \end{center}
    \caption{}
    \label{fig:roadmap:dimacs:mixed:runtime:f}
  \end{subfigure}
  \begin{subfigure}{0.3\textwidth}
    \begin{center}
        \includegraphics[width=\textwidth]{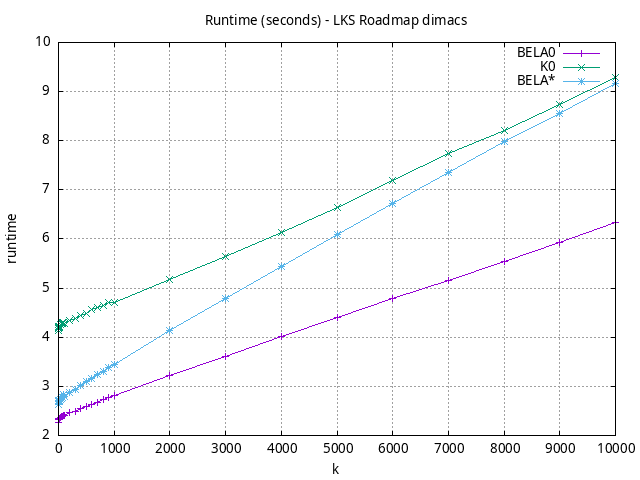}
    \end{center}
    \caption{}
    \label{fig:roadmap:dimacs:mixed:runtime:g}
  \end{subfigure}
  \begin{subfigure}{0.3\textwidth}
    \begin{center}
        \includegraphics[width=\textwidth]{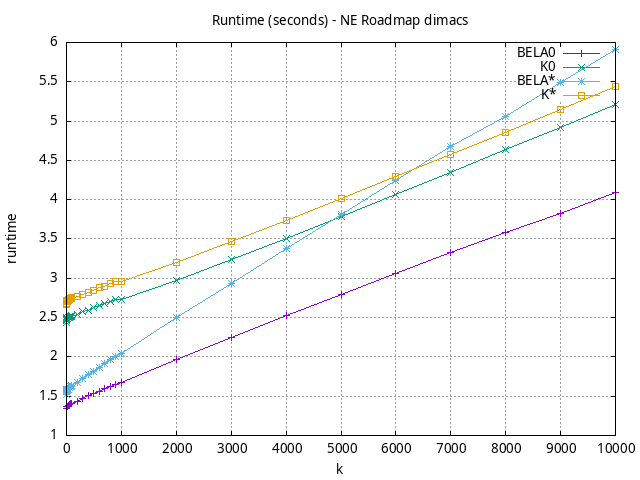}
    \end{center}
    \caption{}
    \label{fig:roadmap:dimacs:mixed:runtime:h}
  \end{subfigure}
  \begin{subfigure}{0.3\textwidth}
    \begin{center}
        \includegraphics[width=\textwidth]{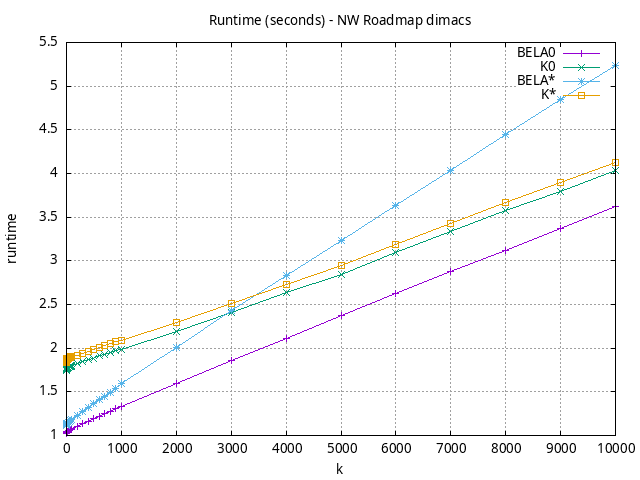}
    \end{center}
    \caption{}
    \label{fig:roadmap:dimacs:mixed:runtime:i}
  \end{subfigure}
  \begin{subfigure}{0.3\textwidth}
    \begin{center}
        \includegraphics[width=\textwidth]{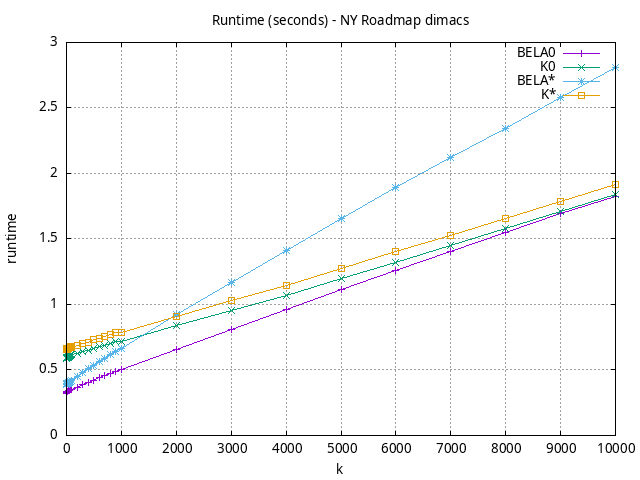}
    \end{center}
    \caption{}
    \label{fig:roadmap:dimacs:mixed:runtime:j}
  \end{subfigure}
  \begin{subfigure}{0.3\textwidth}
    \begin{center}
        \includegraphics[width=\textwidth]{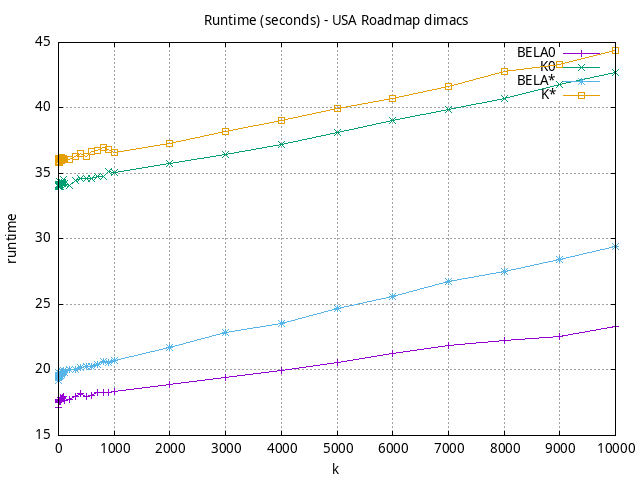}
    \end{center}
    \caption{}
    \label{fig:roadmap:dimacs:mixed:runtime:k}
  \end{subfigure}
  \begin{subfigure}{0.3\textwidth}
    \begin{center}
        \includegraphics[width=\textwidth]{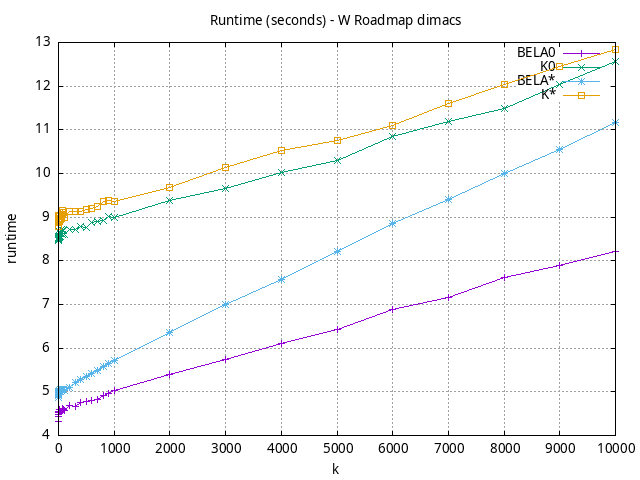}
    \end{center}
    \caption{}
    \label{fig:roadmap:dimacs:mixed:runtime:l}
  \end{subfigure}
  \caption{Runtime (in seconds) in the roadmap (dimacs) domain with mixed search algorithms}
  \label{fig:roadmap:dimacs:mixed:runtime}
\end{figure*}

%% file: roadmap.dimacs.mem.mixed.tex
\begin{figure*}
  \centering
  \begin{subfigure}{0.3\textwidth}
    \begin{center}
        \includegraphics[width=\textwidth]{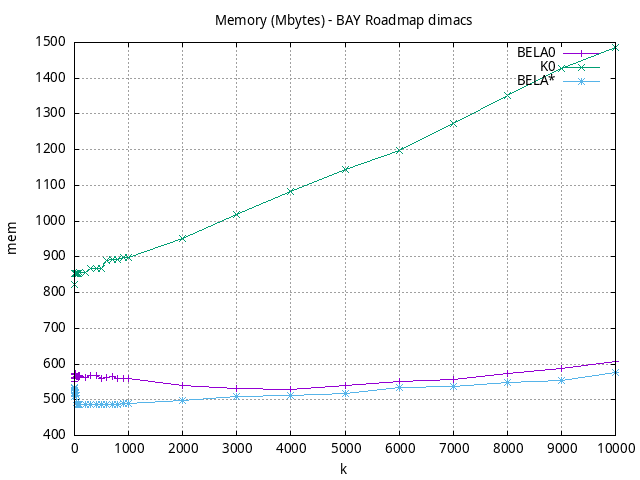}
    \end{center}
    \caption{}
    \label{fig:roadmap:dimacs:mixed:mem:a}
  \end{subfigure}
  \begin{subfigure}{0.3\textwidth}
    \begin{center}
        \includegraphics[width=\textwidth]{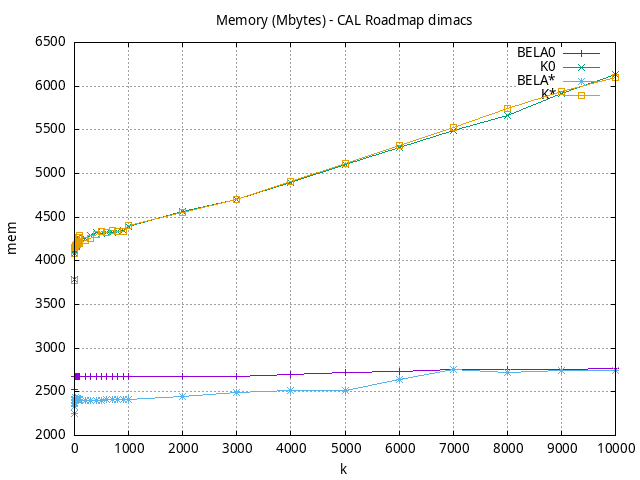}
    \end{center}
    \caption{}
    \label{fig:roadmap:dimacs:mixed:mem:b}
  \end{subfigure}
  \begin{subfigure}{0.3\textwidth}
    \begin{center}
        \includegraphics[width=\textwidth]{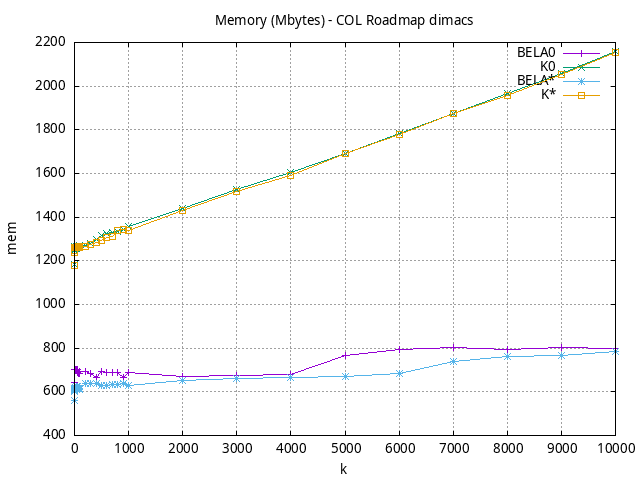}
    \end{center}
    \caption{}
    \label{fig:roadmap:dimacs:mixed:mem:c}
  \end{subfigure}
  \begin{subfigure}{0.3\textwidth}
    \begin{center}
        \includegraphics[width=\textwidth]{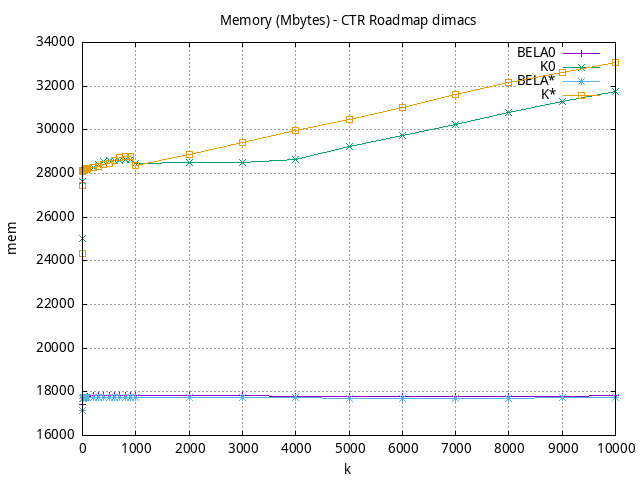}
    \end{center}
    \caption{}
    \label{fig:roadmap:dimacs:mixed:mem:d}
  \end{subfigure}
  \begin{subfigure}{0.3\textwidth}
    \begin{center}
        \includegraphics[width=\textwidth]{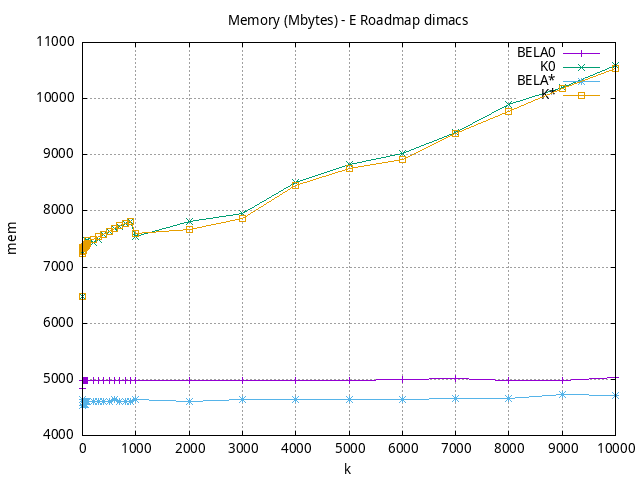}
    \end{center}
    \caption{}
    \label{fig:roadmap:dimacs:mixed:mem:e}
  \end{subfigure}
  \begin{subfigure}{0.3\textwidth}
    \begin{center}
        \includegraphics[width=\textwidth]{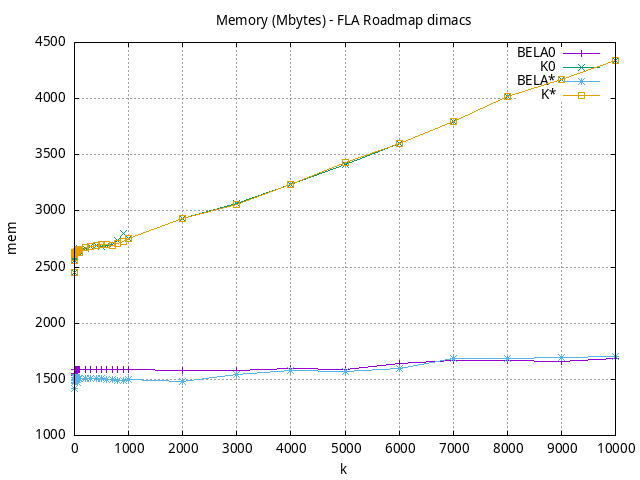}
    \end{center}
    \caption{}
    \label{fig:roadmap:dimacs:mixed:mem:f}
  \end{subfigure}
  \begin{subfigure}{0.3\textwidth}
    \begin{center}
        \includegraphics[width=\textwidth]{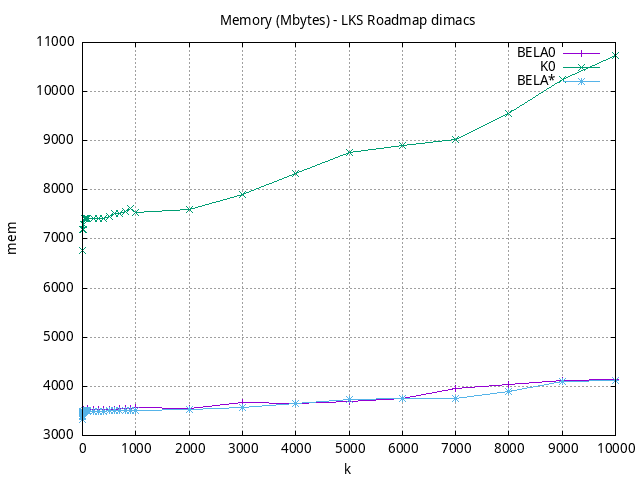}
    \end{center}
    \caption{}
    \label{fig:roadmap:dimacs:mixed:mem:g}
  \end{subfigure}
  \begin{subfigure}{0.3\textwidth}
    \begin{center}
        \includegraphics[width=\textwidth]{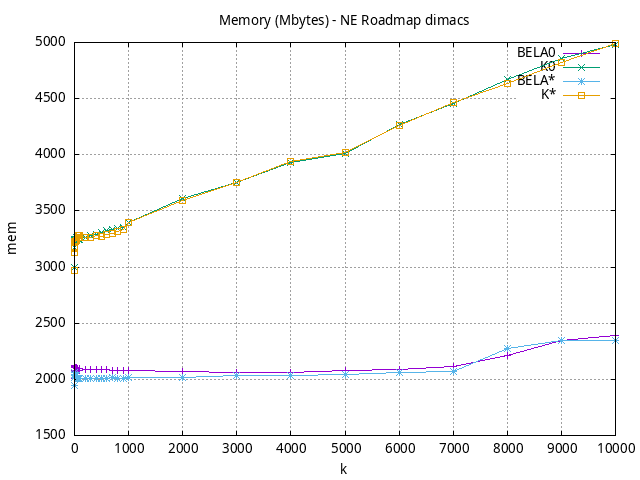}
    \end{center}
    \caption{}
    \label{fig:roadmap:dimacs:mixed:mem:h}
  \end{subfigure}
  \begin{subfigure}{0.3\textwidth}
    \begin{center}
        \includegraphics[width=\textwidth]{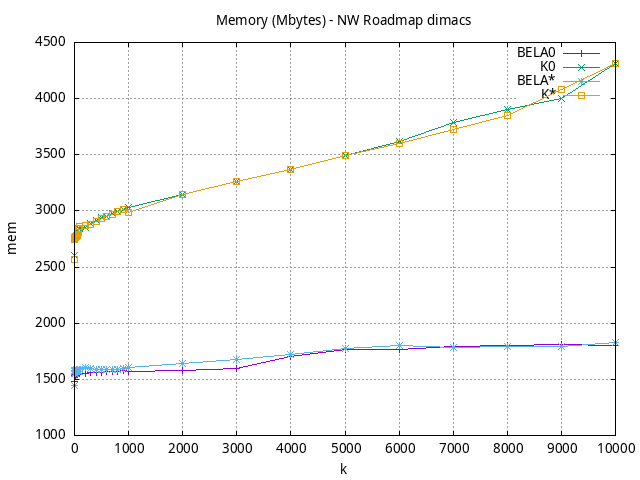}
    \end{center}
    \caption{}
    \label{fig:roadmap:dimacs:mixed:mem:i}
  \end{subfigure}
  \begin{subfigure}{0.3\textwidth}
    \begin{center}
        \includegraphics[width=\textwidth]{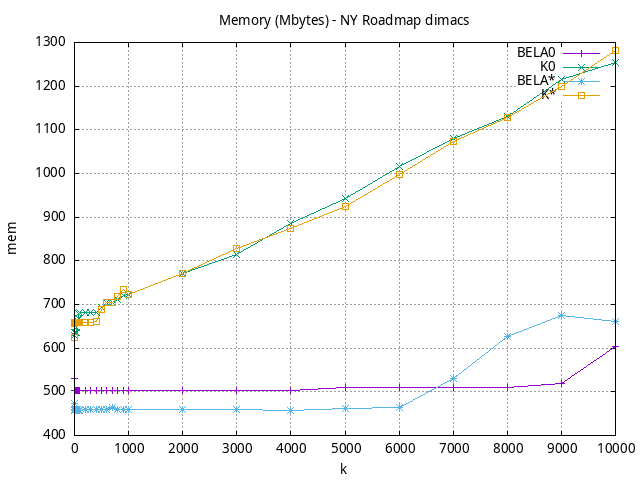}
    \end{center}
    \caption{}
    \label{fig:roadmap:dimacs:mixed:mem:j}
  \end{subfigure}
  \begin{subfigure}{0.3\textwidth}
    \begin{center}
        \includegraphics[width=\textwidth]{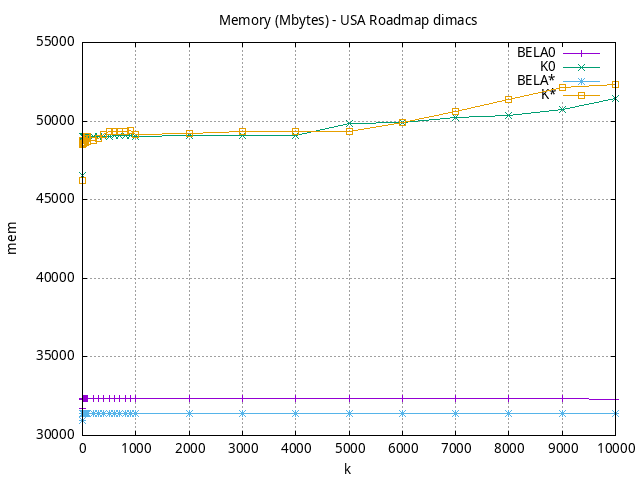}
    \end{center}
    \caption{}
    \label{fig:roadmap:dimacs:mixed:mem:k}
  \end{subfigure}
  \begin{subfigure}{0.3\textwidth}
    \begin{center}
        \includegraphics[width=\textwidth]{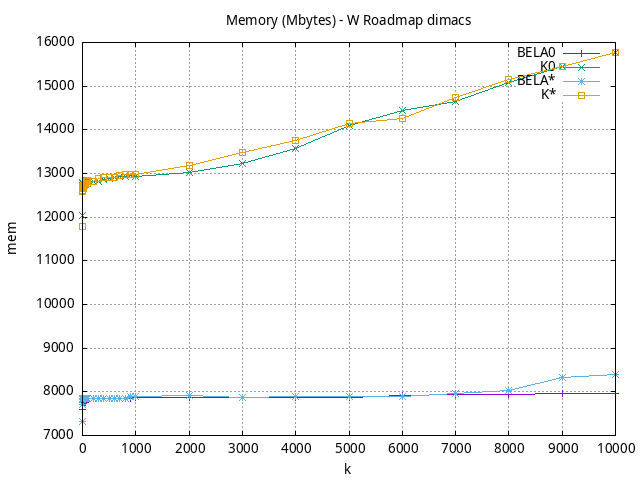}
    \end{center}
    \caption{}
    \label{fig:roadmap:dimacs:mixed:mem:l}
  \end{subfigure}
  \caption{Memory usage (in Mbytes) in the roadmap (dimacs) domain with mixed search algorithms}
  \label{fig:roadmap:dimacs:mixed:mem}
\end{figure*}

%% file: roadmap.dimacs.expansions.mixed.tex
\begin{figure*}
  \centering
  \begin{subfigure}{0.3\textwidth}
    \begin{center}
        \includegraphics[width=\textwidth]{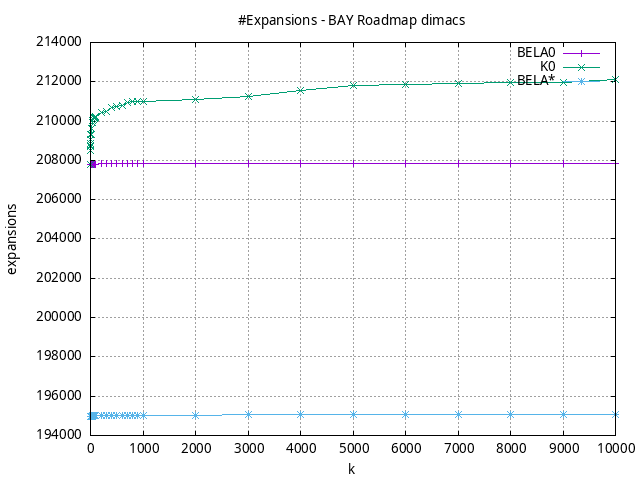}
    \end{center}
    \caption{}
    \label{fig:roadmap:dimacs:mixed:expansions:a}
  \end{subfigure}
  \begin{subfigure}{0.3\textwidth}
    \begin{center}
        \includegraphics[width=\textwidth]{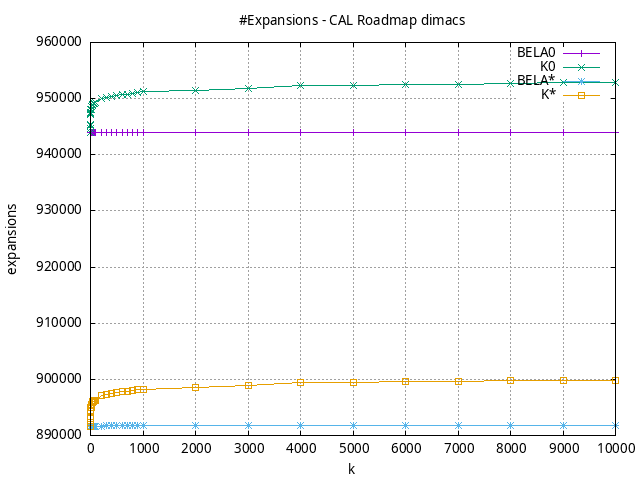}
    \end{center}
    \caption{}
    \label{fig:roadmap:dimacs:mixed:expansions:b}
  \end{subfigure}
  \begin{subfigure}{0.3\textwidth}
    \begin{center}
        \includegraphics[width=\textwidth]{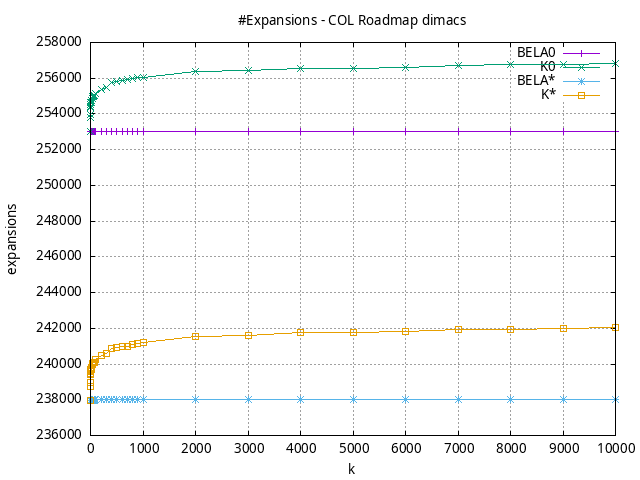}
    \end{center}
    \caption{}
    \label{fig:roadmap:dimacs:mixed:expansions:c}
  \end{subfigure}
  \begin{subfigure}{0.3\textwidth}
    \begin{center}
        \includegraphics[width=\textwidth]{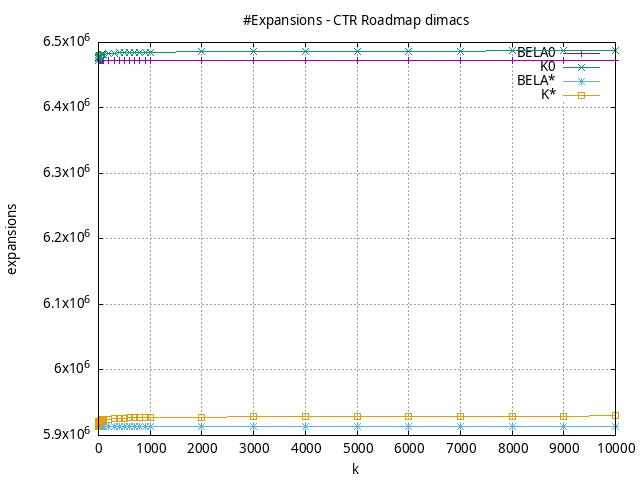}
    \end{center}
    \caption{}
    \label{fig:roadmap:dimacs:mixed:expansions:d}
  \end{subfigure}
  \begin{subfigure}{0.3\textwidth}
    \begin{center}
        \includegraphics[width=\textwidth]{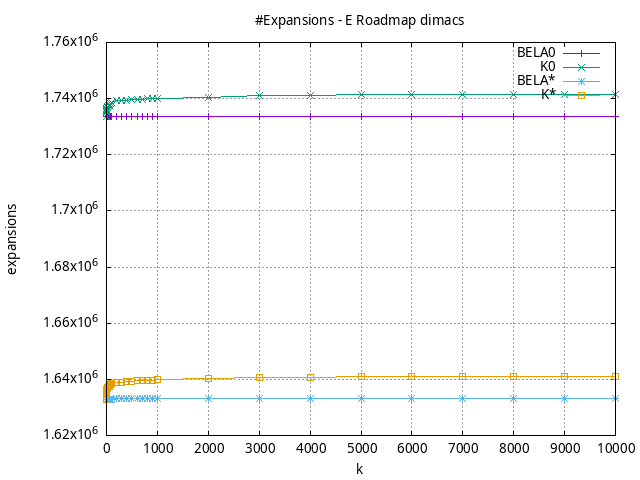}
    \end{center}
    \caption{}
    \label{fig:roadmap:dimacs:mixed:expansions:e}
  \end{subfigure}
  \begin{subfigure}{0.3\textwidth}
    \begin{center}
        \includegraphics[width=\textwidth]{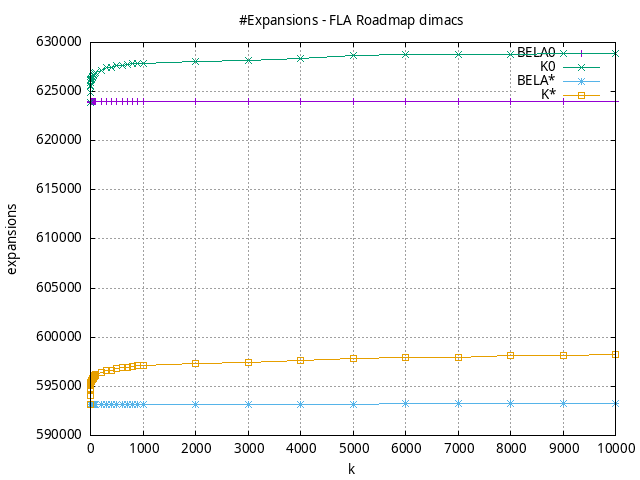}
    \end{center}
    \caption{}
    \label{fig:roadmap:dimacs:mixed:expansions:f}
  \end{subfigure}
  \begin{subfigure}{0.3\textwidth}
    \begin{center}
        \includegraphics[width=\textwidth]{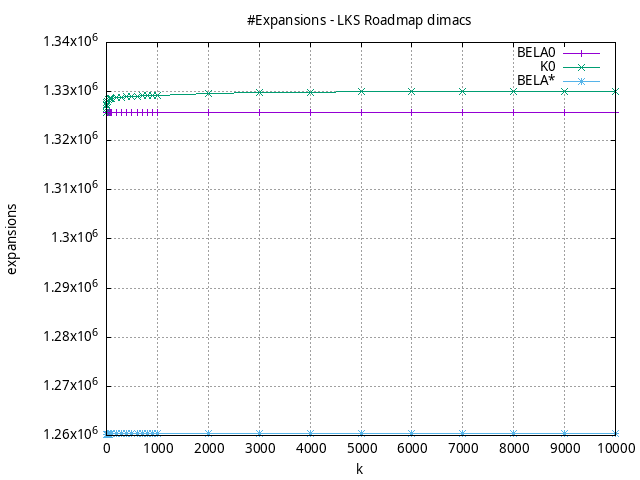}
    \end{center}
    \caption{}
    \label{fig:roadmap:dimacs:mixed:expansions:g}
  \end{subfigure}
  \begin{subfigure}{0.3\textwidth}
    \begin{center}
        \includegraphics[width=\textwidth]{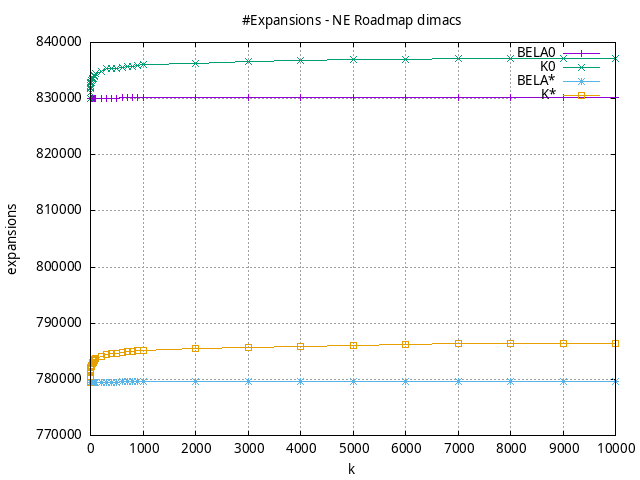}
    \end{center}
    \caption{}
    \label{fig:roadmap:dimacs:mixed:expansions:h}
  \end{subfigure}
  \begin{subfigure}{0.3\textwidth}
    \begin{center}
        \includegraphics[width=\textwidth]{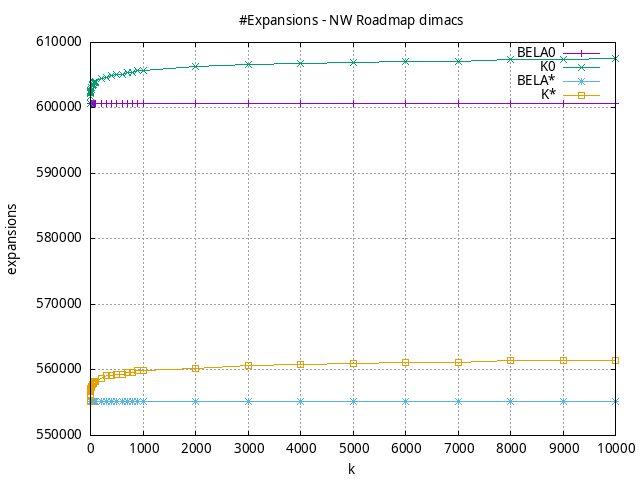}
    \end{center}
    \caption{}
    \label{fig:roadmap:dimacs:mixed:expansions:i}
  \end{subfigure}
  \begin{subfigure}{0.3\textwidth}
    \begin{center}
        \includegraphics[width=\textwidth]{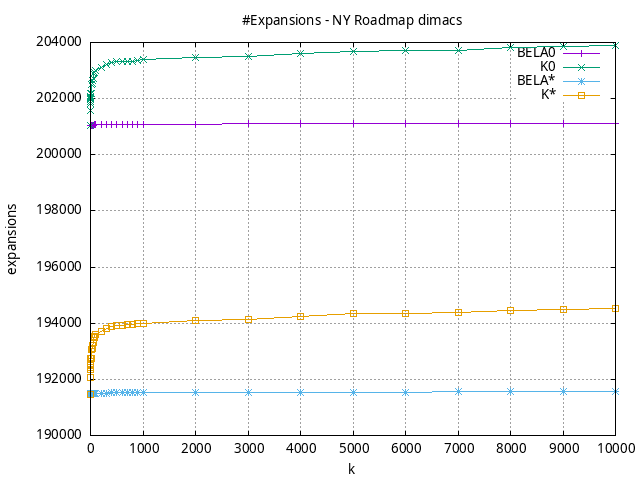}
    \end{center}
    \caption{}
    \label{fig:roadmap:dimacs:mixed:expansions:j}
  \end{subfigure}
  \begin{subfigure}{0.3\textwidth}
    \begin{center}
        \includegraphics[width=\textwidth]{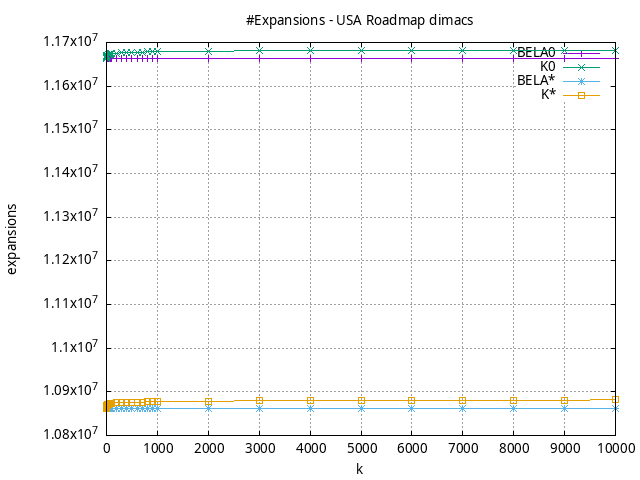}
    \end{center}
    \caption{}
    \label{fig:roadmap:dimacs:mixed:expansions:k}
  \end{subfigure}
  \begin{subfigure}{0.3\textwidth}
    \begin{center}
        \includegraphics[width=\textwidth]{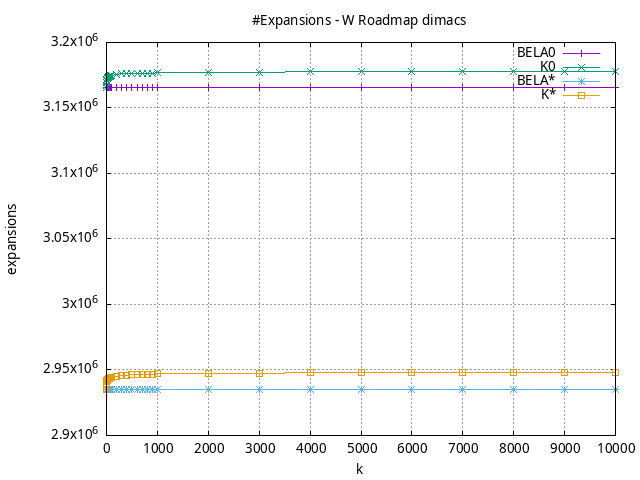}
    \end{center}
    \caption{}
    \label{fig:roadmap:dimacs:mixed:expansions:l}
  \end{subfigure}
  \caption{Number of expansions in the roadmap (dimacs) domain with mixed search algorithms}
  \label{fig:roadmap:dimacs:mixed:expansions}
\end{figure*}

%% file: roadmap.unit.runtime.brute-force.tex
\begin{figure*}
  \centering
  \begin{subfigure}{0.3\textwidth}
    \begin{center}
        \includegraphics[width=\textwidth]{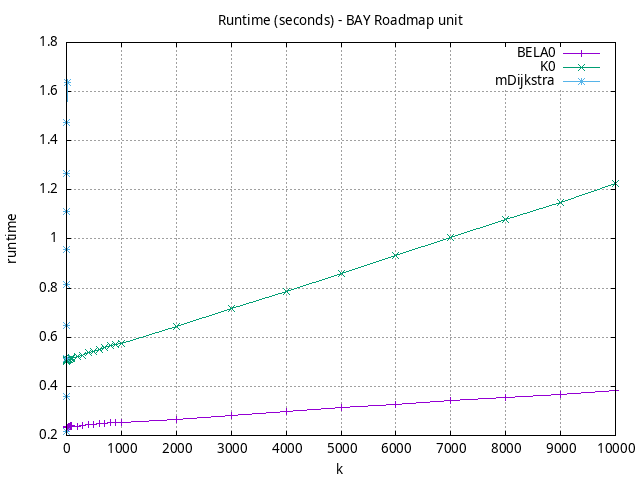}
    \end{center}
    \caption{}
    \label{fig:roadmap:unit:brute-force:runtime:a}
  \end{subfigure}
  \begin{subfigure}{0.3\textwidth}
    \begin{center}
        \includegraphics[width=\textwidth]{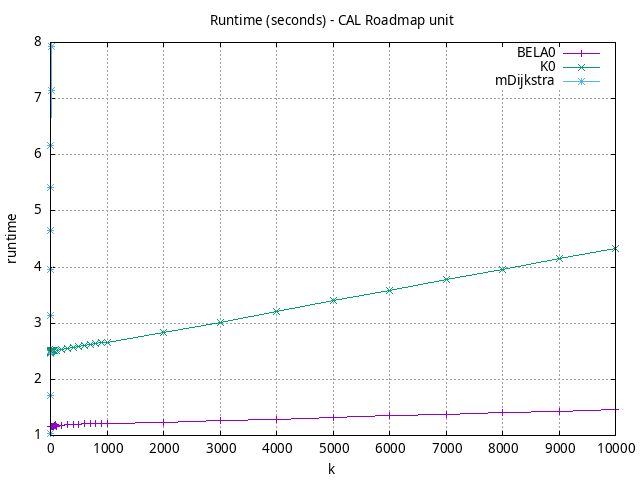}
    \end{center}
    \caption{}
    \label{fig:roadmap:unit:brute-force:runtime:b}
  \end{subfigure}
  \begin{subfigure}{0.3\textwidth}
    \begin{center}
        \includegraphics[width=\textwidth]{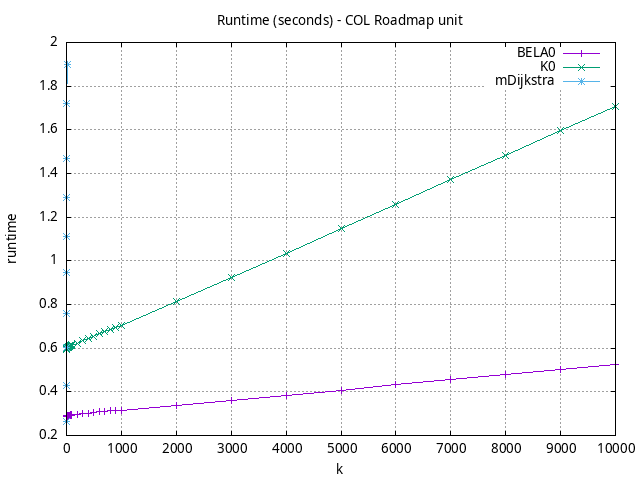}
    \end{center}
    \caption{}
    \label{fig:roadmap:unit:brute-force:runtime:c}
  \end{subfigure}
  \begin{subfigure}{0.3\textwidth}
    \begin{center}
        \includegraphics[width=\textwidth]{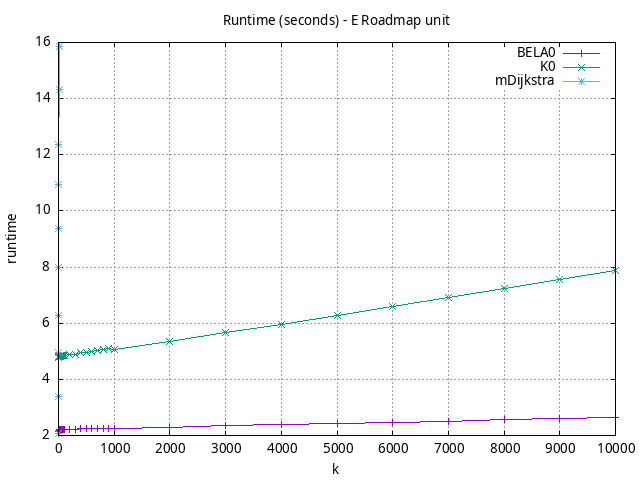}
    \end{center}
    \caption{}
    \label{fig:roadmap:unit:brute-force:runtime:d}
  \end{subfigure}
  \begin{subfigure}{0.3\textwidth}
    \begin{center}
        \includegraphics[width=\textwidth]{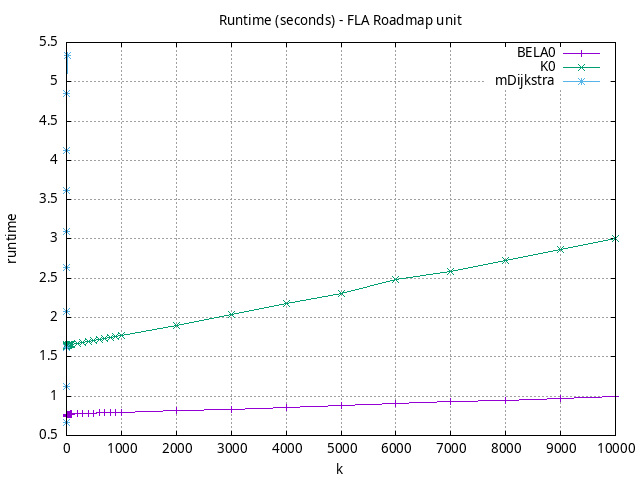}
    \end{center}
    \caption{}
    \label{fig:roadmap:unit:brute-force:runtime:e}
  \end{subfigure}
  \begin{subfigure}{0.3\textwidth}
    \begin{center}
        \includegraphics[width=\textwidth]{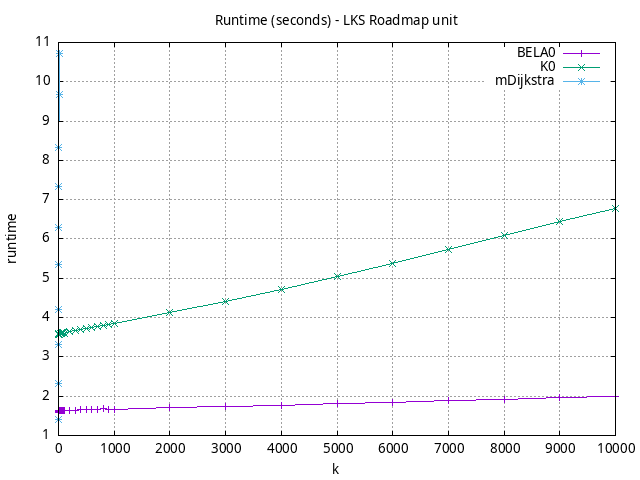}
    \end{center}
    \caption{}
    \label{fig:roadmap:unit:brute-force:runtime:f}
  \end{subfigure}
  \begin{subfigure}{0.3\textwidth}
    \begin{center}
        \includegraphics[width=\textwidth]{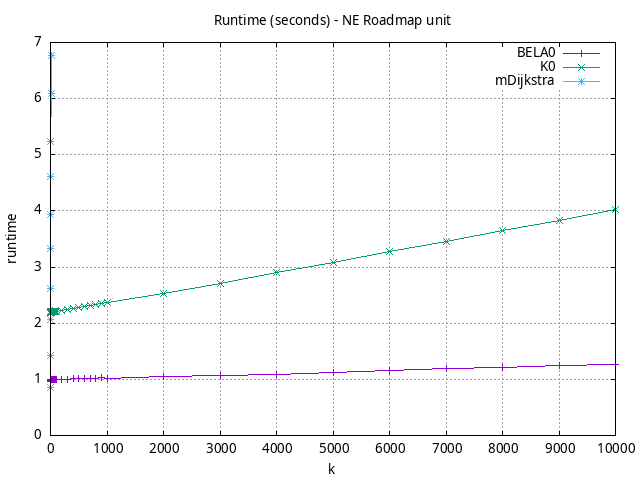}
    \end{center}
    \caption{}
    \label{fig:roadmap:unit:brute-force:runtime:g}
  \end{subfigure}
  \begin{subfigure}{0.3\textwidth}
    \begin{center}
        \includegraphics[width=\textwidth]{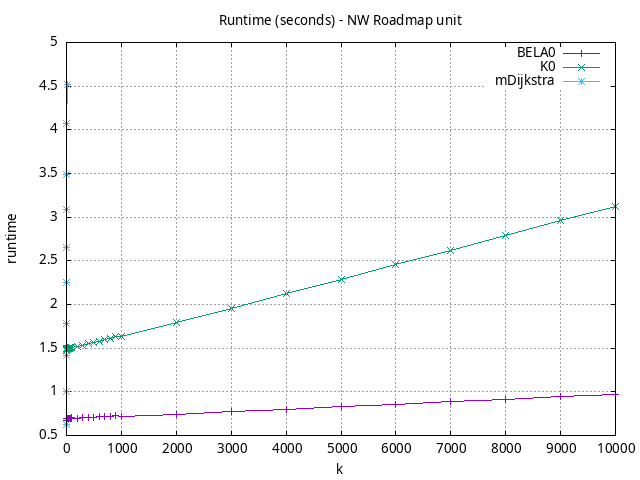}
    \end{center}
    \caption{}
    \label{fig:roadmap:unit:brute-force:runtime:h}
  \end{subfigure}
  \begin{subfigure}{0.3\textwidth}
    \begin{center}
        \includegraphics[width=\textwidth]{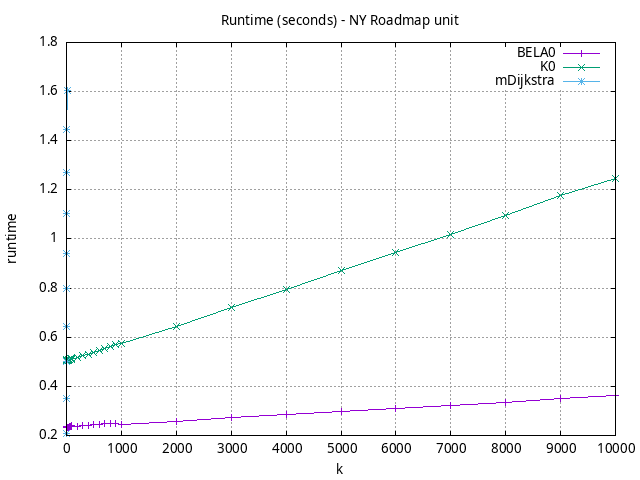}
    \end{center}
    \caption{}
    \label{fig:roadmap:unit:brute-force:runtime:i}
  \end{subfigure}
  \caption{Runtime (in seconds) in the roadmap (unit) domain with brute-force search algorithms}
  \label{fig:roadmap:unit:brute-force:runtime}
\end{figure*}

%% file: roadmap.unit.mem.brute-force.tex
\begin{figure*}
  \centering
  \begin{subfigure}{0.3\textwidth}
    \begin{center}
        \includegraphics[width=\textwidth]{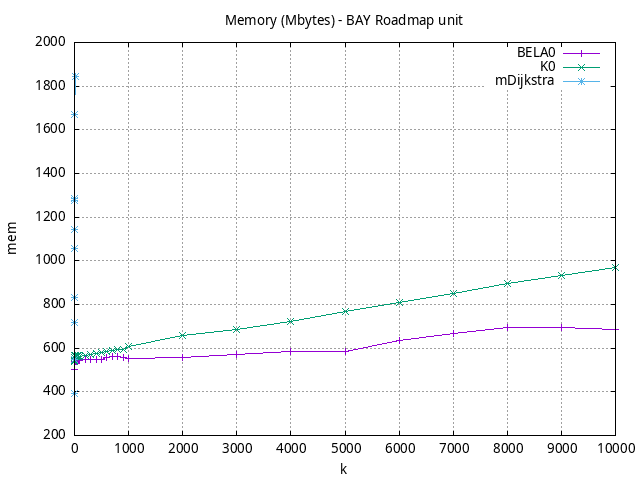}
    \end{center}
    \caption{}
    \label{fig:roadmap:unit:brute-force:mem:a}
  \end{subfigure}
  \begin{subfigure}{0.3\textwidth}
    \begin{center}
        \includegraphics[width=\textwidth]{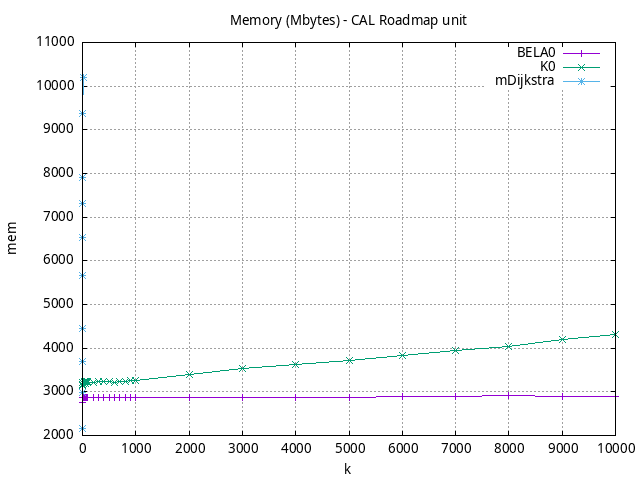}
    \end{center}
    \caption{}
    \label{fig:roadmap:unit:brute-force:mem:b}
  \end{subfigure}
  \begin{subfigure}{0.3\textwidth}
    \begin{center}
        \includegraphics[width=\textwidth]{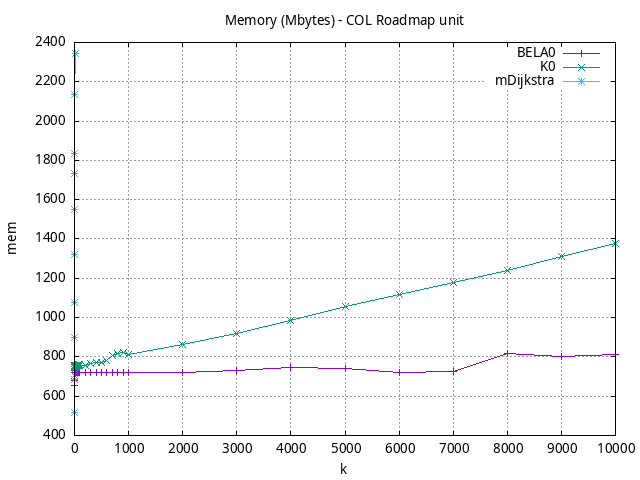}
    \end{center}
    \caption{}
    \label{fig:roadmap:unit:brute-force:mem:c}
  \end{subfigure}
  \begin{subfigure}{0.3\textwidth}
    \begin{center}
        \includegraphics[width=\textwidth]{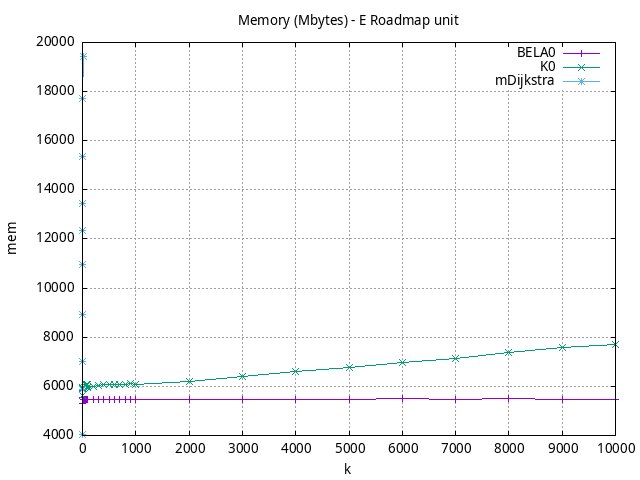}
    \end{center}
    \caption{}
    \label{fig:roadmap:unit:brute-force:mem:d}
  \end{subfigure}
  \begin{subfigure}{0.3\textwidth}
    \begin{center}
        \includegraphics[width=\textwidth]{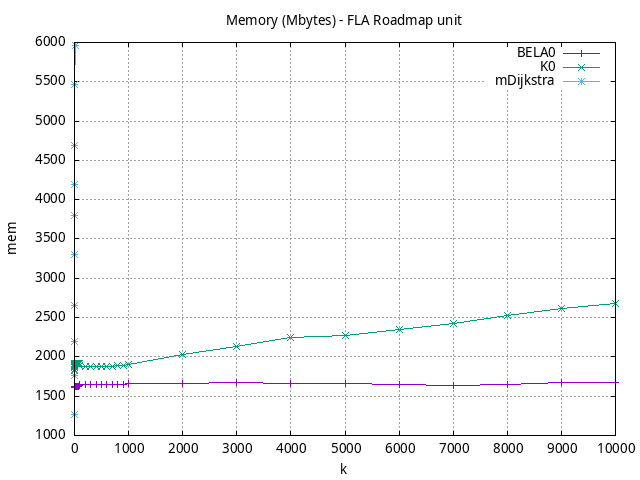}
    \end{center}
    \caption{}
    \label{fig:roadmap:unit:brute-force:mem:e}
  \end{subfigure}
  \begin{subfigure}{0.3\textwidth}
    \begin{center}
        \includegraphics[width=\textwidth]{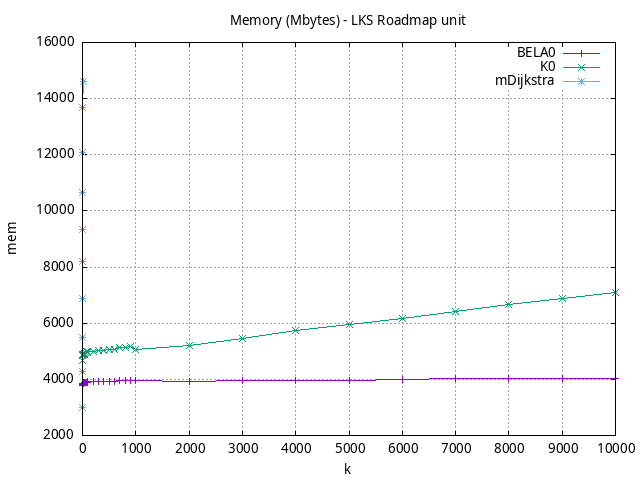}
    \end{center}
    \caption{}
    \label{fig:roadmap:unit:brute-force:mem:f}
  \end{subfigure}
  \begin{subfigure}{0.3\textwidth}
    \begin{center}
        \includegraphics[width=\textwidth]{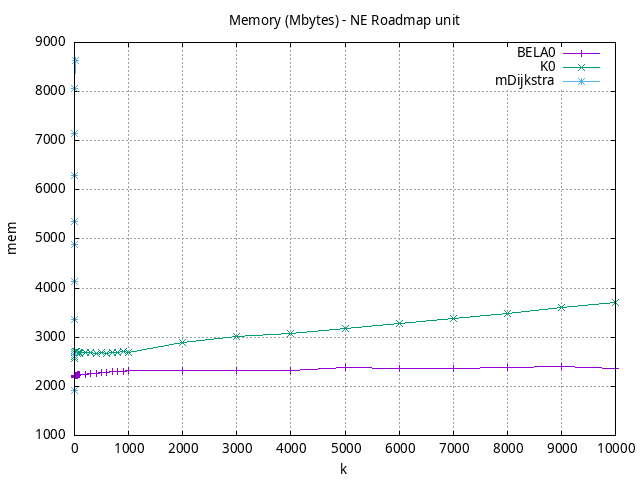}
    \end{center}
    \caption{}
    \label{fig:roadmap:unit:brute-force:mem:g}
  \end{subfigure}
  \begin{subfigure}{0.3\textwidth}
    \begin{center}
        \includegraphics[width=\textwidth]{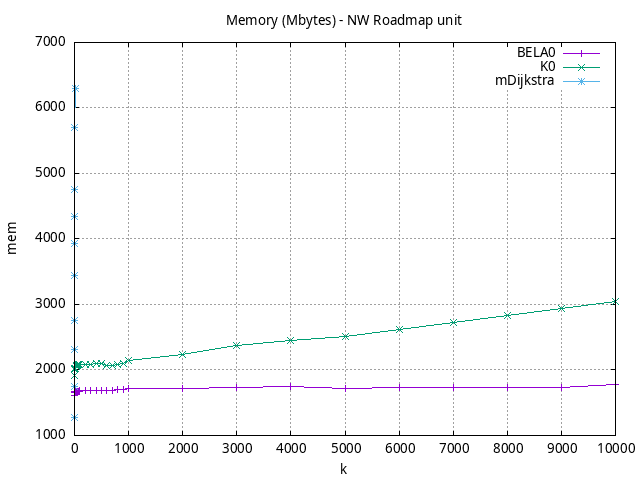}
    \end{center}
    \caption{}
    \label{fig:roadmap:unit:brute-force:mem:h}
  \end{subfigure}
  \begin{subfigure}{0.3\textwidth}
    \begin{center}
        \includegraphics[width=\textwidth]{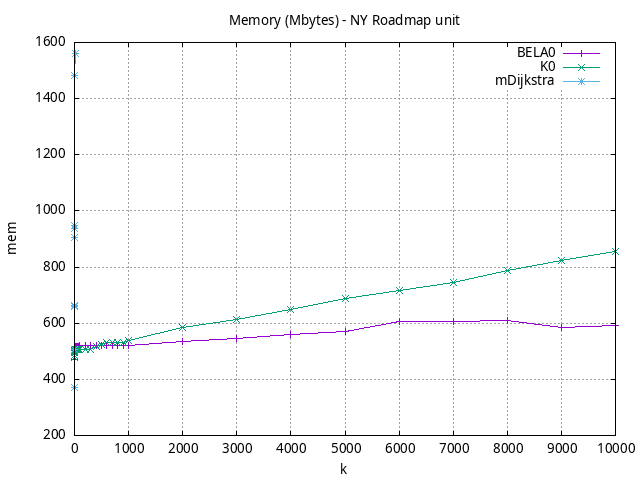}
    \end{center}
    \caption{}
    \label{fig:roadmap:unit:brute-force:mem:i}
  \end{subfigure}
  \caption{Memory usage (in Mbytes) in the roadmap (unit) domain with brute-force search algorithms}
  \label{fig:roadmap:unit:brute-force:mem}
\end{figure*}

%% file: roadmap.unit.expansions.brute-force.tex
\begin{figure*}
  \centering
  \begin{subfigure}{0.3\textwidth}
    \begin{center}
        \includegraphics[width=\textwidth]{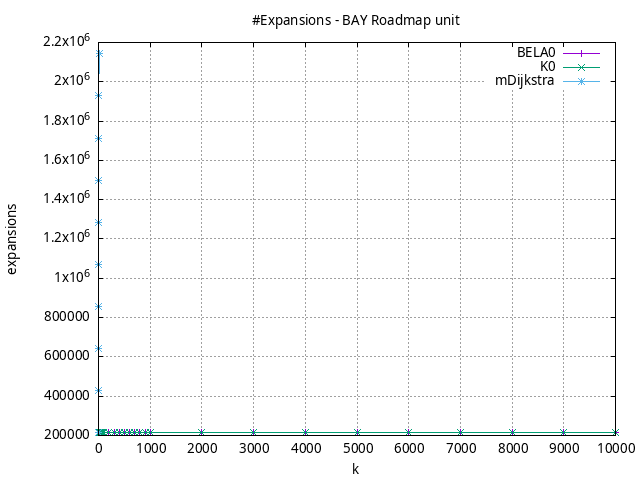}
    \end{center}
    \caption{}
    \label{fig:roadmap:unit:brute-force:expansions:a}
  \end{subfigure}
  \begin{subfigure}{0.3\textwidth}
    \begin{center}
        \includegraphics[width=\textwidth]{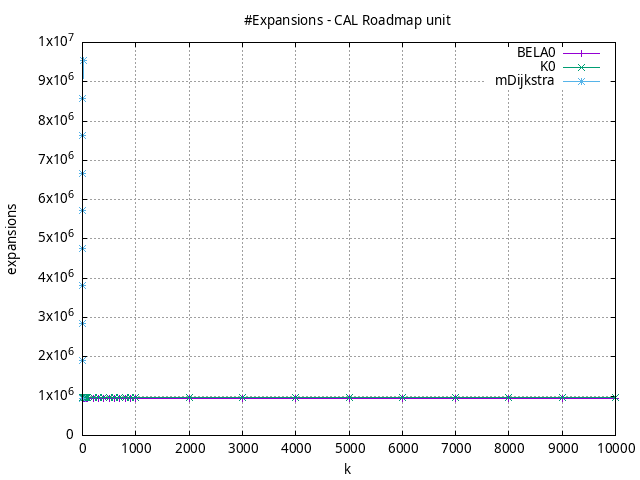}
    \end{center}
    \caption{}
    \label{fig:roadmap:unit:brute-force:expansions:b}
  \end{subfigure}
  \begin{subfigure}{0.3\textwidth}
    \begin{center}
        \includegraphics[width=\textwidth]{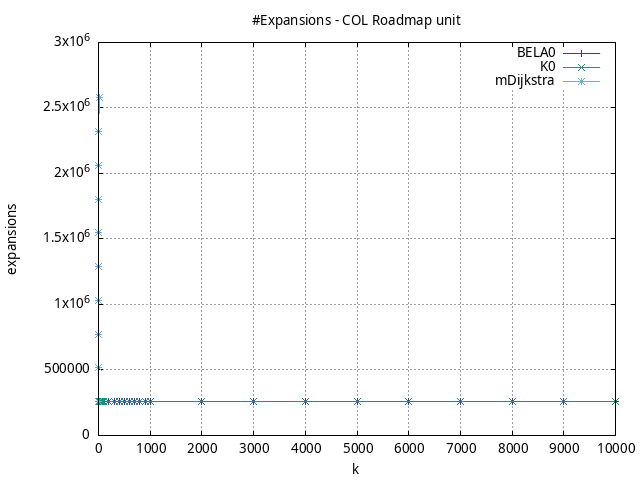}
    \end{center}
    \caption{}
    \label{fig:roadmap:unit:brute-force:expansions:c}
  \end{subfigure}
  \begin{subfigure}{0.3\textwidth}
    \begin{center}
        \includegraphics[width=\textwidth]{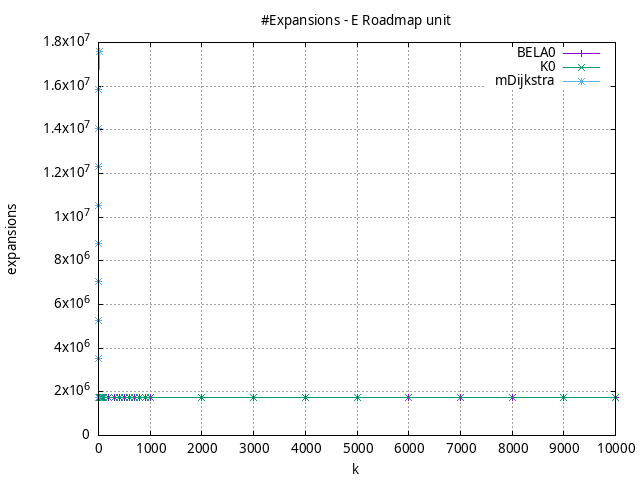}
    \end{center}
    \caption{}
    \label{fig:roadmap:unit:brute-force:expansions:d}
  \end{subfigure}
  \begin{subfigure}{0.3\textwidth}
    \begin{center}
        \includegraphics[width=\textwidth]{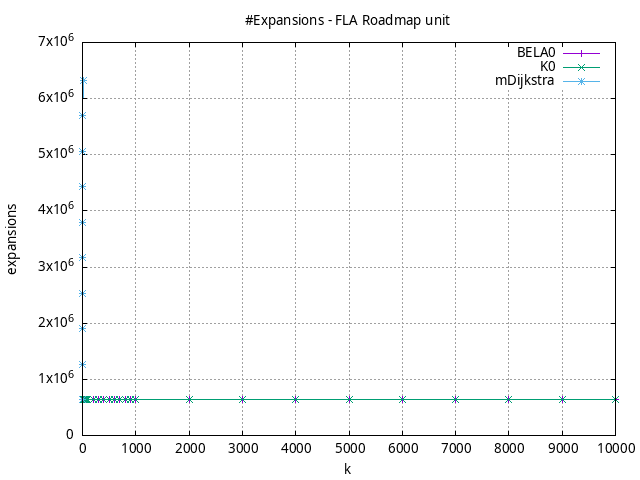}
    \end{center}
    \caption{}
    \label{fig:roadmap:unit:brute-force:expansions:e}
  \end{subfigure}
  \begin{subfigure}{0.3\textwidth}
    \begin{center}
        \includegraphics[width=\textwidth]{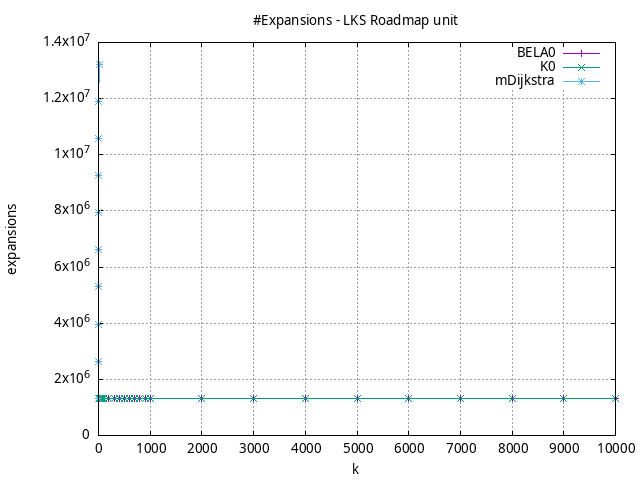}
    \end{center}
    \caption{}
    \label{fig:roadmap:unit:brute-force:expansions:f}
  \end{subfigure}
  \begin{subfigure}{0.3\textwidth}
    \begin{center}
        \includegraphics[width=\textwidth]{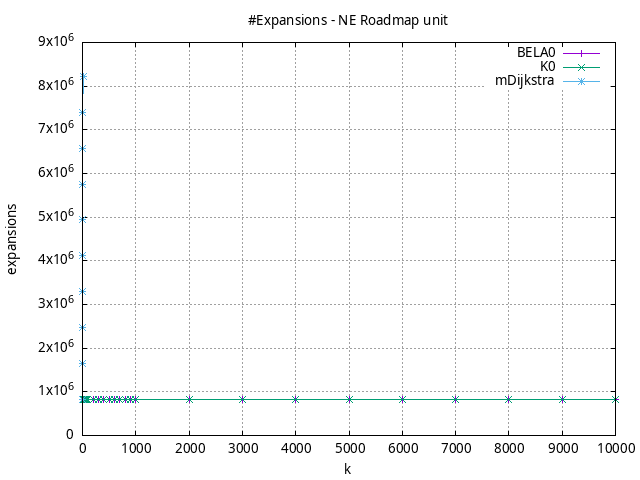}
    \end{center}
    \caption{}
    \label{fig:roadmap:unit:brute-force:expansions:g}
  \end{subfigure}
  \begin{subfigure}{0.3\textwidth}
    \begin{center}
        \includegraphics[width=\textwidth]{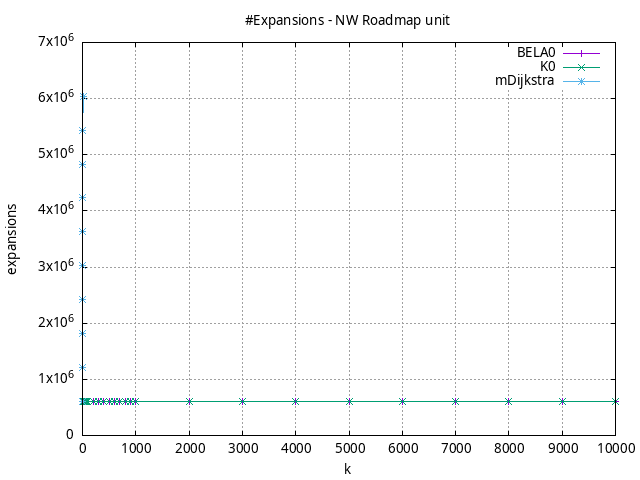}
    \end{center}
    \caption{}
    \label{fig:roadmap:unit:brute-force:expansions:h}
  \end{subfigure}
  \begin{subfigure}{0.3\textwidth}
    \begin{center}
        \includegraphics[width=\textwidth]{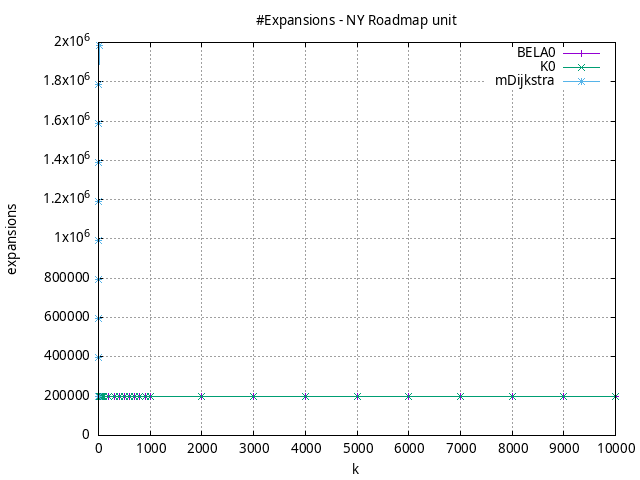}
    \end{center}
    \caption{}
    \label{fig:roadmap:unit:brute-force:expansions:i}
  \end{subfigure}
  \caption{Number of expansions in the roadmap (unit) domain with brute-force search algorithms}
  \label{fig:roadmap:unit:brute-force:expansions}
\end{figure*}

%% file: supplemental6-maps.tex

\subsection{Random maps}
\label{sec:empirical-evaluation:random-maps}

The random map is taken from the 2d Pathfinding \textit{movingai}
benchmark\footnote{\url{https://movingai.com/benchmarks/grids.html}}. Only the
first instance from the random maps benchmark has been used (with 512$\times$512
locations), but considering different percentages of obstruction: 10, 15, 20,
25, 30 and 35, yielding a total of 6 different random maps. For each map, 100
instances were randomly generated where the heuristic distance between the start
and goal state is at least 90\% of the largest possible distance. All
results are averaged over all runs.

\subsubsection{Unit variant}
\label{sec:empirical-evaluation:random-maps:unit-variant}

In the first variant it is only possible to move either horizontally or
vertically, and the cost of all operators is equal to 1. Both brute-force and
heuristic variants of all search algorithms are considered. The heuristic
function used is the Manhattan distance.

Figures~\ref{fig:maps:unit:brute-force:runtime}--\ref{fig:maps:unit:brute-force:expansions}
show the runtime (in seconds), the memory usage (in Mbytes), and the number of
expansions of \bfbela{}, \bfk{}, and mDijkstra. The first observation is that
with the absence of a heuristic function, mDijkstra performs even worse than in
the previous domain, and it only finds $\kappa=4$ paths before using more time
than \bfbela{} takes to output 10,000 different paths. This shows a difference
of several orders of magnitude in runtime. The performance of \kstar{} in this
domain deserves attention. First, it performs much worse than \bfbela{} in all
maps. In fact, \bfk{} was requested only to find $\kappa=1,000$ paths, yet it
always takes significantly longer than \bfbela{} takes to compute
$\kappa=10,000$ paths, even if it expands around the same number of nodes as
shown in Figure~\ref{fig:maps:unit:brute-force:expansions}. This indicates a
difference in runtime of several orders of magnitude. Secondly, as conjectured
in the roadmap domain, \bfk{}'s performance improves as the branching factor is
reduced. As the percentage of obstruction increases, the runtime improves. For
example, it takes roughly 1 second to compute only 1,000 paths when 10\% of the
locations are occupied, but it can find the same number of paths in less than
0.65 seconds when the obstruction percentage gets to its maximum, 35.

When considering the application of the heuristic search algorithms, the
differences between \bfk{} and \bfbela{} become more acute, with a difference in
runtime of one order of magnitude. \kstar{} computes $\kappa=1,000$
different paths in around the same amount of time it takes for \bela{} to find ten times that amount of paths. Even if the
heuristic funtion is not very well informed (in particular, for percentages of
obstruction equal to 25 or larger), the difference in the number of expansions,
shown in Figure~\ref{fig:maps:unit:heuristic:expansions} is of various orders of
magnitude often, in particular in those cases with low percentages of
obstruction. This difference is explained with an increase of the branching
factor, which is conjectured to harm performance of \kstar{} but, more
importantly, by the observation that, in this domain in particular, the number
of paths grow exponentially, so that a single centroid is enough to deliver even
several billions of paths. \bela{} can take advantage of this possibility and,
in the end, it runs various orders of magnitude faster than \kstar{}.

As for \mAstar{}, the consideration of the heuristic function makes it improve
its runtime marginally and it can now find $\kappa=10$ solution paths. The
reason for this low number is, as explained above, because the heuristic
function is not very well informed.

\input{maps.unit.runtime.brute-force.tex}
\input{maps.unit.mem.brute-force.tex}
\input{maps.unit.expansions.brute-force.tex}

\input{maps.unit.runtime.heuristic.tex}
\input{maps.unit.mem.heuristic.tex}
\input{maps.unit.expansions.heuristic.tex}

\subsubsection{Octile variant}
\label{sec:empirical-evaluation:random-maps:octile-variant}

In this variant, in addition to horizontal and vertical moves, it is also possible to move to
cells diagonally adjacent to the current cell, provided they are not marked as inaccessible. This
doubles the branching factor from 4 to 8. In addition, the octile variant is a
non-unit domain because the diagonal moves have a cost equal to 14, whereas
horizontal and vertical moves have a cost equal to 10 units. The heuristic
function used is the octile distance.

This variant is harder than the previous variant for all algorithms.
Again, mDijkstra is only able to find $\kappa=4$ different paths, usually taking
longer than the other algorithms which find either two orders of magnitude or
even four orders of magnitude more paths in the same allotted time, as shown in
Figure~\ref{fig:maps:octile:brute-force:runtime}. This time, \bfk{} is
restricted to find only $\kappa=100$ different paths (10 times less than in the
unit domain) and it consistently takes one order of magnitude more time than
\bfbela{}, which computes $\kappa=10,000$ solution paths. Even if \bfbela{}
also takes longer than it does in the unit variant, it still performs much better than all the
other algorithms, being able to compute up to 10,000 paths in less than a second
(averaged over each map). The difference in runtime between \bfk{} and \bfbela{}
can not be attributed neither to an increase in graph size (since they are the
same than in the unit variant), nor the number of expansions performed by each
algorithm. shown in Figure~\ref{fig:maps:octile:brute-force:expansions}, since
the difference is rather small. The degradation in performance of \bfk{} is
therefore attributed to the increase in the branching factor which forces \bfk{}
to consume more time in building and maintaining the path graph. Regarding
\bfbela{}, its performance does not decrease significantly and, again, it
delivers $\kappa=10,000$ solution paths in less than a second on average across
all maps. One of the reasons for this performance is that in this variant, one
single centroid suffices to deliver all solution paths.

\input{maps.octile.runtime.brute-force.tex}
\input{maps.octile.mem.brute-force.tex}
\input{maps.octile.expansions.brute-force.tex}

\input{maps.octile.runtime.heuristic.tex}
\input{maps.octile.mem.heuristic.tex}
\input{maps.octile.expansions.heuristic.tex}


%% file: maps.unit.runtime.brute-force.tex
\begin{figure*}
  \centering
  \begin{subfigure}{0.3\textwidth}
    \begin{center}
        \includegraphics[width=\textwidth]{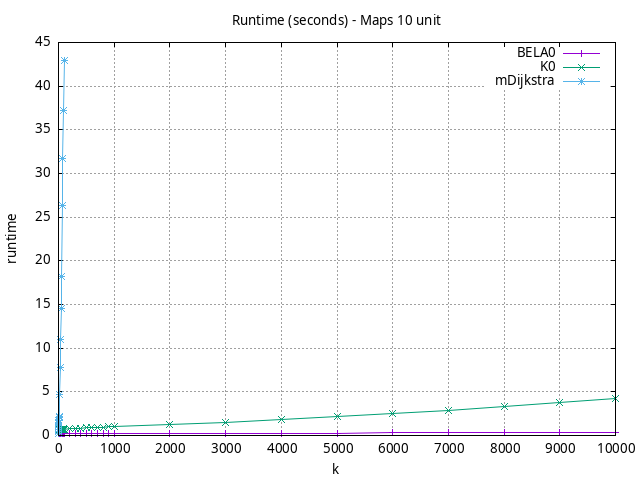}
    \end{center}
    \caption{}
    \label{fig:maps:unit:brute-force:runtime:a}
  \end{subfigure}
  \begin{subfigure}{0.3\textwidth}
    \begin{center}
        \includegraphics[width=\textwidth]{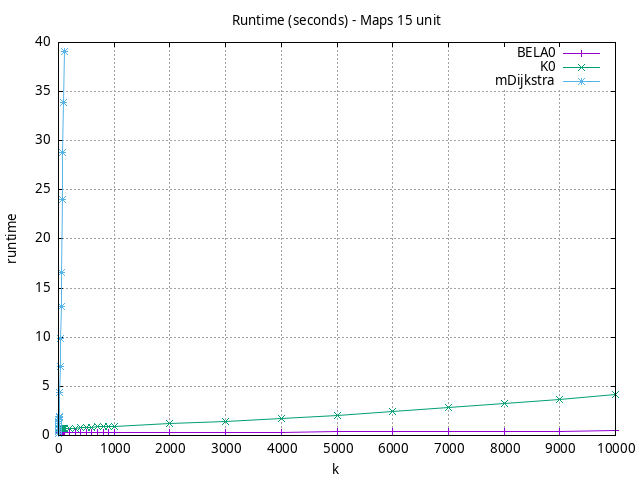}
    \end{center}
    \caption{}
    \label{fig:maps:unit:brute-force:runtime:b}
  \end{subfigure}
  \begin{subfigure}{0.3\textwidth}
    \begin{center}
        \includegraphics[width=\textwidth]{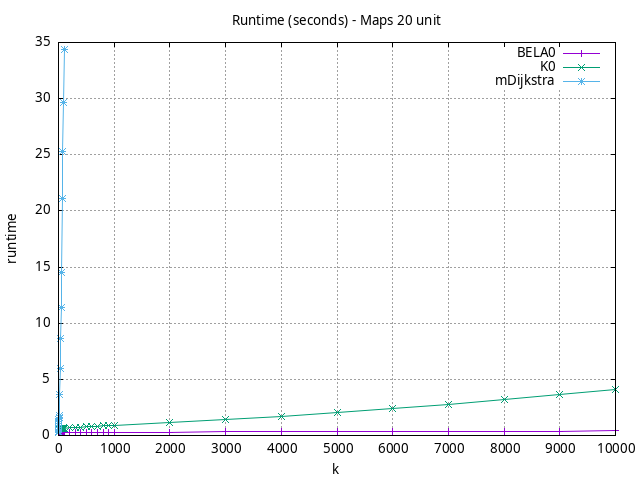}
    \end{center}
    \caption{}
    \label{fig:maps:unit:brute-force:runtime:c}
  \end{subfigure}
  \begin{subfigure}{0.3\textwidth}
    \begin{center}
        \includegraphics[width=\textwidth]{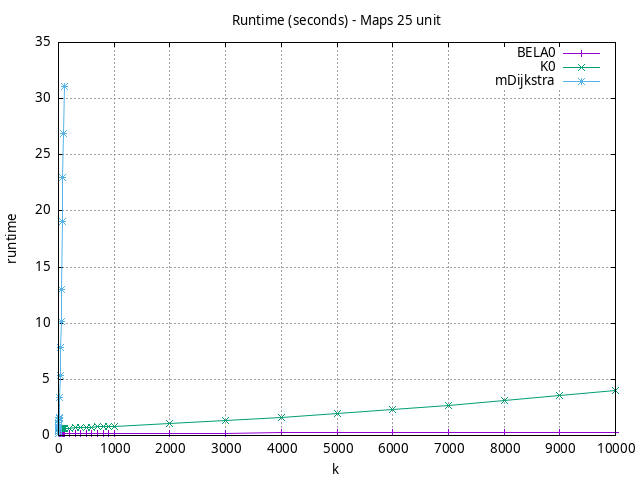}
    \end{center}
    \caption{}
    \label{fig:maps:unit:brute-force:runtime:d}
  \end{subfigure}
  \begin{subfigure}{0.3\textwidth}
    \begin{center}
        \includegraphics[width=\textwidth]{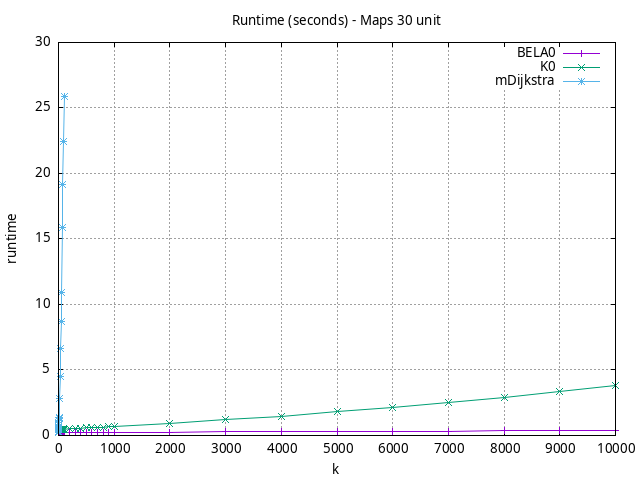}
    \end{center}
    \caption{}
    \label{fig:maps:unit:brute-force:runtime:e}
  \end{subfigure}
  \begin{subfigure}{0.3\textwidth}
    \begin{center}
        \includegraphics[width=\textwidth]{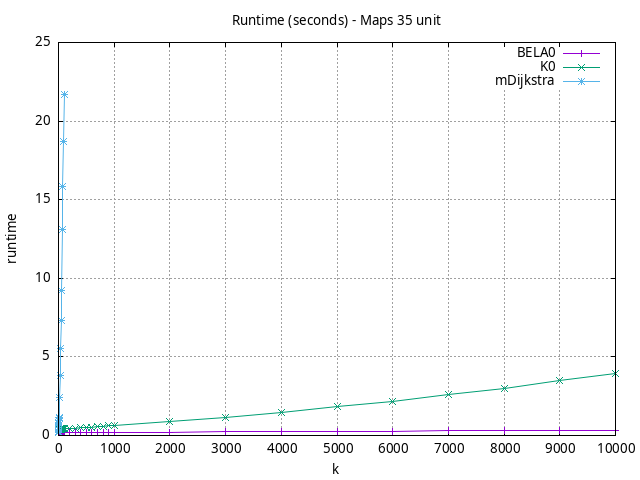}
    \end{center}
    \caption{}
    \label{fig:maps:unit:brute-force:runtime:f}
  \end{subfigure}
  \caption{Runtime (in seconds) in the maps (unit) domain with brute-force search algorithms}
  \label{fig:maps:unit:brute-force:runtime}
\end{figure*}

%% file: maps.unit.mem.brute-force.tex
\begin{figure*}
  \centering
  \begin{subfigure}{0.3\textwidth}
    \begin{center}
        \includegraphics[width=\textwidth]{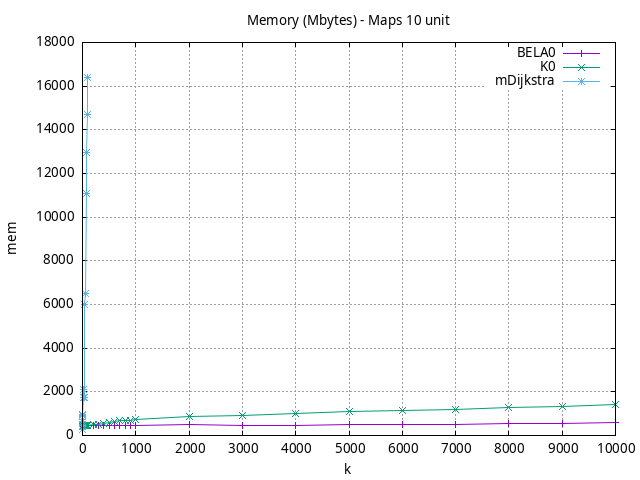}
    \end{center}
    \caption{}
    \label{fig:maps:unit:brute-force:mem:a}
  \end{subfigure}
  \begin{subfigure}{0.3\textwidth}
    \begin{center}
        \includegraphics[width=\textwidth]{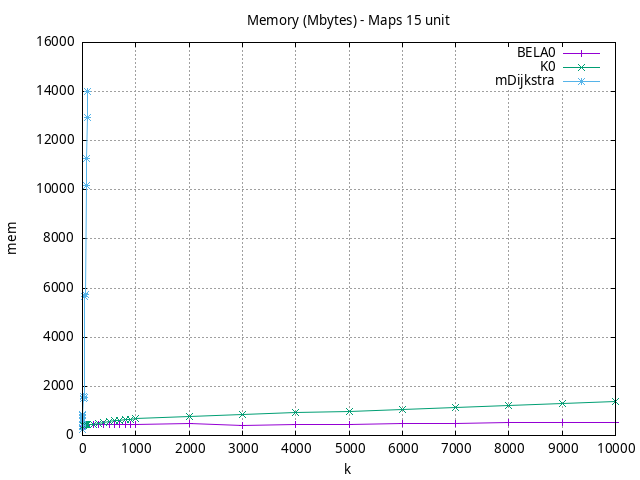}
    \end{center}
    \caption{}
    \label{fig:maps:unit:brute-force:mem:b}
  \end{subfigure}
  \begin{subfigure}{0.3\textwidth}
    \begin{center}
        \includegraphics[width=\textwidth]{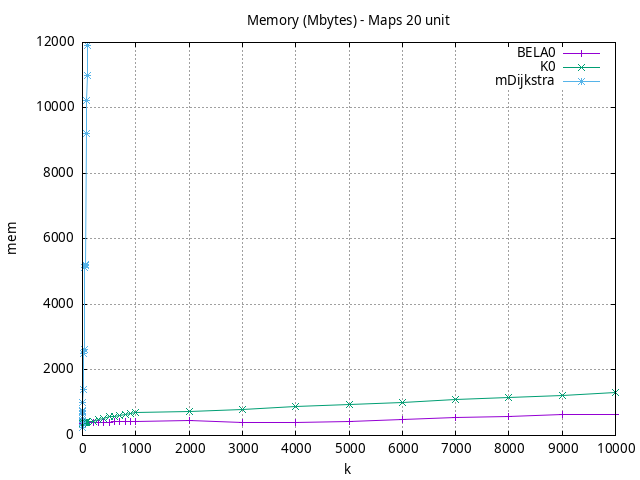}
    \end{center}
    \caption{}
    \label{fig:maps:unit:brute-force:mem:c}
  \end{subfigure}
  \begin{subfigure}{0.3\textwidth}
    \begin{center}
        \includegraphics[width=\textwidth]{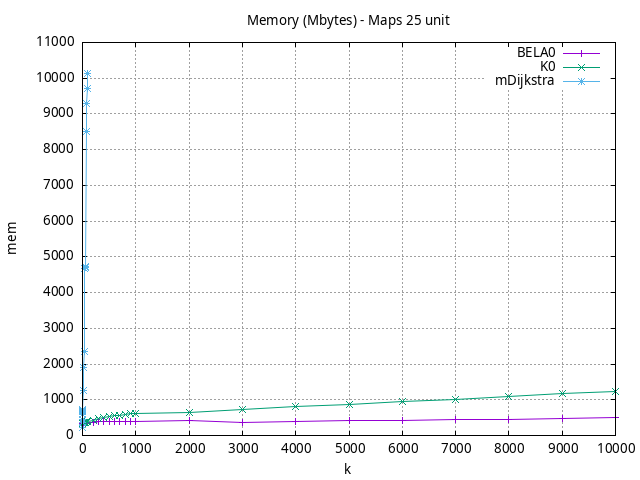}
    \end{center}
    \caption{}
    \label{fig:maps:unit:brute-force:mem:d}
  \end{subfigure}
  \begin{subfigure}{0.3\textwidth}
    \begin{center}
        \includegraphics[width=\textwidth]{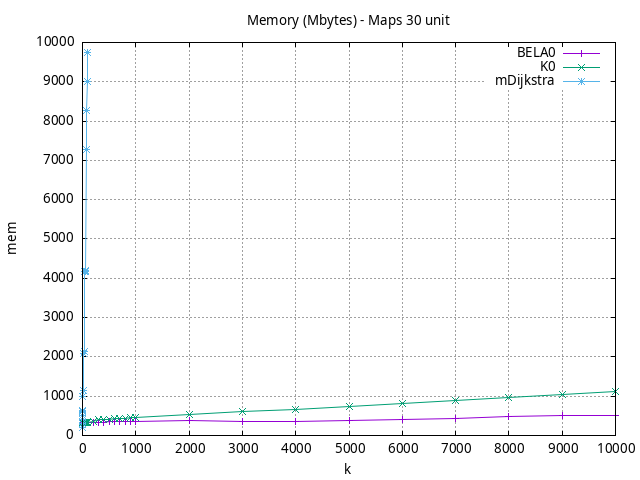}
    \end{center}
    \caption{}
    \label{fig:maps:unit:brute-force:mem:e}
  \end{subfigure}
  \begin{subfigure}{0.3\textwidth}
    \begin{center}
        \includegraphics[width=\textwidth]{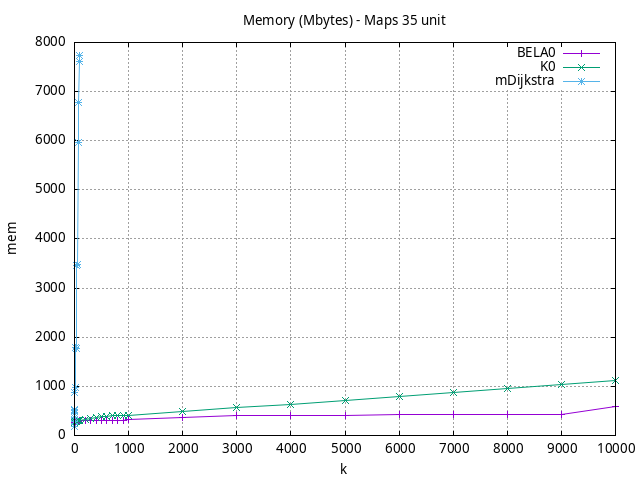}
    \end{center}
    \caption{}
    \label{fig:maps:unit:brute-force:mem:f}
  \end{subfigure}
  \caption{Memory usage (in Mbytes) in the maps (unit) domain with brute-force search algorithms}
  \label{fig:maps:unit:brute-force:mem}
\end{figure*}

%% file: maps.unit.expansions.brute-force.tex
\begin{figure*}
  \centering
  \begin{subfigure}{0.3\textwidth}
    \begin{center}
        \includegraphics[width=\textwidth]{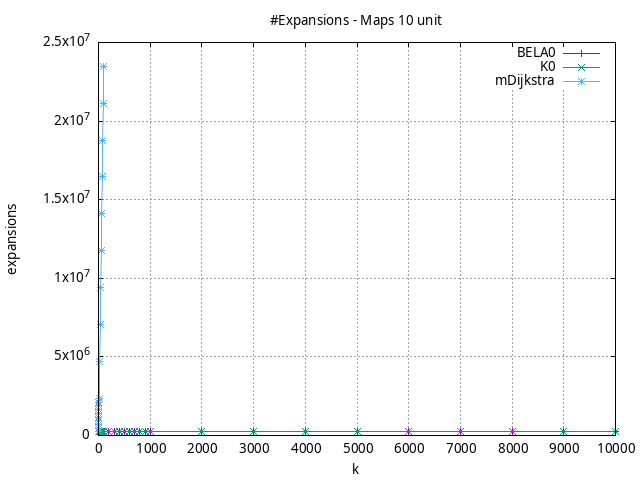}
    \end{center}
    \caption{}
    \label{fig:maps:unit:brute-force:expansions:a}
  \end{subfigure}
  \begin{subfigure}{0.3\textwidth}
    \begin{center}
        \includegraphics[width=\textwidth]{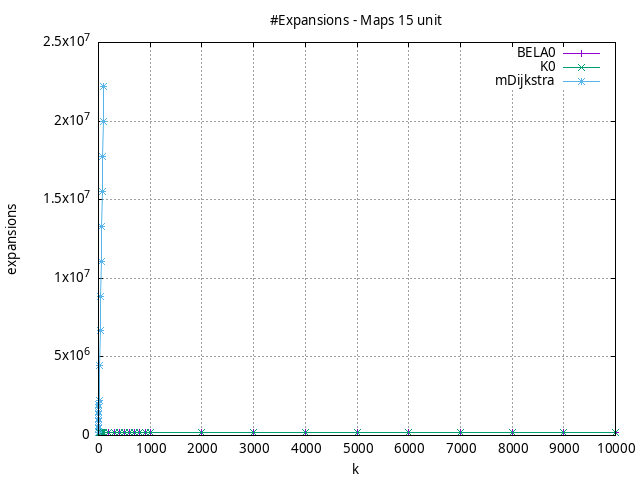}
    \end{center}
    \caption{}
    \label{fig:maps:unit:brute-force:expansions:b}
  \end{subfigure}
  \begin{subfigure}{0.3\textwidth}
    \begin{center}
        \includegraphics[width=\textwidth]{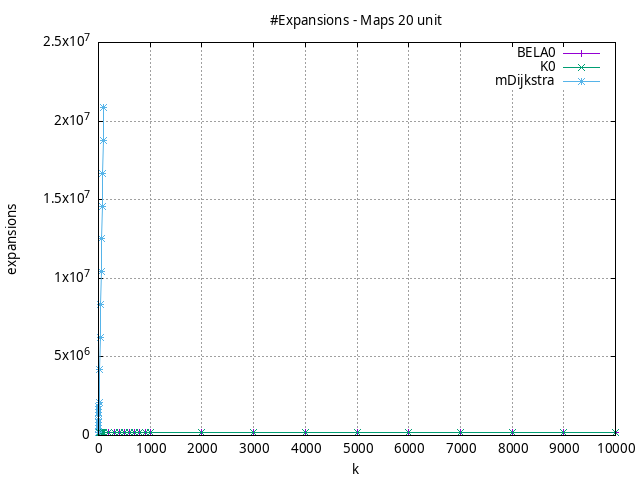}
    \end{center}
    \caption{}
    \label{fig:maps:unit:brute-force:expansions:c}
  \end{subfigure}
  \begin{subfigure}{0.3\textwidth}
    \begin{center}
        \includegraphics[width=\textwidth]{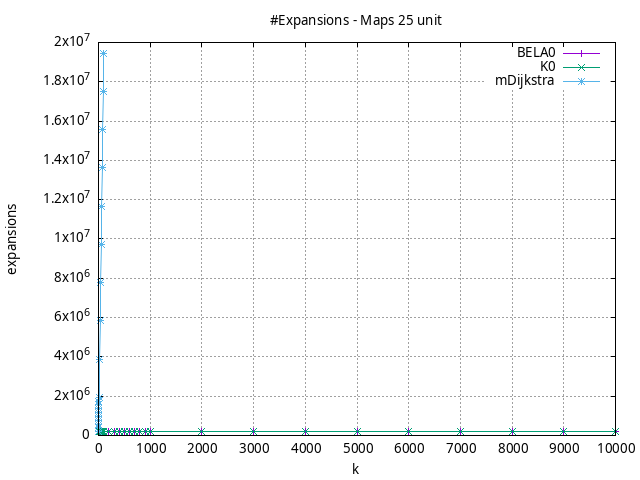}
    \end{center}
    \caption{}
    \label{fig:maps:unit:brute-force:expansions:d}
  \end{subfigure}
  \begin{subfigure}{0.3\textwidth}
    \begin{center}
        \includegraphics[width=\textwidth]{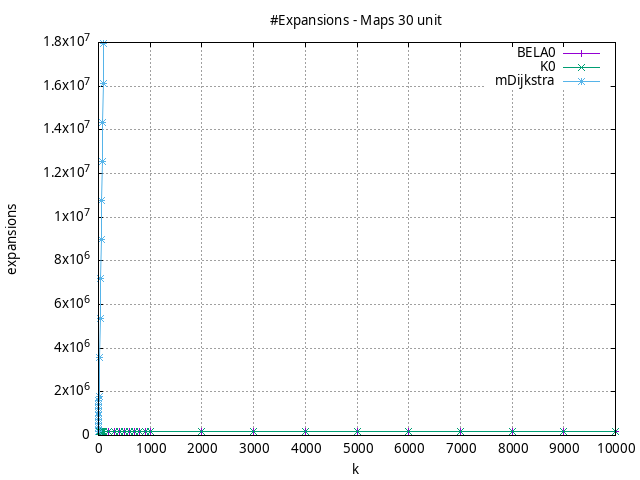}
    \end{center}
    \caption{}
    \label{fig:maps:unit:brute-force:expansions:e}
  \end{subfigure}
  \begin{subfigure}{0.3\textwidth}
    \begin{center}
        \includegraphics[width=\textwidth]{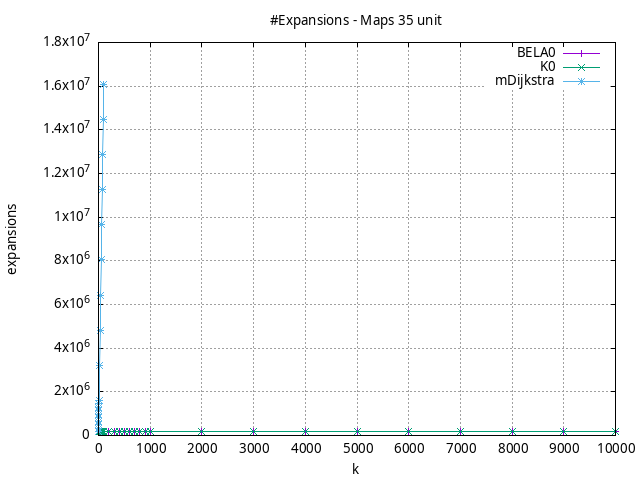}
    \end{center}
    \caption{}
    \label{fig:maps:unit:brute-force:expansions:f}
  \end{subfigure}
  \caption{Number of expansions in the maps (unit) domain with brute-force search algorithms}
  \label{fig:maps:unit:brute-force:expansions}
\end{figure*}

%% file: maps.unit.runtime.heuristic.tex
\begin{figure*}
  \centering
  \begin{subfigure}{0.3\textwidth}
    \begin{center}
        \includegraphics[width=\textwidth]{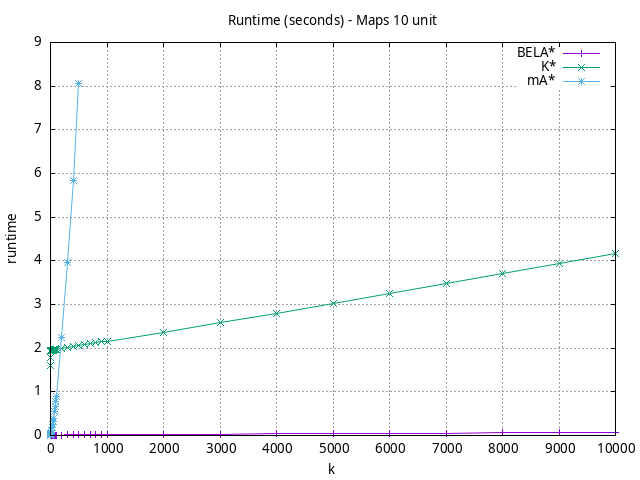}
    \end{center}
    \caption{}
    \label{fig:maps:unit:heuristic:runtime:a}
  \end{subfigure}
  \begin{subfigure}{0.3\textwidth}
    \begin{center}
        \includegraphics[width=\textwidth]{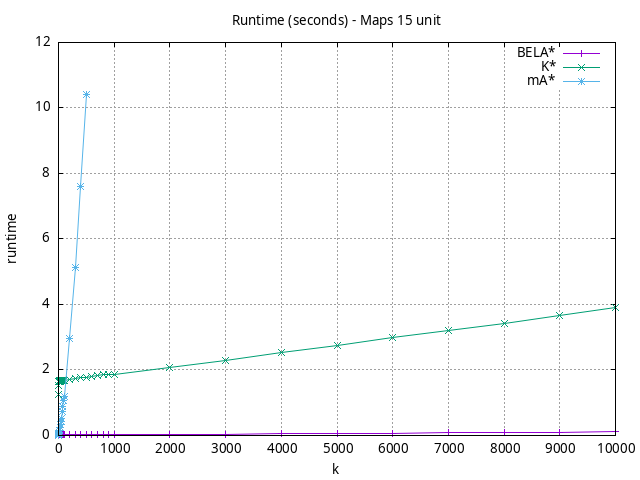}
    \end{center}
    \caption{}
    \label{fig:maps:unit:heuristic:runtime:b}
  \end{subfigure}
  \begin{subfigure}{0.3\textwidth}
    \begin{center}
        \includegraphics[width=\textwidth]{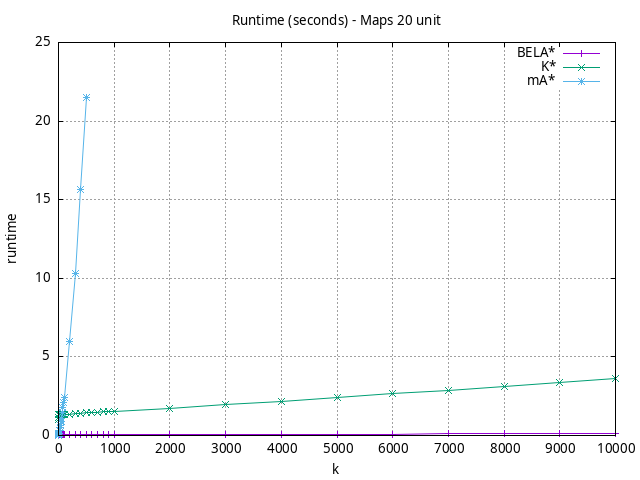}
    \end{center}
    \caption{}
    \label{fig:maps:unit:heuristic:runtime:c}
  \end{subfigure}
  \begin{subfigure}{0.3\textwidth}
    \begin{center}
        \includegraphics[width=\textwidth]{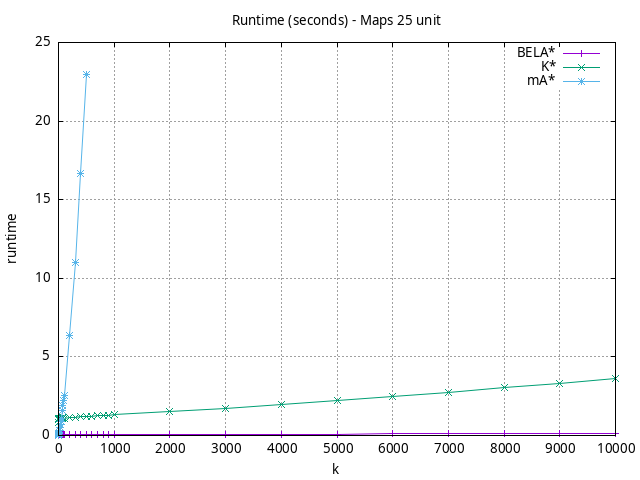}
    \end{center}
    \caption{}
    \label{fig:maps:unit:heuristic:runtime:d}
  \end{subfigure}
  \begin{subfigure}{0.3\textwidth}
    \begin{center}
        \includegraphics[width=\textwidth]{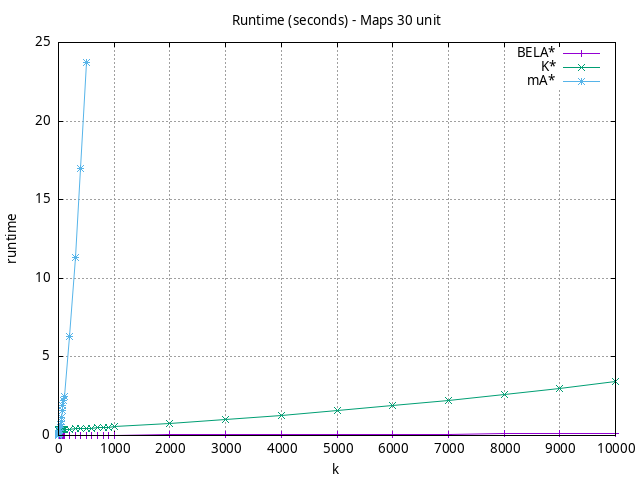}
    \end{center}
    \caption{}
    \label{fig:maps:unit:heuristic:runtime:e}
  \end{subfigure}
  \begin{subfigure}{0.3\textwidth}
    \begin{center}
        \includegraphics[width=\textwidth]{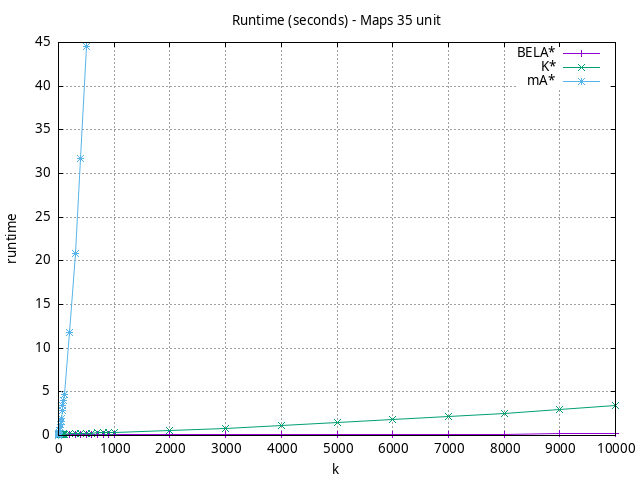}
    \end{center}
    \caption{}
    \label{fig:maps:unit:heuristic:runtime:f}
  \end{subfigure}
  \caption{Runtime (in seconds) in the maps (unit) domain with heuristic search algorithms}
  \label{fig:maps:unit:heuristic:runtime}
\end{figure*}

%% file: maps.unit.mem.heuristic.tex
\begin{figure*}
  \centering
  \begin{subfigure}{0.3\textwidth}
    \begin{center}
        \includegraphics[width=\textwidth]{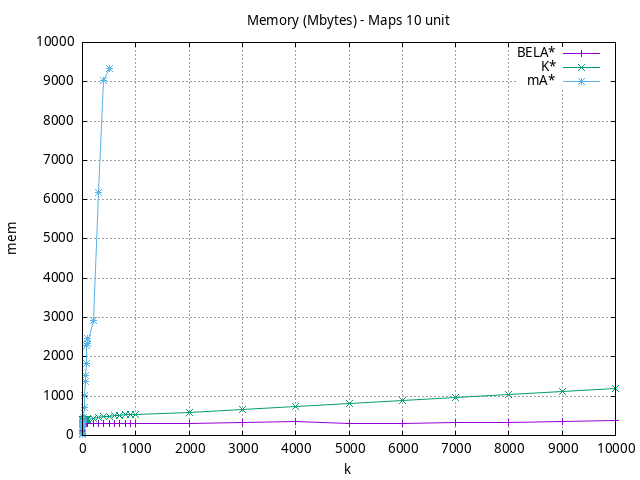}
    \end{center}
    \caption{}
    \label{fig:maps:unit:heuristic:mem:a}
  \end{subfigure}
  \begin{subfigure}{0.3\textwidth}
    \begin{center}
        \includegraphics[width=\textwidth]{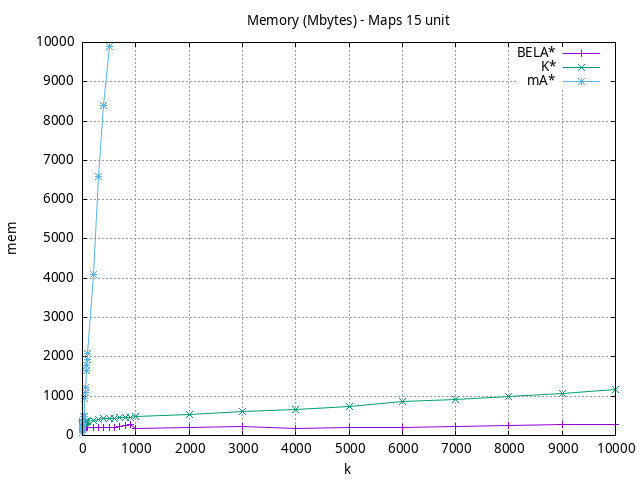}
    \end{center}
    \caption{}
    \label{fig:maps:unit:heuristic:mem:b}
  \end{subfigure}
  \begin{subfigure}{0.3\textwidth}
    \begin{center}
        \includegraphics[width=\textwidth]{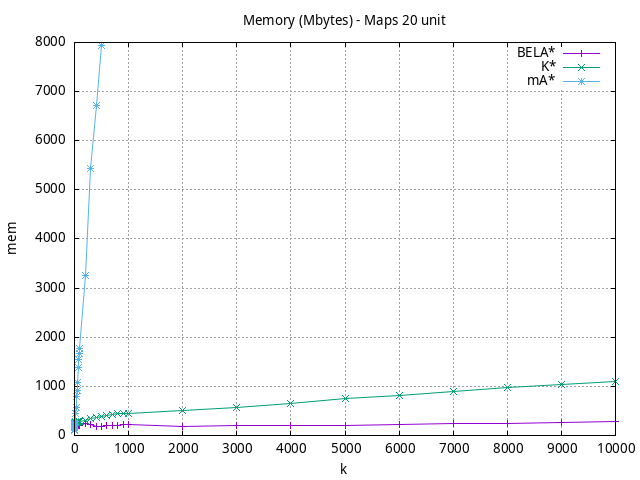}
    \end{center}
    \caption{}
    \label{fig:maps:unit:heuristic:mem:c}
  \end{subfigure}
  \begin{subfigure}{0.3\textwidth}
    \begin{center}
        \includegraphics[width=\textwidth]{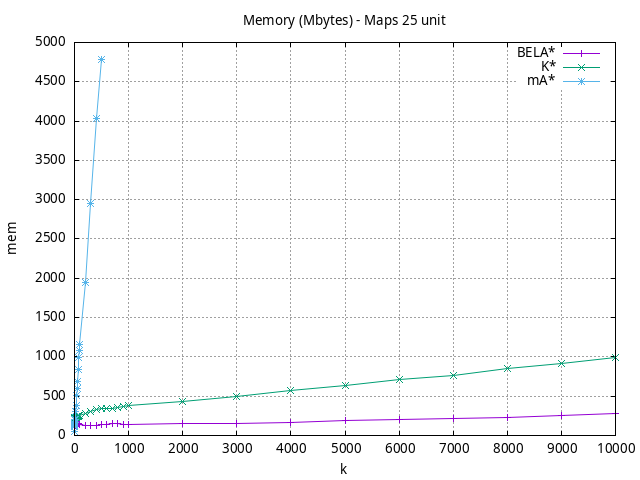}
    \end{center}
    \caption{}
    \label{fig:maps:unit:heuristic:mem:d}
  \end{subfigure}
  \begin{subfigure}{0.3\textwidth}
    \begin{center}
        \includegraphics[width=\textwidth]{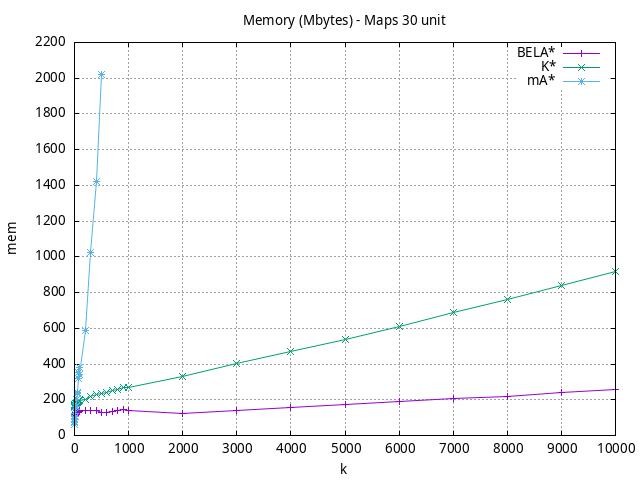}
    \end{center}
    \caption{}
    \label{fig:maps:unit:heuristic:mem:e}
  \end{subfigure}
  \begin{subfigure}{0.3\textwidth}
    \begin{center}
        \includegraphics[width=\textwidth]{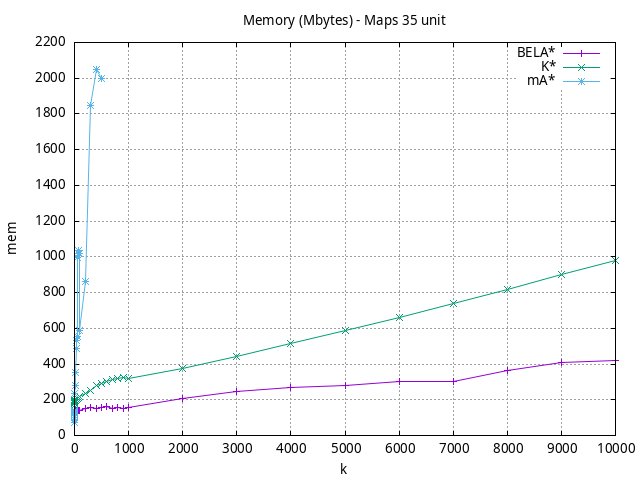}
    \end{center}
    \caption{}
    \label{fig:maps:unit:heuristic:mem:f}
  \end{subfigure}
  \caption{Memory usage (in Mbytes) in the maps (unit) domain with heuristic search algorithms}
  \label{fig:maps:unit:heuristic:mem}
\end{figure*}

%% file: maps.unit.expansions.heuristic.tex
\begin{figure*}
  \centering
  \begin{subfigure}{0.3\textwidth}
    \begin{center}
        \includegraphics[width=\textwidth]{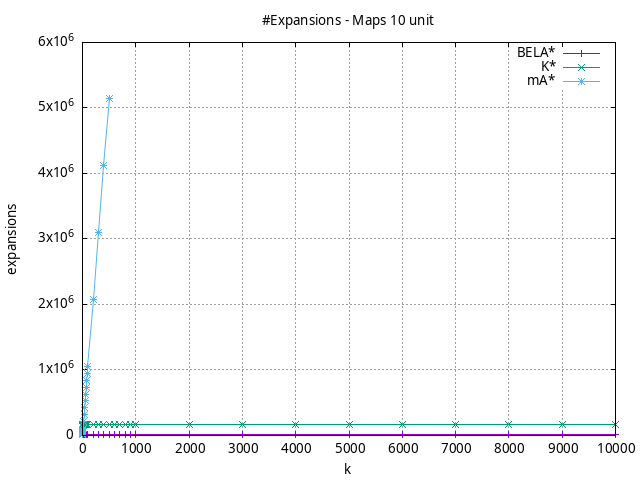}
    \end{center}
    \caption{}
    \label{fig:maps:unit:heuristic:expansions:a}
  \end{subfigure}
  \begin{subfigure}{0.3\textwidth}
    \begin{center}
        \includegraphics[width=\textwidth]{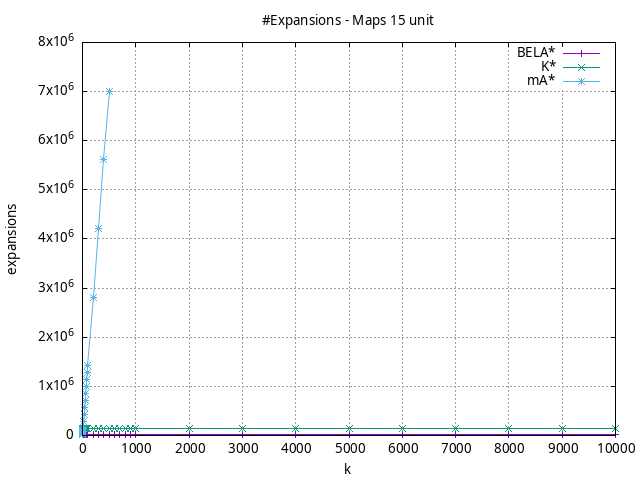}
    \end{center}
    \caption{}
    \label{fig:maps:unit:heuristic:expansions:b}
  \end{subfigure}
  \begin{subfigure}{0.3\textwidth}
    \begin{center}
        \includegraphics[width=\textwidth]{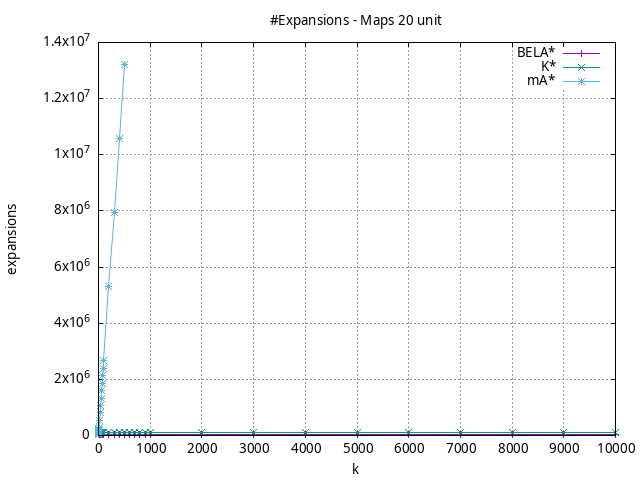}
    \end{center}
    \caption{}
    \label{fig:maps:unit:heuristic:expansions:c}
  \end{subfigure}
  \begin{subfigure}{0.3\textwidth}
    \begin{center}
        \includegraphics[width=\textwidth]{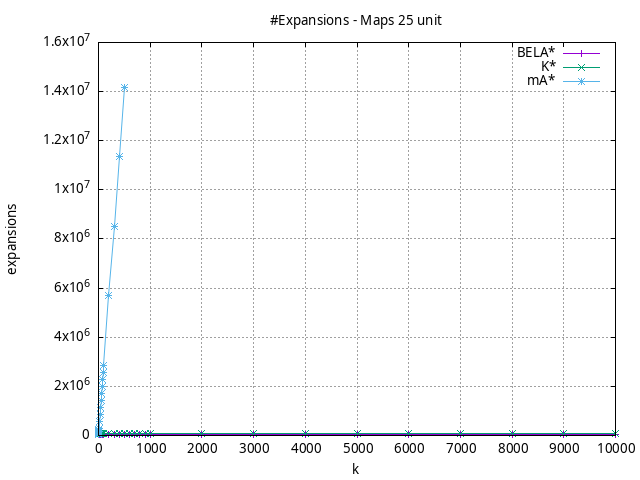}
    \end{center}
    \caption{}
    \label{fig:maps:unit:heuristic:expansions:d}
  \end{subfigure}
  \begin{subfigure}{0.3\textwidth}
    \begin{center}
        \includegraphics[width=\textwidth]{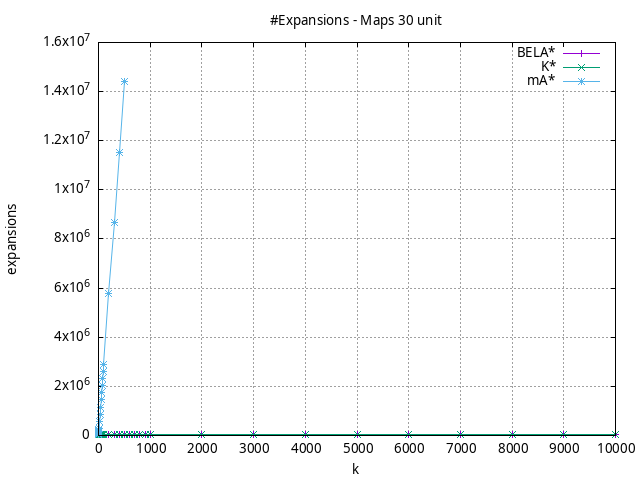}
    \end{center}
    \caption{}
    \label{fig:maps:unit:heuristic:expansions:e}
  \end{subfigure}
  \begin{subfigure}{0.3\textwidth}
    \begin{center}
        \includegraphics[width=\textwidth]{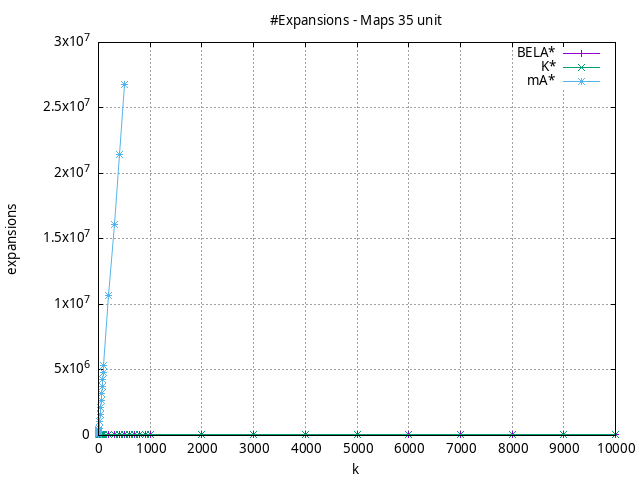}
    \end{center}
    \caption{}
    \label{fig:maps:unit:heuristic:expansions:f}
  \end{subfigure}
  \caption{Number of expansions in the maps (unit) domain with heuristic search algorithms}
  \label{fig:maps:unit:heuristic:expansions}
\end{figure*}

%% file: maps.octile.runtime.brute-force.tex
\begin{figure*}
  \centering
  \begin{subfigure}{0.3\textwidth}
    \begin{center}
        \includegraphics[width=\textwidth]{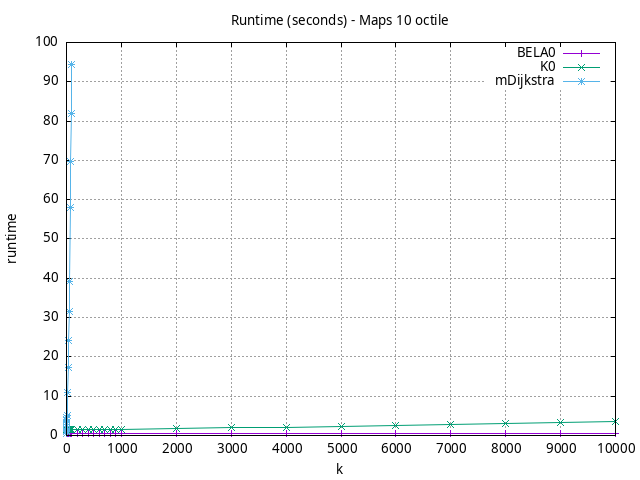}
    \end{center}
    \caption{}
    \label{fig:maps:octile:brute-force:runtime:a}
  \end{subfigure}
  \begin{subfigure}{0.3\textwidth}
    \begin{center}
        \includegraphics[width=\textwidth]{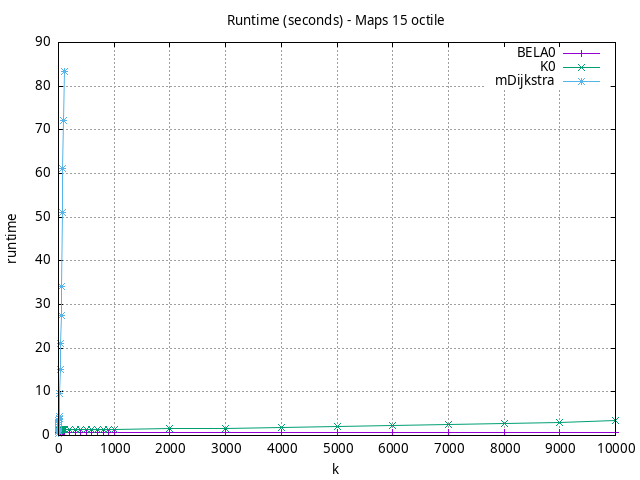}
    \end{center}
    \caption{}
    \label{fig:maps:octile:brute-force:runtime:b}
  \end{subfigure}
  \begin{subfigure}{0.3\textwidth}
    \begin{center}
        \includegraphics[width=\textwidth]{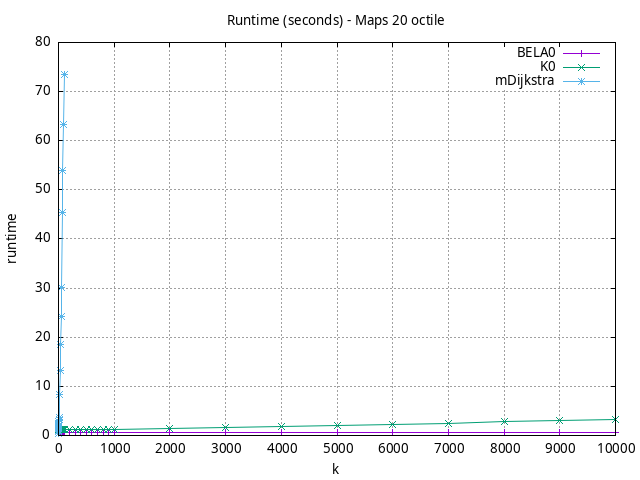}
    \end{center}
    \caption{}
    \label{fig:maps:octile:brute-force:runtime:c}
  \end{subfigure}
  \begin{subfigure}{0.3\textwidth}
    \begin{center}
        \includegraphics[width=\textwidth]{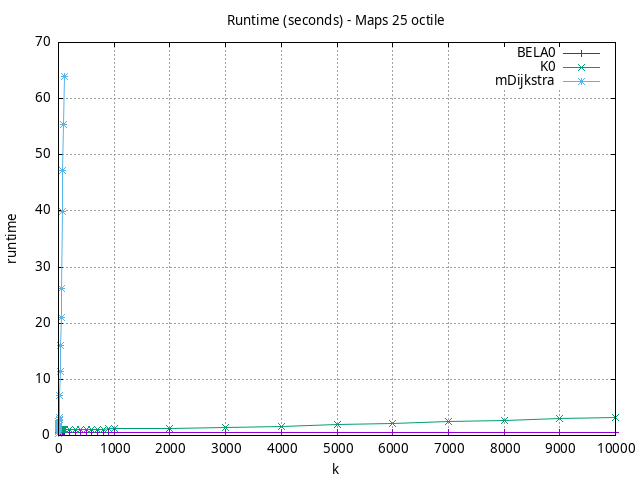}
    \end{center}
    \caption{}
    \label{fig:maps:octile:brute-force:runtime:d}
  \end{subfigure}
  \begin{subfigure}{0.3\textwidth}
    \begin{center}
        \includegraphics[width=\textwidth]{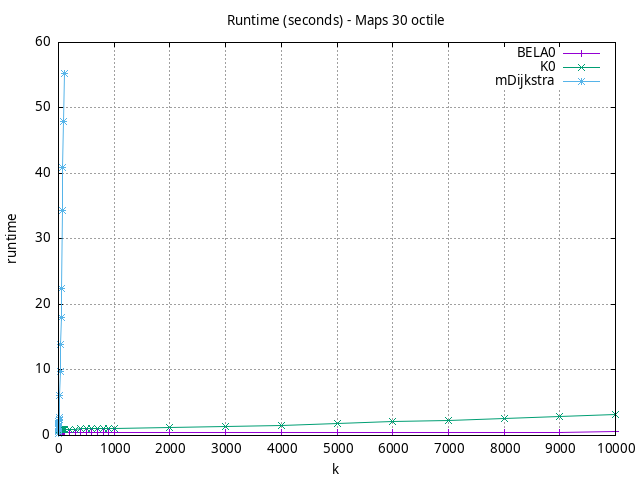}
    \end{center}
    \caption{}
    \label{fig:maps:octile:brute-force:runtime:e}
  \end{subfigure}
  \begin{subfigure}{0.3\textwidth}
    \begin{center}
        \includegraphics[width=\textwidth]{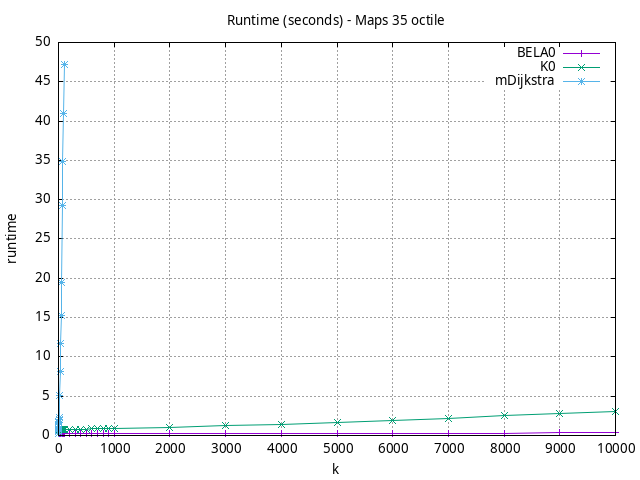}
    \end{center}
    \caption{}
    \label{fig:maps:octile:brute-force:runtime:f}
  \end{subfigure}
  \caption{Runtime (in seconds) in the maps (octile) domain with brute-force search algorithms}
  \label{fig:maps:octile:brute-force:runtime}
\end{figure*}

%% file: maps.octile.mem.brute-force.tex
\begin{figure*}
  \centering
  \begin{subfigure}{0.3\textwidth}
    \begin{center}
        \includegraphics[width=\textwidth]{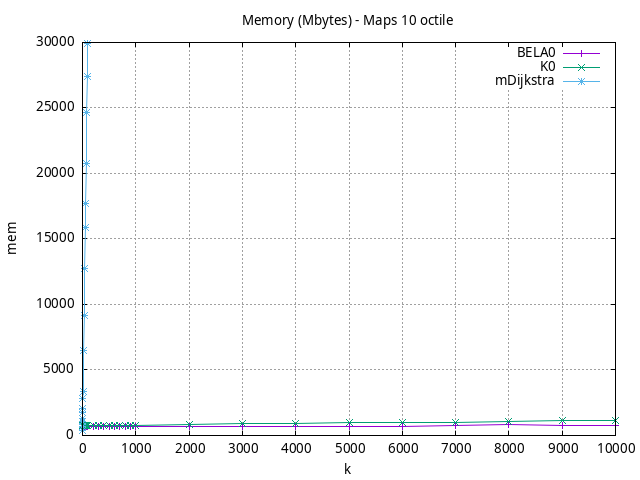}
    \end{center}
    \caption{}
    \label{fig:maps:octile:brute-force:mem:a}
  \end{subfigure}
  \begin{subfigure}{0.3\textwidth}
    \begin{center}
        \includegraphics[width=\textwidth]{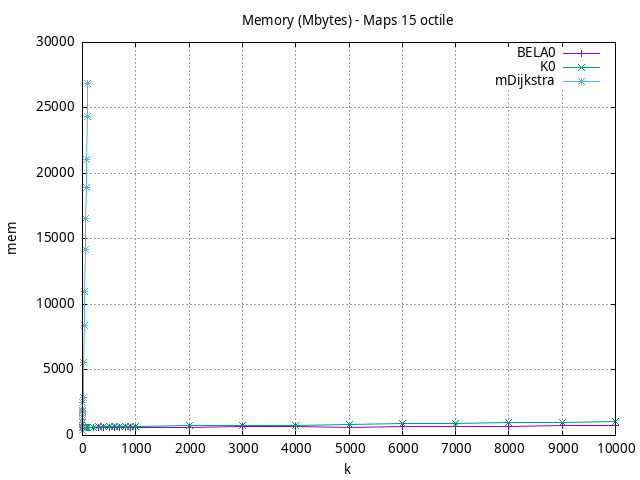}
    \end{center}
    \caption{}
    \label{fig:maps:octile:brute-force:mem:b}
  \end{subfigure}
  \begin{subfigure}{0.3\textwidth}
    \begin{center}
        \includegraphics[width=\textwidth]{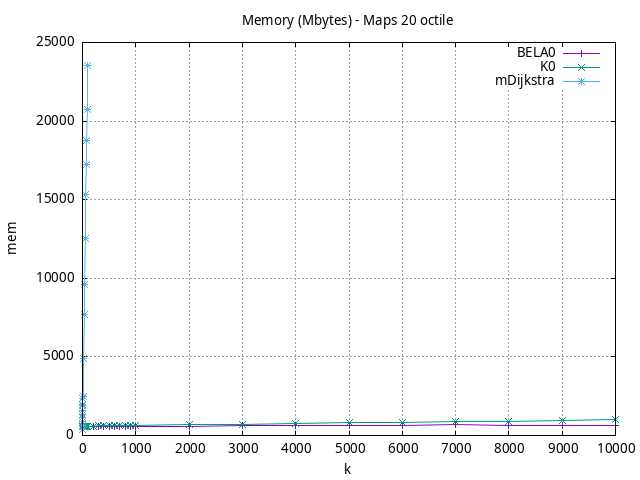}
    \end{center}
    \caption{}
    \label{fig:maps:octile:brute-force:mem:c}
  \end{subfigure}
  \begin{subfigure}{0.3\textwidth}
    \begin{center}
        \includegraphics[width=\textwidth]{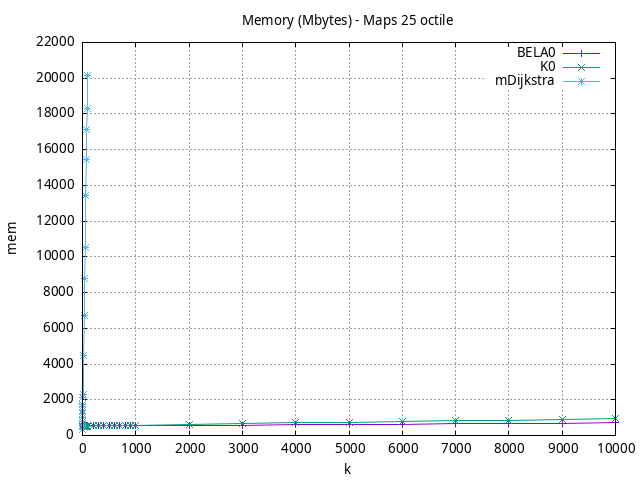}
    \end{center}
    \caption{}
    \label{fig:maps:octile:brute-force:mem:d}
  \end{subfigure}
  \begin{subfigure}{0.3\textwidth}
    \begin{center}
        \includegraphics[width=\textwidth]{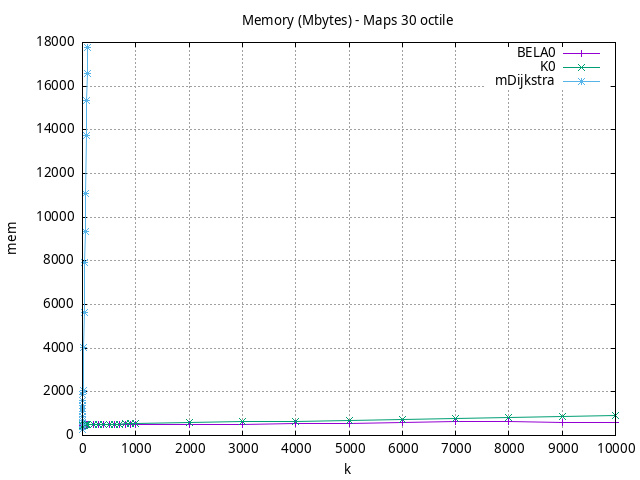}
    \end{center}
    \caption{}
    \label{fig:maps:octile:brute-force:mem:e}
  \end{subfigure}
  \begin{subfigure}{0.3\textwidth}
    \begin{center}
        \includegraphics[width=\textwidth]{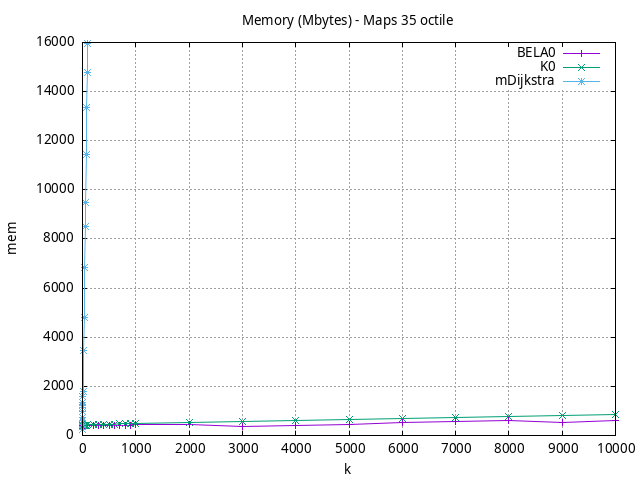}
    \end{center}
    \caption{}
    \label{fig:maps:octile:brute-force:mem:f}
  \end{subfigure}
  \caption{Memory usage (in Mbytes) in the maps (octile) domain with brute-force search algorithms}
  \label{fig:maps:octile:brute-force:mem}
\end{figure*}

%% file: maps.octile.expansions.brute-force.tex
\begin{figure*}
  \centering
  \begin{subfigure}{0.3\textwidth}
    \begin{center}
        \includegraphics[width=\textwidth]{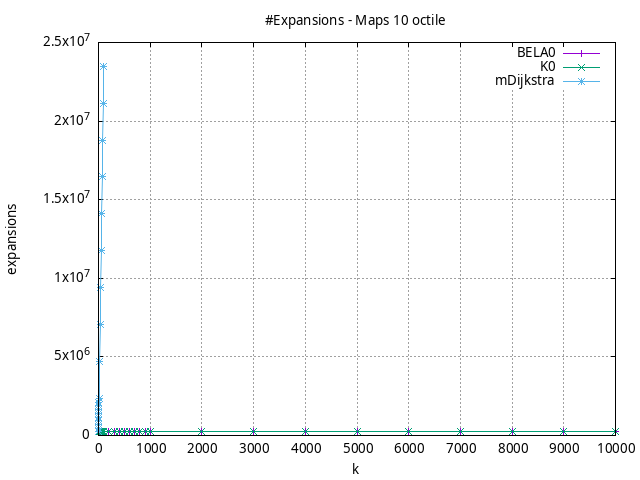}
    \end{center}
    \caption{}
    \label{fig:maps:octile:brute-force:expansions:a}
  \end{subfigure}
  \begin{subfigure}{0.3\textwidth}
    \begin{center}
        \includegraphics[width=\textwidth]{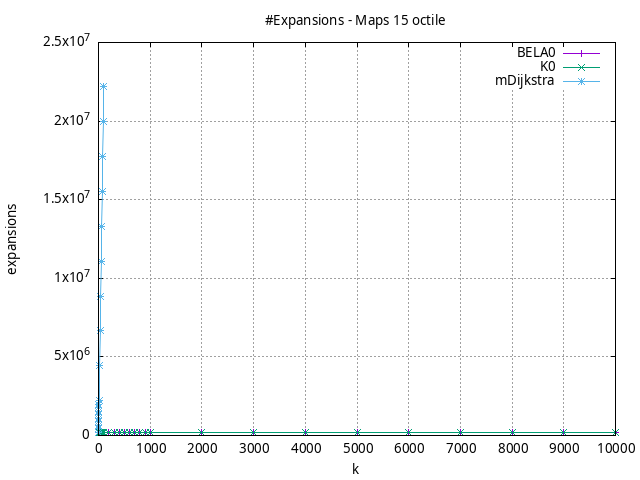}
    \end{center}
    \caption{}
    \label{fig:maps:octile:brute-force:expansions:b}
  \end{subfigure}
  \begin{subfigure}{0.3\textwidth}
    \begin{center}
        \includegraphics[width=\textwidth]{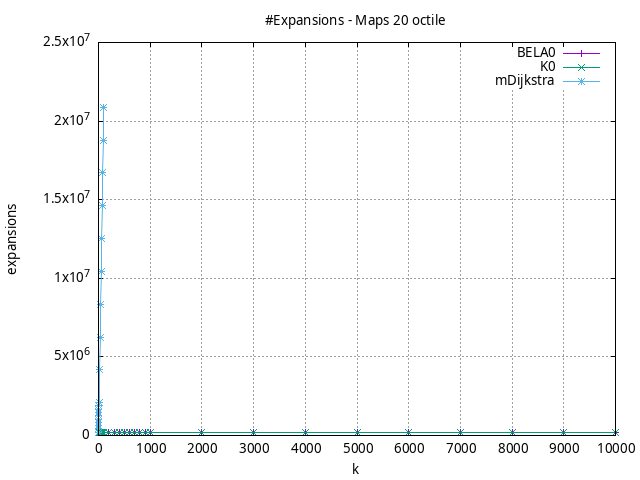}
    \end{center}
    \caption{}
    \label{fig:maps:octile:brute-force:expansions:c}
  \end{subfigure}
  \begin{subfigure}{0.3\textwidth}
    \begin{center}
        \includegraphics[width=\textwidth]{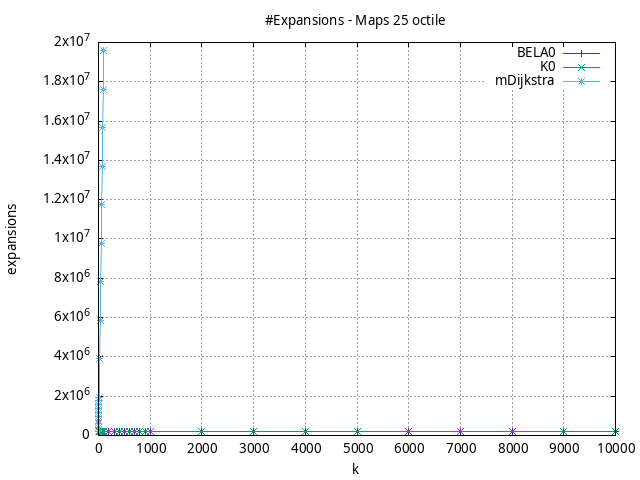}
    \end{center}
    \caption{}
    \label{fig:maps:octile:brute-force:expansions:d}
  \end{subfigure}
  \begin{subfigure}{0.3\textwidth}
    \begin{center}
        \includegraphics[width=\textwidth]{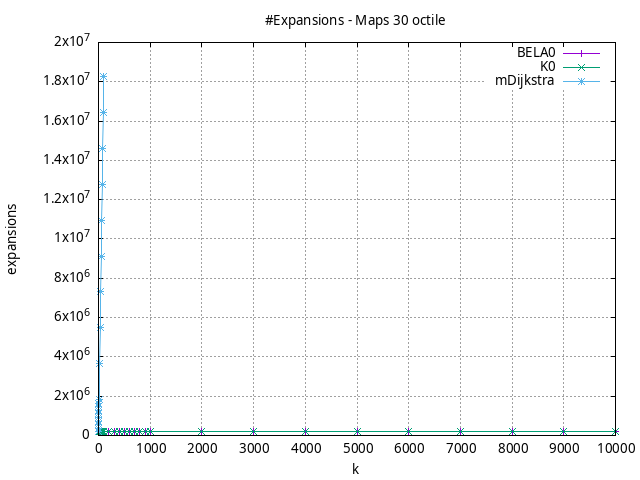}
    \end{center}
    \caption{}
    \label{fig:maps:octile:brute-force:expansions:e}
  \end{subfigure}
  \begin{subfigure}{0.3\textwidth}
    \begin{center}
        \includegraphics[width=\textwidth]{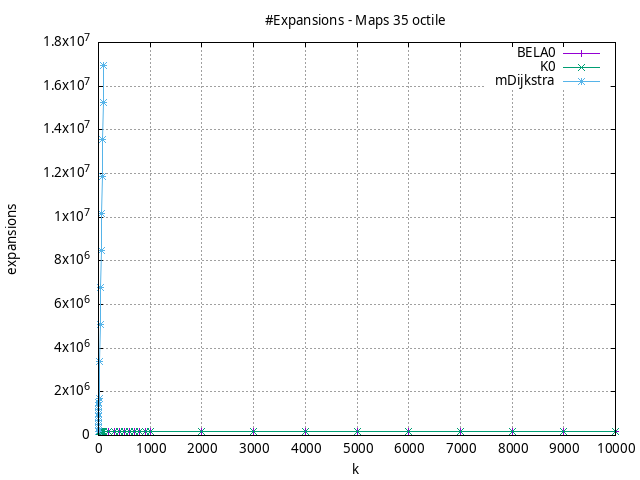}
    \end{center}
    \caption{}
    \label{fig:maps:octile:brute-force:expansions:f}
  \end{subfigure}
  \caption{Number of expansions in the maps (octile) domain with brute-force search algorithms}
  \label{fig:maps:octile:brute-force:expansions}
\end{figure*}

%% file: maps.octile.runtime.heuristic.tex
\begin{figure*}
  \centering
  \begin{subfigure}{0.3\textwidth}
    \begin{center}
        \includegraphics[width=\textwidth]{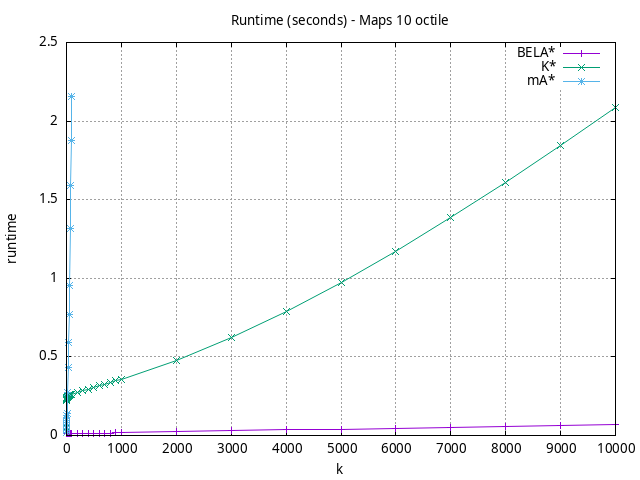}
    \end{center}
    \caption{}
    \label{fig:maps:octile:heuristic:runtime:a}
  \end{subfigure}
  \begin{subfigure}{0.3\textwidth}
    \begin{center}
        \includegraphics[width=\textwidth]{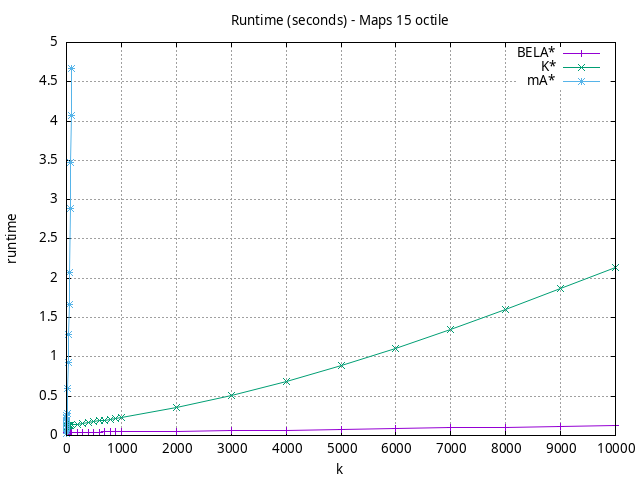}
    \end{center}
    \caption{}
    \label{fig:maps:octile:heuristic:runtime:b}
  \end{subfigure}
  \begin{subfigure}{0.3\textwidth}
    \begin{center}
        \includegraphics[width=\textwidth]{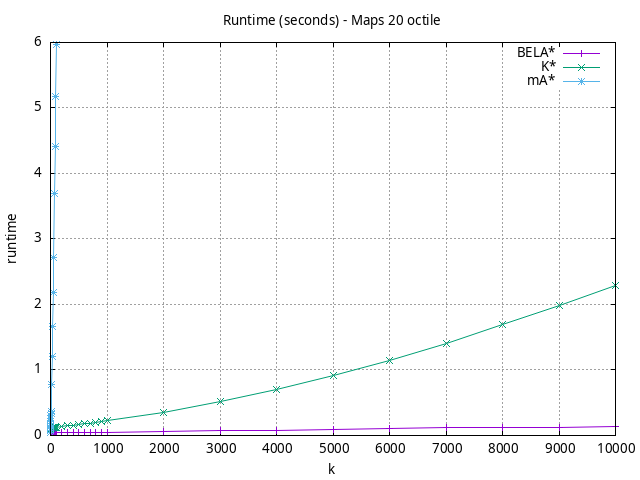}
    \end{center}
    \caption{}
    \label{fig:maps:octile:heuristic:runtime:c}
  \end{subfigure}
  \begin{subfigure}{0.3\textwidth}
    \begin{center}
        \includegraphics[width=\textwidth]{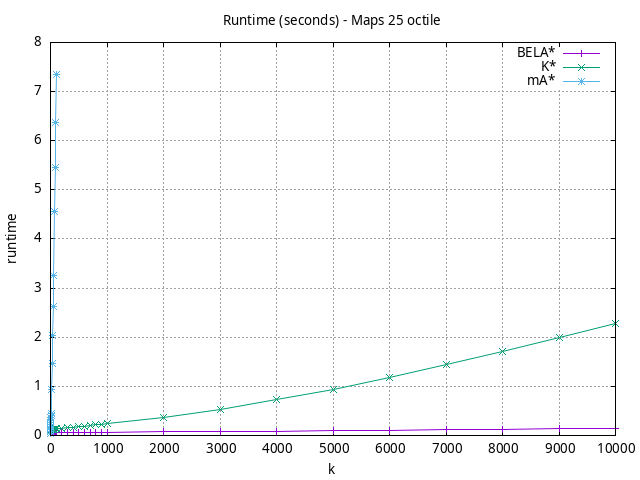}
    \end{center}
    \caption{}
    \label{fig:maps:octile:heuristic:runtime:d}
  \end{subfigure}
  \begin{subfigure}{0.3\textwidth}
    \begin{center}
        \includegraphics[width=\textwidth]{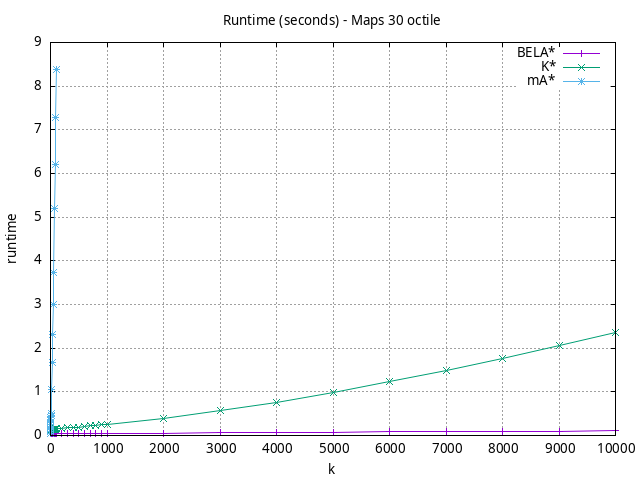}
    \end{center}
    \caption{}
    \label{fig:maps:octile:heuristic:runtime:e}
  \end{subfigure}
  \begin{subfigure}{0.3\textwidth}
    \begin{center}
        \includegraphics[width=\textwidth]{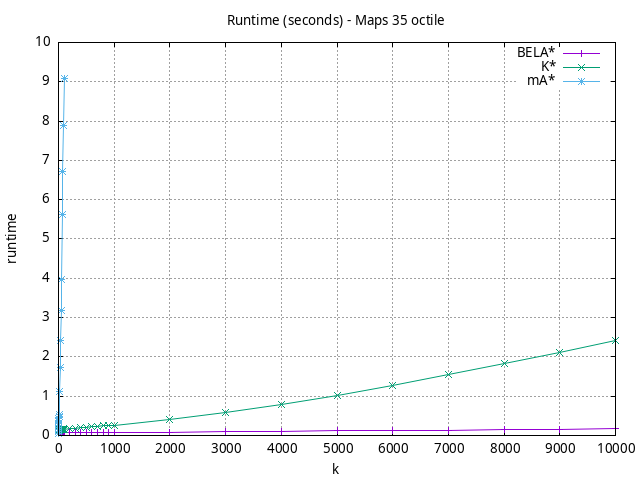}
    \end{center}
    \caption{}
    \label{fig:maps:octile:heuristic:runtime:f}
  \end{subfigure}
  \caption{Runtime (in seconds) in the maps (octile) domain with heuristic search algorithms}
  \label{fig:maps:octile:heuristic:runtime}
\end{figure*}

%% file: maps.octile.mem.heuristic.tex
\begin{figure*}
  \centering
  \begin{subfigure}{0.3\textwidth}
    \begin{center}
        \includegraphics[width=\textwidth]{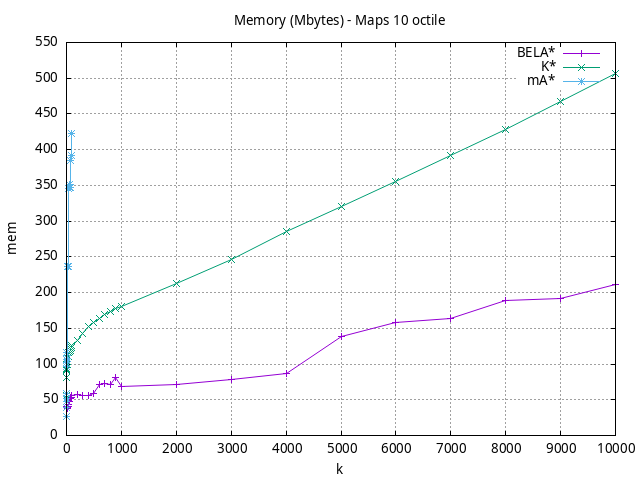}
    \end{center}
    \caption{}
    \label{fig:maps:octile:heuristic:mem:a}
  \end{subfigure}
  \begin{subfigure}{0.3\textwidth}
    \begin{center}
        \includegraphics[width=\textwidth]{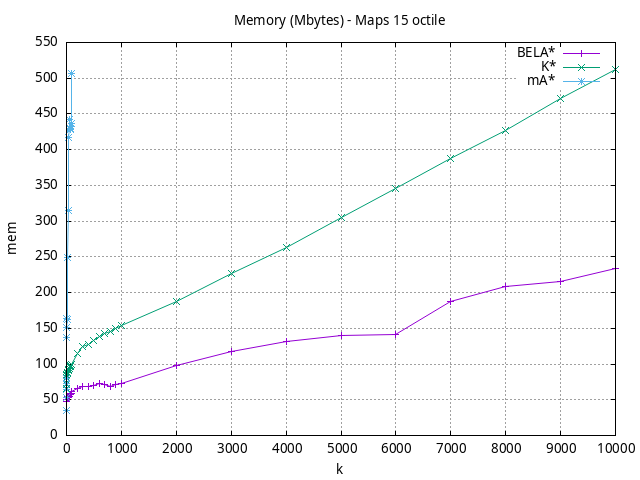}
    \end{center}
    \caption{}
    \label{fig:maps:octile:heuristic:mem:b}
  \end{subfigure}
  \begin{subfigure}{0.3\textwidth}
    \begin{center}
        \includegraphics[width=\textwidth]{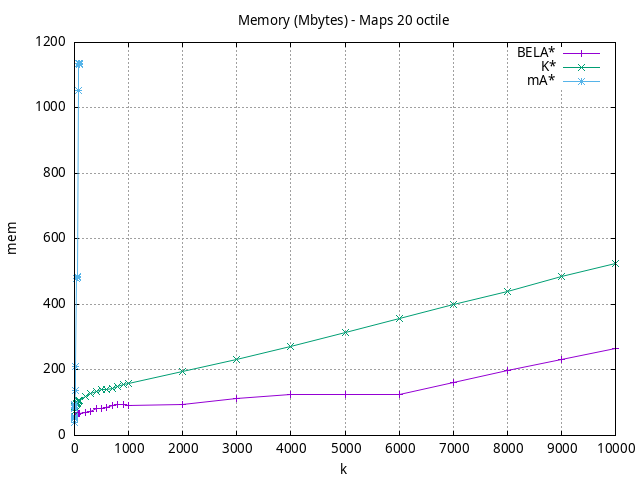}
    \end{center}
    \caption{}
    \label{fig:maps:octile:heuristic:mem:c}
  \end{subfigure}
  \begin{subfigure}{0.3\textwidth}
    \begin{center}
        \includegraphics[width=\textwidth]{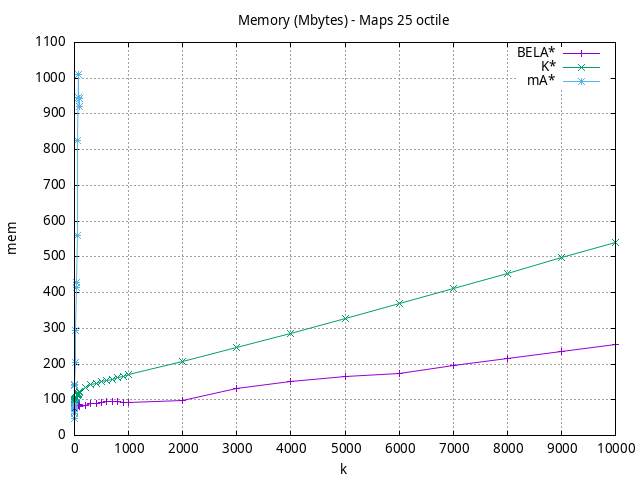}
    \end{center}
    \caption{}
    \label{fig:maps:octile:heuristic:mem:d}
  \end{subfigure}
  \begin{subfigure}{0.3\textwidth}
    \begin{center}
        \includegraphics[width=\textwidth]{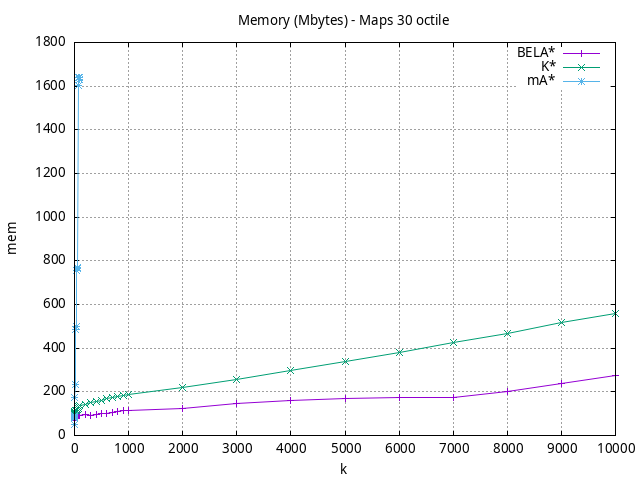}
    \end{center}
    \caption{}
    \label{fig:maps:octile:heuristic:mem:e}
  \end{subfigure}
  \begin{subfigure}{0.3\textwidth}
    \begin{center}
        \includegraphics[width=\textwidth]{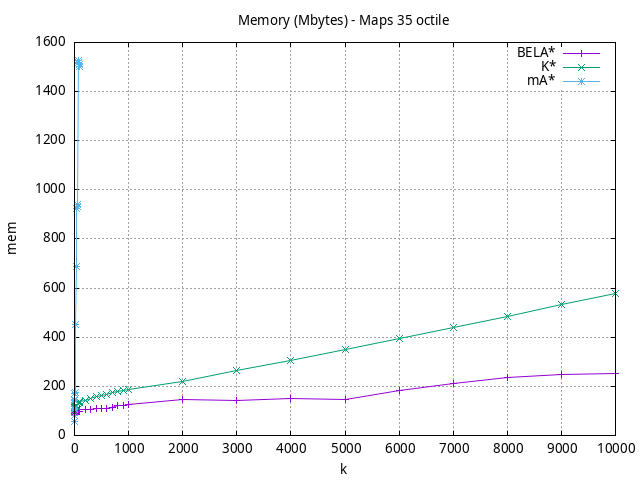}
    \end{center}
    \caption{}
    \label{fig:maps:octile:heuristic:mem:f}
  \end{subfigure}
  \caption{Memory usage (in Mbytes) in the maps (octile) domain with heuristic search algorithms}
  \label{fig:maps:octile:heuristic:mem}
\end{figure*}

%% file: maps.octile.expansions.heuristic.tex
\begin{figure*}
  \centering
  \begin{subfigure}{0.3\textwidth}
    \begin{center}
        \includegraphics[width=\textwidth]{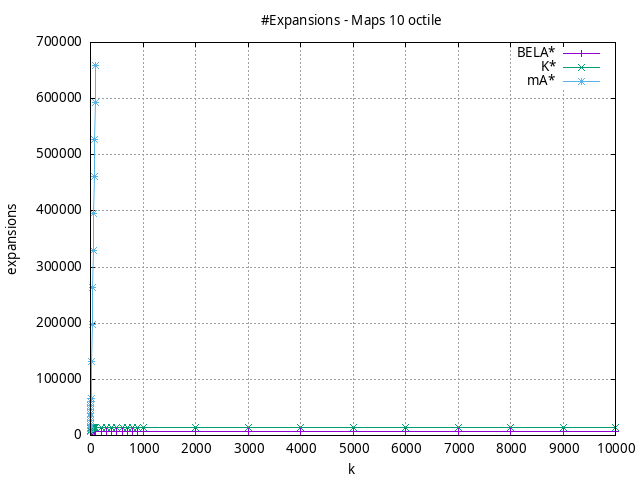}
    \end{center}
    \caption{}
    \label{fig:maps:octile:heuristic:expansions:a}
  \end{subfigure}
  \begin{subfigure}{0.3\textwidth}
    \begin{center}
        \includegraphics[width=\textwidth]{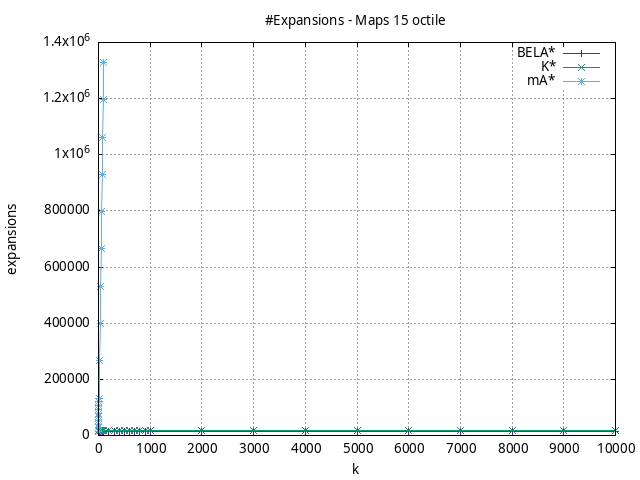}
    \end{center}
    \caption{}
    \label{fig:maps:octile:heuristic:expansions:b}
  \end{subfigure}
  \begin{subfigure}{0.3\textwidth}
    \begin{center}
        \includegraphics[width=\textwidth]{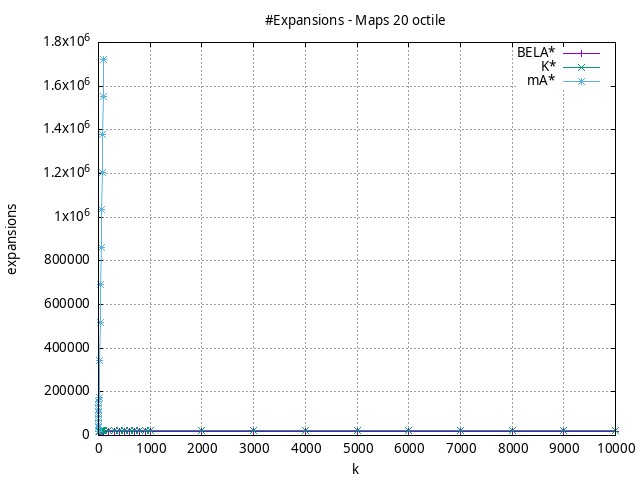}
    \end{center}
    \caption{}
    \label{fig:maps:octile:heuristic:expansions:c}
  \end{subfigure}
  \begin{subfigure}{0.3\textwidth}
    \begin{center}
        \includegraphics[width=\textwidth]{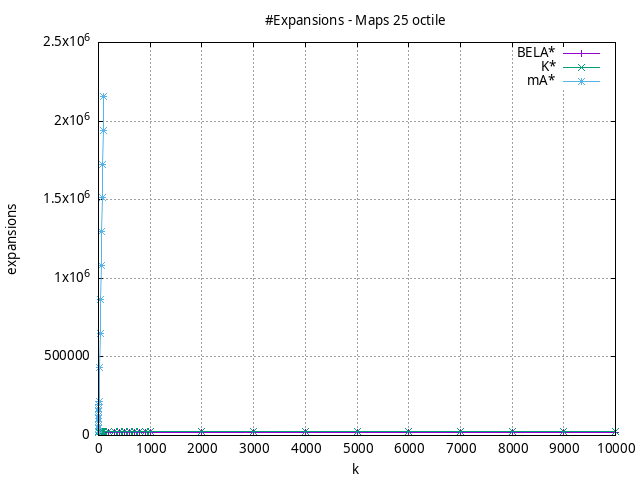}
    \end{center}
    \caption{}
    \label{fig:maps:octile:heuristic:expansions:d}
  \end{subfigure}
  \begin{subfigure}{0.3\textwidth}
    \begin{center}
        \includegraphics[width=\textwidth]{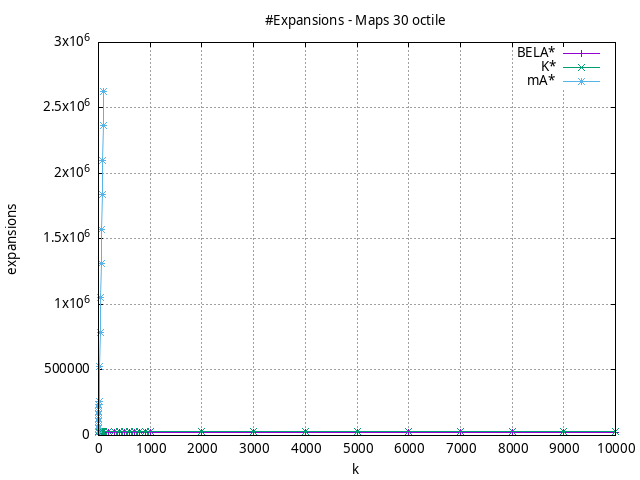}
    \end{center}
    \caption{}
    \label{fig:maps:octile:heuristic:expansions:e}
  \end{subfigure}
  \begin{subfigure}{0.3\textwidth}
    \begin{center}
        \includegraphics[width=\textwidth]{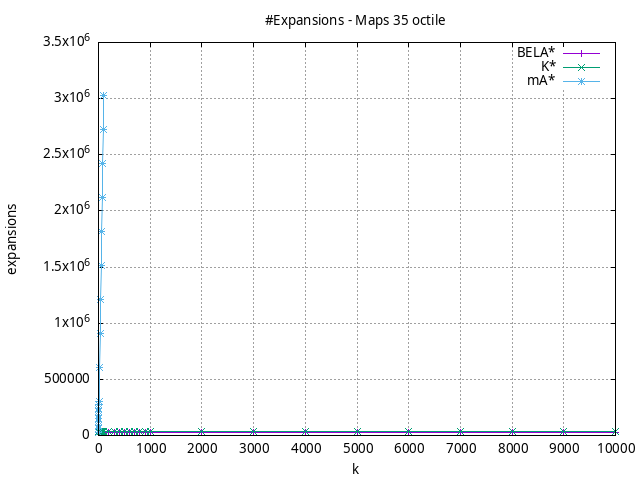}
    \end{center}
    \caption{}
    \label{fig:maps:octile:heuristic:expansions:f}
  \end{subfigure}
  \caption{Number of expansions in the maps (octile) domain with heuristic search algorithms}
  \label{fig:maps:octile:heuristic:expansions}
\end{figure*}

%% file: supplemental6-npancake.tex

\subsection{$N$-Pancake}
\label{sec:empirical-evaluation:n-pancake}

The $N$-Pancake domain defines a permutation state space with a size equal to
$N!$. It is thus a significant challenge, as it also has a large
branching-factor, $N-1$. The heuristic used is the \textsc{gap}
heuristic~\parencite{DBLP:conf/socs/Helmert10}, which is known to be very well
informed in the unit variant discussed next. This allows current state-of-the-art
solvers to solve instances of the 60-Pancake in less than 30 seconds on average
per instance. In all cases, 100 instances were randomly generated and only those
instances where the heuristic distance between the start state and the goal
state was greater than or equal to $(N-2)$ were accepted. All of the points in the following plots
have been averaged over 100 runs each. This domain has never been used, to the
best of the authors' knowledge, as a testbed for algorithms solving the $\kappa$
shortest path problem.

\subsubsection{Unit variant}
\label{sec:empirical-evaluation:n-pancake:unit-variant}

The unit variant is the classic version of the $N$-Pancake
problem~\parencite{dweighter.h:problem}, where an arbitrary permutation of the symbols
$\{1, \ldots, N\}$ has to be transformed into the identity permutation by
performing prefix reversals, all of which have cost 1.

The brute-force variants of the algorithms under consideration were only tested
on the 10-Pancake, because this state space is big enough for them, with
3,628,800 different states. The results are shown in
Figures~\ref{fig:n-pancake:unit:brute-force:runtime}--\ref{fig:n-pancake:unit:brute-force:expansions}.
Only \bfbela{} was able to find 10 different paths in less than 25 seconds on
average. In this domain, mDijkstra performed significantly better than \bfk{},
but only for very low values of $\kappa$. In fact, mDijkstra was requested to
find only $\kappa=3$ different solutions, because finding a fourth solution
exhausts the available memory for some instances. \bfk{} performed much worse
than \bfbela{}, doubling the number of expansions, being almost five times
slower in the end, and taking also five times more memory than it.

Experiments using the \textsc{gap} heuristic were particularly interesting.
\kstar{} is indeed the worst algorithm in this domain. For example, in the
20-Pancake (see Figure~\ref{fig:n-pancake:unit:heuristic:runtime:a}) it takes a
huge amount of time for finding only $\kappa=10$ paths, while \mAstar{} and
\bela{} can find up to 1,000 solutions in much less time. Indeed, both \mAstar{}
and \bela{} are already two orders of magnitude faster with $\kappa=10$, the
maximum value attempted with \kstar{}. In the end, \bela{} is one order of
mangitude faster for finding two orders of magnitude more solutions. This
degradation in the running time of \kstar{} is attributed to two different
factors: On one hand, the large branching factor which forces \kstar{} to spend
much more time updating and maintaining its path graph; secondly, it expands
significantly more nodes than \bela{}, which is likely caused by the swapping
criterion used. For the first time, \mAstar{} seems to be competitive with
\bela{}, even if it consistently performs worse than it over all values of
$\kappa$. This behavior is due to the accuracy of the heuristic function.
Observing the results in the 30 and 40-Pancake (see
Figures~\ref{fig:n-pancake:unit:heuristic:runtime:b}
and~\ref{fig:n-pancake:unit:heuristic:runtime:c}) we can see that that the
difference between \mAstar and \bela{} increases with the growth of $\kappa$.
\mAstar{} seems to be particularly competitive with \bela{} in the 40-pancake
where the latter is roughly 15\% faster only. However, \mAstar{} takes more
memory in this domain (see Figure~\ref{fig:n-pancake:unit:heuristic:mem:c}), and
it exhausts all the available memory with $\kappa=30$ while \bela{} is able to
find solutions up to $\kappa=40$.

Note that in the 30 and 40-Pancake only values for
$\kappa=1$ are given for \kstar{}, which are one order of magnitude worse than
the runtime of the other algorithms.

\input{n-pancake.unit.runtime.brute-force.tex}
\input{n-pancake.unit.mem.brute-force.tex}
\input{n-pancake.unit.expansions.brute-force.tex}

\input{n-pancake.unit.runtime.heuristic.tex}
\input{n-pancake.unit.mem.heuristic.tex}
\input{n-pancake.unit.expansions.heuristic.tex}

\subsubsection{Heavy-cost variant}
\label{sec:empirical-evaluation:n-pancake:heavy-cost-variant}

In the heavy-cost variant, the cost of each prefix reversal is defined as the
size of the disc that becomes first in the permutation after the reversal. This
variant is much harder than the unit version, because the \textsc{gap} heuristic
is not so well informed now, even if a weighted version of the \textsc{gap}
heuristic is being used. In the weighted variant of the \textsc{gap} heuristic,
each gap gets weighted by the size of the smaller disc adjacent to it. As a
result of its hardness, experiments in the octile variant of the $N$-Pancake
were conducted with 32Gb of RAM memory.

Figures~\ref{fig:n-pancake:heavy-cost:brute-force:runtime}--\ref{fig:n-pancake:heavy-cost:brute-force:expansions}
show the results using the brute-force search algorithms. As before, only the
10-Pancake was tested. As shown in
Figure~\ref{fig:n-pancake:heavy-cost:brute-force:runtime:a} \bfbela{} takes an
average time slightly above 30 seconds to find $\kappa=10$ solutions, whereas
mDijkstra can solve instances only with $\kappa\leq 2$ with a much worse average
time than \bfbela{} for $\kappa=2$; \bfk{} behaves also much worse than
\bfbela{} even if it manages to find solutions with up to $\kappa=5$.
Figure~\ref{fig:n-pancake:heavy-cost:brute-force:runtime:a} clearly shows a
trend where \bfbela{} outperforms its contenders by a large margin in running
time. As in the unit variant, \bfk{} expands significantly more nodes than
\bfbela{}. It is important to remark that this figure indicates a general trend
observed in most experiments throughout all domains, this is, that mDijkstra
with $\kappa=1$ is faster than \bfbela{}. This is not surprising at all, since
mDijkstra with $\kappa=1$ becomes vanilla Dijkstra with no significant overhead,
whereas both \bfbela{} and \bfk{} have an overhead necessary for efficiently
solving problems with larger values of $\kappa$. Namely, the runtime cost
originating from maintaining the closed lists in both algorithms. Nevertheless,
just like most experiments conducted, mDijkstra (and also \mAstar) becomes
immediately worse than \bfbela{} (and \bela{}, respectively) for low values of
$\kappa$, even just 2, as shown in
Figure~\ref{fig:n-pancake:heavy-cost:brute-force:runtime:a}.

As a consequence of the degradation in performance of the heuristic function,
experiments with the heavy-cost variant with the informed versions of all
algorithms for values of $N$ larger than 10 took too long and, in many cases
memory was exhausted. For this reason, only experiments in the 10-Pancake were
conducted, though with a larger value of $\kappa$, 100.
Figures~\ref{fig:n-pancake:heavy-cost:heuristic:runtime}--\ref{fig:n-pancake:heavy-cost:heuristic:expansions}
show a dramatic difference in performance. Again, \bela{} is one order of
magnitude faster for finding up to one order of magnitude more solutions when
compared to either \mAstar{} or \kstar{}: While \bela{} finds 100 different
solutions in less than a second on average, both \mAstar{} and \kstar{} take 3
and almost 5 seconds each on average respectively, for computing only 10
solutions.

\input{n-pancake.heavy-cost.runtime.brute-force.tex}
\input{n-pancake.heavy-cost.mem.brute-force.tex}
\input{n-pancake.heavy-cost.expansions.brute-force.tex}

\input{n-pancake.heavy-cost.runtime.heuristic.tex}
\input{n-pancake.heavy-cost.mem.heuristic.tex}
\input{n-pancake.heavy-cost.expansions.heuristic.tex}


%% file: n-pancake.unit.runtime.brute-force.tex
\begin{figure*}
  \centering
  \begin{subfigure}{0.3\textwidth}
    \begin{center}
        \includegraphics[width=\textwidth]{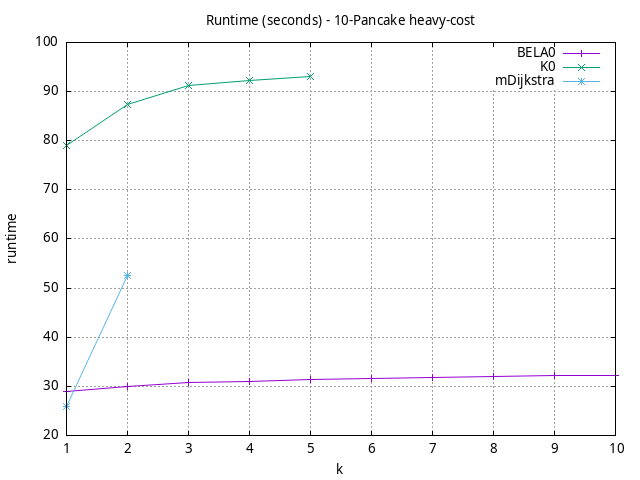}
    \end{center}
    \caption{}
    \label{fig:n-pancake:unit:brute-force:runtime:a}
  \end{subfigure}
  \caption{Runtime (in seconds) in the n-pancake (unit) domain with brute-force search algorithms}
  \label{fig:n-pancake:unit:brute-force:runtime}
\end{figure*}

%% file: n-pancake.unit.mem.brute-force.tex
\begin{figure*}
  \centering
  \begin{subfigure}{0.3\textwidth}
    \begin{center}
        \includegraphics[width=\textwidth]{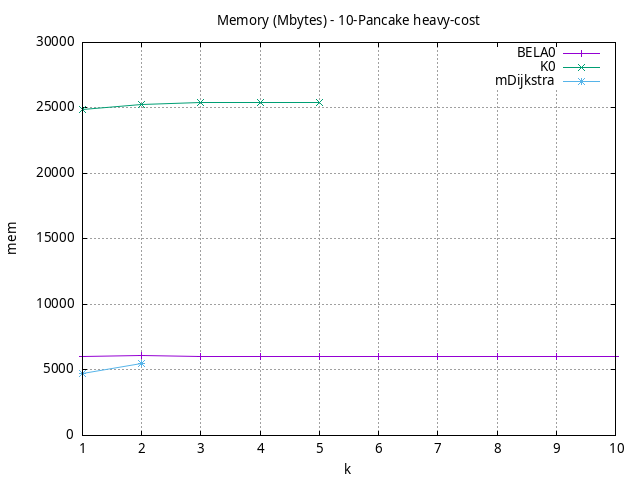}
    \end{center}
    \caption{}
    \label{fig:n-pancake:unit:brute-force:mem:a}
  \end{subfigure}
  \caption{Memory usage (in Mbytes) in the n-pancake (unit) domain with brute-force search algorithms}
  \label{fig:n-pancake:unit:brute-force:mem}
\end{figure*}

%% file: n-pancake.unit.expansions.brute-force.tex
\begin{figure*}
  \centering
  \begin{subfigure}{0.3\textwidth}
    \begin{center}
        \includegraphics[width=\textwidth]{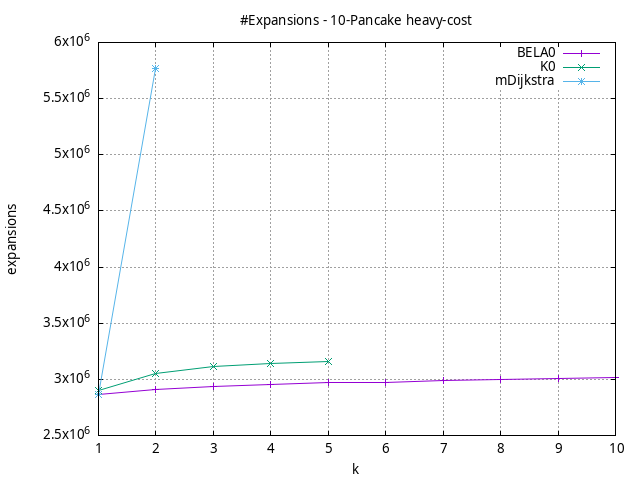}
    \end{center}
    \caption{}
    \label{fig:n-pancake:unit:brute-force:expansions:a}
  \end{subfigure}
  \caption{Number of expansions in the n-pancake (unit) domain with brute-force search algorithms}
  \label{fig:n-pancake:unit:brute-force:expansions}
\end{figure*}

%% file: n-pancake.unit.runtime.heuristic.tex
\begin{figure*}
  \centering
  \begin{subfigure}{0.3\textwidth}
    \begin{center}
        \includegraphics[width=\textwidth]{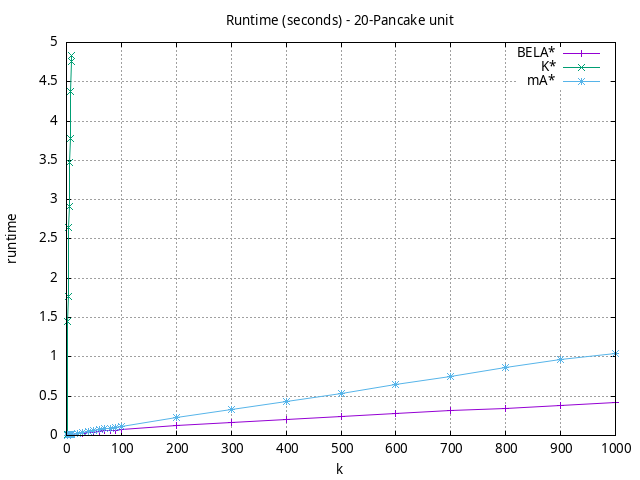}
    \end{center}
    \caption{}
    \label{fig:n-pancake:unit:heuristic:runtime:a}
  \end{subfigure}
  \begin{subfigure}{0.3\textwidth}
    \begin{center}
        \includegraphics[width=\textwidth]{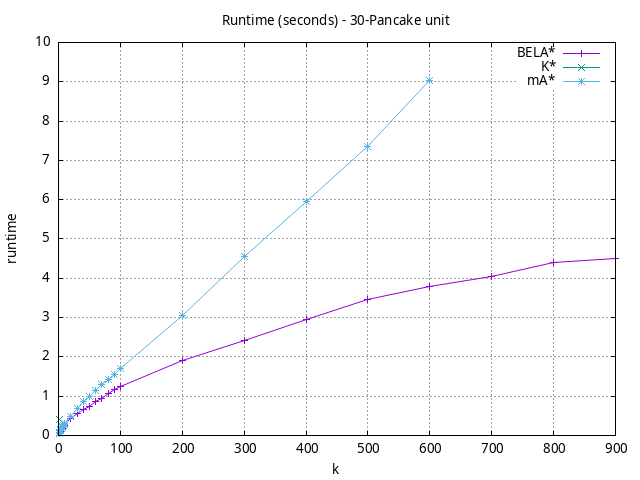}
    \end{center}
    \caption{}
    \label{fig:n-pancake:unit:heuristic:runtime:b}
  \end{subfigure}
  \begin{subfigure}{0.3\textwidth}
    \begin{center}
        \includegraphics[width=\textwidth]{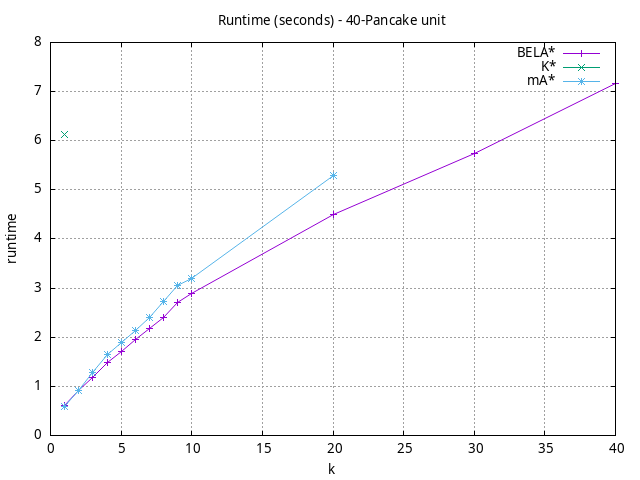}
    \end{center}
    \caption{}
    \label{fig:n-pancake:unit:heuristic:runtime:c}
  \end{subfigure}
  \caption{Runtime (in seconds) in the n-pancake (unit) domain with heuristic search algorithms}
  \label{fig:n-pancake:unit:heuristic:runtime}
\end{figure*}

%% file: n-pancake.unit.mem.heuristic.tex
\begin{figure*}
  \centering
  \begin{subfigure}{0.3\textwidth}
    \begin{center}
        \includegraphics[width=\textwidth]{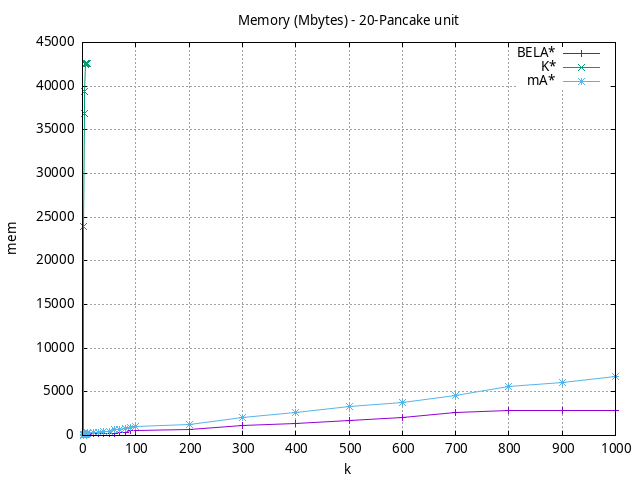}
    \end{center}
    \caption{}
    \label{fig:n-pancake:unit:heuristic:mem:a}
  \end{subfigure}
  \begin{subfigure}{0.3\textwidth}
    \begin{center}
        \includegraphics[width=\textwidth]{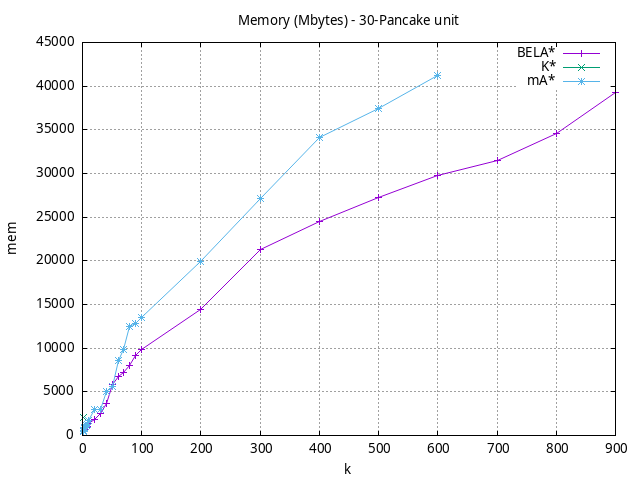}
    \end{center}
    \caption{}
    \label{fig:n-pancake:unit:heuristic:mem:b}
  \end{subfigure}
  \begin{subfigure}{0.3\textwidth}
    \begin{center}
        \includegraphics[width=\textwidth]{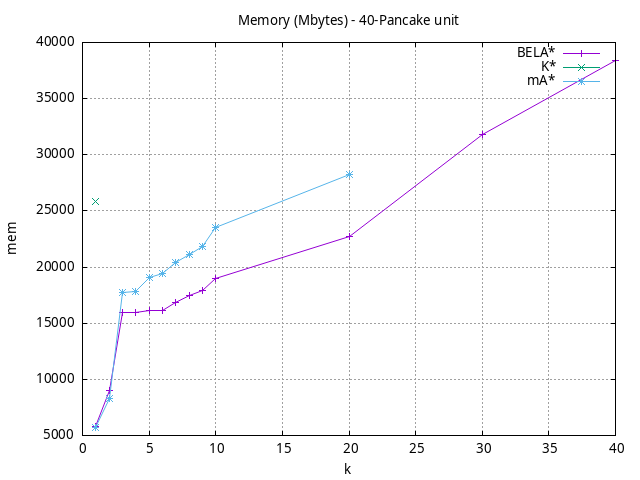}
    \end{center}
    \caption{}
    \label{fig:n-pancake:unit:heuristic:mem:c}
  \end{subfigure}
  \caption{Memory usage (in Mbytes) in the n-pancake (unit) domain with heuristic search algorithms}
  \label{fig:n-pancake:unit:heuristic:mem}
\end{figure*}

%% file: n-pancake.unit.expansions.heuristic.tex
\begin{figure*}
  \centering
  \begin{subfigure}{0.3\textwidth}
    \begin{center}
        \includegraphics[width=\textwidth]{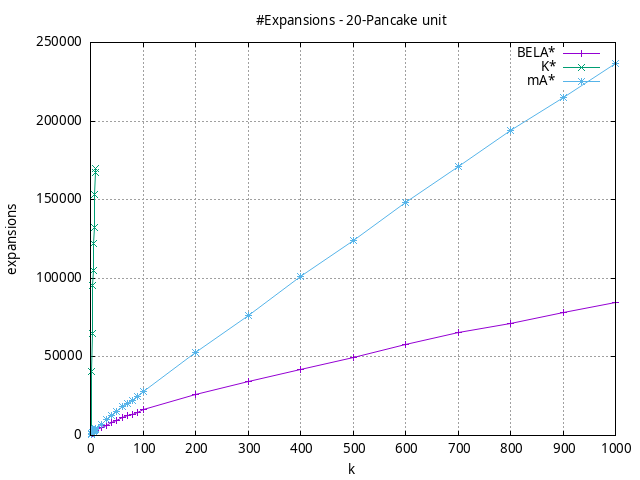}
    \end{center}
    \caption{}
    \label{fig:n-pancake:unit:heuristic:expansions:a}
  \end{subfigure}
  \begin{subfigure}{0.3\textwidth}
    \begin{center}
        \includegraphics[width=\textwidth]{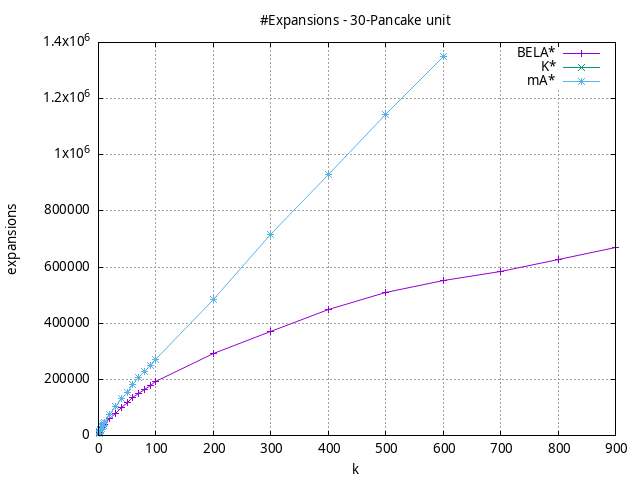}
    \end{center}
    \caption{}
    \label{fig:n-pancake:unit:heuristic:expansions:b}
  \end{subfigure}
  \begin{subfigure}{0.3\textwidth}
    \begin{center}
        \includegraphics[width=\textwidth]{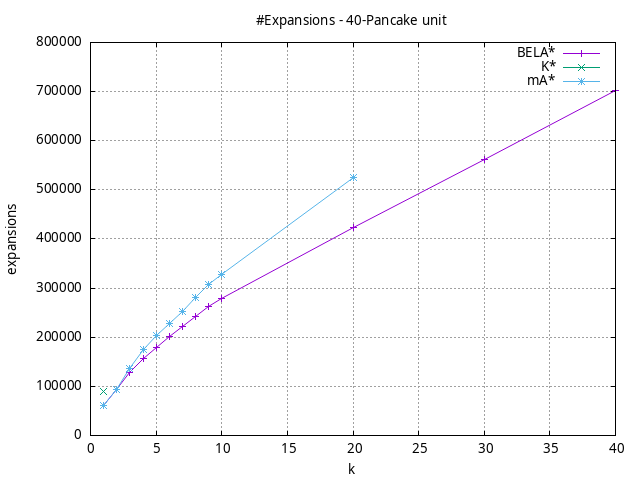}
    \end{center}
    \caption{}
    \label{fig:n-pancake:unit:heuristic:expansions:c}
  \end{subfigure}
  \caption{Number of expansions in the n-pancake (unit) domain with heuristic search algorithms}
  \label{fig:n-pancake:unit:heuristic:expansions}
\end{figure*}

%% file: n-pancake.heavy-cost.runtime.brute-force.tex
\begin{figure*}
  \centering
  \begin{subfigure}{0.3\textwidth}
    \begin{center}
        \includegraphics[width=\textwidth]{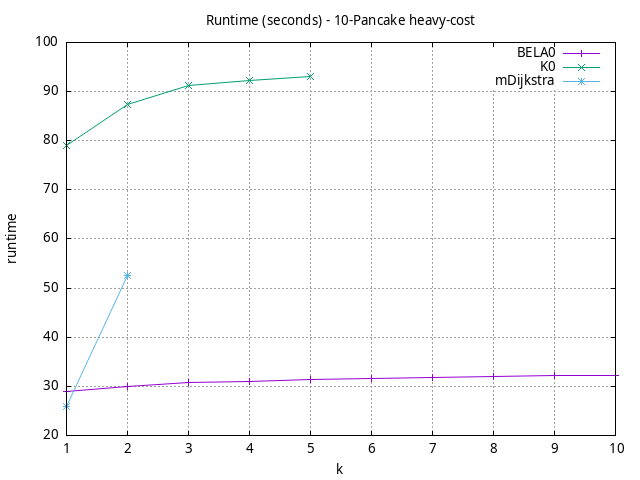}
    \end{center}
    \caption{}
    \label{fig:n-pancake:heavy-cost:brute-force:runtime:a}
  \end{subfigure}
  \caption{Runtime (in seconds) in the n-pancake (heavy-cost) domain with brute-force search algorithms}
  \label{fig:n-pancake:heavy-cost:brute-force:runtime}
\end{figure*}

%% file: n-pancake.heavy-cost.mem.brute-force.tex
\begin{figure*}
  \centering
  \begin{subfigure}{0.3\textwidth}
    \begin{center}
        \includegraphics[width=\textwidth]{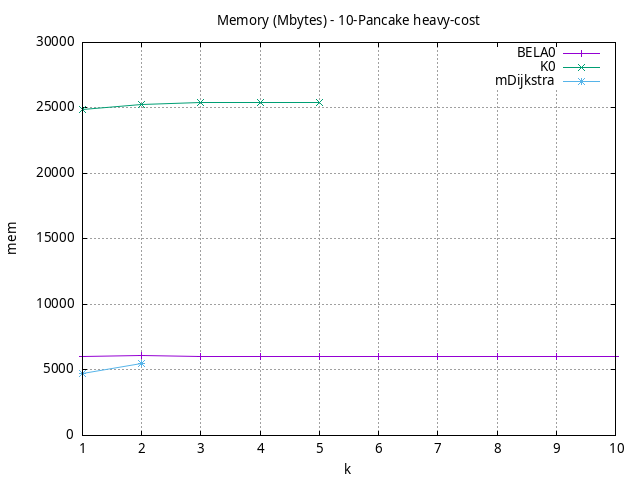}
    \end{center}
    \caption{}
    \label{fig:n-pancake:heavy-cost:brute-force:mem:a}
  \end{subfigure}
  \caption{Memory usage (in Mbytes) in the n-pancake (heavy-cost) domain with brute-force search algorithms}
  \label{fig:n-pancake:heavy-cost:brute-force:mem}
\end{figure*}

%% file: n-pancake.heavy-cost.expansions.brute-force.tex
\begin{figure*}
  \centering
  \begin{subfigure}{0.3\textwidth}
    \begin{center}
        \includegraphics[width=\textwidth]{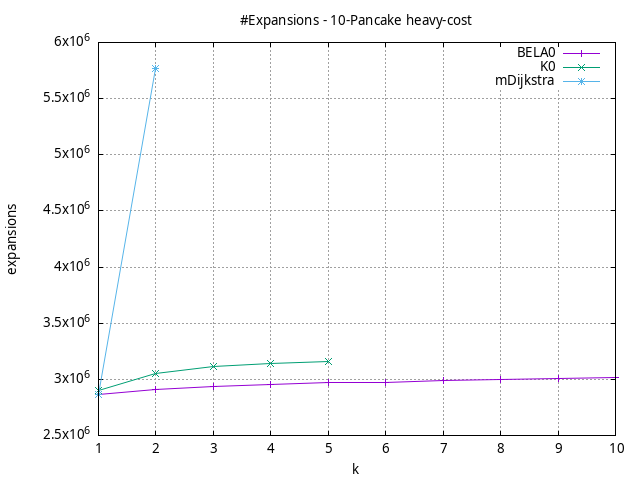}
    \end{center}
    \caption{}
    \label{fig:n-pancake:heavy-cost:brute-force:expansions:a}
  \end{subfigure}
  \caption{Number of expansions in the n-pancake (heavy-cost) domain with brute-force search algorithms}
  \label{fig:n-pancake:heavy-cost:brute-force:expansions}
\end{figure*}

%% file: n-pancake.heavy-cost.runtime.heuristic.tex
\begin{figure*}
  \centering
  \begin{subfigure}{0.3\textwidth}
    \begin{center}
        \includegraphics[width=\textwidth]{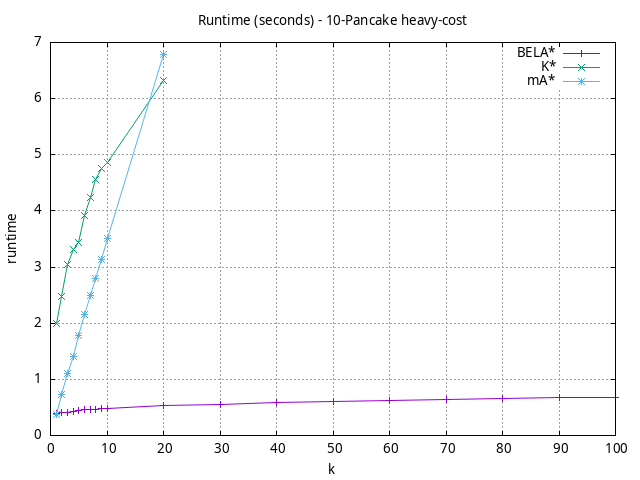}
    \end{center}
    \caption{}
    \label{fig:n-pancake:heavy-cost:heuristic:runtime:a}
  \end{subfigure}
  \caption{Runtime (in seconds) in the n-pancake (heavy-cost) domain with heuristic search algorithms}
  \label{fig:n-pancake:heavy-cost:heuristic:runtime}
\end{figure*}

%% file: n-pancake.heavy-cost.mem.heuristic.tex
\begin{figure*}
  \centering
  \begin{subfigure}{0.3\textwidth}
    \begin{center}
        \includegraphics[width=\textwidth]{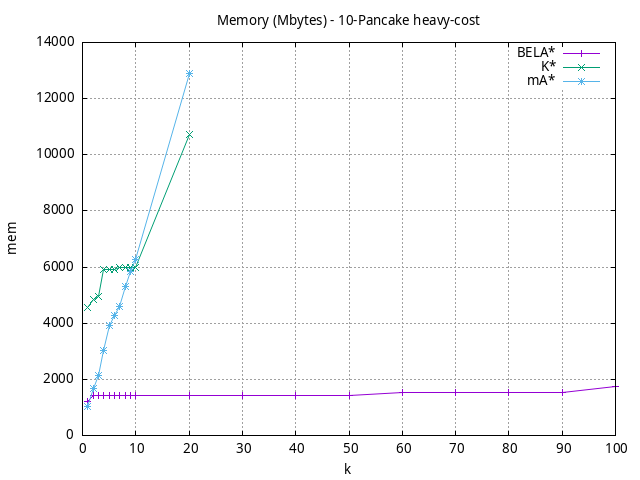}
    \end{center}
    \caption{}
    \label{fig:n-pancake:heavy-cost:heuristic:mem:a}
  \end{subfigure}
  \caption{Memory usage (in Mbytes) in the n-pancake (heavy-cost) domain with heuristic search algorithms}
  \label{fig:n-pancake:heavy-cost:heuristic:mem}
\end{figure*}

%% file: n-pancake.heavy-cost.expansions.heuristic.tex
\begin{figure*}
  \centering
  \begin{subfigure}{0.3\textwidth}
    \begin{center}
        \includegraphics[width=\textwidth]{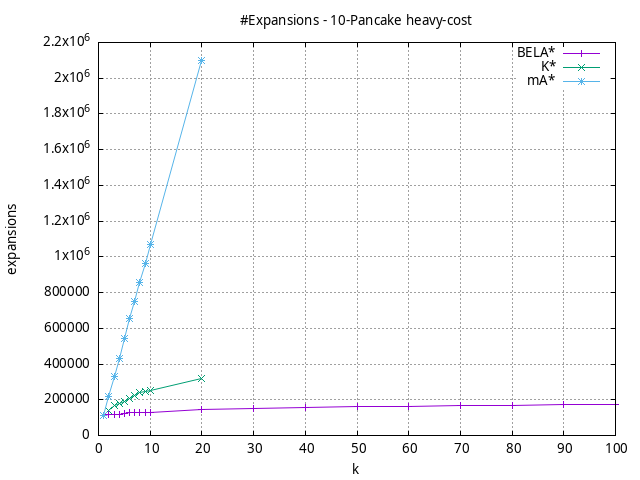}
    \end{center}
    \caption{}
    \label{fig:n-pancake:heavy-cost:heuristic:expansions:a}
  \end{subfigure}
  \caption{Number of expansions in the n-pancake (heavy-cost) domain with heuristic search algorithms}
  \label{fig:n-pancake:heavy-cost:heuristic:expansions}
\end{figure*}

%% file: supplemental6-npuzzle.tex

\subsection{$N$-Puzzle}
\label{sec:empirical-evaluation:n-puzzle}

The $N$-Puzzle is a classical combinatorial task~\parencite{johnson.wa:notes}
that has a state space with $\frac{N^{2}!}{2}$ different states. Up to $N^{2}-1$
different symbols are arranged over a square matrix (though other arrangements
are possible), leaving only one blank position, so that only symbols
horizontally or vertically adjacent to it can swap their locations. The goal is
to re-arrange all symbols into the identity permutation where the blank tile
must be located in the upper-left corner. The 8-Puzzle and the 15-Puzzle were
used for our experiments. In the first case, 100 random instances were randomly
generated, whereas in the 15-Puzzle the 40 easiest instances of the Korf's test
suite were selected~\parencite{korf.re:depth-first}. As a matter of fact, this
test suite is known to extremely difficult for best-first search strategies when
using the Manhattan distance~\parencite{Burns2012}, even without trying to find
$\kappa > 1$ solution paths. As in the case of the $N$-Pancake, this is the
first time this domain is used as a testbed for $\kappa$ shortest path
algorithms to the best of the authors' knowledge.

\subsubsection{Unit variant}
\label{sec:empirical-evaluation:n-puzzle:unit-variant}

In the unit variant, all operators cost the same and thus, they are all equal to
one. There are various heuristic functions for this domain. The current
state-of-the-art uses Additive Pattern Databases~\parencite{Felner2004}.
However, they are known to be inconsistent and thus they have been discarded for
our experimentation, and the Manhattan distance is used instead.

Experiments with the brute-force variants were restricted to the 8-Puzzle, with
181,440 states.
Figures~\ref{fig:n-puzzle:unit:brute-force:runtime}-\ref{fig:n-puzzle:unit:brute-force:expansions}
show the running time, memory usage and number of expansions. In this domain,
\bfk{} is roughly twice as slow as \bfbela{} for $\kappa=10,000$, while
mDijkstra performs very poorly due to the lack of a heuristic function.

Figures~\ref{fig:n-puzzle:unit:heuristic:runtime}--\ref{fig:n-puzzle:unit:heuristic:expansions}
show the results when using heuristic search algorithms, both in the 8-Puzzle
(with $\kappa=10,000$) and the 15-Puzzle ---with $\kappa=100$. The results in
the 15-Puzzle (see Figure~\ref{fig:n-puzzle:unit:heuristic:runtime:a}) show huge
improvements in running time when using \bela{}, which finds the best 100
solutions in roughly 5 seconds on average, whereas both \kstar{} and \mAstar{}
take one order of magnitude more time for very low values of $\kappa$. As
observed in Figures~\ref{fig:n-puzzle:unit:heuristic:mem}
and~\ref{fig:n-puzzle:unit:heuristic:expansions}, the profiles shown in running
time are closely followed by those for memory usage and the number of
expansions.

\input{n-puzzle.unit.runtime.brute-force.tex}
\input{n-puzzle.unit.mem.brute-force.tex}
\input{n-puzzle.unit.expansions.brute-force.tex}

\input{n-puzzle.unit.runtime.heuristic.tex}
\input{n-puzzle.unit.mem.heuristic.tex}
\input{n-puzzle.unit.expansions.heuristic.tex}

\subsubsection{Heavy-cost variant}
\label{sec:empirical-evaluation:n-puzzle:heavy-cost-variant}

In the heavy-cost variant, the cost of a movement is equal to the content of the
tile exchanged with the blank. A weighted variant of the Manhattan distance,
where the distance of each tile is multiplied by its content is used as
our heuristic. The resulting variant is much harder than the previous one, and thus
only experiments with the heuristic versions were conducted.

Results in the 8-Puzzle are almost identical to those in the unit variant ---
compare Figures~\ref{fig:n-puzzle:unit:heuristic:runtime}
and~\ref{fig:n-puzzle:heavy-cost:heuristic:runtime}. The reason is that the
state space of the 8-Puzzle is too small as to pose any significant challenge
when using a heuristic. Things change entirely when considering the 15-Puzzle,
see Figure~\ref{fig:n-puzzle:heavy-cost:heuristic:runtime:a}: \kstar{} takes
almost 20 seconds on average to find a single optimal solution, and \mAstar{} is
able to compute the three best solutions in almost 14 seconds; \bela{}, however,
finds the best 10 solutions in roughly 6 seconds on average. The profiles shown
in Figure~\ref{fig:n-puzzle:heavy-cost:heuristic:runtime:a} show differences of
various orders of magnitude in running time.

\input{n-puzzle.heavy-cost.runtime.heuristic.tex}
\input{n-puzzle.heavy-cost.mem.heuristic.tex}
\input{n-puzzle.heavy-cost.expansions.heuristic.tex}


%% file: n-puzzle.unit.runtime.brute-force.tex
\begin{figure*}
  \centering
  \begin{subfigure}{0.3\textwidth}
    \begin{center}
        \includegraphics[width=\textwidth]{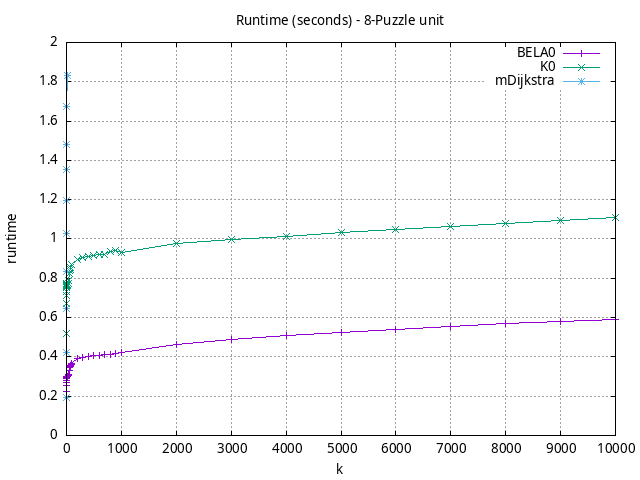}
    \end{center}
    \caption{}
    \label{fig:n-puzzle:unit:brute-force:runtime:a}
  \end{subfigure}
  \caption{Runtime (in seconds) in the n-puzzle (unit) domain with brute-force search algorithms}
  \label{fig:n-puzzle:unit:brute-force:runtime}
\end{figure*}

%% file: n-puzzle.unit.mem.brute-force.tex
\begin{figure*}
  \centering
  \begin{subfigure}{0.3\textwidth}
    \begin{center}
        \includegraphics[width=\textwidth]{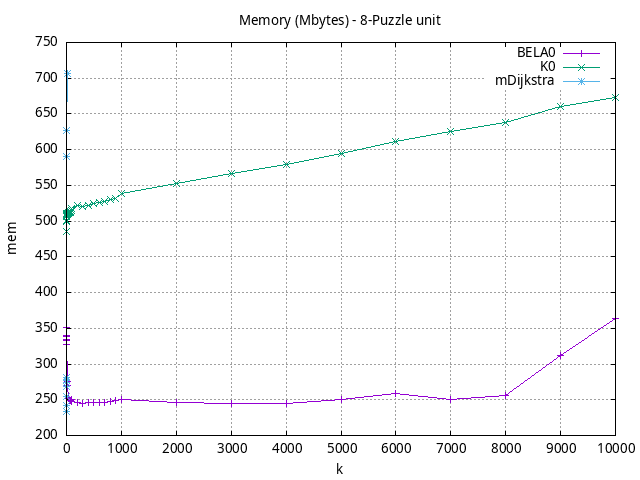}
    \end{center}
    \caption{}
    \label{fig:n-puzzle:unit:brute-force:mem:a}
  \end{subfigure}
  \caption{Memory usage (in Mbytes) in the n-puzzle (unit) domain with brute-force search algorithms}
  \label{fig:n-puzzle:unit:brute-force:mem}
\end{figure*}

%% file: n-puzzle.unit.expansions.brute-force.tex
\begin{figure*}
  \centering
  \begin{subfigure}{0.3\textwidth}
    \begin{center}
        \includegraphics[width=\textwidth]{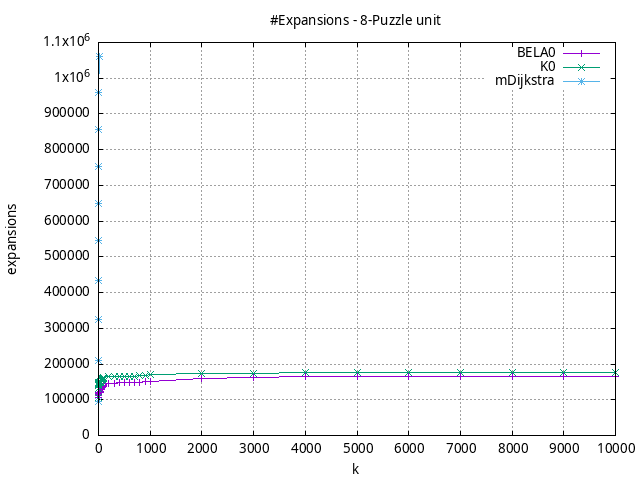}
    \end{center}
    \caption{}
    \label{fig:n-puzzle:unit:brute-force:expansions:a}
  \end{subfigure}
  \caption{Number of expansions in the n-puzzle (unit) domain with brute-force search algorithms}
  \label{fig:n-puzzle:unit:brute-force:expansions}
\end{figure*}

%% file: n-puzzle.unit.runtime.heuristic.tex
\begin{figure*}
  \centering
  \begin{subfigure}{0.3\textwidth}
    \begin{center}
        \includegraphics[width=\textwidth]{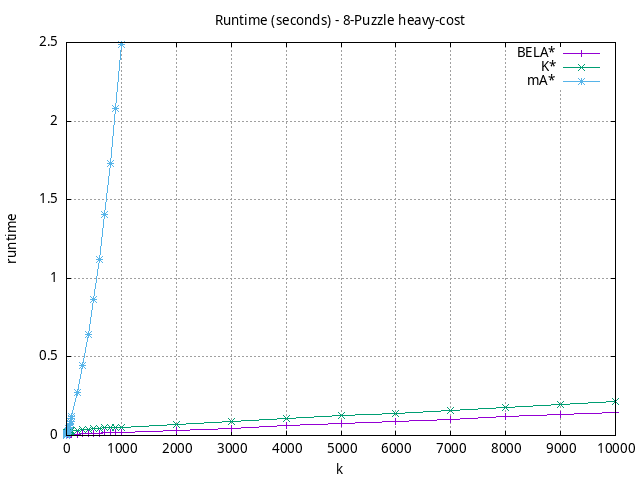}
    \end{center}
    \caption{}
    \label{fig:n-puzzle:unit:heuristic:runtime:b}
  \end{subfigure}
  \begin{subfigure}{0.3\textwidth}
    \begin{center}
        \includegraphics[width=\textwidth]{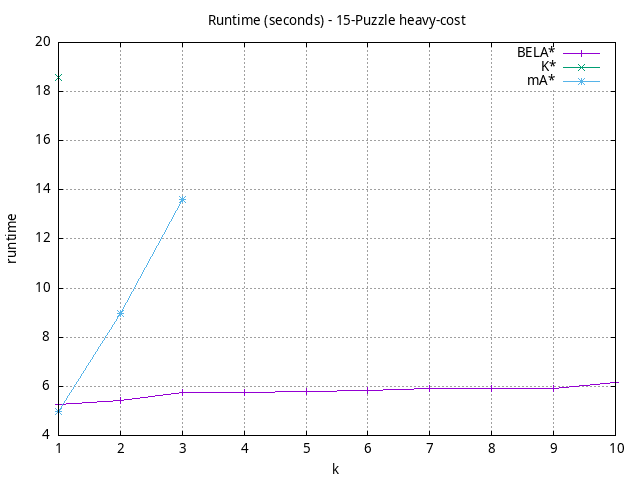}
    \end{center}
    \caption{}
    \label{fig:n-puzzle:unit:heuristic:runtime:a}
  \end{subfigure}
  \caption{Runtime (in seconds) in the n-puzzle (unit) domain with heuristic search algorithms}
  \label{fig:n-puzzle:unit:heuristic:runtime}
\end{figure*}

%% file: n-puzzle.unit.mem.heuristic.tex
\begin{figure*}
  \centering
  \begin{subfigure}{0.3\textwidth}
    \begin{center}
        \includegraphics[width=\textwidth]{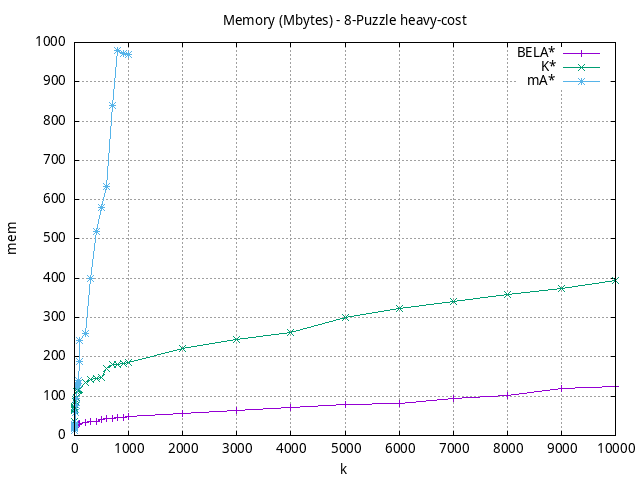}
    \end{center}
    \caption{}
    \label{fig:n-puzzle:unit:heuristic:mem:b}
  \end{subfigure}
  \begin{subfigure}{0.3\textwidth}
    \begin{center}
        \includegraphics[width=\textwidth]{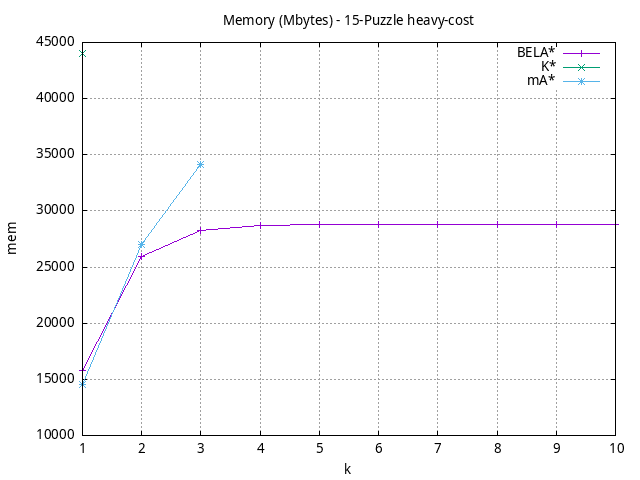}
    \end{center}
    \caption{}
    \label{fig:n-puzzle:unit:heuristic:mem:a}
  \end{subfigure}
  \caption{Memory usage (in Mbytes) in the n-puzzle (unit) domain with heuristic search algorithms}
  \label{fig:n-puzzle:unit:heuristic:mem}
\end{figure*}

%% file: n-puzzle.unit.expansions.heuristic.tex
\begin{figure*}
  \centering
  \begin{subfigure}{0.3\textwidth}
    \begin{center}
        \includegraphics[width=\textwidth]{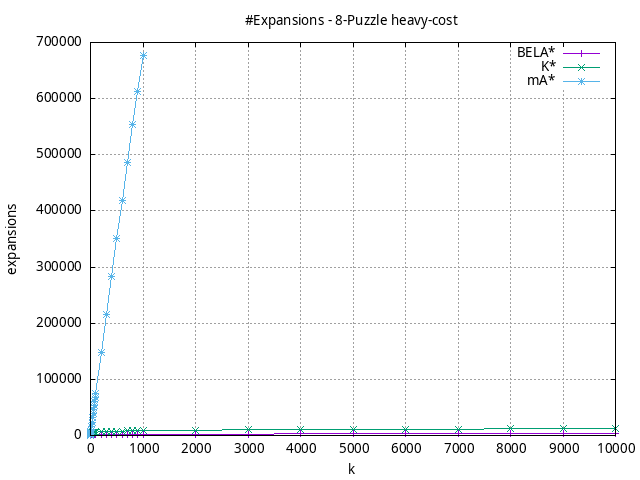}
    \end{center}
    \caption{}
    \label{fig:n-puzzle:unit:heuristic:expansions:b}
  \end{subfigure}
  \begin{subfigure}{0.3\textwidth}
    \begin{center}
        \includegraphics[width=\textwidth]{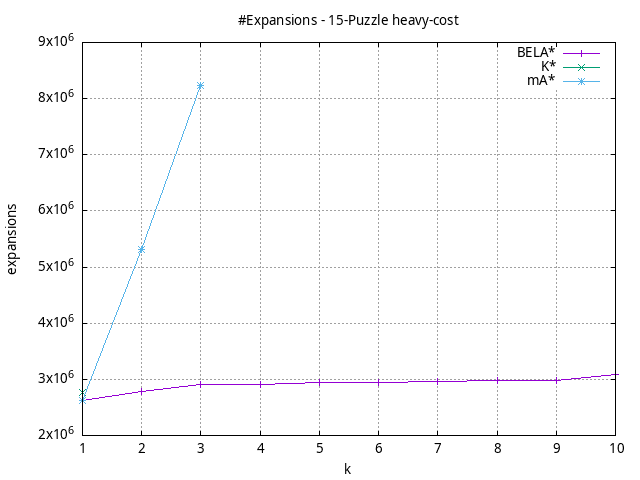}
    \end{center}
    \caption{}
    \label{fig:n-puzzle:unit:heuristic:expansions:a}
  \end{subfigure}
  \caption{Number of expansions in the n-puzzle (unit) domain with heuristic search algorithms}
  \label{fig:n-puzzle:unit:heuristic:expansions}
\end{figure*}

%% file: n-puzzle.heavy-cost.runtime.heuristic.tex
\begin{figure*}
  \centering
  \begin{subfigure}{0.3\textwidth}
    \begin{center}
        \includegraphics[width=\textwidth]{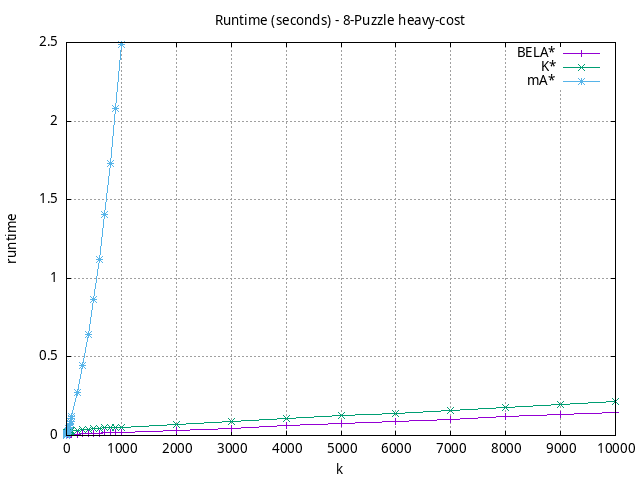}
    \end{center}
    \caption{}
    \label{fig:n-puzzle:heavy-cost:heuristic:runtime:b}
  \end{subfigure}
  \begin{subfigure}{0.3\textwidth}
    \begin{center}
        \includegraphics[width=\textwidth]{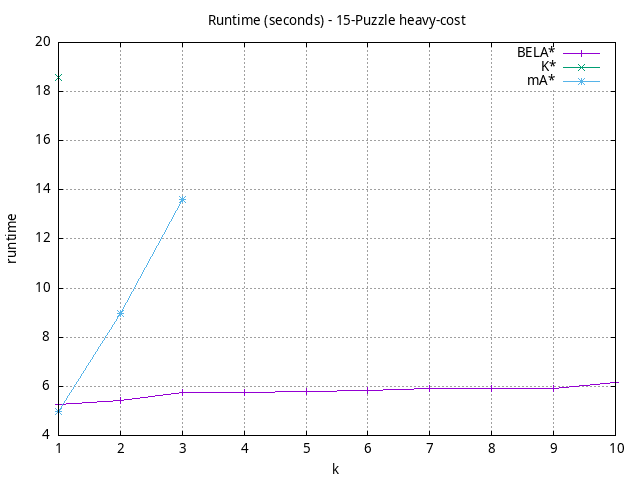}
    \end{center}
    \caption{}
    \label{fig:n-puzzle:heavy-cost:heuristic:runtime:a}
  \end{subfigure}
  \caption{Runtime (in seconds) in the n-puzzle (heavy-cost) domain with heuristic search algorithms}
  \label{fig:n-puzzle:heavy-cost:heuristic:runtime}
\end{figure*}

%% file: n-puzzle.heavy-cost.mem.heuristic.tex
\begin{figure*}
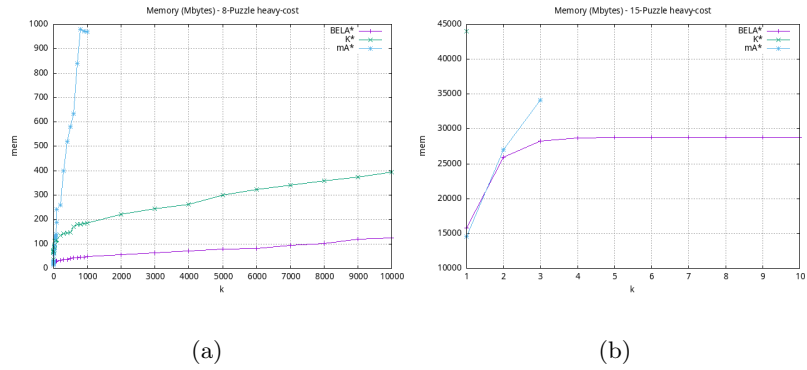

  \centering
  \begin{subfigure}{0.3\textwidth}
    \begin{center}
        \includegraphics[width=\textwidth]{8puzzle.heuristic.mem.png}
    \end{center}
    \caption{}
    \label{fig:n-puzzle:heavy-cost:heuristic:mem:b}
  \end{subfigure}
  \begin{subfigure}{0.3\textwidth}
    \begin{center}
        \includegraphics[width=\textwidth]{15puzzle.heuristic.mem.png}
    \end{center}
    \caption{}
    \label{fig:n-puzzle:heavy-cost:heuristic:mem:a}
  \end{subfigure}
  \caption{Memory usage (in Mbytes) in the n-puzzle (heavy-cost) domain with heuristic search algorithms}
  \label{fig:n-puzzle:heavy-cost:heuristic:mem}
\end{figure*}

%% file: n-puzzle.heavy-cost.expansions.heuristic.tex
\begin{figure*}
  \centering
  \begin{subfigure}{0.3\textwidth}
    \begin{center}
        \includegraphics[width=\textwidth]{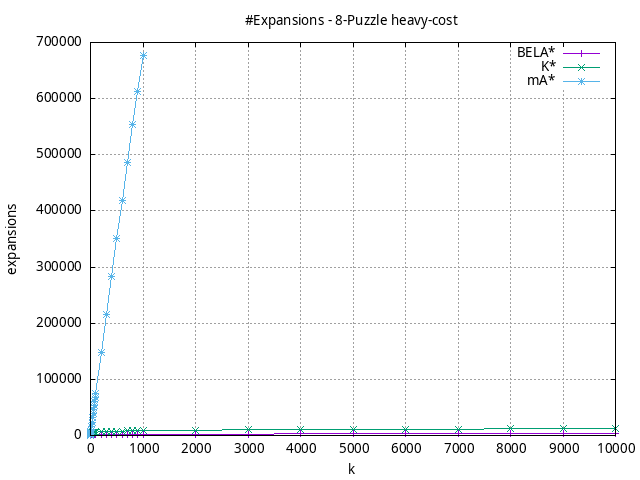}
    \end{center}
    \caption{}
    \label{fig:n-puzzle:heavy-cost:heuristic:expansions:b}
  \end{subfigure}
  \begin{subfigure}{0.3\textwidth}
    \begin{center}
        \includegraphics[width=\textwidth]{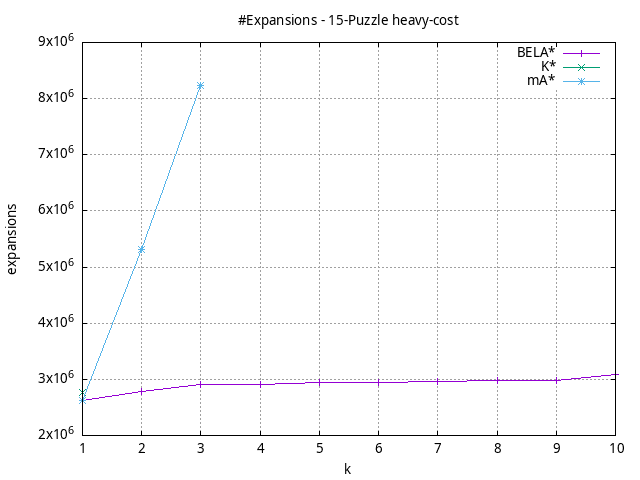}
    \end{center}
    \caption{}
    \label{fig:n-puzzle:heavy-cost:heuristic:expansions:a}
  \end{subfigure}
  \caption{Number of expansions in the n-puzzle (heavy-cost) domain with heuristic search algorithms}
  \label{fig:n-puzzle:heavy-cost:heuristic:expansions}
\end{figure*}

%% file: appendix1.tex

\section{Tables}
\label{cha:tables}

This appendix provides the same information given in
section~\ref{sec:empirical-evaluation} but in tabular form for a selected
collection of $\kappa$ values. Only runtime is provided. In all cases, the best
result is shown in boldface.

\subsection{Roadmap}
\label{cha:tables:roadmap}

Tables~\ref{tab:roadmap:BAY:dimacs:brute-force:runtime}--\ref{tab:roadmap:NY:unit:brute-force:runtime}
summarize the results in the roadmap domain for all variants and graphs. The
size of each graph is given in Table~\ref{tab:dimacs}.

\subsubsection{9th DIMACS Challenge}

The first tables provide all results in tabular form in the \textsc{dimacs}
domain:
Tables~\ref{tab:roadmap:BAY:dimacs:brute-force:runtime}--\ref{tab:roadmap:W:dimacs:brute-force:runtime}
show the runtime of the brute-force variants;
Tables~\ref{tab:roadmap:BAY:dimacs:heuristic:runtime}--\ref{tab:roadmap:W:dimacs:heuristic:runtime}
show the runtime of the heuristics variants, and
Tables~\ref{tab:roadmap:BAY:dimacs:mixed:runtime}--\ref{tab:roadmap:W:dimacs:mixed:runtime}
show a comparison among both brute-force and heuristic variants.

\input{tab.roadmap.dimacs.runtime.brute-force.tex}
\input{tab.roadmap.dimacs.runtime.heuristic.tex}
\input{tab.roadmap.dimacs.runtime.mixed.tex}

\subsubsection{Unit variant}

Next,
Tables~\ref{tab:roadmap:BAY:unit:brute-force:runtime}--\ref{tab:roadmap:NY:unit:brute-force:runtime}
show the results in the unit variant of the roadmap domain.

\input{tab.roadmap.unit.runtime.brute-force.tex}

\subsection{Maps}
\label{cha:tables:maps}

Tables~\ref{tab:maps:10:unit:brute-force:runtime}--\ref{tab:maps:35:octile:heuristic:runtime}
show the runtime of all algorithms in the maps domain.

\subsubsection{Unit variant}

Tables~\ref{tab:maps:10:unit:brute-force:runtime}--\ref{tab:maps:35:unit:brute-force:runtime}
show the results of the brute-force variants being compared, whereas
Tables~\ref{tab:maps:10:unit:heuristic:runtime}--\ref{tab:maps:35:unit:heuristic:runtime}
show the runtime for the heuristic variants, in the unit variant for all sizes of
maps being tested.

\input{tab.maps.unit.runtime.brute-force.tex}
\input{tab.maps.unit.runtime.heuristic.tex}

\subsubsection{Octile variant}

Tables~\ref{tab:maps:10:octile:brute-force:runtime}--\ref{tab:maps:35:octile:brute-force:runtime}
show the runtime of the brute-force variants, and
Tables~\ref{tab:maps:10:octile:heuristic:runtime}--\ref{tab:maps:35:octile:heuristic:runtime}
show the same statistics for the heuristic variants, in the octile variant for
all sizes of maps being tested.

\input{tab.maps.octile.runtime.brute-force.tex}
\input{tab.maps.octile.runtime.heuristic.tex}

\subsection{$N$-Pancake}
\label{cha:tables:npancake}

Tables~\ref{tab:n-pancake:unit:brute-force:runtime}--\ref{tab:n-pancake:heavy-cost:heuristic:runtime}
show the runtime of all algorithms in the $N$-Pancake domain.

\subsubsection{Unit variant}

Table~\ref{tab:n-pancake:unit:brute-force:runtime} shows the runtime of the
brute-force search algorithms tested in the 10-Pancake in the unit variant.
Tables~\ref{tab:n-pancake:20:unit:heuristic:runtime}--\ref{tab:n-pancake:40:unit:heuristic:runtime}
show the runtime of the heuristic search algorithms in the 20-, 30- and
40-Pancake, respectively, in the unit variant.

\input{tab.n-pancake.unit.runtime.brute-force.tex}
\input{tab.n-pancake.unit.runtime.heuristic.tex}

\subsubsection{Heavy-cost variant}

The results in the heavy-cost variant are shown in
Tables~\ref{tab:n-pancake:heavy-cost:brute-force:runtime}--\ref{tab:n-pancake:heavy-cost:heuristic:runtime}.
The first table shows the results of the brute-force variants. The last table
shows the results of the heuristic search algorithms. Due to its difficulty,
only the 10-Pancake was used.

\input{tab.n-pancake.heavy-cost.runtime.brute-force.tex}
\input{tab.n-pancake.heavy-cost.runtime.heuristic.tex}

\subsection{$N$-Puzzle}
\label{cha:tables:npuzzle}

The runtime of all algorithms being tested is shown in
Tables~\ref{tab:n-puzzle:8:unit:brute-force:runtime}--\ref{tab:n-puzzle:15:heavy-cost:heuristic:runtime},
both in the 8- and 15-Puzzle.

\subsubsection{Unit variant}

Table~\ref{tab:n-puzzle:8:unit:brute-force:runtime} shows the runtime of the
brute-force search algorithms in the 8-Puzzle.
Tables~\ref{tab:n-puzzle:8:unit:heuristic:runtime}--\ref{tab:n-puzzle:15:unit:heuristic:runtime}
show the runtime of the heuristic search algorithms in the 8- and the 15-Puzzle
respectively, in the unit variant.

\input{tab.n-puzzle.unit.runtime.brute-force.tex}
\input{tab.n-puzzle.unit.runtime.heuristic.tex}

\subsubsection{Heavy-cost variant}

The heavy-cost variant has being tried only with the heuristic variants.
Tables~\ref{tab:n-puzzle:8:heavy-cost:heuristic:runtime}--\ref{tab:n-puzzle:15:heavy-cost:heuristic:runtime}
show the runtime of all algorithms being tested.

\input{tab.n-puzzle.heavy-cost.runtime.heuristic.tex}


%% file: tab.roadmap.dimacs.runtime.brute-force.tex
\begin{table*}
  \centering
    \input{USA-road-d.BAY.brute-force.runtime.tex}
  \caption{Runtime (in seconds) in the roadmap (dimacs) domain with brute-force search algorithms}
  \label{tab:roadmap:BAY:dimacs:brute-force:runtime}
\end{table*}
  \begin{table*}
  \centering
    \input{USA-road-d.CAL.brute-force.runtime.tex}
  \caption{Runtime (in seconds) in the roadmap (dimacs) domain with brute-force search algorithms}
  \label{tab:roadmap:CAL:dimacs:brute-force:runtime}
\end{table*}
  \begin{table*}
  \centering
    \input{USA-road-d.COL.brute-force.runtime.tex}
  \caption{Runtime (in seconds) in the roadmap (dimacs) domain with brute-force search algorithms}
  \label{tab:roadmap:COL:dimacs:brute-force:runtime}
\end{table*}
  \begin{table*}
  \centering
    \input{USA-road-d.CTR.brute-force.runtime.tex}
  \caption{Runtime (in seconds) in the roadmap (dimacs) domain with brute-force search algorithms}
  \label{tab:roadmap:CTR:dimacs:brute-force:runtime}
\end{table*}
  \begin{table*}
  \centering
    \input{USA-road-d.E.brute-force.runtime.tex}
  \caption{Runtime (in seconds) in the roadmap (dimacs) domain with brute-force search algorithms}
  \label{tab:roadmap:E:dimacs:brute-force:runtime}
\end{table*}
  \begin{table*}
  \centering
    \input{USA-road-d.FLA.brute-force.runtime.tex}
  \caption{Runtime (in seconds) in the roadmap (dimacs) domain with brute-force search algorithms}
  \label{tab:roadmap:FLA:dimacs:brute-force:runtime}
\end{table*}
  \begin{table*}
  \centering
    \input{USA-road-d.LKS.brute-force.runtime.tex}
  \caption{Runtime (in seconds) in the roadmap (dimacs) domain with brute-force search algorithms}
  \label{tab:roadmap:LKS:dimacs:brute-force:runtime}
\end{table*}
  \begin{table*}
  \centering
    \input{USA-road-d.NE.brute-force.runtime.tex}
  \caption{Runtime (in seconds) in the roadmap (dimacs) domain with brute-force search algorithms}
  \label{tab:roadmap:NE:dimacs:brute-force:runtime}
\end{table*}
  \begin{table*}
  \centering
    \input{USA-road-d.NW.brute-force.runtime.tex}
  \caption{Runtime (in seconds) in the roadmap (dimacs) domain with brute-force search algorithms}
  \label{tab:roadmap:NW:dimacs:brute-force:runtime}
\end{table*}
  \begin{table*}
  \centering
    \input{USA-road-d.NY.brute-force.runtime.tex}
  \caption{Runtime (in seconds) in the roadmap (dimacs) domain with brute-force search algorithms}
  \label{tab:roadmap:NY:dimacs:brute-force:runtime}
\end{table*}
  \begin{table*}
  \centering
    \input{USA-road-d.USA.brute-force.runtime.tex}
  \caption{Runtime (in seconds) in the roadmap (dimacs) domain with brute-force search algorithms}
  \label{tab:roadmap:USA:dimacs:brute-force:runtime}
\end{table*}
  \begin{table*}
  \centering
    \input{USA-road-d.W.brute-force.runtime.tex}
  \caption{Runtime (in seconds) in the roadmap (dimacs) domain with brute-force search algorithms}
  \label{tab:roadmap:W:dimacs:brute-force:runtime}
\end{table*}

%% file: tab.roadmap.dimacs.runtime.heuristic.tex
\begin{table*}
  \centering
    \input{USA-road-d.BAY.heuristic.runtime.tex}
  \caption{Runtime (in seconds) in the roadmap (dimacs) domain with heuristic search algorithms}
  \label{tab:roadmap:BAY:dimacs:heuristic:runtime}
\end{table*}
  \begin{table*}
  \centering
    \input{USA-road-d.CAL.heuristic.runtime.tex}
  \caption{Runtime (in seconds) in the roadmap (dimacs) domain with heuristic search algorithms}
  \label{tab:roadmap:CAL:dimacs:heuristic:runtime}
\end{table*}
  \begin{table*}
  \centering
    \input{USA-road-d.COL.heuristic.runtime.tex}
  \caption{Runtime (in seconds) in the roadmap (dimacs) domain with heuristic search algorithms}
  \label{tab:roadmap:COL:dimacs:heuristic:runtime}
\end{table*}
  \begin{table*}
  \centering
    \input{USA-road-d.CTR.heuristic.runtime.tex}
  \caption{Runtime (in seconds) in the roadmap (dimacs) domain with heuristic search algorithms}
  \label{tab:roadmap:CTR:dimacs:heuristic:runtime}
\end{table*}
  \begin{table*}
  \centering
    \input{USA-road-d.E.heuristic.runtime.tex}
  \caption{Runtime (in seconds) in the roadmap (dimacs) domain with heuristic search algorithms}
  \label{tab:roadmap:E:dimacs:heuristic:runtime}
\end{table*}
  \begin{table*}
  \centering
    \input{USA-road-d.FLA.heuristic.runtime.tex}
  \caption{Runtime (in seconds) in the roadmap (dimacs) domain with heuristic search algorithms}
  \label{tab:roadmap:FLA:dimacs:heuristic:runtime}
\end{table*}
  \begin{table*}
  \centering
    \input{USA-road-d.LKS.heuristic.runtime.tex}
  \caption{Runtime (in seconds) in the roadmap (dimacs) domain with heuristic search algorithms}
  \label{tab:roadmap:LKS:dimacs:heuristic:runtime}
\end{table*}
  \begin{table*}
  \centering
    \input{USA-road-d.NE.heuristic.runtime.tex}
  \caption{Runtime (in seconds) in the roadmap (dimacs) domain with heuristic search algorithms}
  \label{tab:roadmap:NE:dimacs:heuristic:runtime}
\end{table*}
  \begin{table*}
  \centering
    \input{USA-road-d.NW.heuristic.runtime.tex}
  \caption{Runtime (in seconds) in the roadmap (dimacs) domain with heuristic search algorithms}
  \label{tab:roadmap:NW:dimacs:heuristic:runtime}
\end{table*}
  \begin{table*}
  \centering
    \input{USA-road-d.NY.heuristic.runtime.tex}
  \caption{Runtime (in seconds) in the roadmap (dimacs) domain with heuristic search algorithms}
  \label{tab:roadmap:NY:dimacs:heuristic:runtime}
\end{table*}
  \begin{table*}
  \centering
    \input{USA-road-d.USA.heuristic.runtime.tex}
  \caption{Runtime (in seconds) in the roadmap (dimacs) domain with heuristic search algorithms}
  \label{tab:roadmap:USA:dimacs:heuristic:runtime}
\end{table*}
  \begin{table*}
  \centering
    \input{USA-road-d.W.heuristic.runtime.tex}
  \caption{Runtime (in seconds) in the roadmap (dimacs) domain with heuristic search algorithms}
  \label{tab:roadmap:W:dimacs:heuristic:runtime}
\end{table*}

%% file: tab.roadmap.dimacs.runtime.mixed.tex
\begin{table*}
  \centering
    \input{USA-road-d.BAY.mixed.runtime.tex}
  \caption{Runtime (in seconds) in the roadmap (dimacs) domain with mixed search algorithms}
  \label{tab:roadmap:BAY:dimacs:mixed:runtime}
\end{table*}
  \begin{table*}
  \centering
    \input{USA-road-d.CAL.mixed.runtime.tex}
  \caption{Runtime (in seconds) in the roadmap (dimacs) domain with mixed search algorithms}
  \label{tab:roadmap:CAL:dimacs:mixed:runtime}
\end{table*}
  \begin{table*}
  \centering
    \input{USA-road-d.COL.mixed.runtime.tex}
  \caption{Runtime (in seconds) in the roadmap (dimacs) domain with mixed search algorithms}
  \label{tab:roadmap:COL:dimacs:mixed:runtime}
\end{table*}
  \begin{table*}
  \centering
    \input{USA-road-d.CTR.mixed.runtime.tex}
  \caption{Runtime (in seconds) in the roadmap (dimacs) domain with mixed search algorithms}
  \label{tab:roadmap:CTR:dimacs:mixed:runtime}
\end{table*}
  \begin{table*}
  \centering
    \input{USA-road-d.E.mixed.runtime.tex}
  \caption{Runtime (in seconds) in the roadmap (dimacs) domain with mixed search algorithms}
  \label{tab:roadmap:E:dimacs:mixed:runtime}
\end{table*}
  \begin{table*}
  \centering
    \input{USA-road-d.FLA.mixed.runtime.tex}
  \caption{Runtime (in seconds) in the roadmap (dimacs) domain with mixed search algorithms}
  \label{tab:roadmap:FLA:dimacs:mixed:runtime}
\end{table*}
  \begin{table*}
  \centering
    \input{USA-road-d.LKS.mixed.runtime.tex}
  \caption{Runtime (in seconds) in the roadmap (dimacs) domain with mixed search algorithms}
  \label{tab:roadmap:LKS:dimacs:mixed:runtime}
\end{table*}
  \begin{table*}
  \centering
    \input{USA-road-d.NE.mixed.runtime.tex}
  \caption{Runtime (in seconds) in the roadmap (dimacs) domain with mixed search algorithms}
  \label{tab:roadmap:NE:dimacs:mixed:runtime}
\end{table*}
  \begin{table*}
  \centering
    \input{USA-road-d.NW.mixed.runtime.tex}
  \caption{Runtime (in seconds) in the roadmap (dimacs) domain with mixed search algorithms}
  \label{tab:roadmap:NW:dimacs:mixed:runtime}
\end{table*}
  \begin{table*}
  \centering
    \input{USA-road-d.NY.mixed.runtime.tex}
  \caption{Runtime (in seconds) in the roadmap (dimacs) domain with mixed search algorithms}
  \label{tab:roadmap:NY:dimacs:mixed:runtime}
\end{table*}
  \begin{table*}
  \centering
    \input{USA-road-d.USA.mixed.runtime.tex}
  \caption{Runtime (in seconds) in the roadmap (dimacs) domain with mixed search algorithms}
  \label{tab:roadmap:USA:dimacs:mixed:runtime}
\end{table*}
  \begin{table*}
  \centering
    \input{USA-road-d.W.mixed.runtime.tex}
  \caption{Runtime (in seconds) in the roadmap (dimacs) domain with mixed search algorithms}
  \label{tab:roadmap:W:dimacs:mixed:runtime}
\end{table*}

%% file: tab.roadmap.unit.runtime.brute-force.tex
\begin{table*}
  \centering
    \input{USA-road-d.BAY.brute-force.runtime.tex}
  \caption{Runtime (in seconds) in the roadmap (unit) domain with brute-force search algorithms}
  \label{tab:roadmap:BAY:unit:brute-force:runtime}
\end{table*}
  \begin{table*}
  \centering
    \input{USA-road-d.CAL.brute-force.runtime.tex}
  \caption{Runtime (in seconds) in the roadmap (unit) domain with brute-force search algorithms}
  \label{tab:roadmap:CAL:unit:brute-force:runtime}
\end{table*}
  \begin{table*}
  \centering
    \input{USA-road-d.COL.brute-force.runtime.tex}
  \caption{Runtime (in seconds) in the roadmap (unit) domain with brute-force search algorithms}
  \label{tab:roadmap:COL:unit:brute-force:runtime}
\end{table*}
  \begin{table*}
  \centering
    \input{USA-road-d.E.brute-force.runtime.tex}
  \caption{Runtime (in seconds) in the roadmap (unit) domain with brute-force search algorithms}
  \label{tab:roadmap:E:unit:brute-force:runtime}
\end{table*}
  \begin{table*}
  \centering
    \input{USA-road-d.FLA.brute-force.runtime.tex}
  \caption{Runtime (in seconds) in the roadmap (unit) domain with brute-force search algorithms}
  \label{tab:roadmap:FLA:unit:brute-force:runtime}
\end{table*}
  \begin{table*}
  \centering
    \input{USA-road-d.LKS.brute-force.runtime.tex}
  \caption{Runtime (in seconds) in the roadmap (unit) domain with brute-force search algorithms}
  \label{tab:roadmap:LKS:unit:brute-force:runtime}
\end{table*}
  \begin{table*}
  \centering
    \input{USA-road-d.NE.brute-force.runtime.tex}
  \caption{Runtime (in seconds) in the roadmap (unit) domain with brute-force search algorithms}
  \label{tab:roadmap:NE:unit:brute-force:runtime}
\end{table*}
  \begin{table*}
  \centering
    \input{USA-road-d.NW.brute-force.runtime.tex}
  \caption{Runtime (in seconds) in the roadmap (unit) domain with brute-force search algorithms}
  \label{tab:roadmap:NW:unit:brute-force:runtime}
\end{table*}
  \begin{table*}
  \centering
    \input{USA-road-d.NY.brute-force.runtime.tex}
  \caption{Runtime (in seconds) in the roadmap (unit) domain with brute-force search algorithms}
  \label{tab:roadmap:NY:unit:brute-force:runtime}
\end{table*}

%% file: tab.maps.unit.runtime.brute-force.tex
\begin{table*}
  \centering
    \input{random512-10.brute-force.unit.runtime.tex}
  \caption{Runtime (in seconds) in the maps (unit) domain with brute-force search algorithms}
  \label{tab:maps:10:unit:brute-force:runtime}
\end{table*}
  \begin{table*}
  \centering
    \input{random512-15.brute-force.unit.runtime.tex}
  \caption{Runtime (in seconds) in the maps (unit) domain with brute-force search algorithms}
  \label{tab:maps:15:unit:brute-force:runtime}
\end{table*}
  \begin{table*}
  \centering
    \input{random512-20.brute-force.unit.runtime.tex}
  \caption{Runtime (in seconds) in the maps (unit) domain with brute-force search algorithms}
  \label{tab:maps:20:unit:brute-force:runtime}
\end{table*}
  \begin{table*}
  \centering
    \input{random512-25.brute-force.unit.runtime.tex}
  \caption{Runtime (in seconds) in the maps (unit) domain with brute-force search algorithms}
  \label{tab:maps:25:unit:brute-force:runtime}
\end{table*}
  \begin{table*}
  \centering
    \input{random512-30.brute-force.unit.runtime.tex}
  \caption{Runtime (in seconds) in the maps (unit) domain with brute-force search algorithms}
  \label{tab:maps:30:unit:brute-force:runtime}
\end{table*}
  \begin{table*}
  \centering
    \input{random512-35.brute-force.unit.runtime.tex}
  \caption{Runtime (in seconds) in the maps (unit) domain with brute-force search algorithms}
  \label{tab:maps:35:unit:brute-force:runtime}
\end{table*}

%% file: tab.maps.unit.runtime.heuristic.tex
\begin{table*}
  \centering
    \input{random512-10.heuristic.unit.runtime.tex}
  \caption{Runtime (in seconds) in the maps (unit) domain with heuristic search algorithms}
  \label{tab:maps:10:unit:heuristic:runtime}
\end{table*}
  \begin{table*}
  \centering
    \input{random512-15.heuristic.unit.runtime.tex}
  \caption{Runtime (in seconds) in the maps (unit) domain with heuristic search algorithms}
  \label{tab:maps:15:unit:heuristic:runtime}
\end{table*}
  \begin{table*}
  \centering
    \input{random512-20.heuristic.unit.runtime.tex}
  \caption{Runtime (in seconds) in the maps (unit) domain with heuristic search algorithms}
  \label{tab:maps:20:unit:heuristic:runtime}
\end{table*}
  \begin{table*}
  \centering
    \input{random512-25.heuristic.unit.runtime.tex}
  \caption{Runtime (in seconds) in the maps (unit) domain with heuristic search algorithms}
  \label{tab:maps:25:unit:heuristic:runtime}
\end{table*}
  \begin{table*}
  \centering
    \input{random512-30.heuristic.unit.runtime.tex}
  \caption{Runtime (in seconds) in the maps (unit) domain with heuristic search algorithms}
  \label{tab:maps:30:unit:heuristic:runtime}
\end{table*}
  \begin{table*}
  \centering
    \input{random512-35.heuristic.unit.runtime.tex}
  \caption{Runtime (in seconds) in the maps (unit) domain with heuristic search algorithms}
  \label{tab:maps:35:unit:heuristic:runtime}
\end{table*}

%% file: tab.maps.octile.runtime.brute-force.tex
\begin{table*}
  \centering
    \input{random512-10.brute-force.octile.runtime.tex}
  \caption{Runtime (in seconds) in the maps (octile) domain with brute-force search algorithms}
  \label{tab:maps:10:octile:brute-force:runtime}
\end{table*}
  \begin{table*}
  \centering
    \input{random512-15.brute-force.octile.runtime.tex}
  \caption{Runtime (in seconds) in the maps (octile) domain with brute-force search algorithms}
  \label{tab:maps:15:octile:brute-force:runtime}
\end{table*}
  \begin{table*}
  \centering
    \input{random512-20.brute-force.octile.runtime.tex}
  \caption{Runtime (in seconds) in the maps (octile) domain with brute-force search algorithms}
  \label{tab:maps:20:octile:brute-force:runtime}
\end{table*}
  \begin{table*}
  \centering
    \input{random512-25.brute-force.octile.runtime.tex}
  \caption{Runtime (in seconds) in the maps (octile) domain with brute-force search algorithms}
  \label{tab:maps:25:octile:brute-force:runtime}
\end{table*}
  \begin{table*}
  \centering
    \input{random512-30.brute-force.octile.runtime.tex}
  \caption{Runtime (in seconds) in the maps (octile) domain with brute-force search algorithms}
  \label{tab:maps:30:octile:brute-force:runtime}
\end{table*}
  \begin{table*}
  \centering
    \input{random512-35.brute-force.octile.runtime.tex}
  \caption{Runtime (in seconds) in the maps (octile) domain with brute-force search algorithms}
  \label{tab:maps:35:octile:brute-force:runtime}
\end{table*}

%% file: tab.maps.octile.runtime.heuristic.tex
\begin{table*}
  \centering
    \input{random512-10.heuristic.octile.runtime.tex}
  \caption{Runtime (in seconds) in the maps (octile) domain with heuristic search algorithms}
  \label{tab:maps:10:octile:heuristic:runtime}
\end{table*}
  \begin{table*}
  \centering
    \input{random512-15.heuristic.octile.runtime.tex}
  \caption{Runtime (in seconds) in the maps (octile) domain with heuristic search algorithms}
  \label{tab:maps:15:octile:heuristic:runtime}
\end{table*}
  \begin{table*}
  \centering
    \input{random512-20.heuristic.octile.runtime.tex}
  \caption{Runtime (in seconds) in the maps (octile) domain with heuristic search algorithms}
  \label{tab:maps:20:octile:heuristic:runtime}
\end{table*}
  \begin{table*}
  \centering
    \input{random512-25.heuristic.octile.runtime.tex}
  \caption{Runtime (in seconds) in the maps (octile) domain with heuristic search algorithms}
  \label{tab:maps:25:octile:heuristic:runtime}
\end{table*}
  \begin{table*}
  \centering
    \input{random512-30.heuristic.octile.runtime.tex}
  \caption{Runtime (in seconds) in the maps (octile) domain with heuristic search algorithms}
  \label{tab:maps:30:octile:heuristic:runtime}
\end{table*}
  \begin{table*}
  \centering
    \input{random512-35.heuristic.octile.runtime.tex}
  \caption{Runtime (in seconds) in the maps (octile) domain with heuristic search algorithms}
  \label{tab:maps:35:octile:heuristic:runtime}
\end{table*}

%% file: tab.n-pancake.unit.runtime.brute-force.tex
\begin{table*}
  \centering
    \input{10pancake.brute-force.runtime.tex}
  \caption{Runtime (in seconds) in the n-pancake (unit) domain with brute-force search algorithms}
  \label{tab:n-pancake:unit:brute-force:runtime}
\end{table*}

%% file: tab.n-pancake.unit.runtime.heuristic.tex
\begin{table*}
  \centering
    \input{20pancake.heuristic.runtime.tex}
  \caption{Runtime (in seconds) in the n-pancake (unit) domain with heuristic search algorithms}
  \label{tab:n-pancake:20:unit:heuristic:runtime}
\end{table*}
  \begin{table*}
  \centering
    \input{30pancake.heuristic.runtime.tex}
  \caption{Runtime (in seconds) in the n-pancake (unit) domain with heuristic search algorithms}
  \label{tab:n-pancake:30:unit:heuristic:runtime}
\end{table*}
  \begin{table*}
  \centering
    \input{40pancake.heuristic.runtime.tex}
  \caption{Runtime (in seconds) in the n-pancake (unit) domain with heuristic search algorithms}
  \label{tab:n-pancake:40:unit:heuristic:runtime}
\end{table*}

%% file: tab.n-pancake.heavy-cost.runtime.brute-force.tex
\begin{table*}
  \centering
    \input{10pancake.brute-force.runtime.tex}
  \caption{Runtime (in seconds) in the n-pancake (heavy-cost) domain with brute-force search algorithms}
  \label{tab:n-pancake:heavy-cost:brute-force:runtime}
\end{table*}

%% file: tab.n-pancake.heavy-cost.runtime.heuristic.tex
\begin{table*}
  \centering
    \input{10pancake.heuristic.runtime.tex}
  \caption{Runtime (in seconds) in the n-pancake (heavy-cost) domain with heuristic search algorithms}
  \label{tab:n-pancake:heavy-cost:heuristic:runtime}
\end{table*}

%% file: tab.n-puzzle.unit.runtime.brute-force.tex
\begin{table*}
  \centering
    \input{8puzzle.brute-force.runtime.tex}
  \caption{Runtime (in seconds) in the n-puzzle (unit) domain with brute-force search algorithms}
  \label{tab:n-puzzle:8:unit:brute-force:runtime}
\end{table*}

%% file: tab.n-puzzle.unit.runtime.heuristic.tex
\begin{table*}
  \centering
    \input{8puzzle.heuristic.runtime.tex}
  \caption{Runtime (in seconds) in the n-puzzle (unit) domain with heuristic search algorithms}
  \label{tab:n-puzzle:8:unit:heuristic:runtime}
\end{table*}
\begin{table*}
  \centering
    \input{15puzzle.heuristic.runtime.tex}
  \caption{Runtime (in seconds) in the n-puzzle (unit) domain with heuristic search algorithms}
  \label{tab:n-puzzle:15:unit:heuristic:runtime}
\end{table*}

%% file: 15puzzle.heuristic.runtime.tex
\begin{tabular}{c|cccccc}\toprule
\multicolumn{7}{c}{Runtime (seconds) - 15-Puzzle heavy-cost}\\ \midrule
Algorithm & k=1 & k=2 & k=3 & k=4 & k=5 & k=10 \\ \midrule
BELA$^*$ & 5.26 & \textbf{5.43} & \textbf{5.75} & \textbf{5.74} & \textbf{5.81} & \textbf{6.17} \\
K$^*$ & 18.58 & -- & -- & -- & -- & -- \\
mA$^*$ & \textbf{4.98} & 8.95 & 13.59 & -- & -- & -- \\ \bottomrule 
\end{tabular}

%% file: tab.n-puzzle.heavy-cost.runtime.heuristic.tex
\begin{table*}
  \centering
    \input{8puzzle.heuristic.runtime.tex}
  \caption{Runtime (in seconds) in the n-puzzle (heavy-cost) domain with heuristic search algorithms}
  \label{tab:n-puzzle:8:heavy-cost:heuristic:runtime}
\end{table*}
\begin{table*}
  \centering
    \input{15puzzle.heuristic.runtime.tex}
  \caption{Runtime (in seconds) in the n-puzzle (heavy-cost) domain with heuristic search algorithms}
  \label{tab:n-puzzle:15:heavy-cost:heuristic:runtime}
\end{table*}